\documentclass[twocolumn]{aastex631}

\usepackage{tabularx}
\usepackage{xcolor}

\received{Mar 23, 2021}
\revised{Apr 16, 2021}
\accepted{Sep 28, 2021}
\submitjournal{ApJ}

\shorttitle{Mid-IR imaging survey of stars within 5\,pc}
\shortauthors{Gauza et al.}

\begin{document}

\title{GTC/CanariCam Deep Mid-Infrared Imaging Survey of Northern Stars within 5\,pc}

\correspondingauthor{Bartosz Gauza}
\email{b.gauza@herts.ac.uk}

\author[0000-0001-5452-2056]{Bartosz Gauza}
\affiliation{Centre for Astrophysics Research, School of Physics, Astronomy and Mathematics \\
University of Hertfordshire College Lane, Hatfield AL10 9AB, UK},
\affiliation{Janusz Gil Institute of Astronomy, University of Zielona G\'ora, Lubuska 2, 65-265 Zielona G\'ora, Poland}

\author{V\'ictor J.~S. B\'ejar}
\affiliation{Instituto de Astrof\'isica de Canarias (IAC), Calle V\'ia L\'actea s/n, 38200 La Laguna, Tenerife, Spain}
\affiliation{Departamento de Astrof\'isica, Universidad de La Laguna (ULL), 38205 La Laguna, Tenerife, Spain}

\author{Rafael Rebolo}
\affiliation{Instituto de Astrof\'isica de Canarias (IAC), Calle V\'ia L\'actea s/n, 38200 La Laguna, Tenerife, Spain}
\affiliation{Departamento de Astrof\'isica, Universidad de La Laguna (ULL), 38205 La Laguna, Tenerife, Spain}
\affiliation{Consejo Superior de Investigaciones Cient\'ificas (CSIC), Madrid, Spain}

\author{Carlos \'Alvarez}
\affiliation{W. M. Keck Observatory, 65-1120 Mamalahoa Highway, Kamuela, HI 96743, USA}

\author{Mar\'ia Rosa Zapatero Osorio}
\affiliation{Centro de Astrobiolog\'ia (CSIC-INTA), Ctra. Ajalvir km 4, E-28850 Torrej\'on de Ardoz, Madrid, Spain}

\author{Gabriel Bihain}
\affiliation{Max Planck Institute for Gravitational Physics (Albert Einstein Institute), Callinstrasse 38, 30167 Hannover, Germany}
\affiliation{Leibniz Universit\"at Hannover, D-30167 Hannover, Germany}

\author{Jos\'e A. Caballero}
\affiliation{Centro de Astrobiolog\'ia (CSIC-INTA), Ctra. Ajalvir km 4, E-28850 Torrej\'on de Ardoz, Madrid, Spain}

\author{David J. Pinfield}
\affiliation{Centre for Astrophysics Research, School of Physics, Astronomy and Mathematics \\
University of Hertfordshire College Lane, Hatfield AL10 9AB, UK}

\author{Charles M. Telesco}
\affiliation{Department of Astronomy, University of Florida, Gainesville, FL 32611, USA}

\author{Christopher Packham}
\affiliation{Department of Physics and Astronomy, University of Texas at San Antonio, San Antonio, TX 78249, USA}
\affiliation{National Astronomical Observatory of Japan, Mitaka, Tokyo 181-8588, Japan}



\begin{abstract}
In this work we present the results of a direct imaging survey for brown dwarf companions around the nearest 
stars at the mid-infrared 10 micron range ($\lambda_{c}$\,=\,8.7\,$\mu$m, $\Delta\lambda$\,=\,1.1\,$\mu$m) using 
the CanariCam instrument at the 10.4\,m Gran Telescopio Canarias (GTC). We imaged the 25 nearest stellar 
systems within 5\,pc of the Sun at declinations $\delta$\,$>$\,$-25^{\circ}$ (at least half have planets 
from radial velocity), reaching a mean detection limit of 11.3$\,\pm$\,0.2\,mag (1.5\,mJy) in the Si-2 
8.7\,$\mu$m band over a range of angular separations from 1 to 10\,arcsec. This would have allowed us to 
uncover substellar companions at projected orbital separations between $\sim$2 and 50\,au, with effective 
temperatures down to 600\,K and masses greater than 30\,$M_{\rm Jup}$ assuming an average age of 5\,Gyr
and down to the deuterium-burning mass limit for objects with ages $<$1\,Gyr. From the non-detection of such 
companions, we determined upper limits on their occurrence rate at depths and orbital separations yet 
unexplored by deep imaging programs. For the M dwarfs, main components of our sample, we found with a 90\% 
confidence level that less than 20\% of these low-mass stars have L and T-type brown dwarf companions with 
$m$\,$\gtrsim$\,30\,$M_{\rm Jup}$ and $T_{\rm eff}$\,$\gtrsim$\,600\,K at $\sim$3.5--35\,au projected 
orbital separations.
\end{abstract}



\section{Introduction} \label{sec:intro}
High contrast imaging surveys of stars constitute one of the foremost methods to find and study brown dwarfs 
and extrasolar planets. Their results complement our knowledge on these populations drawn from the objects 
detected by transit, radial velocity and other techniques. Modern imaging searches extend the accessible range 
of parameter space of cool companions to wider separations and longer orbital periods ($a$\,$>$\,5\,au, 
$P$\,$>$\,10\,yr) than those presently explored by radial velocities (RVs) or transits methods 
\cite[e.g.,][]{2007ApJ...670.1367L, 2013ApJ...777..160B, 2015A&A...573A.127C, 2015MNRAS.453.2533M, 
2016AA...594A..63G, 2017AA...603A...3V, 2018AJ....156..286S, 2019AJ....158..187B, 2021AA...651A..72V,
2020AA...635A.162L}. Moreover, directly detected brown dwarfs and planets provide a unique opportunity for 
spectroscopic characterisation and thereby for detailed study of their fundamental physical properties, in 
particular their atmospheres \cite[e.g.,][]{2015ApJ...804...96G, 2016ApJ...817..166S, 2016A&A...587A..58B, 
2018AA...618A..63B, 2021AA...645A..17C}. 

A large number of imaging programs have thus far focused on young stars (less than 1\,Gyr old) from the solar 
neighbourhood ($d\lesssim$\,50\,pc), e.g., \citet{2013ApJ...777..160B, 2016AA...596A..83L, 2019AJ....158...13N, 
2021AA...651A..72V}. Young nearby stars are ideal targets for direct imaging surveys because their potential 
substellar companions, at the early stages of evolution are still warm and relatively bright, favouring their 
detection. As a consequence of both the limitations in sensitivity to the coldest companions at field ages, 
and enhanced chances of detection at young ages, the most successful searches by means of confirmed detections 
are so far those carried on known young stars.

The first large imaging program sensitive to substellar companions that targeted the nearest stars was the 
one led by \cite{2001AJ....121.2189O}. They observed a sample of 163 northern stars in 111 star systems, 
located within 8\,pc from the Sun using the Adaptive Optics Coronograph on the Palomar 1.5\,m telescope for 
the optical imaging and the Cassegrain Infrared Camera on the Palomar 5\,m Hale Telescope for the near-IR. 
For about 80\% of the surveyed stars, companions more massive than 40\,$M_{\rm Jup}$ at an age of 5\,Gyr 
would have been detected at separations between 40 and 120\,au. Among the most sensitive imaging surveys of 
the nearest stars are those by \cite{2012AJ....144...64D} and \cite{2011ApJ...743..141C}. The first group 
employed the NICMOS on the Hubble Space Telescope to obtain high resolution images of 255 stars within 
$\sim$10\,pc, while the second used the IRAC on the Spitzer Space Telescope and observed 117 targets at 
distances from 1.3 to 43.8\,pc.

Despite several remarkable discoveries like the methane brown dwarf Gliese 229B \citep{1995Natur.378..463N} or 
the planetary system around HR 8799 \citep{2008Sci...322.1348M}, gas giant planets and brown dwarf companions 
at orbital separations beyond a few astronomical units were found to be rare both by the surveys of young stars 
and the nearest stars. In contrast to thousands of discoveries from RV and transit methods, only a few tens of 
companions in or around the planetary mass regime were found by direct imaging surveys sensitive enough to 
detect them. \cite{2012AJ....144...64D}, from analysis of their sub-sample of 138 M dwarfs, calculated a 
multiplicity fraction of $2.3^{+5.0}_{-0.7}$\% for L and T-type companions to M dwarfs at orbital separations 
of 10--70\,au. The IRAC/Spitzer search performed by \cite{2011ApJ...743..141C} had the ability to detect 
600--1100\,K brown dwarf companions at semimajor axes $\gtrsim$\,35\,au and 500--600\,K companions beyond 
60\,au. Using Monte Carlo simulations they estimated a 600--1100\,K T dwarf companion fraction of $<$\,3.4\% 
at 35--1200\,au and $<$\,12.4\% for 500--600\,K companions at 60--1000\,au. Due to limitations in spatial 
resolution, contrast and sensitivities achieved by available instruments, the orbital separations of less 
than 10--15\,au remained largely unexplored for the presence of massive planets and brown dwarfs by the past 
imaging surveys.

\begin{deluxetable*}{llrrcccrr}
\tablenum{1}
\tablecaption{CanariCam target sample\label{CCSample}}
\tablewidth{0pt}
\tabletypesize{\scriptsize}
\tablehead{
\colhead{Star} & 
\colhead{Other Name} & 
\colhead{RA (J2000)} & 
\colhead{Dec (J2000)} & 
\colhead{Spectral} & 
\colhead{$d$} & 
\colhead{$\pi$} & 
\colhead{$\mu_{\alpha}\cos{\delta}$} & 
\colhead{$\mu_{\delta}$} \\
\colhead{} & 
\colhead{} & 
\colhead{(hh:mm:sss)} & 
\colhead{(dd:mm:ss)} & 
\colhead{Type} & 
\colhead{(pc)} & 
\colhead{(mas)} & 
\colhead{(mas/yr)} & 
\colhead{(mas/yr)}
}
\startdata
GJ 699  	& Barnard's Star       	& 17:57:48.499	& +04:41:36.11	& M3.5		& 1.827\,$\pm$\,0.0010 & 547.45\,$\pm$\,0.29 & -802.80\,$\pm$\,0.64 & 10362.54\,$\pm$\,0.36\\				
GJ 406  	& CN Leo	       	& 10:56:28.826	& +07:00:52.34	& M6.0		& 2.409\,$\pm$\,0.0004 & 415.18\,$\pm$\,0.07 & -3866.34\,$\pm$\,0.08 & -2699.22\,$\pm$\,0.07\\
GJ 411  	& Lalande 21185  	& 11:03:20.194	& +35:58:11.57	& M2.0		& 2.546\,$\pm$\,0.0002 & 392.75\,$\pm$\,0.03 & -580.06\,$\pm$\,0.03 & -4776.59\,$\pm$\,0.03\\
GJ 244  	& Sirius               	& 06:45:08.917	& -16:42:58.02	& A1.0		& 2.670\,$\pm$\,0.0017 & 374.49\,$\pm$\,0.23 & -461.57\,$\pm$\,0.28 & -914.52\,$\pm$\,0.33\\
GJ 65\,AB	& BL Cet\,+\,UV Cet    	& 01:39:01.453	& -17:57:02.04	& M5.5, M6.0    & 2.720\,$\pm$\,0.0055 & 367.71\,$\pm$\,0.74 & 3385.32\,$\pm$\,0.67 & 544.39\,$\pm$\,0.38\\
GJ 729  	& V1216 Sgr            	& 18:49:49.364	& -23:50:10.45	& M3.5		& 2.976\,$\pm$\,0.0003 & 336.03\,$\pm$\,0.03 & 639.37\,$\pm$\,0.04 & -193.96\,$\pm$\,0.03\\
GJ 905  	& HH And               	& 23:41:55.036	& +44:10:38.82	& M5.0		& 3.160\,$\pm$\,0.0004 & 316.48\,$\pm$\,0.04 & 112.53\,$\pm$\,0.04 & -1591.65\,$\pm$\,0.03\\
GJ 144  	& $\epsilon$ Eridani   	& 03:32:55.845	& -09:27:29.73	& K2.0		& 3.220\,$\pm$\,0.0014 & 310.58\,$\pm$\,0.14 & -974.76\,$\pm$\,0.16 & 20.88\,$\pm$\,0.12\\
GJ 447  	& FI Vir               	& 11:47:44.397	& +00:48:16.40	& M4.0		& 3.375\,$\pm$\,0.0003 & 296.31\,$\pm$\,0.03 & 607.30\,$\pm$\,0.03 & -1223.03\,$\pm$\,0.02\\
GJ 866\,(AC)B  & EZ Aqr             	& 22:38:36.081	& -15:17:23.89	& M5.0		& 3.406\,$\pm$\,0.0105 & 293.60\,$\pm$\,0.90 & 2314.8\,$\pm$\,8.0 & 2295.3\,$\pm$\,8.0\\ 				
GJ 820\,A  	& 61 Cyg A 		& 21:06:53.940	& +38:44:57.90	& K5.0		& 3.497\,$\pm$\,0.0012 & 285.95\,$\pm$\,0.10 & 4164.17\,$\pm$\,0.19 & 3249.99\,$\pm$\,0.25\\
GJ 820\,B  	& 61 Cyg B 		& 21:06:55.264	& +38:44:31.36	& K7.0		& 3.495\,$\pm$\,0.0007 & 286.15\,$\pm$\,0.06 & 4105.79\,$\pm$\,0.09 & 3155.76\,$\pm$\,0.10\\
GJ 280  	& Procyon  		& 07:39:18.119	& +05:13:29.95	& F5	        & 3.507\,$\pm$\,0.0079 & 285.17\,$\pm$\,0.64 & -714.59\,$\pm$\,2.06 & -1036.80\,$\pm$\,1.15\\			
GJ 725\,A 	& HD 173739  		& 18:42:46.705	& +59:37:49.41	& M3.0		& 3.523\,$\pm$\,0.0003 & 283.84\,$\pm$\,0.02 & -1311.68\,$\pm$\,0.03 & 1792.33\,$\pm$\,0.03\\
GJ 725\,B 	& HD 173740  		& 18:42:46.894	& +59:37:36.72	& M3.5		& 3.523\,$\pm$\,0.0004 & 283.84\,$\pm$\,0.03 & -1400.26\,$\pm$\,0.04 & 1862.53\,$\pm$\,0.03\\
GJ 15 A  	& GX And  		& 00:18:22.885	& +44:01:22.64	& M1.5		& 3.562\,$\pm$\,0.0003 & 280.71\,$\pm$\,0.02 & 2891.52\,$\pm$\,0.02 & 411.83\,$\pm$\,0.01\\
GJ 15 B  	& GQ And  		& 00:18:25.824	& +44:01:38.09	& M3.5		& 3.563\,$\pm$\,0.0004 & 280.69\,$\pm$\,0.03 & 2862.80\,$\pm$\,0.02 & 336.43\,$\pm$\,0.02\\
GJ 1111  	& DX Cnc  		& 08:29:49.353	& +26:46:33.63	& M6.5		& 3.581\,$\pm$\,0.0008 & 279.25\,$\pm$\,0.06 & -1113.69\,$\pm$\,0.06 & -612.19\,$\pm$\,0.05\\
GJ 71  		& $\tau$ Cet  		& 01:44:04.091	& -15:56:14.93	& G8.5		& 3.652\,$\pm$\,0.0023 & 273.81\,$\pm$\,0.17 & -1721.73\,$\pm$\,0.18 & 854.96\,$\pm$\,0.09\\
GJ 54.1  	& YZ Cet  		& 01:12:30.637	& -16:59:56.36	& M4.5		& 3.717\,$\pm$\,0.0005 & 269.06\,$\pm$\,0.03 & 1205.07\,$\pm$\,0.04 & 637.55\,$\pm$\,0.05\\
GJ 273  	& Luyten's Star		& 07:27:25.093	& +05:12:35.63	& M3.5		& 3.786\,$\pm$\,0.0006 & 264.13\,$\pm$\,0.04 & 571.23\,$\pm$\,0.05 & -3691.49\,$\pm$\,0.04\\
SO 0253+16& Teegarden's Star	        & 02:53:00.891	& +16:52:52.64	& M7.0		& 3.832\,$\pm$\,0.0014 & 260.99\,$\pm$\,0.09 & 3429.08\,$\pm$\,0.09 & -3805.54\,$\pm$\,0.08\\
GJ 860\,AB  	& Kruger 60 AB 		& 22:27:59.557	& +57:41:42.08	& M3.0, M4.0    & 4.010\,$\pm$\,0.0026 & 249.39\,$\pm$\,0.16 & -725.23\,$\pm$\,0.54 & -223.46\,$\pm$\,0.35\\
GJ 83.1  	& TZ Ari  		& 02:00:12.956	& +13:03:07.02	& M4.0		& 4.470\,$\pm$\,0.0014 & 223.73\,$\pm$\,0.07 & 1096.46\,$\pm$\,0.07 & -1771.53\,$\pm$\,0.06\\
GJ 687  	& LHS 450  		& 17:36:25.899	& +68:20:20.90	& M3.0		& 4.550\,$\pm$\,0.0004 & 219.79\,$\pm$\,0.02 & -320.68\,$\pm$\,0.02 & -1269.89\,$\pm$\,0.03\\
GJ 1245\,ABC & LHS 3494  		& 19:53:54.482	& +44:24:51.34	&M5.5,\,M,\,M6  & 4.660\,$\pm$\,0.0010 & 214.57\,$\pm$\,0.05 & 349.36\,$\pm$\,0.06 & -480.32\,$\pm$\,0.05\\
GJ 876  	& IL Aqr  		& 22:53:16.732	& -14:15:49.30	& M3.5		& 4.672\,$\pm$\,0.0008 & 214.04\,$\pm$\,0.04 & 957.72\,$\pm$\,0.04 & -673.60\,$\pm$\,0.03\\
GJ 1002  	& LHS 2  		& 00:06:43.197	& -07:32:17.02	& M5.0		& 4.846\,$\pm$\,0.0011 & 206.35\,$\pm$\,0.05 & -811.57\,$\pm$\,0.06 & -1893.25\,$\pm$\,0.03
\enddata
\end{deluxetable*}

We used the CanariCam instrument at the Gran Telescopio Canarias (GTC) to carry out deep, high spatial resolution 
mid-infrared imaging in the 10 micron window, targeting the nearest known stars visible from a northern site 
($\delta$\,$>$\,$-25^{\circ}$), to search for ultra-cool brown dwarfs and massive planets. We aimed to detect 
companions at 1--10\,arcsec separations, which translate to 2.0--50\,au, or orbital periods typically longer than 
10 years. Therefore, our search extends to orbital separations and periods not explored yet by previous imaging 
or RV surveys. Since stars in the solar vicinity are typically old, at ages over 1\,Gyr, any low-mass substellar 
companion will have cooled down to $T_{\rm eff}$ well below 1000\,K. At such temperatures, the maximum flux 
emission shifts from the near- to mid-IR. Hence, CanariCam at GTC provided an opportunity to perform a competitive 
search with respect to direct imaging surveys completed at the optical or near-IR using adaptive optics and 
coronographic systems, by means of sensitivity to the coolest companions. In this work we present the generic 
results of the program. The rest of the paper is structured as follows. Section~2 sets out the observed sample 
of stars, Section~3 describes the observations and data processing steps, and Section~4 presents the analysis 
and results: relative astrometry of resolved binaries, contrast curves and sensitivities, constraints on physical 
parameters of detectable companions, and the upper limit estimates on the occurrence rate of companions. Section 5 
contains comparison of our results to other surveys and a discussion regarding the stellar binaries and known 
planets hosts in the sample. Conclusions and final remarks are presented in Section~6.

\section{The sample} \label{sec:sample}

The sample of CanariCam targets consists of the nearest known stars from the northern sky, visible from the 
Roque de los Muchachos Observatory, that is, with declinations $\delta$\,$>$\,$-25^{\circ}$. We used the One 
Hundred Nearest Star Systems list provided by the Research Consortium On Nearby Stars (RECONS; 
\citealt{1997AJ....114..388H, 2006AJ....132.2360H, 2018AJ....155..265H}), complemented 
with the astrometric data from the Gaia DR2 and EDR3 \citep{2018A&A...616A...1G, 2021AA...649A...1G}
where available, starting from the nearest star in the Northern Hemisphere, GJ~699 (Barnard's Star) and 
moving to more distant ones.

\begin{deluxetable*}{lrrrrrrrr}
\tablenum{2}
\tablecaption{Near- and mid-infrared photometry (from 2MASS, WISE and Akari S9W) of stars in the sample\label{SamplePhot}}
\tablewidth{0pt}
\tabletypesize{\scriptsize}
\tablehead{
\colhead{Star} & 
\colhead{$J$} & 
\colhead{$H$} & 
\colhead{$K_{s}$} & 
\colhead{$W$1} & 
\colhead{$W$2} & 
\colhead{$W$3} & 
\colhead{$W$4} & 
\colhead{S9W} 
\\ 
\colhead{} & 
\colhead{(mag)} & 
\colhead{(mag)} &
\colhead{(mag)} & 
\colhead{(mag)} & 
\colhead{(mag)} & 
\colhead{(mag)} & 
\colhead{(mag)} & 
\colhead{(Jy)}
}
\startdata
GJ 699  &  5.244\,$\pm$\,0.020  & 4.834\,$\pm$\,0.034  & 4.524\,$\pm$\,0.020  & 4.386\,$\pm$\,0.073  & 3.600\,$\pm$\,0.062  & 4.036\,$\pm$\,0.016  & 3.921\,$\pm$\,0.025  & ... \\
GJ 406  &  7.085\,$\pm$\,0.024  & 6.482\,$\pm$\,0.042  & 6.084\,$\pm$\,0.017  & 5.807\,$\pm$\,0.055  & 5.487\,$\pm$\,0.031  & 5.481\,$\pm$\,0.015  & 5.310\,$\pm$\,0.031  & 0.38\,$\pm$\,0.02 \\ 
GJ 411  &  4.203\,$\pm$\,0.242  & 3.640\,$\pm$\,0.202  & 3.254\,$\pm$\,0.306  & 3.239\,$\pm$\,0.136  & 2.360\,$\pm$\,0.071  & 3.045\,$\pm$\,0.010  & 2.934\,$\pm$\,0.024  & 3.40\,$\pm$\,0.02 \\
GJ 244  &  -1.391\,$\pm$\,0.109  & -1.391\,$\pm$\,0.184  & -1.390\,$\pm$\,0.214  & 2.387\,$\pm$\,0.059  & 0.786\,$\pm$\,0.112  & 0.497\,$\pm$\,0.018  & -1.330\,$\pm$\,0.005  & 198.0\,$\pm$\,6.9 \\
GJ 65\,AB  & 6.283\,$\pm$\,0.019  & 5.690\,$\pm$\,0.029  & 5.343\,$\pm$\,0.021  & 5.053\,$\pm$\,0.072  & 4.575\,$\pm$\,0.041  & 4.762\,$\pm$\,0.015  & 4.616\,$\pm$\,0.025  & 0.65\,$\pm$\,0.01 \\
GJ 729  &  6.222\,$\pm$\,0.018  & 5.655\,$\pm$\,0.034  & 5.370\,$\pm$\,0.016  & 5.164\,$\pm$\,0.062  & 4.754\,$\pm$\,0.033  & 4.911\,$\pm$\,0.014  & 4.715\,$\pm$\,0.026  & 0.60\,$\pm$\,0.01 \\
GJ 905  &  6.884\,$\pm$\,0.026  & 6.247\,$\pm$\,0.027  & 5.929\,$\pm$\,0.020  & 5.694\,$\pm$\,0.056  & 5.410\,$\pm$\,0.029  & 5.393\,$\pm$\,0.015  & 5.254\,$\pm$\,0.031  & 0.38\,$\pm$\,0.01 \\
GJ 144  &  2.228\,$\pm$\,0.298  & 1.880\,$\pm$\,0.276  & 1.776\,$\pm$\,0.286  & 2.970\,$\pm$\,0.215  & 2.285\,$\pm$\,0.055  & 1.770\,$\pm$\,0.006  & 1.288\,$\pm$\,0.005  & 12.86\,$\pm$\,0.05 \\
GJ 447  &  6.505\,$\pm$\,0.023  & 5.945\,$\pm$\,0.024  & 5.654\,$\pm$\,0.024  & 5.457\,$\pm$\,0.064  & 5.012\,$\pm$\,0.034  & 5.176\,$\pm$\,0.013  & 5.027\,$\pm$\,0.031  & 0.53\,$\pm$\,0.01 \\
GJ 866\,(AC)B &  6.553\,$\pm$\,0.019  & 5.954\,$\pm$\,0.031  & 5.537\,$\pm$\,0.020  & 5.314\,$\pm$\,0.062  & 4.889\,$\pm$\,0.035  & 5.006\,$\pm$\,0.015  & 4.877\,$\pm$\,0.030  & 0.56\,$\pm$\,0.02 \\
GJ 820\,A  & 3.114\,$\pm$\,0.268  & 2.540\,$\pm$\,0.198  & 2.248\,$\pm$\,0.318  & 2.822\,$\pm$\,0.317  & 2.120\,$\pm$\,0.080  & 2.334\,$\pm$\,0.009  & 2.206\,$\pm$\,0.011  & 7.00\,$\pm$\,0.09 \\
GJ 820\,B  & 3.546\,$\pm$\,0.278  & 2.895\,$\pm$\,0.218  & 2.544\,$\pm$\,0.328  & 6.224\,$\pm$\,0.010  & 2.884\,$\pm$\,0.001  & 2.595\,$\pm$\,0.009  & 2.529\,$\pm$\,0.013  & 5.95\,$\pm$\,0.12 \\
GJ 280  &  -0.498\,$\pm$\,0.151  & -0.666\,$\pm$\,0.270  & -0.658\,$\pm$\,0.322  & 2.147\,$\pm$\,0.397  & 0.625\,$\pm$\,0.255  & 1.148\,$\pm$\,0.022  & -0.646\,$\pm$\,0.003  & 109.4\,$\pm$\,1.7 \\
GJ 725\,A  & 5.189\,$\pm$\,0.017  & 4.741\,$\pm$\,0.036  & 4.432\,$\pm$\,0.020  & 4.498\,$\pm$\,0.226  & 3.520\,$\pm$\,0.157  & 4.070\,$\pm$\,0.014  & 3.937\,$\pm$\,0.018  & 2.09\,$\pm$\,0.03 \\
GJ 725\,B  & 5.721\,$\pm$\,0.020  & 5.197\,$\pm$\,0.024  & 5.000\,$\pm$\,0.023  & 5.014\,$\pm$\,0.325  & 4.309\,$\pm$\,0.206  & 4.588\,$\pm$\,0.016  & 4.464\,$\pm$\,0.025  & 2.09\,$\pm$\,0.03 \\
GJ 15\,A  & 5.252\,$\pm$\,0.264  & 4.476\,$\pm$\,0.200  & 4.018\,$\pm$\,0.020  & 3.853\,$\pm$\,0.099  & 3.130\,$\pm$\,0.074  & 3.707\,$\pm$\,0.015  & 3.595\,$\pm$\,0.022  & 1.84\,$\pm$\,0.02 \\
GJ 15\,B  & 6.789\,$\pm$\,0.024  & 6.191\,$\pm$\,0.016  & 5.948\,$\pm$\,0.024  & 5.745\,$\pm$\,0.045  & 5.419\,$\pm$\,0.028  & 5.463\,$\pm$\,0.015  & 5.303\,$\pm$\,0.030  & 1.84\,$\pm$\,0.02 \\
GJ 1111  & 8.235\,$\pm$\,0.021  & 7.617\,$\pm$\,0.018  & 7.260\,$\pm$\,0.024  & 7.030\,$\pm$\,0.031  & 6.819\,$\pm$\,0.020  & 6.630\,$\pm$\,0.015  & 6.467\,$\pm$\,0.058  & 0.14\,$\pm$\,0.01 \\
GJ 71  &  2.149\,$\pm$\,0.310  & 1.800\,$\pm$\,0.234  & 1.794\,$\pm$\,0.274  & 2.444\,$\pm$\,0.510  & 1.846\,$\pm$\,0.163  & 2.071\,$\pm$\,0.011  & 1.671\,$\pm$\,0.010  & 12.37\,$\pm$\,0.07 \\
GJ 54.1  & 7.258\,$\pm$\,0.020  & 6.749\,$\pm$\,0.033  & 6.420\,$\pm$\,0.017  & 6.167\,$\pm$\,0.044  & 5.929\,$\pm$\,0.021  & 5.888\,$\pm$\,0.014  & 5.719\,$\pm$\,0.036  & 0.27\,$\pm$\,0.01 \\
GJ 273  &  5.714\,$\pm$\,0.032  & 5.219\,$\pm$\,0.063  & 4.857\,$\pm$\,0.023  & 4.723\,$\pm$\,0.074  & 4.108\,$\pm$\,0.041  & 4.461\,$\pm$\,0.016  & 4.325\,$\pm$\,0.027  & 0.93\,$\pm$\,0.01 \\
SO 0253+16  & 8.394\,$\pm$\,0.027  & 7.883\,$\pm$\,0.040  & 7.585\,$\pm$\,0.046  & 7.322\,$\pm$\,0.027 & 7.057\,$\pm$\,0.020  & 6.897\,$\pm$\,0.017  & 6.718\,$\pm$\,0.076  & 0.10\,$\pm$\,0.01 \\
GJ 860\,AB  & 5.575\,$\pm$\,0.027  & 5.038\,$\pm$\,0.034  & 4.777\,$\pm$\,0.029  & 4.690\,$\pm$\,0.075  & 4.089\,$\pm$\,0.037  & 4.299\,$\pm$\,0.014  & 4.122\,$\pm$\,0.025  & 1.10\,$\pm$\,0.02 \\
GJ 83.1  & 7.514\,$\pm$\,0.017  & 6.970\,$\pm$\,0.027  & 6.648\,$\pm$\,0.017  & 6.438\,$\pm$\,0.042  & 6.162\,$\pm$\,0.021  & 6.100\,$\pm$\,0.014  & 5.964\,$\pm$\,0.043  & 0.22\,$\pm$\,0.02 \\
GJ 687  &  5.335\,$\pm$\,0.021  & 4.766\,$\pm$\,0.033  & 4.548\,$\pm$\,0.021  & 4.397\,$\pm$\,0.094  & 3.763\,$\pm$\,0.061  & 4.182\,$\pm$\,0.015  & 4.064\,$\pm$\,0.018  & 1.16\,$\pm$\,0.01 \\
GJ 1245\,AC  & 7.791\,$\pm$\,0.023  & 7.194\,$\pm$\,0.016  & 6.854\,$\pm$\,0.016  & 6.600\,$\pm$\,0.065  & 6.379\,$\pm$\,0.025  & 6.244\,$\pm$\,0.016  & 6.076\,$\pm$\,0.051  & ... \\
GJ 1245\,B  & 8.275\,$\pm$\,0.026  & 7.728\,$\pm$\,0.031  & 7.387\,$\pm$\,0.018  & 7.178\,$\pm$\,0.066  & 6.968\,$\pm$\,0.029  & 6.853\,$\pm$\,0.022  & 6.765\,$\pm$\,0.089  & ... \\
GJ 876  &  5.934\,$\pm$\,0.019  & 5.349\,$\pm$\,0.049  & 5.010\,$\pm$\,0.021  & 4.844\,$\pm$\,0.077  & 4.374\,$\pm$\,0.046  & 4.635\,$\pm$\,0.014  & 4.538\,$\pm$\,0.026  & 0.79\,$\pm$\,0.03 \\
GJ 1002  & 8.323\,$\pm$\,0.019  & 7.792\,$\pm$\,0.034  & 7.439\,$\pm$\,0.021  & 7.176\,$\pm$\,0.028 & 6.993\,$\pm$\,0.020  & 6.860\,$\pm$\,0.016  & 6.766\,$\pm$\,0.080  & 0.14\,$\pm$\,0.02
\enddata
\end{deluxetable*}

In total we have observed 33 individual stars within 5\,pc arranged in 25 systems, five of which are double: 
GJ~820\,A+B, GJ~15\,A+B, GJ~65\,AB, GJ~725\,A+B and GJ~860\,AB, and two of which are triple: GJ~866\,ABC and 
GJ~1245\,ABC. We count here GJ~866\,ABC as two stars, since its individual components AC were not 
resolved and (AC)B were marginally resolved in our observations. Additionally, two stars, Sirius and 
Procyon, have known white dwarf companions. The notation ``A+B'' signifies that the components were observed 
individually as separate CanariCam targets, and ``AB'' that both components were observed simultaneously as a 
single target. The sample includes one A, one F and one G type star, three K stars and 27 M dwarfs. Such 
distribution of spectral types implies that our statistical results will be significant only for M dwarfs. 
The sample is a volume limited sample complete up to 4.0\,pc. We have imaged all of the 19 known stellar 
systems at $\delta$\,$>$\,$-25^{\circ}$ within this distance, and 6 out of 15 known systems between 4 and 5\,pc 
observable from our site. Due to observational limitations from target brightness, substellar objects were not 
considered as target primaries. Hence, the Y2-type brown dwarf WISE 0855-0714 located at 2.23\,$\pm$\,0.04\,pc 
\citep{2014ApJ...786L..18L, 2016AJ....152...78L} and the $\sim$500~K brown dwarf UGPS J072227.51-054031.2 at 
4.12\,$\pm$\,0.04\,pc \citep{2012ApJ...748...74L} were not included in our sample. The remaining 9 objects 
between 4 and 5\,pc were not observed because of the limited telescope time available for the programme.

Table \ref{CCSample} lists the observed stars, including compiled information on their equatorial coordinates 
at J2000 epoch (proper motions taken into account), spectral types, trigonometric parallax, distance and proper 
motions. Table~\ref{SamplePhot} list their near- and mid-IR photometry. Because all known stars in the 
solar vicinity have large and well determined proper motions, our survey was designed to find common proper 
motion companions. Any additional source detected within the field of view would have been considered as a 
potential companion, without any criteria based on photometric colours. Its companionship could be easily 
verified through second epoch observations.

We have searched through the literature to gather the available information on the planets discovered around 
our sample stars by RVs, transits and other methods, as well as constraints on the substellar companions from 
other surveys or signs of RV or astrometric trends indicating a possible presence of a distant companion. 
Notes with selected essential information regarding each star are compiled in Table~\ref{known_exoplanets} 
in the Appendix. 

\section{Observations and data processing} \label{sec:observations}
The program was carried out in queue mode observations, starting in 2012 and completed in 2015. We used the 
mid-infrared camera CanariCam \citep{2008SPIE.7014E..0RT} operating at the Nasmyth-A focal station of the 
10.4\,m Gran Telescopio Canarias (GTC) at the Roque de los Muchachos Observatory on the island of La Palma 
(Spain). CanariCam was designed to reach the diffraction limit of the GTC at mid-IR wavelengths (7.5--25\,$\mu$m). 
The instrument uses a Raytheon 320$\times$240 Si:As detector with a pixel scale of 79.8\,$\pm$\,0.2\,mas, which 
covers a field of view of 25.6\,$\times$\,19.2\,arcsec on the sky. We imaged our targets in the 10 micron 
window, using a medium-band silicate filter centred at $\lambda$\,=\,8.7\,$\mu$m ($\delta\lambda$\,=\,1.1\,$\mu$m). 
The choice of this particular bandpass was a compromise between the instrument performance, in particular filter 
transmissivity, and the sky background contribution, which is significantly higher at the $N$ broad-band and 
other narrow-band filters than at the Si-2 filter. Si-2 is also favoured by a better spatial resolution, since 
the diffraction disc is larger at the other available narrow-band filters at longer wavelengths. Observations 
were executed under the following restricted atmospheric conditions: spectroscopic (clear sky with possible 
thin cirrus) or better, i.e., photometric/clear sky transparency, precipitable water vapor (PWV) at the level 
of 5--12\,mm and image quality of $<$\,0\farcs3, corresponding to a seeing of $\sim$\,0\farcs8 in 
the $R$ band.

Observations were performed with the standard chopping and nodding technique used in the mid-IR to remove the 
sky emission and radiative offset. Chopping consists of switching the telescope secondary mirror at a typical 
frequency of a few (2--5)\,Hz between the position of the source (on-source) and the nearby sky (off-source). 
This rapid movement of the secondary mirror allows subtraction of the sky background emission that is varying 
in time at frequencies below the chop frequency. Movement of the secondary mirror changes the optical 
configuration of the telescope, resulting in two different emission patterns seen by the camera and producing 
a spurious signal termed the radiative offset seen in the chop-differenced images. To remove the radiative 
offset, the telescope is moved between two nod positions to swap over on- and off-source positions.

\startlongtable
\begin{deluxetable*}{lccccccccccc}
\tablenum{3}
\tablecaption{CanariCam observation log\label{obslog}}
\tablewidth{0pt}
\tabletypesize{\scriptsize}
\tablehead{
\colhead{Star} & 
\colhead{OB} & 
\colhead{Observation} &  
\colhead{MJD} & 
\colhead{Saveset} & 
\colhead{On-source} & 
\colhead{Instr.} & 
\colhead{Chop} & 
\colhead{Nod} & 
\colhead{Readout} & 
\colhead{Sky} & 
\colhead{PWV} 
\\
\colhead{} & 
\colhead{\#} & 
\colhead{date} &  
\colhead{} & 
\colhead{(s)} & 
\colhead{(s)} & 
\colhead{PA} & 
\colhead{PA} & 
\colhead{PA} & 
\colhead{type} & 
\colhead{} & 
\colhead{(mm)}
}
\startdata
\hline
\multicolumn{12}{l}{GTC4-12BGCAN, Semester: 2012B} \\
\hline
GJ 1111 & 01 & 2012-12-26 & 56287.267541 & 5.96 & 3$\times$404 & 0 & 90 & -90 & S1R1-CR & Ph & $\sim$7.1 \\
        & 02 & 2012-12-28 & 56289.211395 & 5.96 & 3$\times$404 & 300 & 150 & -30 & S1R1-CR & Ph & $\sim$4.2 \\
GJ 71   & 03 & 2012-09-29 & 56199.097535 & 5.96 & 3$\times$404 & 1.83 & 90 & -90 & S1R1-CR & Ph & $<$9.1 \\
        & 04a & 2012-09-29 & 56199.138148 & 5.96 & 1$\times$404 & 1.83 & 90 & -90 & S1R1-CR & Ph & $<$9.1 \\
        & 04a1 & 2012-10-05 & 56205.036337 & 5.96 & 3$\times$404 & 300 & 150 & -30 & S1R1-CR & L.Cs. & 9.0 \\
        & 04a2 & 2012-12-03 & 56264.907141 & 5.96 & 1$\times$404 & 0 & 150 & -30 & S1R1-CR & Cl & 6.3 \\
GJ 406  & 05 & 2012-12-04 & 56265.186042 & 1.49 & 3$\times$454 & 0 & 90 & -90 & S1R1-CR & Cl & 6.0 \\
        & 06 & 2013-01-29 & 56321.090712 & 1.49 & 3$\times$454 & 300 & 150 & -30 & S1R1-CR & Ph & 5.5--5.7\\
        & 06a1 & 2012-12-04 & 56265.283756 & 1.49 & 2$\times$454 & 0 & 150 & -30 & S1R1-CR & Cl & 6.7 \\
        & 06a2 & 2012-12-28 & 56289.257245 & 5.96 & 3$\times$404 & 300 & 150 & -30 & S1R1-CR & Ph & $<$5.0\\
GJ 144  & 07 & 2012-09-29 & 56199.208941 & 5.96, 1.49 & 404, 545, 378 & 1.83 & 90& -90 & S1R1-CR & Ph & $<$9.1 \\
        & 08 & 2012-10-05 & 56205.085058 & 1.49 & 4$\times$454 & 300 & 150 & -30 & S1R1-CR & L.Cs. & 9.0 \\
\hline
\multicolumn{12}{l}{GTC9-12AGCAN, Semesters: 2012AB, 2013AB} \\
\hline
GJ 820 A & 01  & 2013-09-05  & 56540.063970 & 1.55 & 3$\times$432 & 0 & 90   & -90 & S1R3  & Ph & 7.7--8.9 \\
         & 02  & 2013-09-05  & 56540.114497 & 1.55 & 3$\times$432 & 300 & 30   & -150 & S1R3  & Ph & 8.3--8.6 \\
GJ 699   & 05  & 2012-07-29  & 56137.973895 & 5.96 & 3$\times$404 & 0   & -90  & 90  & S1R1-CR & Ph & 8.6--9.3 \\
         & 06  & 2012-07-30  & 56138.020220 & 5.96 & 3$\times$404 & 90  & -180 & 0  & S1R1-CR &  Ph & 8.6--9.3 \\
         & 19  & 2013-06-09  & 56452.160365 & 1.55 & 3$\times$432 & 0   & 90   & -90 & S1R3  & Ph & 6.7 \\
         & 20  & 2013-06-10  & 56453.185434 & 1.55 & 3$\times$360 & 300 & 0    & 180 & S1R3  & Ph & 6.7 \\
GJ 729   & 09  & 2013-09-05  & 56540.897378 & 1.55 & 3$\times$432 & 0   & 90   & -90 & S1R3  & Ph & 8.7--9.2 \\
         & 10  & 2013-09-14  & 56549.853825 & 1.55 & 3$\times$432 & 330 & 60   & -120 & S1R3 &  Cl &  9.9--9.2 \\
GJ 905   & 11  & 2013-06-07  & 56450.190631 & 6.21 & 3$\times$417 & 0 & 90   & -90 & S1R3  & Ph & 6.3 \\
         & 12  & 2013-06-08  & 56451.191418 & 6.21 & 3$\times$417 & 300 & -180 & 0 & S1R3   & Ph & 7.2 \\
GJ 15 A  & 15  & 2012-12-27  & 56288.852112 & 5.96 & 3$\times$404 & 0 & 90   & -90  & S1R1-CR & Ph & $\sim$4.6 \\
         & 16  & 2012-12-27  & 56288.916597 & 5.96 & 1211         & 90  & -180 & 0  & S1R1-CR & Ph & 4.6 \\
GJ 15 B  & 17  & 2012-12-28  & 56289.825174 & 5.96 & 3$\times$404 & 0 & 90   & -90  & S1R1-CR & Ph & $<$5.5 \\
         & 18  & 2012-12-28  & 56289.867014 & 5.96 & 3$\times$404 & 0 & -180 & 0  & S1R1-CR &  Ph & $<$6.0 \\
GJ 54.1  & 21  & 2013-08-30  & 56534.128715 & 6.21 & 3$\times$417 & 0   & 90   & -90 & S1R3  & Ph & 4.7--5.0 \\
         & 22  & 2013-08-30  & 56534.173160 & 6.21 & 4$\times$417 & 0   & 90   & -90 & S1R3  & Ph & 5.0--5.2 \\
GJ 65 AB & 26  & 2013-09-15  & 56550.091308 & 1.55 & 3$\times$432 & 0   & 90   & -90 & S1R3 &  Cl & $<$12 \\
         & 27  & 2013-09-15  & 56550.160683 & 1.55 & 3$\times$432 & 330 & 60   & -120 & S1R3 & Cl & 9.9--12 \\
GJ 866 (AC)B  & 29  & 2013-09-08  & 56543.051701 & 1.55 & 3$\times$432 & 330 & 60   & -120 & S1R3 & Cl & 9.0--10.3 \\ 
GJ 280   & 30  & 2014-01-04  & 56661.065336 & 1.55 & 3$\times$432 & 0   & 90   & -90 & S1R3 &  Cl & 7.9--9.1 \\
         & 31  & 2014-01-04  & 56661.143403 & 1.55 & 3$\times$432 & 330 & 60   & -120 & S1R3 &  Cl  & 8.3--9.4 \\
GJ 725 A & 32  & 2013-09-05  & 56540.951082 & 1.55 & 3$\times$432 & 0 & 90   & -90 & S1R3  & Ph & 8.7--9.2 \\
         & 33  & 2013-09-05  & 56540.996568 & 1.55 & 3$\times$432 & 330 & 60   & -120 & S1R3 & Ph & 8.4--9.1 \\
GJ 725 B & 34  & 2013-09-08  & 56543.926146 & 1.55 & 3$\times$432 & 0 & 90   & -90 & S1R3 &  Cl & 8.1--9.1 \\
         & 35  & 2013-09-08  & 56543.976256 & 1.55 & 3$\times$432 & 330 & 60   & -120 & S1R3 &  Cl & 8.7--9.0 \\
\hline
\multicolumn{12}{l}{GTC8-14AGCAN, Semesters: 2014AB, 2015AB} \\
\hline
GJ 411   & 01 & 2015-02-01 & 57054.289022 & 1.55 & 2$\times$432, 360 & 0 & 90 & -90 & S1R3 & Sp & $\sim$8 \\
         & 02 & 2015-02-03 & 57056.005301 & 1.55 & 3$\times$432 & 330 & 60 & -120 & S1R3 &  Cl &  6.0 \\
         & 39 & 2015-06-03 & 57176.938229 & 1.55 & 3$\times$432 & 0 & 90 & -90 & S1R3 &  Cl & 10.8--12.2 \\
GJ 244   & 03 & 2015-02-06 & 57059.926707 & 1.55 & 3$\times$432 & 0 & 90 & -90 & S1R3 &  Sp & 13.3 \\
         & 04 & 2015-02-01 & 57054.923368 & 1.55 & 2$\times$432, 360 & 330 & 60 & -120 & S1R3 &  L.Cs. & 7.0 \\
GJ 447   & 05 & 2014-05-10 & 56787.896094 & 6.21 & 3$\times$417 & 0 & 90 & -90 & S1R3 &  Cl &  6.1--8.5 \\
         & 06a & 2016-01-05 & 57392.188420 & 6.21 & 3$\times$417 & 30 & 60 & -120 & S1R3 &  Cl &  5.2--6.6 \\
GJ 15 A  & 09 & 2014-09-08  & 56908.079543 & 1.55 & 3$\times$432 & 0 & -90 & 90 & S1R3 &  Ph & 7.5 \\
         & 10 & 2014-09-08  & 56908.129653 & 1.55 & 3$\times$432 & 330 & -120 & 60 & S1R3 &  Ph & 7.5 \\
GJ 15 B  & 11 & 2014-09-23  & 56923.122396 & 1.55 & 4$\times$432 & 0 & 90 & -90 & S1R3 &  Cl &  7.0 \\
         & 11a & 2014-09-08 & 56908.178374 & 1.55 & 2$\times$432 & 0 & -90 & 90 & S1R3 &  Ph & 7.5 \\
         & 12 & 2014-09-23  & 56923.191435 & 1.55 & 360, 2$\times$432 & 330 & 60 & -120 & S1R3 & Cl &  7.0 \\
GJ 65 AB & 13 & 2014-12-02  & 56993.926748 & 1.55 & 3$\times$432 & 0 & 90 & -90 & S1R3 &  Ph & 4.7 \\
         & 14 & 2014-12-03  & 56994.886817 & 1.55 & 2$\times$432, 360 & 330 & 60 & -120 & S1R3 & Ph & 6.2--7.0 \\
GJ 820 B & 15 & 2014-06-11  & 56819.217245 & 1.55 & 2$\times$432, 360 & 0 & 90 & -90 & S1R3 & Ph & 10.8--11.1 \\
         & 16 & 2014-06-12  & 56820.185712 & 1.55 & 3$\times$432 & 330 & 60 & -120 & S1R3 &  Ph & 12.8--13.6 \\
GJ 866 (AC)B  & 17 & 2014-09-22 & 56922.996505 & 1.55 & 3$\times$432 & 0 & 90 & -90 & S1R3 &  Cl &  $\sim$6.0 \\
         & 18 & 2014-09-23  & 56923.051458 & 1.55 & 3$\times$432, 360 & 330 & 60 & -120 & S1R3 &  Cl &  7.0 \\
GJ 144   & 19 & 2014-10-04  & 56934.128079 & 1.55 & 2$\times$432, 360 & 0 & 90 & -90 & S1R3 & Ph & 9.7--10.6 \\
GJ 729   & 20 & 2014-07-10  & 56848.043125 & 1.55 & 3$\times$432 & 330 & 60 & -120 & S1R3 & Ph & 4.5--5.8 \\
GJ 273   & 21 & 2014-03-13  & 56729.891626 & 1.55 & 3$\times$432 & 0 & 90 & -90 & S1R3 &  Sp & $<$10 \\
         & 22 & 2014-03-13  & 56729.997425 & 1.55 & 3$\times$432 & 330 & 60 & -120 & S1R3 &  Sp & $<$10 \\
GJ 860 AB& 23 & 2014-09-02  & 56902.981678 & 6.21 & 417, 348, 209 & 0 & 90 & -90 & S1R3 &  Cl & 12--14 \\
         & 24 & 2014-09-03  & 56903.016580 & 6.21 & 278, 2$\times$417, 487 & 0 & 60 & -120 & S1R3 & Cl & 12--14 \\
SO0253+13& 30 & 2015-02-02  & 57055.942853 & 1.55 & 2$\times$432, 360 & 0 & 60 & -120 & S1R3 &  T.Cs. & 6.0 \\
         & 29\_2 & 2015-08-28  & 57262.166314 & 6.21 & 209, 2$\times$487, 417 & 0 & 90 & -90 & S1R3 & Cl &  6.5--7.4 \\
         & 30\_2 & 2015-09-02  & 57267.227269 & 6.21 & 2$\times$417, 348 & 0 & 60 & -120 & S1R3 &  Cl &  n.a. \\
GJ 83.1  & 47 & 2015-08-25  & 57259.131956 & 6.21 & 3$\times$417 & 0 & 90 & -90 & S1R3 &  Cl &  9.5. \\
         & 48a & 2015-08-25 & 57259.180666 & 6.21 & 3$\times$417 & 330 & 60 & -120 & S1R3 &  Cl &  9.3--9.9 \\
         & 48 & 2015-08-27  & 57261.200226 & 6.21 & 3$\times$417 & 330 & 60 & -120 & S1R3 &  Cl &  5.5 \\
GJ 687   & 49 & 2015-08-22  & 57256.935787 & 1.55 & 3$\times$432 & 0 & 90 & -90 & S1R3 &  Ph & 8.4 \\
         & 50a & 2015-08-22 & 57256.986562 & 1.55 & 3$\times$432 & 330 & 60 & -120 & S1R3 &  Ph & 7.8--8.7 \\
         & 50 & 2015-08-24  & 57258.972818 & 1.55 & 2$\times$432, 360 & 330 & 60 & -120 & S1R3 &  Cl &  7.6--9.7 \\
GJ 1245  & 51 & 2015-08-19  & 57253.955602 & 6.21 & 3$\times$417 & 0 & 90 & -90 & S1R3 &  Cl & 10.1--10.8 \\
         & 52 & 2015-08-19  & 57254.003559 & 6.21 & 3$\times$417 & 330 & 60 & -120 & S1R3 &  Cl &  10.8--12.2 \\
GJ 876   & 53 & 2015-08-25  & 57259.022564 & 1.55 & 2$\times$432, 360 & 0 & 90 & -90 & S1R3 &  Cl & 8.0--9.0 \\
         & 54 & 2015-08-25  & 57259.067002 & 1.55 & 2$\times$432, 360 & 330 & 60 & -120 & S1R3 &  Cl &  8.2--9.5 \\
GJ 1002  & 55 & 2015-08-07  & 57241.139595 & 6.21 & 4$\times$417 & 0 & 90 & -90 & S1R3 &  Cl &  7.8 \\
         & 56 & 2015-09-16  & 57281.062135 & 6.21 & 3$\times$417 & 330 & 60 & -120 & S1R3 &  Ph & 7.9--8.5
\enddata
\tablecomments{Sky conditions: Ph -- photometric, Cl -- clear, Sp -- spectroscopic, L.Cs. -- light cirrus, T.Cs. -- thick cirrus}
\end{deluxetable*}

We used an ABBA nodding sequence and ``on-chip'' chopping and nodding, with a chop-throw and nod offset of 
8\,arcsec, a chopping frequency of 1.93 or 2.01\,Hz and a nod settle time of about 45\,s. On chip method is 
recommended whenever the scientific target is point-like, since both on-source and off-source chop positions 
contain the signal of the target inside the detector field of view and can be aligned and combined. Individual 
frames of 26 and 19\,ms exposures were co-added by CanariCam control software to savesets of 1.6 and 6\,s 
depending on the brightness of the source. We used an on-source integration time of 40\,min in total, divided 
into two observing blocks (OBs) of 20\,min. Each block contained three data cube files composed of a set of 
individual images (savesets) at subsequent chopping and nodding positions. For the two observing blocks we set 
the instrument at two different position angles to rotate the field of view typically by 30\,deg (60 and 90\,deg 
rotations were also used in some cases), and adjusted the configuration of chop and nod position angles so as 
to maintain the chop/nod parallel to the longer axis of the detector. The use of two different orientations of 
instrument position angle was a way to initially check the reliability of potential faint sources, distinguish 
from bad pixels and explore the region along the horizontal axis of detector, where the cross-talk of the star 
in the 16 channels is more evident and the areas otherwise obscured by the negative off-source chops. 
A detailed observation log is presented in Table~\ref{obslog}.

\begin{figure}
\centering
\includegraphics[width=1.0\columnwidth]{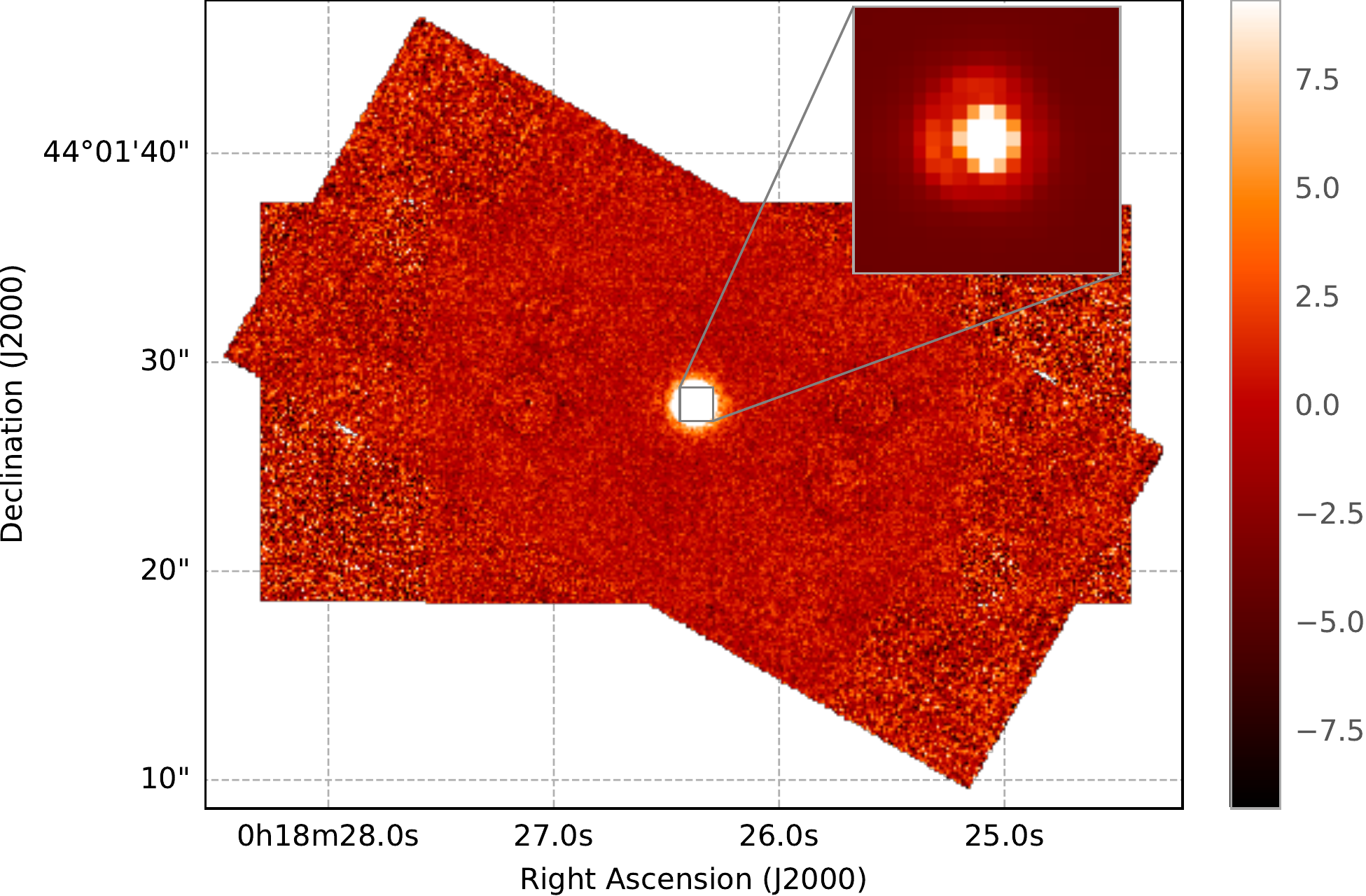}
\caption{The final 8.7\,$\mu$m image mosaic of one of the target stars, GJ\,15A, encompassing 
the full area covered by the CanariCam data. Counts are in linear scale and in the range 
$\pm$7$\sigma$ relative to the zero background. The deepest central region, where the stacks 
area overlap, is a rectangle of $\sim25\arcsec\times19\arcsec$. A zoomed in inset of 
$1.5\arcsec\times1.5\arcsec$ shows the core of the PSF with the first Airy disk visible. 
North is up and east is to the left.
\label{CCimgs}}
\end{figure}

CanariCam images are stored in the standard multi-extension FITS files, with a structure of [320, 240, 2, M][N], 
where 320 and 240 are the image pixel dimensions, 2 is the number of chop positions, M of savesets and N of 
nod positions. The data were processed using a set of dedicated {\tt\string IRAF/PyRAF}\footnote{{\tt PyRAF} 
is a product of the Space Telescope Science Institute, which is operated by AURA for NASA.} scripts developed 
within our group.

\begin{figure*}
\centering
\includegraphics[height=2.8cm,keepaspectratio]{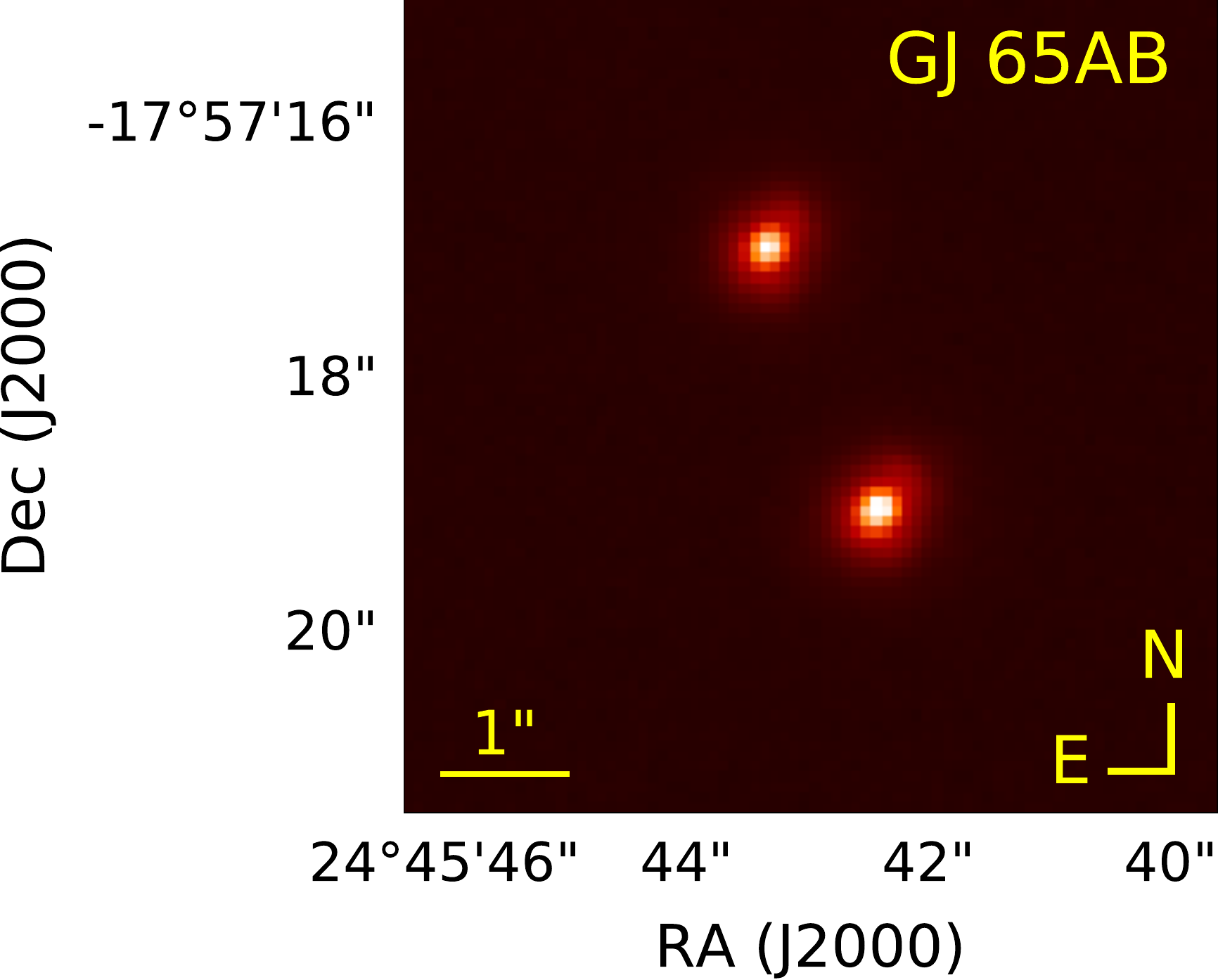}
\includegraphics[clip=true,trim=30pt 0pt 0pt 0pt,height=2.8cm,keepaspectratio]{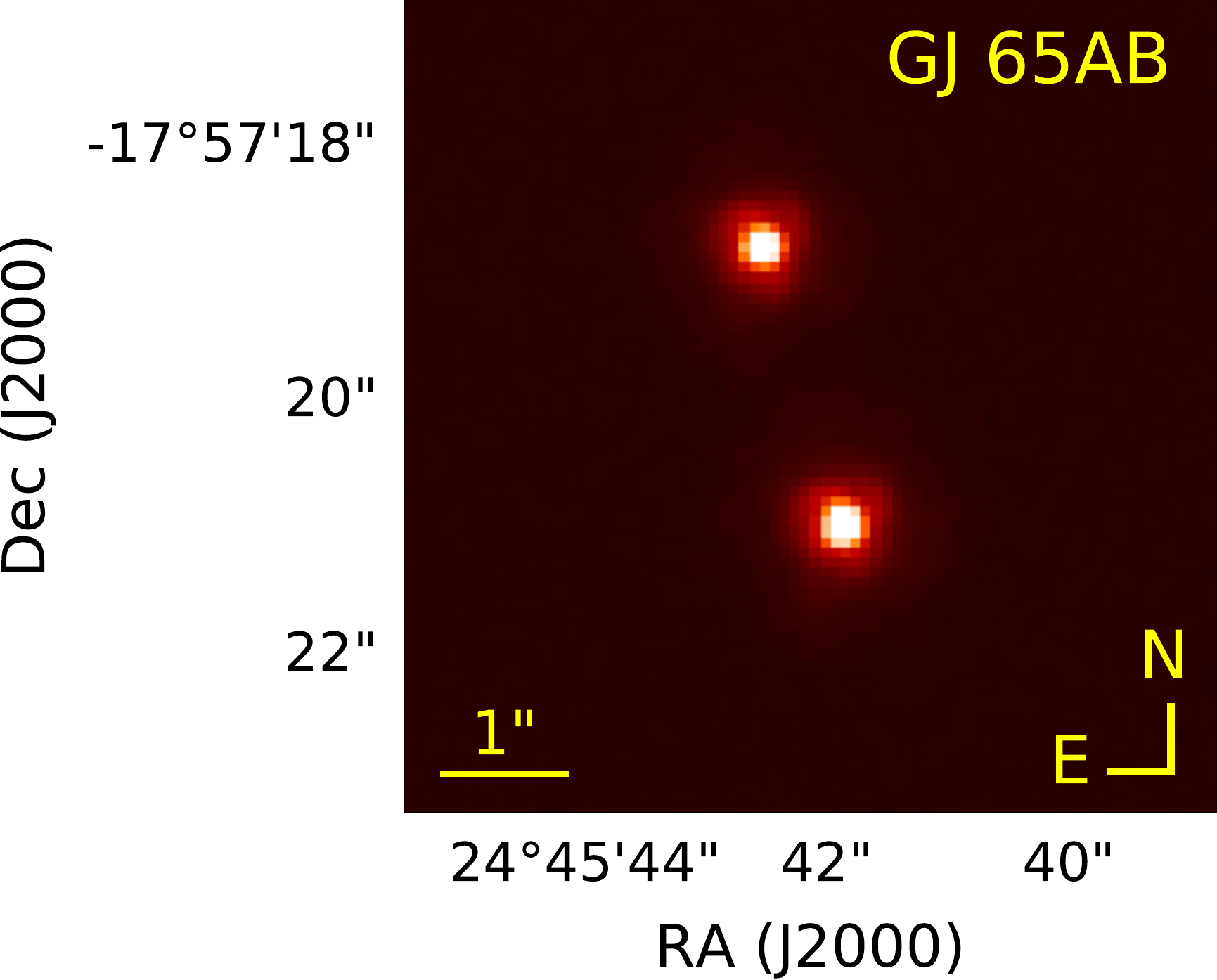}
\includegraphics[clip=true,trim=30pt 0pt 0pt 0pt,height=2.8cm,keepaspectratio]{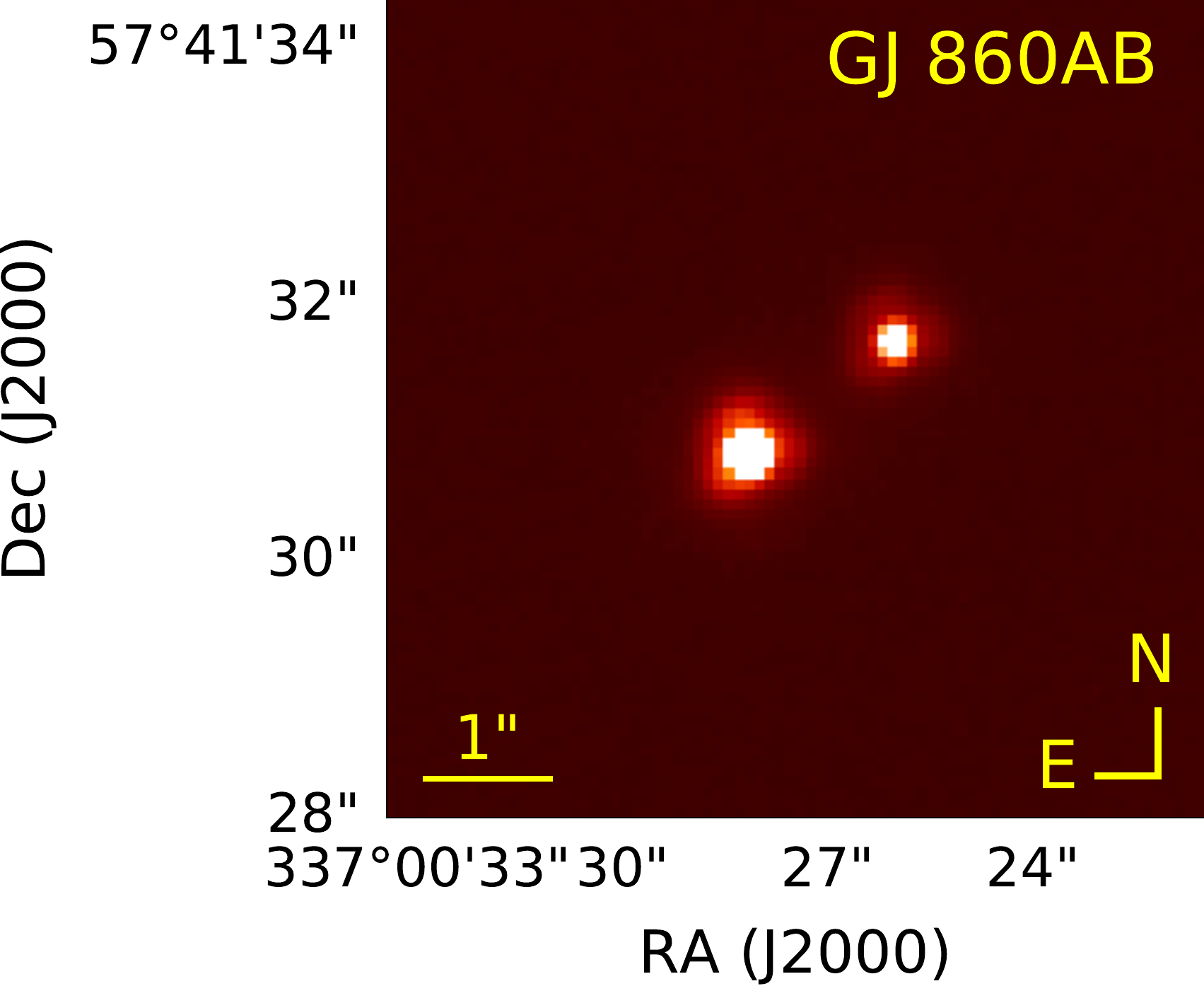}
\includegraphics[clip=true,trim=30pt 0pt 0pt 0pt,height=2.8cm,keepaspectratio]{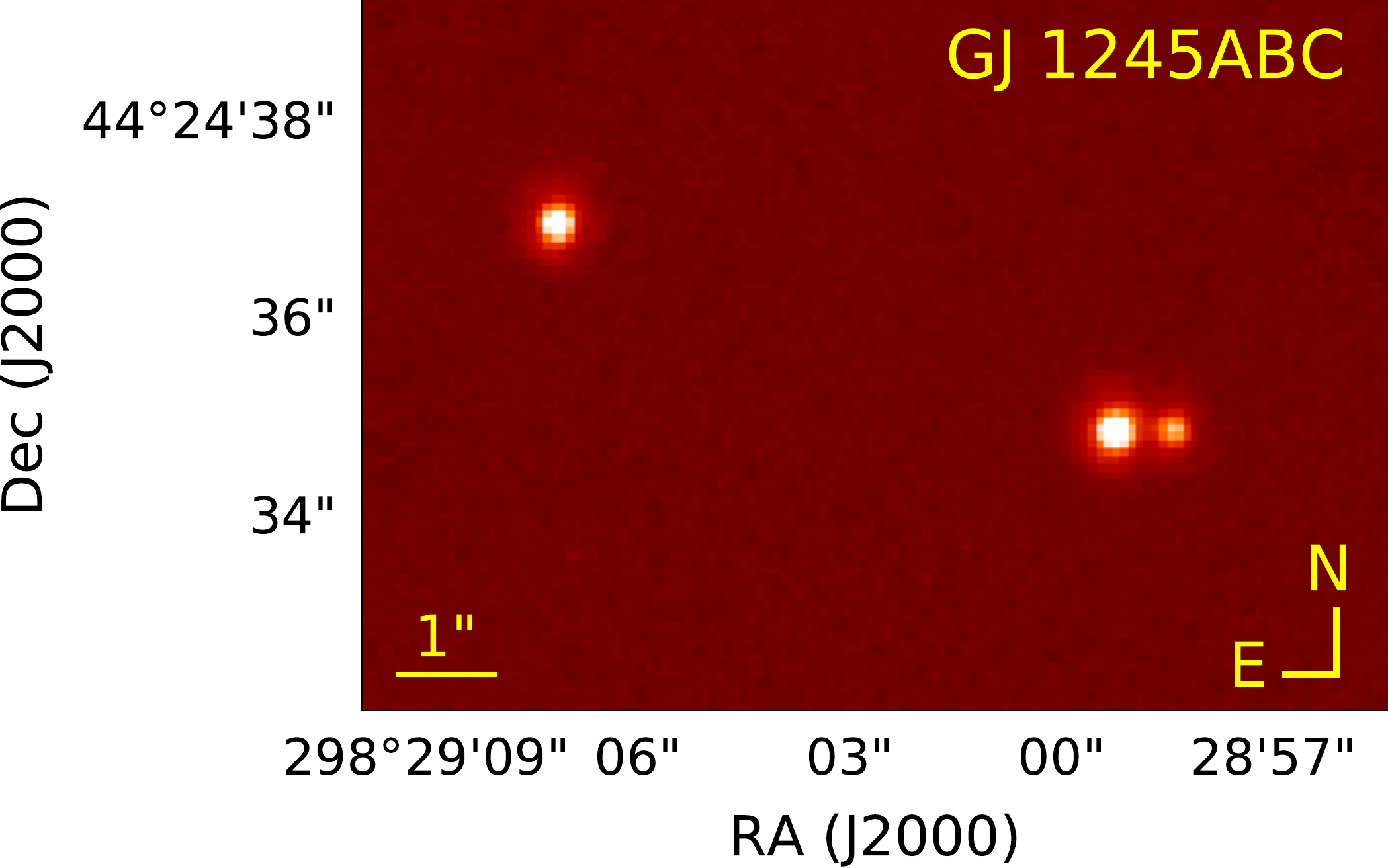}
\includegraphics[clip=true,trim=30pt 0pt 0pt 0pt,height=2.8cm,keepaspectratio]{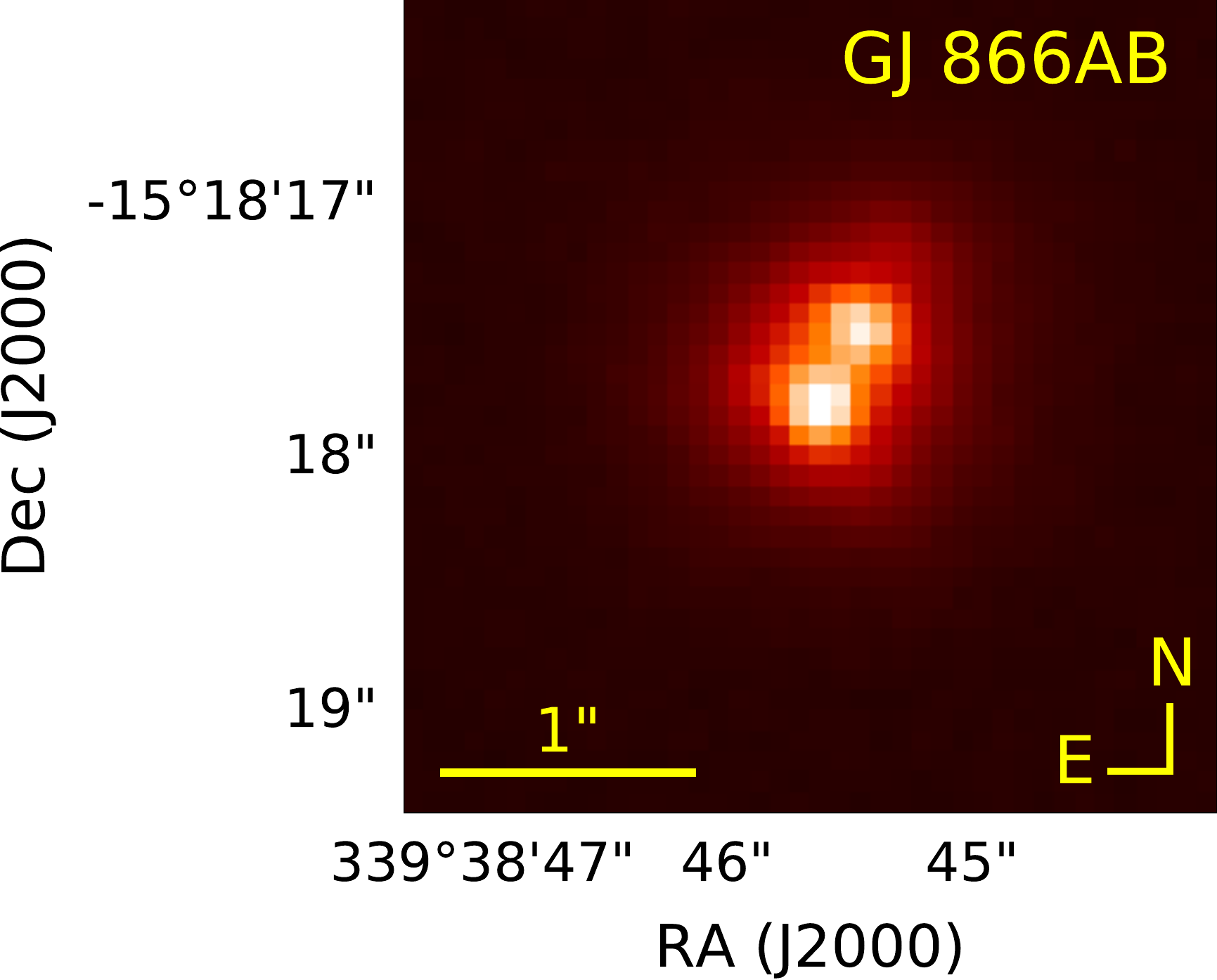}
\caption{Images of the resolved binary and multiple stars: GJ~65AB (at two different epochs), 
GJ~860AB, GJ~1245ABC and GJ~866(AC)B at the first epoch, 2013 Sep. 8. A 1\,arcsec scale 
is indicated. North is up and East is to the left. 
\label{CCimgs_binaries}}
\end{figure*}

As a first step, the off-source savesets, where the star is not located at the centre of the detector, were 
subtracted from the corresponding on-source savesets, for respective nod beam position. These chop-/sky-subtracted 
frames, where the star is located at the centre of the detector, were then aligned to correct for relatively small
misalignments (typically of less than five pixels) with respect to the preliminary shifts computed from the chop and 
nod pointing offsets. Then each pair of frames corresponding to the A and B nod positions was combined, to subtract 
the radiative offset. The sky-subtracted frames were multiplied by $-$1 to recover the negative contributions of 
the star (off-source position of the secondary mirror). 

Because the negatives in the A and B nod positions do not overlap, being at opposite sides and at 8\,arcsec of 
the on-source central location, they were also combined to subtract the radiative offset before they were 
aligned. Residual detector levels constant along single columns or lines but varying across these remained in 
both the positive and negative chop- and nod-subtracted frames; these were background fitted (masking the target) 
and subtracted. 

The alignment itself was applied at once, to all (positive and negative) images of consecutive repetitions of 
an OB, relative to a same reference image, and so that the centroids of the target in all images including the 
reference image were shifted to a same, integer pixel position value. Therefore, the ulterior alignment and 
combination, even with other epochs, were simplified to integer pixel shifts, obviating the need of reinterpolation. 
Before aligning, the images were copied into larger ones to avoid the trimming of outer data regions. Then the 
frames were average-combined per repetition using a shallow sigma upper and lower clipping to discard occasional 
short transients and sharp outliers.

Each combination involved masking the negative counts of the target. Horizontal patterns of cross-talk 
features, apparent for the brightest targets were removed. Repetitions from OBs acquired with position 
angles differing from the North-up East-left orientation were resampled to common orientation using the 
{\tt\string mscimage} task. For the combination of the stacks of the different OBs, the repetitions were 
flux-scaled according to their zero-point magnitude -- as measured on the target -- and weighted inversely 
proportional to the scaled variance of their background noise and the square of the full-width half-maximum 
(FWHM) of the target. 

A final processed image mosaic of one of the target stars -- GJ~15A, with a total on-source time of 86\,min is 
displayed in Figure~\ref{CCimgs}. The counts are represented in linear scale and within $\pm7\sigma$ of the 
background level, where $\sigma$ is the standard deviation of the background noise. The central region with the 
highest sensitivity, where the sky area of stacked frames overlap, covers a rectangle of approximately 
$25\arcsec\times19\arcsec$ ($\sim$89$\times$67\,au at the distance of GJ~15A). A collection of the processed 
CanariCam images of all the observed stars is presented in Figs.~\ref{CCallimages} and \ref{CCallimages_fullFOV} 
in the Appendix. Fig.~\ref{CCallimages} contains a set of common size ($25\farcs6\times19\farcs2$) central 
parts of the images, with the highest sensitivity and smooth background levels, whereas 
Fig.~\ref{CCallimages_fullFOV} includes full FOV mosaics. All the reduced image stacks are available at:
\url{https://cloud.iac.es/index.php/s/kT3cCdP9Wxw92gZ}.

\section{Analysis and results}
\subsection{Relative astrometry of resolved binaries}

As part of our CanariCam observations, we resolved the components of the binary stars GJ~65AB and GJ~860AB, 
and the triple systems GJ~1245ABC and GJ~866(AC)B. Cut-out images of these systems are displayed in 
Figure~\ref{CCimgs_binaries}. We measured the relative angular separations and position angles 
($\rho$, $\theta$) of the resolved components. Observations of GJ~65AB and GJ~866(AC)B were repeated 
at two separate epochs, about 1.2 and 1.0\,yr apart, respectively. In case of GJ~866, the A and C pair is 
a spectroscopic binary \citep{1999AA...350L..39D, 2000AA...353..253W} at $\rho\sim0.01\arcsec$, and thus 
only the B component was resolved in the first epoch. In the second epoch the components got closer by 
their orbital motion, to a separation below the angular resolution of 0.298\arcsec achieved on the images. 
Sirius B was marginally detected at $\rho\sim$10.3\arcsec, $\theta\sim$\,77.7\,deg, but its low signal-to-noise
(S/N$\sim$4) precludes more accurate measurements.

\begin{deluxetable}{lccc}
\tablenum{4}
\tablecaption{Relative astrometry of resolved binary/multiple stars\label{astrometry}}
\tablewidth{0pt}
\tabletypesize{\scriptsize}
\tablehead{
\colhead{Star} & 
\colhead{Epoch (MJD)} & 
\colhead{$\rho$ ($\arcsec$)} & 
\colhead{$\theta$ ($^{\circ}$)} }
\startdata
GJ 65AB	    & 56550.125995  & 2.240\,$\pm$\,0.012 & 22.94\,$\pm$\,0.39\\
            & 56994.406782  & 2.267\,$\pm$\,0.010 & 16.27\,$\pm$\,0.35\\
GJ 860AB	& 56902.999129  & 1.4338\,$\pm$\,0.0036 & 307.61\,$\pm$\,0.33\\
GJ 866(AC)B & 56543.051701  & 0.33\,$\pm$\,0.02 & 331.65\,$\pm$\,1.90\\
            & 56922.996505  & $<$0.298 & ... \\
GJ 1245AB	& 57253.979581  & 6.0063\,$\pm$\,0.0080 & 69.63\,$\pm$\,0.16\\
GJ 1245AC   & 57253.979581  &  0.582\,$\pm$\,0.007 & 271.26\,$\pm$\,0.73\\
\enddata
\end{deluxetable}

We obtained centroids of the sources using the {\tt\string IRAF imcentroid} task, and transformed the $X$ and 
$Y$ pixel coordinates with their respective errors into angular separations and North-to-East position angles 
using the CanariCam pixel scale of 79.8\,$\pm$\,0.2\,mas ({\url{http://www.gtc.iac.es/instruments/canaricam/canaricam.php}}) 
and the orientation given by the instrument position angle in the image header. We checked that the precision 
of this orientation is better than 0.3\,deg by inspecting the alignment of the chopping-nodding throw with 
the detector $X$ axis. We did not perform any calibration observations. Therefore, in the determination of 
the relative astrometry we relied on the default internal calibration of the instrument position angle and 
pixel scale. Measured values are listed in Table~\ref{astrometry}.

\subsection{Contrast curves and achieved sensitivities} \label{subsec:contr}

\begin{figure*}
\centering
\includegraphics[scale=0.55,keepaspectratio=true,clip=true,trim=0pt 0pt 0pt 0pt]{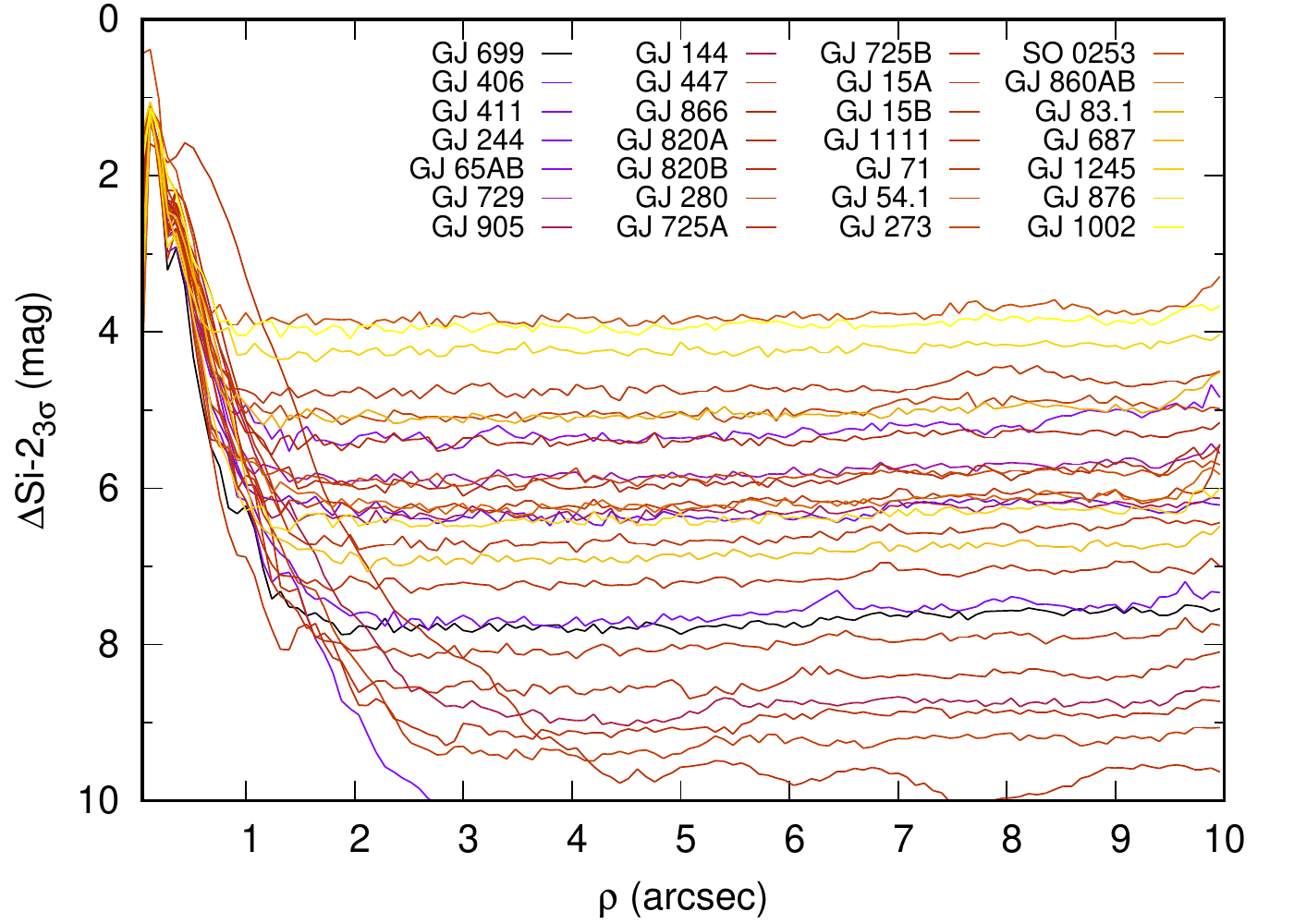}
\includegraphics[scale=0.55,keepaspectratio=true,clip=true,trim=39pt 0pt 0pt 0pt]{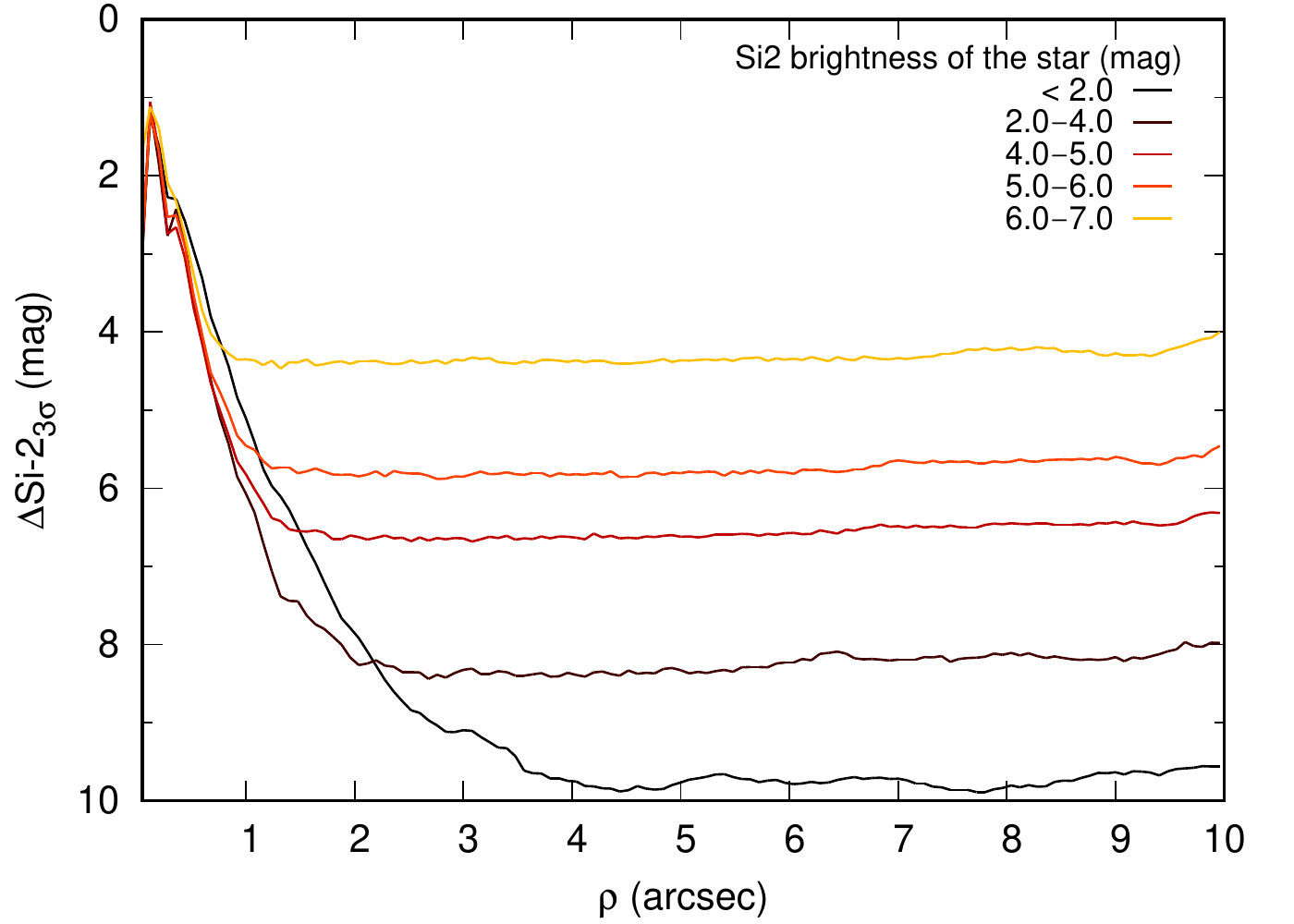}
\caption{{\it Left:} Si-2 (8.7\,$\mu$m) contrast curves at the 3$\sigma$ level
for all individual targets. The curves are ordered and color coded by distance 
of the star (as in Table~\ref{detlims}). {\it Right:} the mean contrast curves 
per given interval of brightness of the stars.
\label{contrasts}}
\end{figure*}

\begin{figure*}
\centering
\includegraphics[scale=0.55,keepaspectratio=true,clip=true,trim=0pt 0pt 0pt 0pt]{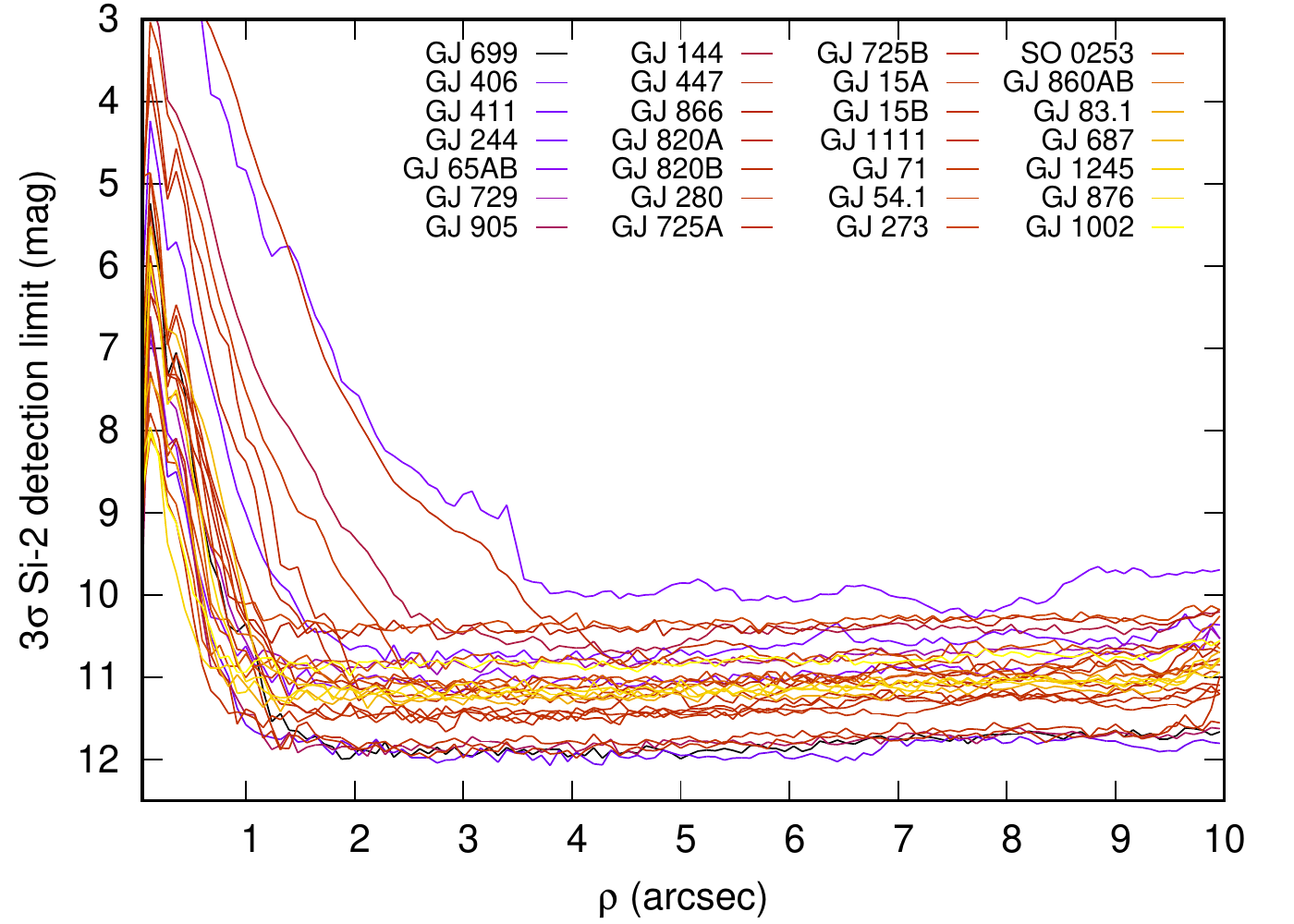}
\includegraphics[scale=0.55,keepaspectratio=true,clip=true,trim=39pt 0pt 0pt 0pt]{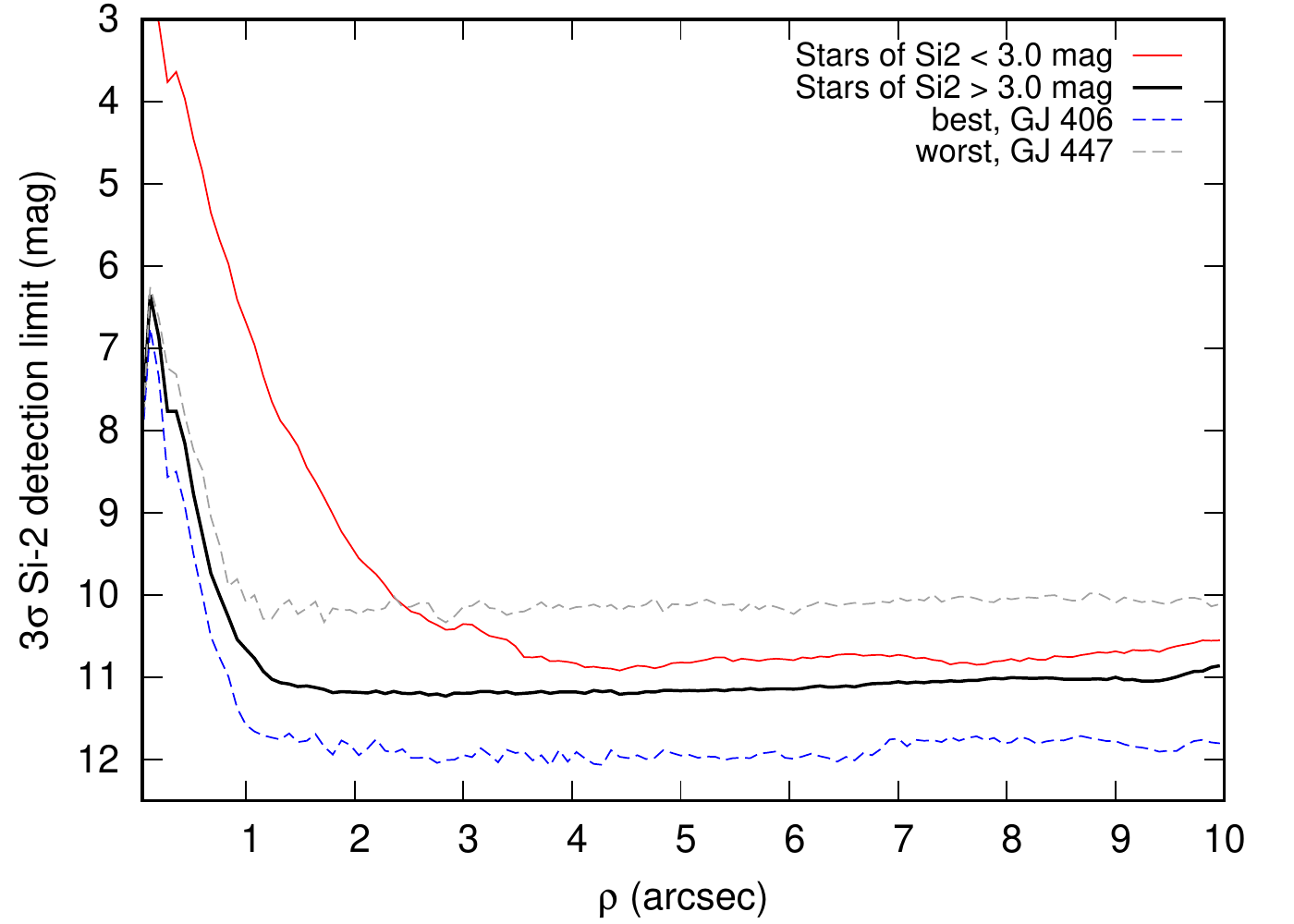}
\caption{CanariCam Si-2 band (8.7\,$\mu$m) detection limits. For each individual 
star on the left panel, and on the right panel, average curves for stars brighter 
and fainter than Si-2\,=\,3.0\,mag, and the best (GJ 406) and the worst (GJ 447) cases.
\label{detlimits}}
\end{figure*}

Since our observations have a bright star at the centre, by aligning the individual chop-subtracted frames we 
could improve the resulting FWHM as compared with straight stacking of images done by default by the automatic 
data reduction pipeline of the instrument provided by the observatory, and to some extent compensate the lack 
of fast-guiding mode of the telescope. The mean FWHM of the PSFs of all images is of 0.28\arcsec (3.5\,pix), 
with the best and the worst values of 0.23\arcsec (2.8\,pix) and 0.55\arcsec (6.9\,pix), respectively. The 
quality of our data is close to the theoretical FWHM of the diffraction-limited PSF, which for GTC is 
0.23\arcsec at 8.7\,$\mu$m.

We measured the detection limits on the deepest region of the final images of each target in the survey 
by using the ratio of the peak counts of the star to 3 times the background noise ($\sigma$). To determine the 
3$\sigma$ limiting magnitudes of the images we estimated the magnitudes of the sample stars in the Si-2 filter, 
using the {\it JHKs} photometry from the 2MASS and WISE W1, W2, W3, W4 photometry from the All-Sky and AllWISE 
Source Catalogs \citep{2010AJ....140.1868W}. In cases when the mid-IR photometry from WISE was highly 
affected by saturation, we used other measurements available in the literature from Spitzer/IRAC and Akari 
S09W and L18W bands \citep{2007PASJ...59S.369M}. We converted the 2MASS and WISE magnitudes into fluxes 
using the corresponding Vega zero points and their errors for each band from \cite{2003AJ....126.1090C} and
\cite{2011ApJ...735..112J}. Then, we fitted a power funcion $f(\lambda) = c_1\lambda^{c_2}+c_3$ to the available 
measurements via a least squares method and used the obtained parameters to estimate the average flux at 
$\lambda$\,=\,8.7\,$\mu$m. This approach does not take into account the different spectral types. However, 
we checked that the effect on the estimated magnitudes at this wavelength is negligible relative to overall
uncertainties. The Si-2 magnitude of each star was calculated using the Vega system zero point determined for 
this CanariCam filter. For Sirius A (GJ 244) we used directly the well-calibrated Si-2 flux within the standard 
filters from Gemini/T-ReCS observations \citep{2011ApJ...730...53S}. For tight binary stars in the sample 
resolved by CanariCam but unresolved by WISE, the magnitudes of individual components were descomposed from the 
integrated magntiudes and peak-to-peak flux ratios. The values of FWHM, and Si-2 3$\sigma$ detection limits 
for individual targets are listed in Table~\ref{detlims}.

To measure the sensitivity as a function of angular separation from the central star, we computed the background 
noise, $\sigma$, as a function of radial separation from the star, by measuring the standard deviation in 1 pixel 
wide concentric annuli around it. The 3$\sigma$ noise counts were converted to contrast (difference in magnitude 
between the primary star and the measured quantity, noise in this case) by relating to the peak pixel value of 
the star's PSF. Results of this method were found to be consistent with a more realistic procedure based on 
inclusion of artificial sources, as in the analysis of the Barnard's Star data described in \cite{2015MNRAS.452.1677G}. 
Then, the sensitivity limit was calculated using the corresponding Si-2 magnitude of a given star.

\begin{figure}
 \centering
 \includegraphics[width=1.0\columnwidth,keepaspectratio=true]{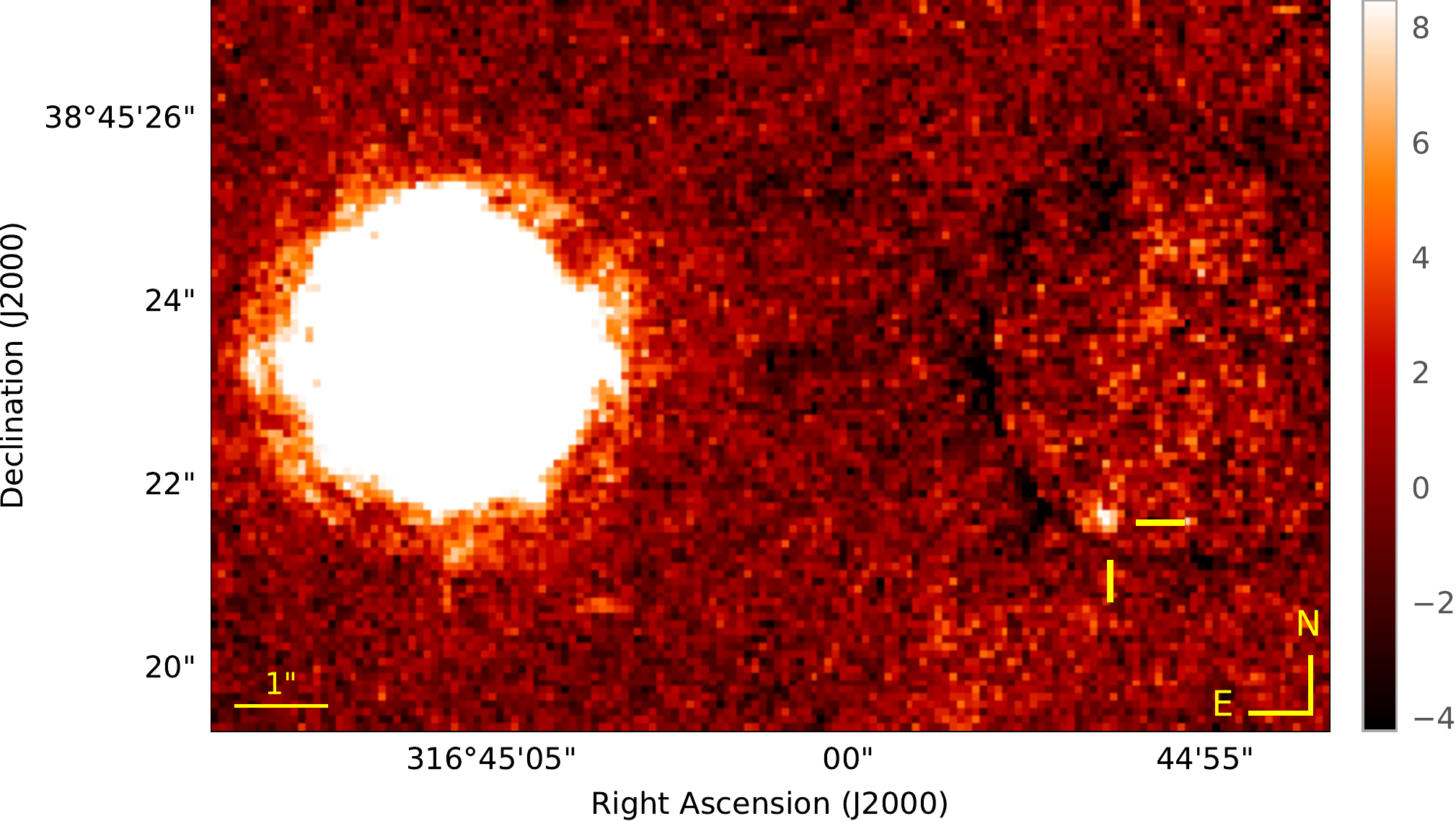}
 \caption{CanariCam image of GJ 820A and the faint (Si-2$\sim$10.4\,mag) 
 source detected at $\rho=7\farcs20\pm0\farcs04$, $\theta=256.2\pm0.3$\,deg 
 (epoch J2013.75), indicated by two yellow lines. The faint source is the 
 background star TYC 3168-590-1. 
 \label{gj820A}}
\end{figure}

The set of graphs of Figs.~\ref{contrasts} and \ref{detlimits} comprises the 3$\sigma$ contrast curves and 
detection limit curves of the survey. The left plot of Fig.~\ref{contrasts} collects the contrast curves 
($\Delta$Si-2\,mag as a function of angular separation $\rho$) for all observed stars individually. On the right 
graph are plotted the average achieved contrast versus separation, for a given brightness range of the target 
stars. For the brightest stars with Si-2\,$<$\,2.0\,mag the maximum dynamic range, of about 10\,mag, 
is achieved at $\sim$3\,arcsec. For stars fainter than 4.0\,mag (20 stars of the sample) the maximum contrast 
is reached at 1.0--1.5\,arcsec. The detection limit curves for each observed star are plotted on the left 
panel of Figure \ref{detlimits}. On the right panel, we plot the mean detectability curves for stars brighter 
and fainter than 3.0\,mag, and also the best and the worst cases, which were on GJ 699 and GJ 273, respectively. 
For most of the observed stars ($\sim$80\%), i.e., those with Si-2\,$<$\,3.0\,mag, the detectability limit of 
Si-2\,=\,11.3\,$\pm$\,0.2\,mag on average, was reached at $\rho\gtrsim$\,1.0--1.5\,arcsec separation.

For eight targets (GJ~699, GJ~65AB, GJ~729, GJ~144, GJ~447, GJ~866, GJ~15A and GJ~15B), observations were 
performed at two epochs separated by $\sim$1.0--1.7\,yr. In these cases, the orbital motion of companions 
may not be negligible. We estimate, considering a 20--40\,$M_{\rm Jup}$ companion in a circular, face-on 
orbit, that the angular shift induced by orbital motion becomes significant, i.e., exceeds the average 
spatial resolution of the images, for $a\lesssim8.0-16.5$\,au orbits, corresponding to 
$\rho\lesssim2\farcs0-6\farcs5$ angular separations at $d$\,=\,2.5--3.5\,pc. Stacking of images of well 
separated epochs would result either in a smearing or point-source doubling for any potential close-in 
faint companion. In any case, we searched over these closest separations around that stars in the stacks 
of two epochs observations both combined all together and of each epoch separately. The constrast 
curves and detection limits reported for these targets were computed on single-epochs images stacks.

\subsection{Constraints on substellar companions} \label{sec:frequencies}
We did not find any new companion to the stars imaged in this survey. All the final processed images were 
examined by us for the presence of faint, point-like sources directly by a visual inspection. We searched for 
candidate objects both on images combined of all the available CanariCam data, as well as on images stacked 
separately at different observing epochs or at different orientations of instrument position angle (typically 
contained in two separate observing blocks of a half of the total on-source time available). We also looked 
for candidates in the regions masked by the negatives by using the complementing images. We did not employ 
any PSF subtraction method because adjusting the image display contrast suffice to efficiently inspect the 
immediate surroundings of the target stars even to one FWHM separation of the stellar PSF.

\begin{figure}
\centering
\includegraphics[width=0.8\columnwidth,keepaspectratio=true,clip=true,trim=0pt 10pt 10pt 0pt]{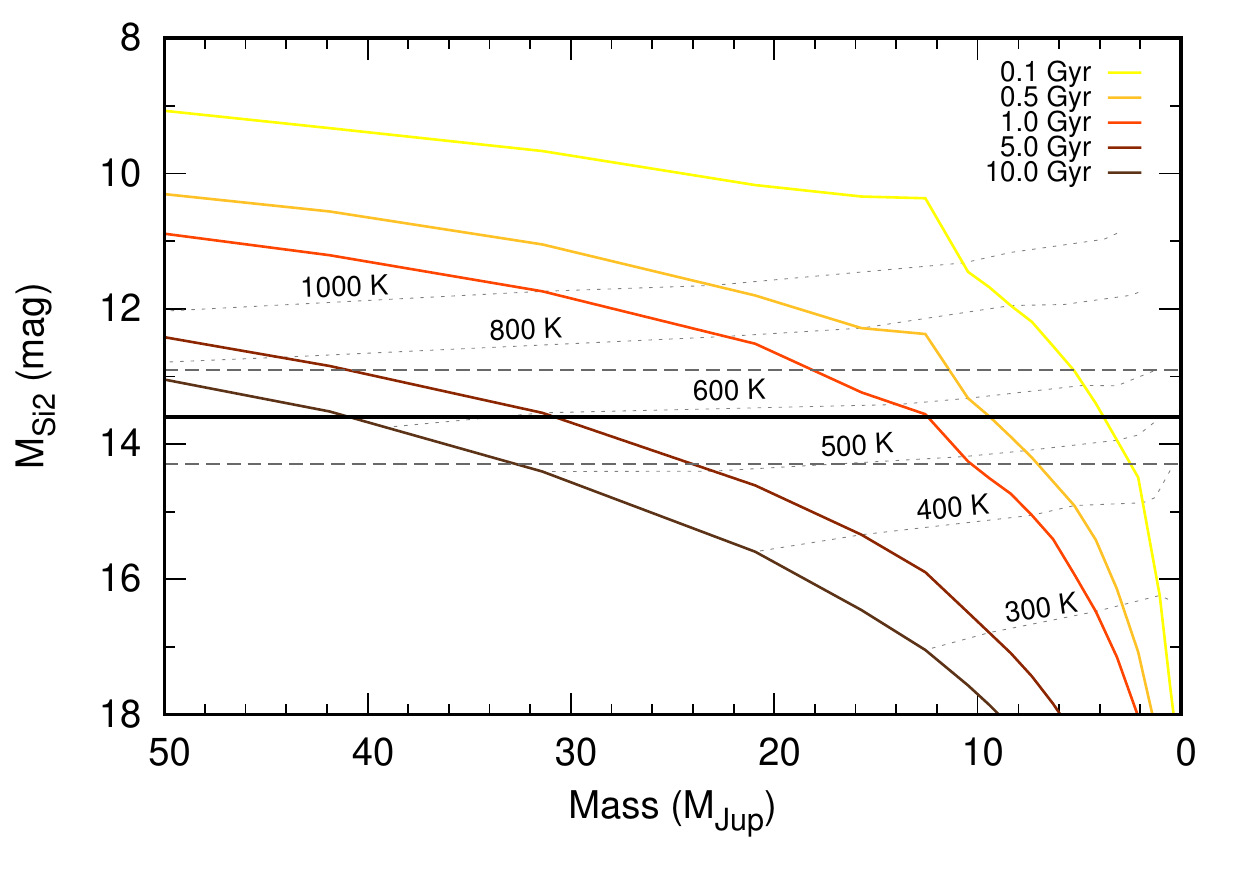}
\caption{Theoretical absolute magnitudes versus mass of brown dwarfs and 
giant planets at Si-2 8.7\,$\mu$m band obtained using the solar abundance 
Ames-COND models at ages of 0.1, 0.5, 1, 5 and 10\,Gyr. Several corresponding 
isotherms are plotted with dotted lines. The solid and dashed horizontal 
lines mark the mean 3$\sigma$ detection limit range of the survey and a 
1$\sigma$ dispersion: $M_{\rm Si2}$\,=\,13.6\,$\pm$\,0.7\,mag.
\label{massesfig}}
\end{figure}

\begin{figure*}
\centering
\includegraphics[width=0.85\columnwidth,keepaspectratio=true,clip=true,trim=0pt 0pt 0pt 0pt]{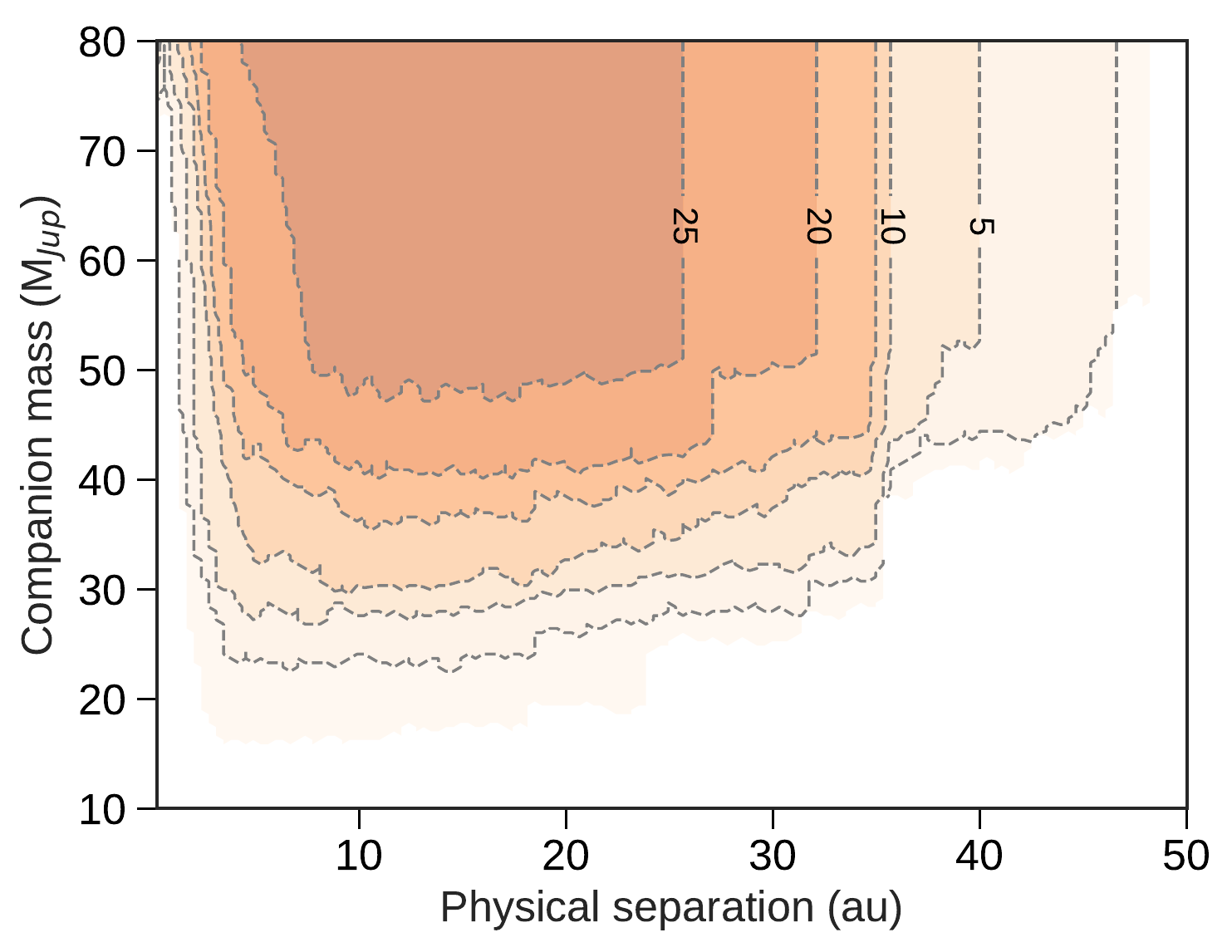}
\includegraphics[width=0.85\columnwidth,keepaspectratio=true,clip=true,trim=0pt 0pt 0pt 0pt]{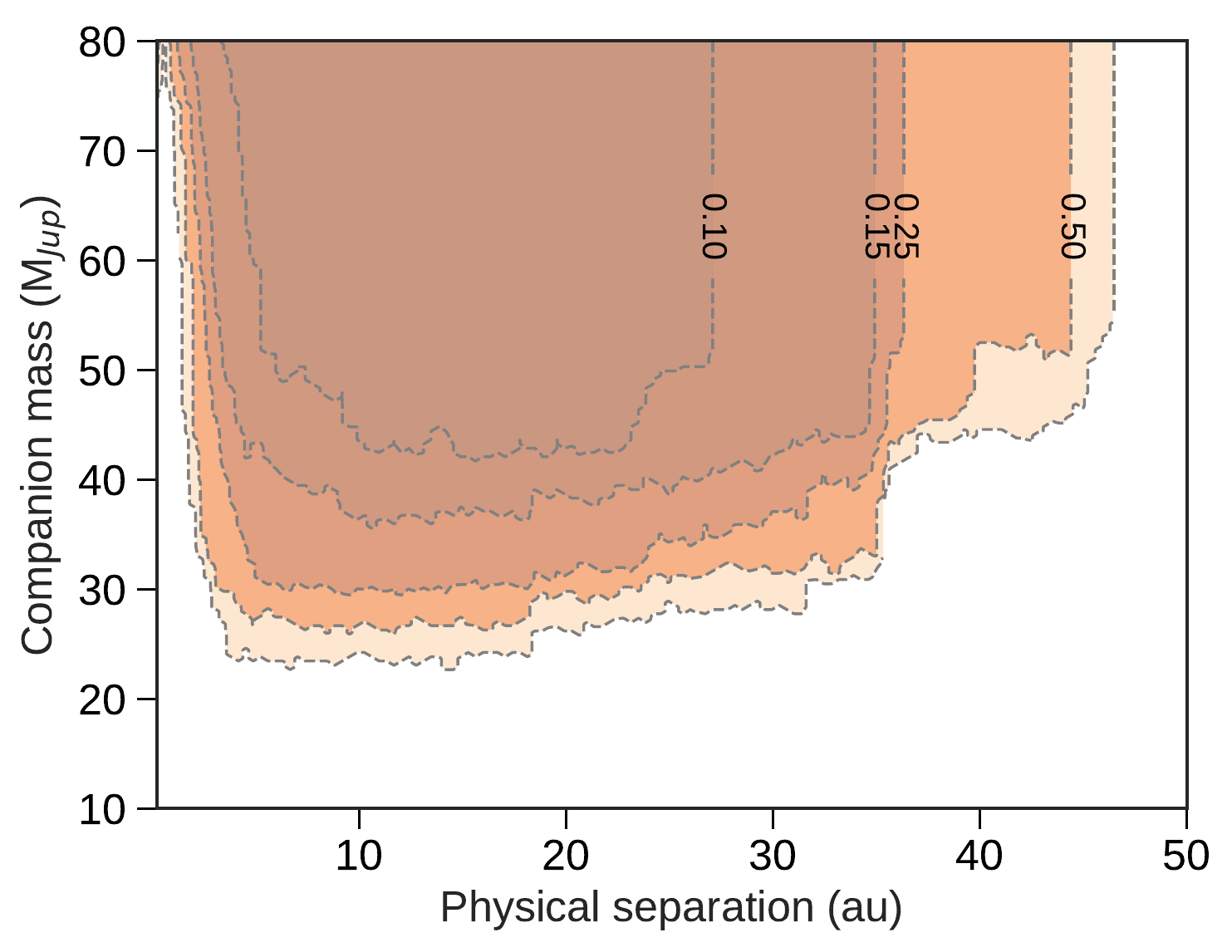}
\caption{{\it Left panel:} Overall completeness map of the survey. The contour lines illustrate the number of stars around
which the survey is sensitive to substellar companions as a function of companion mass and projected orbital separation.
{\it Right panel:} Upper limits constraints on substellar companions occurrence frequencies, in the same range of masses and 
projected separations.
\label{completeness_map}}
\end{figure*}

The only one additional faint source was detected within the field of view on the images of GJ~820A 
(displayed in Fig.~\ref{gj820A}). It was detected in both OBs taken with different instrument position 
angle (North to East sky orientation rotated by 60\,deg), with an apparent magnitude of 10.4\,$\pm$\,0.3 
in the Si-2 band, at an angular separation and position angle of $\rho\,=\,7.20\pm0.04$\,arcsec and 
$\theta\,=\,256.2\pm0.3$\,deg, respectively. We calculated its equatorial coordinates at the epoch of the 
CanariCam observation based on the precise proper motion of GJ~820A and the measured $\rho$ and $\theta$. 
Using available archival observations and catalogues of this area of the sky, we found that the source is 
the background star TYC 3168-590-1 (2MASS J21065820+3845411) with $V_T$=10.737$\pm$0.066, $J$=10.669$\pm$0.024, 
$H$=10.466$\pm$0.016, $K_{\rm s}$=10.405$\pm$0.013\,mag. It is also catalogued in Gaia EDR3 with 
$G$=11.517$\pm$0.003\,mag and a parallax of 2.51$\pm$0.01\,mas. Apart from this source, no other additional 
sources were identified. In this section we translate the detection limits of each star to constraints on 
the physical properties of detectable substellar companions, specifically to their masses and effective 
temperatures.


Because the substellar objects continuously cool down as they evolve, it is not possible to determine their 
mass applying unique relations independent of age, such as the mass--luminosity or mass--effective temperature 
relations for the main-sequence stars. Therefore, in this case one needs to rely on theoretical models providing 
a grid of luminosities, effective temperatures and synthetic photometry for different substellar masses as a 
function of age. In this work, to estimate the minimum masses and temperatures of companions that would have 
been detected, we used the Ames-COND models \citep{2001ApJ...556..357A, 2012EAS....57....3A, 2003A&A...402..701B} 
for solar metallicity. The COND models are valid up to $T_{\rm eff}$ of 1300\,K and extend down to 100\,K. 
They include the formation of dust in the atmospheres, but dust grains are considered to settle below the 
photosphere and are not included in the photospheric opacity. To compute the synthetic magnitudes for the 
Si-2 8.7\,$\mu$m band we used the {\tt\string PHOENIX} Star, Brown Dwarf \& Planet Simulator available 
online\footnote{\url{http://phoenix.ens-lyon.fr/simulator/index.faces}}. For input we used the transmission 
file of the Si-2 filter and obtained the isochrones for a set of five different ages: 0.1, 0.5, 1, 5, and 10\,Gyr.

In Fig.~\ref{massesfig} are plotted the synthetic Si-2 absolute magnitudes versus masses obtained using the 
COND models, for objects at these five ages, jointly with several isotherms in the $T_{\rm eff}$ range between 
300 and 1000\,K. The solid and two dashed horizontal lines mark the mean detectability level reached in this 
program, with the 1$\sigma$ dispersion taken as uncertainty: $M_{\rm Si2}$\,=\,13.6\,$\pm$\,0.7\,mag. For an 
age of 1\,Gyr these limits would extend to objects at the deuterium-burning mass limit and, for 10\,Gyr, to 
about 40\,$M_{\rm Jup}$. Assuming an age of 5\,Gyr as a typical age expected for stars in the solar vicinity, 
this sensitivity limit translates to an average minimum mass and temperature of companions that would have 
been detected of $m$\,=\,30\,$\pm$\,3\,$M_{\rm Jup}$ and $T_{\rm eff}$\,=\,600\,$\pm$\,40\,K. The derived 
constraints on companions masses and temperatures around each observed star, assuming a 5\,Gyr age, are listed 
in Table~\ref{detlims}. 

We converted each detection limit curve into mass limits using the COND evolutionary models and considering 
the nominal 5\,Gyr age, and counted the number of stars for which a companion at a given mass and projected
orbital separation would be detectable. The resulting contour map representing the overall depth of our search for
the observed sample over a grid of companion masses and separations is presented in Fig.~\ref{completeness_map}.

\begin{deluxetable*}{lcccccccc}
\tablenum{5}
\tablecaption{Detection limits of the CanariCam search\label{detlims}}
\tablewidth{0pt}
\tabletypesize{\footnotesize}
\tablehead{
\colhead{Star} & 
\colhead{FWHM} & 
\colhead{FWHM} & 
\colhead{Determined} & 
\colhead{Detection limit} & 
\colhead{$M_{\rm Si2}$} & 
\colhead{$M_{\rm min,comp}$} & 
\colhead{$T_{\rm eff,min,comp}$} & 
\colhead{$s$} \\
	       & 
\colhead{(pix)} & 
\colhead{('')} & 
\colhead{Si-2 (mag)} & 
\colhead{Si-2 (mag)} & 
\colhead{(mag)} & 
\colhead{($M_{\rm Jup}$)} & 
\colhead{(K)} & 
\colhead{(au)} 
          }
\startdata
GJ 699	&	  3.07 & 0.246 & 4.12\,$\pm$\,0.19 & 11.68\,$\pm$\,0.19 & 15.37\,$\pm$\,0.19 & 15.0\,$\pm$\,2.0 & 400\,$\pm$\,30 & 3--18 \\
GJ 406	&	  3.25 & 0.260 & 5.59\,$\pm$\,0.23 & 11.97\,$\pm$\,0.23 & 15.08\,$\pm$\,0.24 & 19.5\,$\pm$\,2.0 & 450\,$\pm$\,30 & 3--24 \\
GJ 411	&	  3.24 & 0.259 & 3.03\,$\pm$\,0.13 & 10.75\,$\pm$\,0.13 & 13.72\,$\pm$\,0.14 & 29.0\,$\pm$\,3.0 & 570\,$\pm$\,40 & 5--25 \\
GJ 244	&	  3.24 & 0.259 & -1.32\,$\pm$\,0.10& 10.05\,$\pm$\,0.14 & 12.95\,$\pm$\,0.15 & 40.0\,$\pm$\,3.0 & 740\,$\pm$\,50 & 18--26 \\
GJ 65AB	&	  3.38 & 0.270 & 5.69\,$\pm$\,0.19 & 11.26\,$\pm$\,0.19 & 14.13\,$\pm$\,0.21 & 25.0\,$\pm$\,2.0 & 520\,$\pm$\,30 & 3--27 \\
GJ 729	&	  3.47 & 0.278 & 4.97\,$\pm$\,0.19 & 10.73\,$\pm$\,0.19 & 13.37\,$\pm$\,0.20 & 35.0\,$\pm$\,4.0 & 650\,$\pm$\,20 & 4--30 \\
GJ 905	&	  2.99 & 0.239 & 5.49\,$\pm$\,0.19 & 11.87\,$\pm$\,0.19 & 14.37\,$\pm$\,0.19 & 23.0\,$\pm$\,2.0 & 490\,$\pm$\,25 & 3--32 \\
GJ 144	&	  3.91 & 0.313 & 1.67\,$\pm$\,0.19 & 10.70\,$\pm$\,0.19 & 13.16\,$\pm$\,0.19 & 37.0\,$\pm$\,3.0 & 680\,$\pm$\,50 & 7--32 \\
GJ 447	&	  3.96 & 0.317 & 5.01\,$\pm$\,0.22 & 10.20\,$\pm$\,0.22 & 12.57\,$\pm$\,0.23 & 46.5\,$\pm$\,5.0 & 830\,$\pm$\,40 & 3--34 \\
GJ 866(AC)B&  4.40 & 0.352 & 5.13\,$\pm$\,0.22 & 11.13\,$\pm$\,0.22 & 13.43\,$\pm$\,0.25 & 32.5\,$\pm$\,4.0 & 600\,$\pm$\,60 & 4--35 \\
GJ 820A	&	  2.95 & 0.236 & 2.37\,$\pm$\,0.08 & 11.50\,$\pm$\,0.08 & 13.79\,$\pm$\,0.08 & 28.0\,$\pm$\,1.5 & 560\,$\pm$\,20 & 7--35 \\
GJ 820B	&	  3.15 & 0.252 & 2.67\,$\pm$\,0.21 & 11.24\,$\pm$\,0.21 & 13.52\,$\pm$\,0.21 & 31.0\,$\pm$\,3.5 & 600\,$\pm$\,50 & 6--35 \\
GJ 280	&	  6.90 & 0.552 & 1.07\,$\pm$\,0.25 & 11.01\,$\pm$\,0.25 & 13.29\,$\pm$\,0.26 & 36.0\,$\pm$\,4.0 & 670\,$\pm$\,60 & 25--35 \\
GJ 725A	&	  2.99 & 0.239 & 4.21\,$\pm$\,0.14 & 11.45\,$\pm$\,0.14 & 13.72\,$\pm$\,0.15 & 29.0\,$\pm$\,2.0 & 570\,$\pm$\,25 & 4--35 \\
GJ 725B	&	  3.15 & 0.252 & 4.71\,$\pm$\,0.17 & 11.45\,$\pm$\,0.17 & 13.71\,$\pm$\,0.18 & 29.0\,$\pm$\,2.5 & 570\,$\pm$\,30 & 5--35 \\
GJ 15A	&	  2.83 & 0.226 & 3.80\,$\pm$\,0.06 & 11.88\,$\pm$\,0.06 & 14.11\,$\pm$\,0.06 & 25.0\,$\pm$\,1.0 & 520\,$\pm$\,15 & 4--36 \\
GJ 15B	&	  3.00 & 0.240 & 5.55\,$\pm$\,0.09 & 11.84\,$\pm$\,0.09 & 14.07\,$\pm$\,0.10 & 25.5\,$\pm$\,1.5 & 530\,$\pm$\,15 & 4--36 \\
GJ 1111	&	  3.79 & 0.303 & 6.70\,$\pm$\,0.05 & 11.50\,$\pm$\,0.06 & 13.70\,$\pm$\,0.08 & 29.0\,$\pm$\,3.0 & 570\,$\pm$\,45 & 3--36 \\
GJ 71	&	  3.58 & 0.286 & 1.78\,$\pm$\,0.26 & 11.18\,$\pm$\,0.26 & 13.37\,$\pm$\,0.26 & 35.0\,$\pm$\,5.0 & 650\,$\pm$\,70 & 10--37 \\
GJ 54.1	&	  3.72 & 0.298 & 5.96\,$\pm$\,0.23 & 11.05\,$\pm$\,0.23 & 13.20\,$\pm$\,0.25 & 36.5\,$\pm$\,4.0 & 680\,$\pm$\,60 & 3--37 \\
GJ 273	&	  3.23 & 0.258 & 4.48\,$\pm$\,0.11 & 10.50\,$\pm$\,0.11 & 12.63\,$\pm$\,0.12 & 47.0\,$\pm$\,2.5 & 850\,$\pm$\,40 & 4--19 \\
SO 0253+16&	  4.22 & 0.338 & 6.96\,$\pm$\,0.03 & 10.83\,$\pm$\,0.03 & 12.90\,$\pm$\,0.04 & 40.0\,$\pm$\,1.5 & 740\,$\pm$\,20 & 3--39 \\
GJ 860AB&	  3.22 & 0.258 & 4.82\,$\pm$\,0.10 & 11.08\,$\pm$\,0.10 & 13.05\,$\pm$\,0.11 & 38.5\,$\pm$\,2.0 & 710\,$\pm$\,30 & 5--40 \\
GJ 83.1	&	  4.19 & 0.335 & 6.17\,$\pm$\,0.28 & 11.26\,$\pm$\,0.28 & 13.02\,$\pm$\,0.31 & 38.5\,$\pm$\,5.0 & 710\,$\pm$\,75 & 4--44 \\
GJ 687	&	  3.30 & 0.264 & 4.31\,$\pm$\,0.13 & 11.23\,$\pm$\,0.13 & 12.95\,$\pm$\,0.14 & 40.0\,$\pm$\,3.0 & 740\,$\pm$\,50 & 5--45 \\
GJ 1245ABC&	  3.40 & 0.272 & 6.90\,$\pm$\,0.14 & 11.10\,$\pm$\,0.14 & 12.81\,$\pm$\,0.15 & 41.5\,$\pm$\,3.5 & 760\,$\pm$\,55 & 5--45 \\
GJ 876	&	  3.22 & 0.258 & 4.76\,$\pm$\,0.14 & 11.20\,$\pm$\,0.14 & 12.86\,$\pm$\,0.15 & 40.5\,$\pm$\,3.0 & 740\,$\pm$\,40 & 5--47 \\
GJ 1002	&	  3.98 & 0.318 & 6.89\,$\pm$\,0.04 & 10.88\,$\pm$\,0.04 & 12.52\,$\pm$\,0.08 & 47.0\,$\pm$\,2.5 & 850\,$\pm$\,35 & 4--47    
\enddata
\tablecomments{M$_{\rm Si2}$ -- absolute Si2 magnitude of an object corresponding to the detection limit, 
$M_{\rm min,comp}$, $T_{\rm eff,min,comp}$ -- lower limits on mass and effective temperature of detectable 
companions, $s$ -- range of projected physical separations at which detection limits and minimum 
$M_{\rm comp}$ and $T_{\rm eff}$ apply.}
\end{deluxetable*}

In Table~\ref{tab_ages}, we show the estimations of ages for our sample from the literature and those derived 
using rotation periods obtained from the literature and the gyrochronology relations from \cite{2007ApJ...669.1167B}, 
\cite{2008ApJ...687.1264M}, and \cite{2015MNRAS.450.1787A}. Most of the stars have an estimated age in the 
range of 1--10\,Gyr, which justifies the selection of the 5\,Gyr isochrone. However, some of them are below 
the 1\,Gyr age and, in these cases, our estimations of detection mass limits were conservative.

We excluded the presence of brown dwarf companions with masses $m$\,$\gtrsim$\,40\,$M_{\rm Jup}$ and 
$T_{\rm eff}$ higher than $\sim$750\,K around 24 of the observed stellar systems in the solar vicinity at 
distances within 4.6\,pc, at the range of angular separations from 1.0--3.0\,arcsec depending on the 
brightness of the target star, and up to 10\,arcsec. For an average distance of 3.5\,pc these angual 
separations correspond to projected orbital separations in the range from 3.5--10.5 to 35\,au. The 
non-detection of substellar companions throughout this search allowed us to determine an upper limit to 
the real fraction of companions at the distances and masses mentioned above. We did not assume any 
shape of underlying distribution of the population of companions in terms of their masses and semi-major 
axes, which is equivalent to a uniform, linear flat distribution. Considering that the number of objects 
with companions can be well represented by a Poisson distribution, the probability of having no companions 
is given by the formula:

\begin{equation}
P\left[k\equiv0\right]=\left(e^{-\lambda}\frac{\lambda^k}{k!}\right)_{k\equiv0}=e^{-\lambda}
\end{equation}
\noindent
where $\lambda$ is the Poisson parameter, $k$ is the number of occurrences, $\lambda$\,=\,$np$, with $n$ being 
the number of events (observed stars) and $p$ the real fraction of companions. For a certain confidence level 
($\gamma$), where $P$\,=\,1$-\gamma$, the upper limit to the real frequency of substellar companions is 
$p$\,=\,$-$ln($P$)/$n$. Considering this general sample of 24 stars, the frequency of such companions 
is below 9.6\% at a confidence level of 90\%.

Among these 24 stars, 18 stars are M dwarfs. For them, we were sensitive to substellar companions with 
$m$\,$\gtrsim$\,40\,$M_{\rm Jup}$ and $T_{\rm eff}$\,$\gtrsim$\,750\,K at 1--10\,arcsec separations, which 
give 1.8--18\,au, $\sim$5--50\,au and 3.5--35\,au projected orbital separations for the closest, furthest 
and average distance of stars in our sample. Excluding the 10\% of the nearest and the furthest observed stars, 
which set the minimum lower limit and the maximum upper limit of the projected physical separations, we 
covered from 2.5 to 45\,au in 90\% of this sample, i.e. our survey is complete from 2.5 to 45\,au with a 
confidence level of 90\%. Previous imaging programs that targeted the nearest stars \citep{2001AJ....121.2189O, 
2011ApJ...743..141C, 2012AJ....144...64D} could already detect companions with such masses. However, they 
explored wider orbital separations, beyond 10--30\,au. The CanariCam survey of nearby M dwarfs allowed us to 
study for the first time the occurrence of substellar companions with $m$\,$>$\,40\,$M_{\rm Jup}$ at orbits 
below 10\,au. With a 90\% level of confidence we set an upper limit of 12.8\% on their frequency. 
This value is consistent with constraints derived by other imaging surveys at larger orbital separations, 
indicating that brown dwarf companions around low-mass M-type stars of the solar vicinity are rare, both 
at close orbits between 2.5 and 10\,au and at wider ones. 

For 11 of the observed M dwarfs we were able to detect companions at 1--10\,arcsec separations with 
$m$\,$\gtrsim$\,30\,$M_{\rm Jup}$ and $T_{\rm eff}$\,$\gtrsim$\,600\,K, equivalent to masses and temperatures 
of the expected T/Y dwarf boundary. We were thus able to establish for the first time a constrain on 
the upper limit on the frequency of the L and T type companions around M dwarfs, at a range of projected 
orbital separations between 2--3.5 and 35\,au. With a 90\% confidence level, we found that frequency of such 
companions is less than 20\% around M stars. Such range of masses and temperatures was explored by imaging 
surveys so far only in young nearby stars, at wider orbital separations, typically beyond 30--50\,au. This 
imaging search is the first one that probes the presence of L and T companions to M stars at orbital 
separations around and below 10\,au, down to 2.0--3.5\,au.

\begin{deluxetable*}{p{0.25\textwidth} p{0.3\textwidth} p{0.16\textwidth} p{0.06\textwidth} p{0.09\textwidth} p{0.04\textwidth}}
\tablenum{6}
\tablecaption{Frequencies of substellar companions from imaging and RV surveys\label{surveys}}
\tablecolumns{6}
\tabletypesize{\scriptsize}
\tablehead{
\colhead{Survey,} \vspace{-0.2cm} &  
\colhead{Sample} & 
\colhead{Occurrence} & 
\colhead{Masses} & 
\colhead{Separations,} & 
\colhead{Conf.} \\
\colhead{reference} &  
\colhead{} & 
\colhead{rate} & 
\colhead{($M_{\rm Jup}$)} & 
\colhead{orb. periods} &
\colhead{level}
}
\startdata 
SHINE (VLT/SPHERE, \citealt{2021AA...651A..72V})         &  subset of 150 stars in young moving groups, typically $\sim$10--150\,Myr  &  12.6 (+12.9, -7.1)\% for M dwarfs &  1--75  &  5--300 au  &  \\
GPIES (Gemini South/GPI, \citealt{2019AJ....158...13N})  &  300 stars, young ($<$1 Gyr old), BA and FGK type 0.2--5.0\,$M_{\odot}$, (M dwarfs not included) &  4.7 (+4.4, -2.7)\% \hspace{0.5cm} \newline 0.8 (+0.8, -0.5)\%  &   5--80 \newline 13--80 &  10--100 au  &  95\%  \\
VLT/NaCo large program (\citealt{2017AA...603A...3V})    &  young nearby stars, mostly $<$1\,Gyr, $d$\,$\le$\,100\,pc, 199 individual stars, FGK types (not M dwarfs) a few stars at $>$1\,Gyr &  2.45\% (0.25--5.55\%)  &  5--75  &  5--500 au  &  95\%   \\
Spitzer/IRAC (\citealt{2016ApJ...824...58D})$^a$         &  73 young stars and 48 exoplanet host stars &  $<$9\%  &  0.5--13  &  100--1000 au  &  95\%  \\ 
IDPS survey (\citealt{2016AA...594A..63G})               &  292 young nearby stars, median age 120 Myr, 5, 107, 63, 24, 44, 49 B, A, F, G, K, M stars &  1.05 (+2.8, -0.7)\%  \newline for M dwarfs: $<$9.15\%  &  0.5--14 \newline 1--13 &  20--300 au \newline 10--200 au  &  95\%   \\  
\cite{2010ApJ...717..878N} analysis                      &  118 stars, majority young stars, only a few at $>$1\,Gyr age  & for FGKM stars: $<$20\% \newline for M stars: $<$20\%  &  $\ge$4  &  22--507 au  \newline 9--207 au  & 95\%   \newline 68\% \\
GEMINI/NICI campaign,\newline \cite{2013ApJ...777..160B} &  80 members of young moving groups, 23 K and 33 M stars & $<$18\% ($<$6\%)$^b$  \newline $<$21\% ($<$7\%)$^b$ &  1--20   &  10--150\,au  \newline  10--50\,au &  95.4\%  \\
SEEDS survey, \citealt{2017AJ....153..106U}              &  68 young stellar objects ($<$10\,Myr)   &  $\sim$2.9\%  &  1--70   &  50--400\,au   &   \\
VLT/NaCo {\it L}'-band imaging \cite{2016AA...596A..83L} &  58 young and nearby M dwarfs  &  4.4 (+3.2, -1.3)\%   &  $>$2   &   8--400 au   &   68\%    \\
PALMS (Keck/NIRC2, Subaru/HiCIAO, \citealt{2015ApJS..216....7B})    &   122 nearby ($<$40\,pc) young M dwarfs, 78 single M dwarfs, 90\% of stars younger than the Hyades (620\,Myr)  &  $<$6.0\% ($<$9.9\%)$^c$  \newline  4.5 (+3.1, -2.1)\%  &   5--13  \newline  13--75  &  10--100 au  \newline  10--200 au &  \\
\cite{2019AJ....158..187B} analysis, AO and seeing-limited imaging combined   &  344 members of nearby young associations, $\sim$120 M dwarfs  & 2.6 (+7, -1)\%  &  1--20  &  20--5000\,au  & 95\%  \\
\cite{2016PASP..128j2001B} analysis                      &   384 unique and single young ($\sim$5--300\,Myr) stars, stellar masses between 0.1 and 3.0\,$M_{\odot}$  &   for M stars: $<$4.2\%   &  5--13   &  10--100 au   &   95\%   \\
\cite{2014ApJ...794..159B} analysis                      &  merged samples, 248 unique stars, SpT from late B to mid M, $d$\,$\sim$\,5--130\,pc   &  0.52--4.9\%   &  5--70  &  10--100\,au  &  95\%\\
TRENDS \cite{2014ApJ...781...28M}, RV+imaging survey,    &  RVs of 111 M-dwarfs within 16\,pc, imaging follow-up of 4 targets with RV drift   &  6.5\,$\pm$\,3.0\%  &  1--13  &  $<$20 au  &  \\ 
\cite{2014ApJ...791...91C, 2016ApJ...819..125C} analysis, RV, microlensing, imaging &  synthesis of various samples of M dwarfs &  3.8 (+1.9, -2.0)\%  &  1--13   &  1--10$^5$\,days &   \\
HARPS (RV, \citealt{2013AA...549A.109B})                 &  102 nearby ($<$11\,pc) M dwarfs  & 4 (+5, -1)\%  \newline  $<$1\%  &  $\sim$0.3--3  \newline $\sim$3--30$^d$  & 10$^3$--10$^4$\,days \newline $<$10$^4$\,days & \\
AAPS survey (RV, \citealt{2016ApJ...819...28W, 2020MNRAS.492..377W})  & 203 solar type stars (FGK) &  6.7 (+2, -1)\%  &   0.3--13   &  $\sim$3--8\,au  & 68.7\% \\
CLS survey (RV, \citealt{2021ApJS..255....8R, 2021ApJS..255...14F})   &  719 nearby FGKM stars  &  14.1 (+2.0, -1.8)\%  \newline  8.9 (+3.0, -2.4)\%   & $\ge$0.1$^d$   &  2--8\,au  \newline 8--32\,au & \\
CARMENES survey (RV, \citealt{2021arXiv210703802S})      & subsample of 71 M dwarfs  &  6 (+4, -3)\%   &  $\ge$0.3$^d$   &   $<$10$^3$\,days  & \\
This work                                                & 18 nearest M dwarfs at $\delta$\,$>$\,$-25^{\circ}$ &  $<$12.8\% \newline $<$20\% &  $\ge$40 \newline $\ge$30 & 3.5--35\,au  &  90\% 
\enddata
\tablecomments{ $a$) -- re-analysis of archival IRAC data; $b$) -- determined applying DUSTY and COND models, respectively; $c$) -- assuming a hot-start (cold-start) formation scenario; $d$) -- minimum masses, $m\sin i$}
\end{deluxetable*}

\section{Discussion} \label{sec:discussion}

\subsection{Comparison to other surveys}
We attempt to compare the determined frequency limits of stars harboring substellar companions to results 
of previous works. Of the numerous programs aiming to detect giant planets and brown dwarf companions, we 
focus mainly on large, high-contrast imaging surveys including general field surveys probing wide
separations and the most recent surveys (see references in Table~\ref{surveys}) that placed constraints 
on substellar companions around M dwarfs. The essential outcomes of many of these programs can be found 
summarized in, e.g., Table~1 of \cite{2015AA...573A.127C}, \cite{2016PASP..128j2001B} and \cite{2021AA...651A..72V}. 
We also consider the results from several RV programs and studies that combine the results from various 
techniques.

Table~\ref{surveys} summarizes the frequency constraints and the explored intervals of masses and orbital 
separations or periods, determined by each of these surveys. It is not straightforward to compare the 
results of these surveys to one another and to our results, because the domains of probed parameter 
ranges are not the same and different teams make different assumptions regarding the distributions laws 
of the substellar mass companions. In general, all the studies agree that the occurrence rates of substellar 
objects are below 20\%, regardless of the companions masses/separations intervals in question and masses 
of the primaries. The majority of surveys points to a maximum frequency of 12\% or below, typically of 
a few percent ($\sim$2--5\%).

As for the low-mass stars, high-contrast imaging searches targeting specifically the M dwarfs, or 
those including them as part of the target lists, explored objects at young ages of a few tens to few 
hundred million years. The most sensitive ones were capable of detecting Jupiter-mass planets at separations
down to 10\,au, and, most recently, even to 5\,au by the SHINE survey \citep{2021AA...651A..72V}. 
On the other side, Doppler measurements restrict the frequencies of planetary companions with minimum 
masses ($m\sin i$) as small as a few tens Earth masses and are starting to explore orbital periods of
up to 10$^4$--10$^5$ days, equivalent to approximately 5--8\,au \citep{2021ApJS..255...14F, 2021arXiv210703802S}.

In this context, our results bridge the explored separation ranges between wide-orbit imaging constraints 
and those from RVs approaching the snow line. Our limits on the frequency of substellar companions 
are compatible with previous studies confirming that their presence around M dwarfs is rare. In 
comparison with recent high contrast imaging programs, our survey is more sensitive to companions of 
somewhat higher masses, but at more advanced ages of a few Gyr, and hence of significantly lower 
$T_{\rm eff}$, extending down to 600\,K. 

\subsection{Stellar binaries in the sample}
From theory it is expected that a stellar companion alters the formation processes of exoplanets within 
a protoplanetary disc around the host star (see e.g. a review by \citealt{2015pes..book..309T}). The 
binary component will also introduce a parameter space of dynamical instability in which we would not 
expect planets or brown dwarf companions to persist on long timescales \citep{1999AJ....117..621H, 
2015enas.book.1942H}.

A non-negligible part of stars in our targets sample has one or more stellar companions known, including 
five stellar binaries, two triple systems and two stars with a white dwarf companion. We measure a 
multiplicity frequency (which quantifies the number of multiple systems within our sample) and a 
companion frequency (which quantifies the total number of companions) of 36$\pm$14\% and 44$\pm$16\%, 
respectively, which are consistent within wide uncertainties with the comprehensive determinations 
by \cite{2021AA...650A.201R} in the 10 pc sample and by \cite{2021MNRAS.tmp.2160B} for the M dwarfs.
For the statistical analysis, we considered the relatively close systems ($s$\,$\lesssim$\,50\,au;
Sirius, GJ 65, GJ 866, Procyon, GJ 860 and GJ 1245) as individual systems and those having components at 
wider average orbital separations ($s$\,$\gtrsim$\,50\,au; GJ 820AB, GJ 725AB and GJ 15AB) as two 
individual stars. 

We add a note of caution that such aproach introduces a potential physical bias in the interpretation 
of our analysis. Our search probes both circumstellar and circumbinary companions depending 
on the separation of the binaries, which are very different science cases to one another and to a search 
around single stars only. We recored a null companion detection in these systems. However, any secondary 
component introduces a region of instability at certain range of physical separations for both S- 
and P-type orbits \citep{1999AJ....117..621H, 2007AA...462..345D}, in which we would not expect an 
additional body to be found. Thus, an element of bias is inherent in the statistical result derived 
from a combined single and binary star sample.

Several works have concluded that binarity has a minimal effect on overall planet frequency 
\citep{2007AA...468..721B, 2013MNRAS.428..182B, 2015ApJ...814..148P, 2020AA...635A..74S}. 
On the other hand, some recent observational studies demonstrate an excess of wide stellar companions
to stars which host high-mass hot Jupiters and brown dwarf companions on short-period orbits
\citep{2016ApJ...827....8N, 2019MNRAS.485.4967F, 2019arXiv191201699M, 2021FrASS...8...16F}.
These conclude that certain types of binaries may support the proccess of formation of close-in 
high-mass planetary and substellar companions. Although early research on this front have been carried, 
e.g. by the SPOTS \citep{2016AA...593A..38B, 2018AA...619A..43A} and the VIBES \citep{2020AA...643A..98H} 
surveys, a more detailed analysis of the effect multiplicity has on the occurrence of substellar 
companions in general requires a larger, dedicated sample that is complete to both single stars 
and binaries.

\subsection{Known planets hosts}
As many as 12 of the 33 stars in the sample have at least one small planet of a few to a few tens Earth masses 
on a close orbit at around 1\,au or less. All of these cases, except the G8.5V-type $\tau$~Cet, are M dwarf 
stars. In contrast, only two of these stars were found to host more massive planets -- K2.0V $\epsilon$ Eri 
(b: $m\sin i$\,=\,0.78\,$M_{\rm Jup}$ at 3.48$\pm$0.02\,au) and M3.5V GJ 876 (b: $m\sin i$\,=\,2.27\,$M_{\rm Jup}$ 
at $a$\,=\,0.21\,au; c: $m\sin i$\,=\,0.71\,$M_{\rm Jup}$ at $a$\,=\,0.13\,au). This goes in line with results 
of surveys that noticed initial indications that frequencies of more massive giant planets scale positively 
with the host star mass \cite[e.g.,][]{2010PASP..122..905J, 2016PASP..128j2001B}.

There are no theoretical premises indicating that such close-in planets will preclude the formation of more 
massive companions in wider orbits. Instead, observational studies showed that hot Jupiters host stars tend to 
have far-away companions \cite[e.g.,][]{2014ApJ...785..126K, 2014AA...569A.120L, 2015ApJ...806..248W}. 
The dynamical interactions between multiple planets or a distant brown dwarf companion may affect the 
final orbital configuration of the system. A variety of mechanisms, such as planet-planet scattering, the 
Kozai-Lidov effect or secular gravitational interactions \citep{2015MNRAS.448.1044H, 2016ARAA..54..441N},
have been invoked for smaller planets to explain inwards migration to short orbital periods 
\citep{2003ApJ...588..494M}.

\section{Conclusions and final remarks} \label{sec:conclusions}
We completed a deep, high spatial resolution imaging search of substellar companions around the nearest 
northern stars in the mid-IR at 8.7\,$\mu$m using the CanariCam instrument at the 10.4\,m GTC telescope. 
Our target sample included 25 stellar systems composed of 33 individual stars, with declinations 
$\delta$\,$>$\,$-$25$^{\circ}$ within 5\,pc of the Sun. No previously undetected companions were identified 
in our survey. 

We explored the angular separations between 1--3 and 10\,arcsec with sensitivities sufficient to detect 
companions with masses and temperatures higher than 40\,$M_{\rm Jup}$ and 750\,K for 24 of the 
observed stars, and as low as 30\,$M_{\rm Jup}$ and 600\,K for 11 M-type stars. Considering an 
average distance of 3.5\,pc of our sample, 3.5--35\,au projected orbital separations were probed for 
faint companions. The non-detections enabled us to determine upper limits for the fraction of substellar 
companions. At a 90\% confidence level, we found that less than 9.6\% of the nearby stars have 
companions with $m$\,$\gtrsim$\,40\,$M_{\rm Jup}$ and $T_{\rm eff}$\,$\gtrsim$\,750\,K, and that less 
than 20\% of the closest M dwarfs have L and T companions with $m$\,$>$\,30\,$M_{\rm Jup}$ and 
$T_{\rm eff}$\,$\gtrsim$\,600\,K within the range of explored physical separations. This is one the 
first imaging programs capable to detect mature substellar companions in this range of masses and 
temperatures below 10\,au separations and provides evidence that substellar companions to low-mass 
M dwarfs are rare also at such closer orbits. Concurrently, extending the constraints beyond 5\,au, 
our results are complementary to the evidences brought by RV programmes \cite[e.g.,][]{2013AA...549A.109B, 
2015ARAA..53..409W, 2019Sci...365.1441M, 2021arXiv210703802S}, which find occurrence of giant planets 
around M stars to be less than 3\% at orbital periods up to 25--30\,yr ($\sim$5\,au).

This work demonstrates that the modern ground-based mid-IR imaging instruments operating on 10\,m class 
telescopes can reach angular resolutions and sensitivity limits as good as and, in certain cases (e.g. 
nearby, relatively old stars), better than adaptive optics systems in the optical or near-IR or space 
telescopes. This technique presents a high potential for direct imaging detection and studies of brown 
dwarfs and exoplanets. Our survey was not extensive enough to determine more precisely the true fraction 
of L and T brown dwarf companions at close orbits. Nonetheless results on the observed sample provide 
valuable constraints for next generation facilities, such as the James Webb Space Telescope or the 
Extremely Large Telescope, which will allow for detection and accurate characterization of the coldest 
companions of stars \citep{2015IJAsB..14..279Q, 2018AJ....156..276D}.

\begin{acknowledgments}
We thank the anonymous referee for a careful review of our manuscript and his/her constructive comments 
which substantially helped improving the quality of the paper. We are grateful to the GTC staff for 
performing the CanariCam observations. Based on observations made with the Gran Telescopio Canarias (GTC), 
installed in the Spanish Observatorio del Roque de los Muchachos of the Instituto de Astrof\'isica de 
Canarias, in the island of La Palma. 
%
%
BG and DJP acknowledge support from the UK Science and Technology Facilities Council (STFC) via the  
Consolidated Grant ST/R000905/1. VJSB, MRZO and JAC acknowledge financial support from the Agencia 
Estatal de Investigaci\'on of the Ministerio de Ciencia, Innovaci\'on y Universidades and the 
European Regional Development Fund through projects PID2019-109522GB-C5[1,3]
%
%
We acknowledge the use of Carmencita, the CARMENES input catalogue (\citealt{2016csss.confE.148C}).
%
%
%
%
%
%
This research has made use of the SIMBAD database, operated at CDS, Strasbourg, France.
\end{acknowledgments}

\newpage
\appendix
\restartappendixnumbering

\section{Known planetary/substellar companions of the sample stars}

\startlongtable
\begin{deluxetable*}{l l p{0.7\textwidth}}
\tablewidth{0pt}
\tabletypesize{\footnotesize}
\tablecolumns{3}
\tablecaption{Notes on the currently known exoplanets and substellar companions of the sample stars \label{known_exoplanets}}
\tablehead{
\colhead{Star} & 
\colhead{} &
\colhead{Notes}
}
\startdata
GJ 699  	& Barnard's Star   &   \cite{2018Natur.563..365R}: RV super-Earth planet near the snow line; from CARMENES, HARPS, HARPS-N 20\,yr monitoring;
                                   $m\sin i$\,=\,3.23$\pm$0.44\,$M_{\earth}$, $a$\,=\,0.404$\pm$0.018\,au, $P_{\rm orb}$\,=\,232.8\,d, $\rho$\,=\,221$\pm$10\,mas;
                                   evidence for a second, longer period signal at $\sim$6600\,d, the presence of an outer planet cannot be ruled out. 
                                   The fit would suggest an object of $\gtrsim$15\,$M_{\earth}$ on a $\sim$4\,au orbit.\\  
GJ 406  	& CN Leo  	       &   Monitored with HARPS \citep{2013AA...549A.109B}, HIRES \citep{2017AJ....153..208B} and CARMENES \citep{2015AA...577A.128A}. 
                                   Relatively active flare star \citep{2007AA...466L..13R}. No planets reported yet. \\
GJ 411  	& Lalande 21185    &   One RV planet, confirmed status. Discovered by \cite{2019AA...625A..17D}, confirmed by \cite{2020AA...643A.112S}
                                   using HIRES, SOPHIE and CARMENES data; $m\sin i$\,=\,2.69$\pm$0.25\,$M_{\earth}$, 0.0789$\pm$0.0007\,au, $P_{\rm orb}$\,=\,12.946$\pm$0.005\,d\\
GJ 244  	& Sirius           &   HST astrometry rules out companions larger than a small brown dwarf or large exoplanet \citep{2017ApJ...840...70B}: 
                                   $m$\,$>$\,0.033\,$M_{\odot}$ orbiting in 0.5\,yr, and $m$\,$>$\,0.014\,$M_{\odot}$ in 2\,yr. No planets reported yet.\\
GJ 65 AB	& BL\,Cet\,+\,UV\,Cet& M5.5V\,+\,M6V binary at $a$\,=\,2.1--8.8\,au, $P_{\rm orb}$\,=\,26.52\,yr, $e$\,=\,0.62;
                                   No planets reported yet.\\
GJ 729  	& V1216 Sgr        &   Monitored with HARPS \citep{2013AA...549A.109B}; on the CARMENES GTO target list \citep{2018AA...612A..49R}.
                                   No planets reported yet. \\
GJ 905  	& HH And           &   SPIRou Input Catalog \citep{2018MNRAS.475.1960F} - a slow rotator star $P_{\rm rot}$\,=\,99.58\,d, rejected as a spectroscopic binary candidate.
                                   On the CARMENES GTO target list \citep{2018AA...612A..49R}. No planets reported yet. \\
GJ 144  	& $\epsilon$ Eridani & One confirmed planet ($\epsilon$ Eri b) + one unconfirmed candidate planet ($\epsilon$ Eri c) \newline
                                   $\epsilon$ Eri b: 0.78\,$M_{\rm Jup}$, 3.48$\pm$0.02\,au, $P_{\rm orb}$\,7.37$\pm$0.07\,yr \citep{2019AJ....157...33M}, 
                                   using Ms band (4.7\,$\mu$m) Keck/NIRC2 imaging and RV data (HIRES, HARPS, others) over 30\,yr. \newline
                                   Direct imaging constraints, for 200, 400 and 800\,Myr age:\newline
                                   3.0, 4.5, 6.5\,$M_{\rm Jup}$ at 1\,au \newline
                                   1.5, 1.7, 2.5\,$M_{\rm Jup}$ at 2\,au \newline
                                   0.8, 1.7, 5.0\,$M_{\rm Jup}$ at 3\,au \newline
                                   The putative planet ``c'' should be orbiting at around 40\,au, to shape the dust disk.
                                   Limits from Spitzer imaging \citep{2015AA...574A.120J}:
                                   $m$\,=\,0.5--2.5\,$M_{\rm Jup}$ at 20--140\,au separations (at\,800 Myr)\\
GJ 447  	& FI Vir           &   An exo-Earth planet discovered in HARPS RVs \citep{2018AA...613A..25B}, 
                                   $m\sin i$\,=\,1.35$\pm$0.2\,$M_{\earth}$, $a$\,=\,0.049$\pm$0.002\,au, $P_{\rm orb}$\,=\,9.86$\pm$0.01\,d.
                                   CARMENES GTO target star \citep{2018AA...612A..49R}.\\
GJ 866 ABC  & EZ Aqr           &   A compact triple system of M dwarfs \citep{1999AA...350L..39D, 2000AA...353..253W}\newline
                                   AC - spectroscopic binary 0.012--0.016\,au, $\rho\sim$0.01'', $P_{\rm orb}\sim$3.8\,d \newline
                                   AC-B - 0.41--0.77\,au, $\rho\sim$0.36'', $P_{\rm orb}$\,=\,823\,d \newline
                                   Well determined dynamical masses: A 0.1216, B 0.1161, C 0.0957\,$M_{\odot}$.
                                   No planets reported yet.\\
GJ 820 A  	& 61 Cyg A 	       &   Limits from RVs: 0.09--0.98\,$M_{\rm Jup}$ between 0.05--5.2\,au, by \cite{2006AJ....132..177W}.
                                   No planets reported yet.\\
GJ 820 B  	& 61 Cyg B 	       &   Possibly a third body orbiting the B component, from proper motion anomalies in Gaia DR2 \citep{2019AA...623A..72K},
                                   hipothetically a companion of $m$\,=\,2.95\,[$M_{\rm Jup}$ au$^{(-1/2)}$] normalized to $r$\,=\,1\,au orbit explains the anomaly.\newline
                                   Limits from RVs: 0.07--0.8\,$M_{\rm Jup}$ between 0.05--5.2\,au, from \cite{2006AJ....132..177W}. No  planets reported yet.\\
GJ 280  	& Procyon  	       &   White dwarf (DQZ) companion Procyon B, $P_{\rm orb}$\,=\,40.8\,yr, $e$\,=\,0.4, average $a$\,=\,15\,au (8.9--21.0\,au) \cite{2015ApJ...813..106B}.
                                   HST astrometry excludes $m$\,$\gtrsim$\,10\,$M_{\rm Jup}$ companions at $>$1.5\,yr orbital periods \citep{2015ApJ...813..106B}. No planets reported yet.\\
GJ 725 A 	& HD 173739	       &   Monitored using HARPS-N \citep{2016MNRAS.459.3551B}. On the CARMENES GTO target list \citep{2018AA...612A..49R}. 
                                   No planets reported yet.\\
GJ 725 B 	& HD 173740  	   &   A candidate planet proposed by \cite{2016MNRAS.459.3551B} from HARPS-N RV: 
                                   $m\sin i$\,=\,1.2\,$M_{\earth}$, $P_{\rm orb}$\,=\,2.7$\pm$0.3\,d, $a$=0.025\,au (S-type orbit).
                                   CARMENES GTO target star \citep{2018AA...612A..49R}.\\
GJ 15 A  	& GX And  	       &   Two planets detected: \newline
                                   b:~$m\sin i$\,=\,3.03\,$\pm$\,0.4\,$M_{\earth}$, 0.07\,au, $P_{\rm orb}$\,=\,11.4\,d, (HIRES, \citealt{2014ApJ...794...51H})\newline
                                   c:~$m\sin i$\,=\,36 (+25,-18)\,$M_{\earth}$, 5.4\,au, $P_{\rm orb}\,\sim$\,7600\,d, longest period Neptune-mass exoplanet, (CARMENES, \citealt{2018AA...609A.117T})
                                   both planets confirmed with more data: HIRES, CARMENES, HARPS-N, \cite{2018AA...617A.104P}\\
GJ 15 B  	& GQ And  	       &   On the CARMENES GTO target list \citep{2018AA...612A..49R}. No planets reported yet.\\
GJ 1111  	& DX Cnc  	       &   On the CARMENES target list.\newline 
                                   No planets reported yet.\\
GJ 71  		& $\tau$ Ceti  	   &   A 4-planet system found using $>$9000 HARPS and HIRES RV measurements from 2003 to 2013 \citep{2013AA...551A..79T, 2017AJ....154..135F}\newline
                                   g: $m\sin i$\,=\,1.75 [+0.25, -0.40]\,$M_{\earth}$, $a$\,=\,0.13\,au, $P_{\rm orb}$\,=\,20.0\,d \newline
                                   h: $m\sin i$\,=\,1.83 [+0.68, -0.26]\,$M_{\earth}$, $a$\,=\,0.24\,au, $P_{\rm orb}$\,=\,49.4\,d \newline
                                   e: $m\sin i$\,=\,3.93 [+0.86, -0.64]\,$M_{\earth}$, $a$\,=\,0.54\,au, $P_{\rm orb}$\,=\,162.9\,d \newline
                                   f: $m\sin i$\,=\,3.93 [+1.05, -1.37]\,$M_{\earth}$, $a$\,=\,1.33\,au, $P_{\rm orb}$\,=\,636.1\,d \newline
                                   Planets originally denoted b, c, d were not confirmed. All 4 planets tightly packed at orbits $<$1.5\,au. 
                                   Tangential velocity anomaly indicates a possible presence of a Jupiter-analogue planet, by at 
                                   most 5\,$M_{\rm Jup}$ if orbiting between 3 and 20\,au \citep{2019AA...623A..72K}.\\
GJ 54.1  	& YZ Cet  	       &   A 3-planet system in compact orbits confirmed (HARPS RVs, \citealt{2017AA...605L..11A}). Small, Earth-mass planets: 
                                   $P_{\rm orb}$\,=\,1.97, 3.06, 4.66\,d; $m\sin i$\,=\,0.75\,$\pm$\,0.13, 0.98\,$\pm$\,0.14, and 1.14\,$\pm$\,0.17\,$M_{\earth}$.
                                   Fourth planet candidate discarded by \cite{2020AA...636A.119S}.\\
GJ 273  	& Luyten's Star	   &   Two planets detected using $\sim$13\,yr HARPS RV monitoring \citep{2017AA...602A..88A}\newline
                                   c: $m\sin i$\,=\,1.18\,$\pm$\,0.16\,$M_{\rm \earth}$, $a$\,=\,0.0365\,au, $P_{\rm orb}$\,=\,4.7234\,$\pm$\,0.0004\,d\newline
                                   b: $m\sin i$\,=\,2.89\,$\pm$\,0.26\,$M_{\rm \earth}$, $a$\,=\,0.09110\,$\pm$\,0.00002\,au, $P_{\rm orb}$\,=\,18.65\,$\pm$\,0.01\,d\newline
                                   GJ 273b is a super-Earth within the habitable zone \citep{2016ApJ...819...84K}. \newline
                                   On the CARMENES GTO target list \citep{2018AA...612A..49R}.\\
SO 0253+1652& Teegarden's Star &   Two Earth-mass planets discovered with CARMENES \citep{2019AA...627A..49Z}\newline
                                   b: $m\sin i$\,=\,1.05\,$M_{\rm \earth}$, $a$\,=\,0.025\,au, $P_{\rm orb}$\,=\,4.91\,d\newline
                                   c: $m\sin i$\,=\,1.11\,$M_{\rm \earth}$, $a$\,=\,0.044\,au, $P_{\rm orb}$\,=\,11.41\,d\\
GJ 860 AB  	& Kruger 60 AB     &   M3V\,+\,M4V binary ($a$\,=\,5.5--13.5\,au, $P_{\rm orb}$\,=\,44.6\,yr, $e$\,=\,0.41, nearly face-on orbit $i$\,=\,167.2\,deg, \citealt{2018JDSO...14....3K}).
                                   Astrometric observations rule out companions $\geq$\,0.45 and $\geq$\,0.37\,$M_{\rm Jup}$ at $\geq$\,1\,au 
                                   orbits around A and B components, respectively \citep{2009MNRAS.400..406H}. RVs monitored with HIRES \citep{2017AJ....153..208B};
                                   No planets reported yet.\\
GJ 83.1  	& TZ Ari  	       &   Two RV planets in 3:1 mean motion resonance, \cite{2020ApJS..250...29F} using HARPS and HIRES data\newline
                                   b: $m\sin i$\,=\,12.8--42.8\,$M_{\rm \earth}$, $a$\,=\,0.40\,au, $P_{\rm orb}$\,=\,243.1\,d \newline
                                   c: $m\sin i$\,=\,96.0--174\,$M_{\rm \earth}$,  $a$\,=\,0.87\,au, $P_{\rm orb}$\,=\,773.4\,d \newline
                                   On the CARMENES GTO target list \citep{2018AA...612A..49R}.\\
GJ 687  	& LHS 450  	       &   Two RV planets reported \citep{2014ApJ...789..114B, 2020ApJS..250...29F}\newline
                                   b: $m\sin i$\,=\,18.4\,$M_{\rm \earth}$, $a$\,=\,0.16\,au, $P_{\rm orb}$\,=\,38.1\,d \newline
                                   c: $m\sin i$\,=\,16.0\,$M_{\rm \earth}$, $a$\,=\,1.16\,au, $P_{\rm orb}$\,=\,692--796\,d, $e$=0.4 \newline
                                   On the CARMENES GTO target list \citep{2018AA...612A..49R}.\\
GJ 1245 ABC & LHS 3494 	       &   Triple system composed of M dwarfs: M5\,+\,M5\,+\,M8 (AC-B: 33\,au, A-C: 8\,au; \citealt{1988ApJ...333..943M, 2014ApJ...797..121H}),
                                   Kepler light curves of A and B analyzed, $P_{\rm rot}$\,=\,0.26 and 0.71\,d of A and B, resp. \citep{2015ApJ...800...95L}
                                   HIRES and HPF RV planet search \citep{2020ApJ...897..125R}. 
                                   No planets reported yet.\\
GJ 876  	& IL Aqr  	       &   A 4-planet system currently known (RVs), first two planets discovered in 1998 and 2001 \citep{1998ApJ...505L.147M, 2001ApJ...556..296M}\newline
                                   d: $m\sin i$\,=\,6.83\,$M_{\rm \earth}$, $a$\,=\,0.021\,au, $P_{\rm orb}$\,=\,1.94\,d \newline
                                   c: $m\sin i$\,=\,0.714\,$M_{\rm Jup}$,  $a$\,=\,0.129\,au, $P_{\rm orb}$\,=\,30.01\,d \newline
                                   b: $m\sin i$\,=\,2.275\,$M_{\rm Jup}$,  $a$\,=\,0.208\,au, $P_{\rm orb}$\,=\,61.12\,d \newline
                                   e: $m\sin i$\,=\,14.6\,$M_{\rm \earth}$, $a$\,=\,0.334\,au, $P_{\rm orb}$\,=\,124.26\,d \newline
                                   Mean motion resonance 1:2:4 of the outermost planets. A triple conjunction between the outer three planets 
                                   once per every orbit of the outer planet,``e'' \citep{2010ApJ...719..890R, 2018AJ....155..106M}. 
                                   GJ 876 is a candidate parent system for the 1I/'Oumuamua object \citep{2018AA...610L..11D}.\\
GJ 1002  	& LHS 2  	       &   Surveyed by HARPS and CARMENES \citep{2013AA...549A.109B, 2018AA...612A..49R}. \newline
                                   No planets reported yet.
\enddata
\end{deluxetable*}

\begin{deluxetable*}{lccccc}[h!]
\tablewidth{0pt}
\tabletypesize{\footnotesize}
\tablecolumns{6}
\tablecaption{Ages of the stars in the sample \label{tab_ages}}
\tablehead{
\colhead{Star} & 
\colhead{Age lit.} &
\colhead{Ref.} &
\colhead{$P_{\rm rot}$} & 
\colhead{Ref.} &
\colhead{Age} \\
\colhead{} &
\colhead{(Gyr)} &
\colhead{} &
\colhead{(d)} &
\colhead{} &
\colhead{(Gyr)}
}
\startdata
GJ 699       & 7--10            & Rib18  & 148.6\,$\pm$\,0.1     & SM15   & $>$10     \\
GJ 406       & 0.1--1           & Pav06  & 2.704\,$\pm$\,0.003   & DA19   & 0.01-0.08 \\
GJ 411       & 5--10            & ...    & 48.0                  & KS07   & 2--10     \\
GJ 244       & 0.228$\pm$0.010  & Bond17 & ...                   & ...    & ...       \\
GJ 65 AB     & 5.0              & MacD18 & 0.243\,$\pm$\,0.0005  & Bar17  & ...       \\
GJ 729$^a$   & $<$1.0           & War08  & 2.8502                & CC21   & 0.01-0.08 \\
GJ 905       & ...              & ...    & 106.0\,$\pm$\,6.0     & DA19   & $>$10     \\
GJ 144       & 0.4--0.8         & Jan15   & 11.1\,$\pm$\,0.03     & GB95   & 0.3--3.2  \\
GJ 447       & 9.45$\pm$0.60    & Man15  & 163\,$\pm$\,3         & DA19   & $>$10     \\
GJ 866 (AC)  & ...              & ...    & ...                   & ...    & ...       \\
GJ 820 A     & 6.1\,$\pm$\,1    & Ker08  & 35.37\,$\pm$\,10      & Don96   & 1--10     \\
GJ 820 B     & 6.1\,$\pm$\,1    & Ker08  & 37.84\,$\pm$\,10      & Don96   & 1--10     \\
GJ 280       & 1.87\,$\pm$\,0.13& Lie13  & 23\,$\pm$\,2          & Ayr91  & ...       \\
GJ 725 A     & 3.0              & Man15  & ...                   & ...    & ...       \\
GJ 725 B     & 2.4              & Man15  & ...                   & ...    & ...       \\
GJ 15A       & 3.02             & Man15  & 45\,$\pm$\,4.4        & SM18   & 1--10     \\
GJ 15B       & 2.754            & Man15  & ...                   & ...    & ...       \\
GJ 1111$^a$  & 0.2--0.4         & Les06  & 0.459\,$\pm$\,0.00001 & DA19   & ...       \\
GJ 71        & 5.8              & Mam08  & 34                    & Bal96  & 3.5--10   \\
GJ 54.1      & 5.0              & ...    & 69.2\,$\pm$\,2.4      & DA19   & $>$10     \\
GJ 273       & $>$8.0           & Poz20  & 93.5\,$\pm$\,16       & SM17   & $>$10     \\
SO 0253+1652 & $>$8.0           & Zech19 & $\sim$100             & Zech19 & $>$10     \\
GJ 860 AB    & ...              & ...    & ...                   & ...    & ...       \\
GJ 83.1      & $\sim$5          & Yee17  & ...                   & ...    & ...       \\
GJ 687       & ...              & ...    & $\sim$60              & Bur14  & 1.5--10   \\
GJ 1245 AC   & ...              & ...    & 0.2632\,$\pm$\,0.0001 & Haw14  & ...       \\
GJ 1245 B    & $\sim$\,0.3      & ...    & 0.709\,$\pm$\,0.001   & Haw14  & ...       \\
GJ 876$^b$   & 0.1--5           & Cor10  & 81.0\,$\pm$\,0.8      & DA19   & $>$10     \\
GJ 1002      & ...              & ...    & ...                   & ...    & ...       
\enddata
\tablecomments{
$^a$ -- Castor YMG member, $^b$ -- slow rotator. 
{\sc References:}
Rib18: \citet{2018Natur.563..365R},
SM15: \citet{2015MNRAS.452.2745S},
DA19: \citet{2019AA...621A.126D},
Pav06: \citet{2006AA...447..709P},
KS07: \citet{2007AcA....57..149K},
Bond17: \citet{2017ApJ...840...70B},
MacD18: \citet{2018ApJ...860...15M},
Bar17: \citet{2017MNRAS.471..811B},
War18 \citet{2008ApJ...676..610W},
CC21: Cortes-Contreras et al. 2021 in prep,
Jan15: \citet{2015AA...574A.120J},
GB15: \citet{1995ApJ...441..436G},
Man15: \citet{2015ApJ...804...64M},
Ker08: \citet{2008AA...488..667K},
Don96: \citet{1996ApJ...466..384D},
Lie13: \citet{2005ApJ...630L..69L},
Ayr91: \citet{1991ApJ...375..704A},
SM18: \citet{2018AA...612A..89S},
Les06: \citet{2006AA...460..733L},
Mam08: \citet{2008ApJ...687.1264M},
Bal96: \citet{1995ApJ...441..436G},
Poz20: \citet{2020AA...641A..23P},
SM17: \citet{2017MNRAS.468.4772S},
Zech19: \citet{2019AA...627A..49Z},
Yee17: \citet{2017ApJ...836...77Y},
Bur14: \citet{2014ApJ...789..114B},
Cor10: \citet{2010AA...511A..21C}.
}
\end{deluxetable*}

\section{A full set of the final CanariCam images}
\begin{figure*}[h!]
\gridline{
          \fig{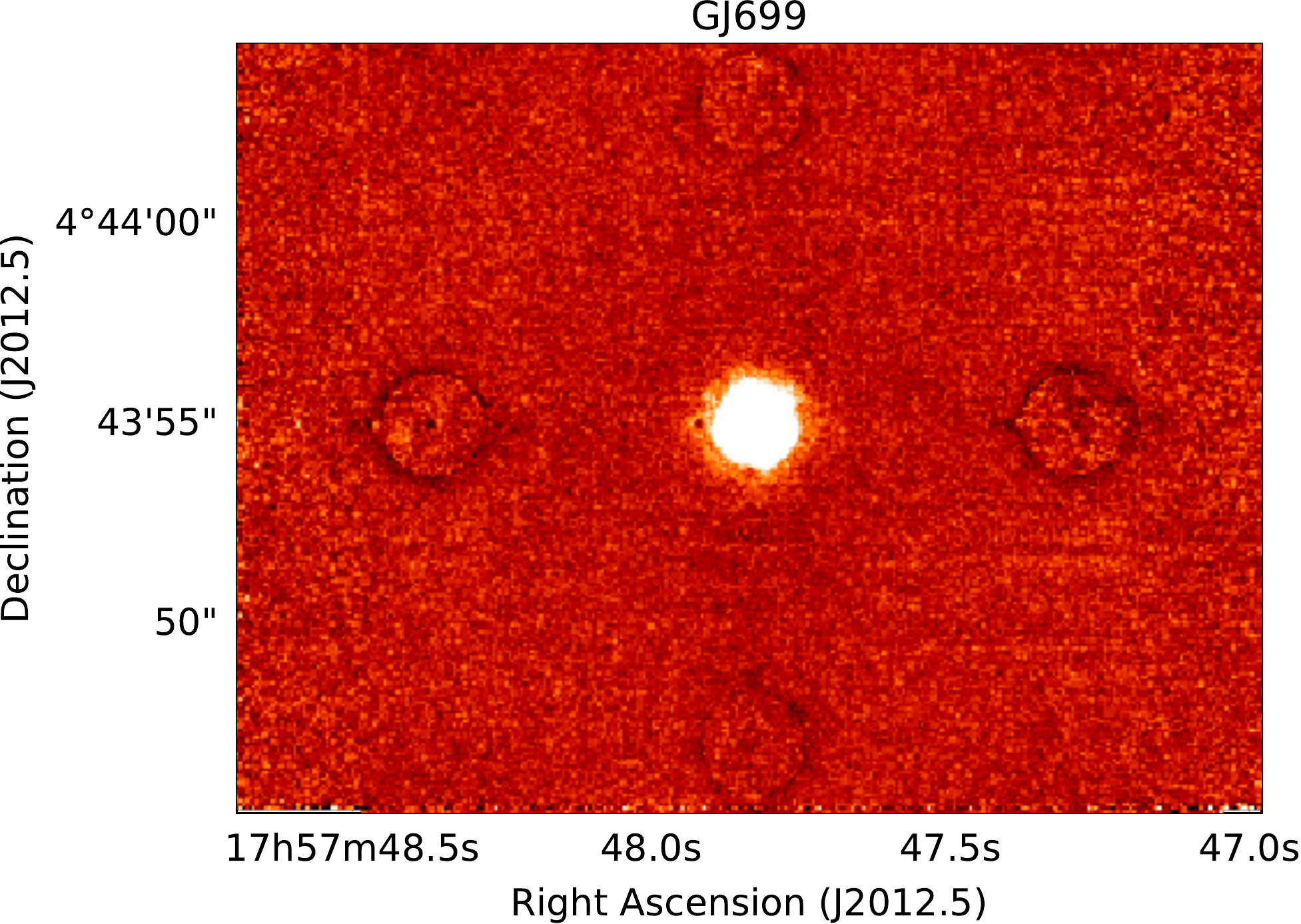}{0.31\textwidth}{}
          \fig{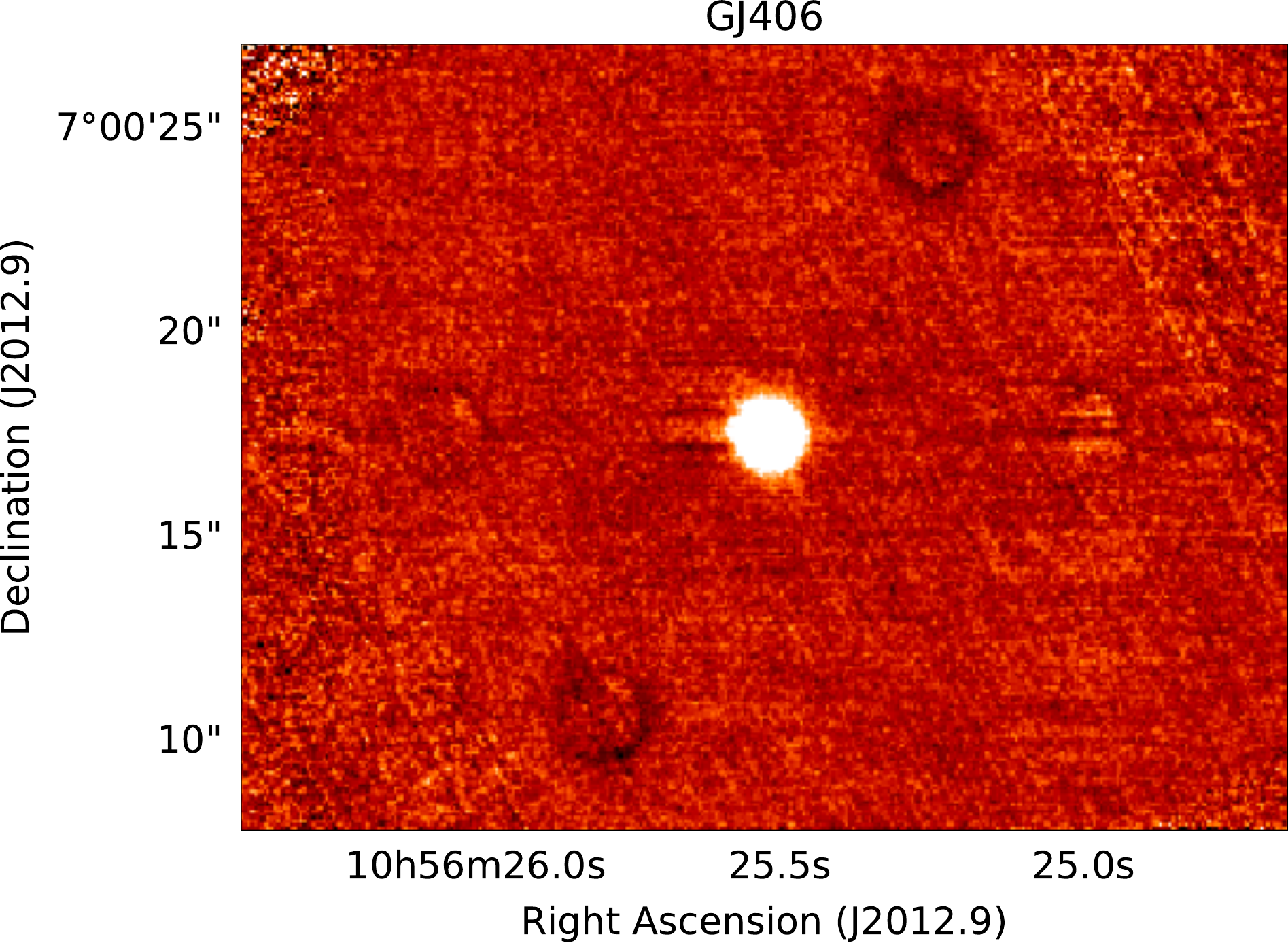}{0.31\textwidth}{}
          \fig{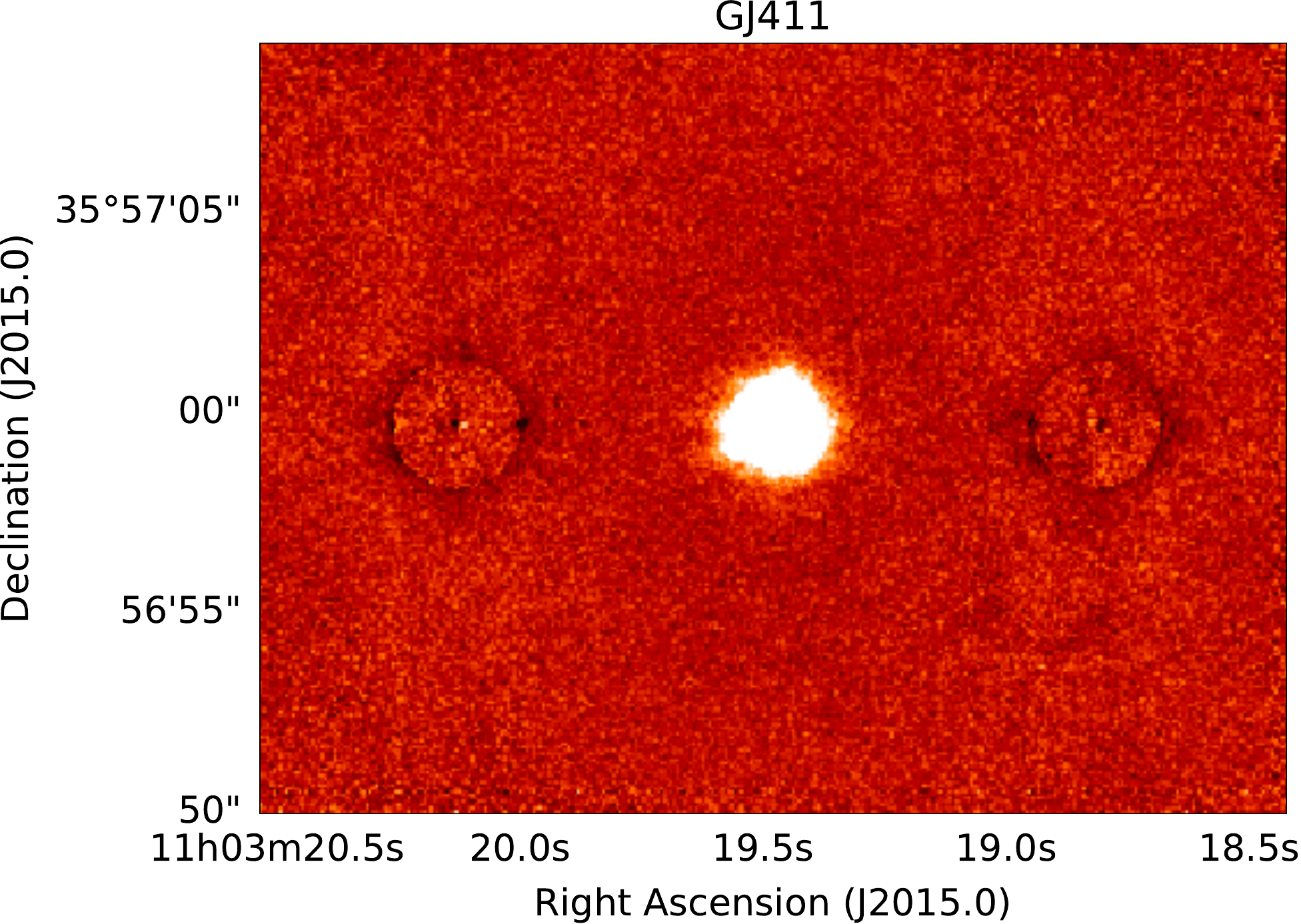}{0.31\textwidth}{}
          }
\vspace*{-0.3cm}
\gridline{\fig{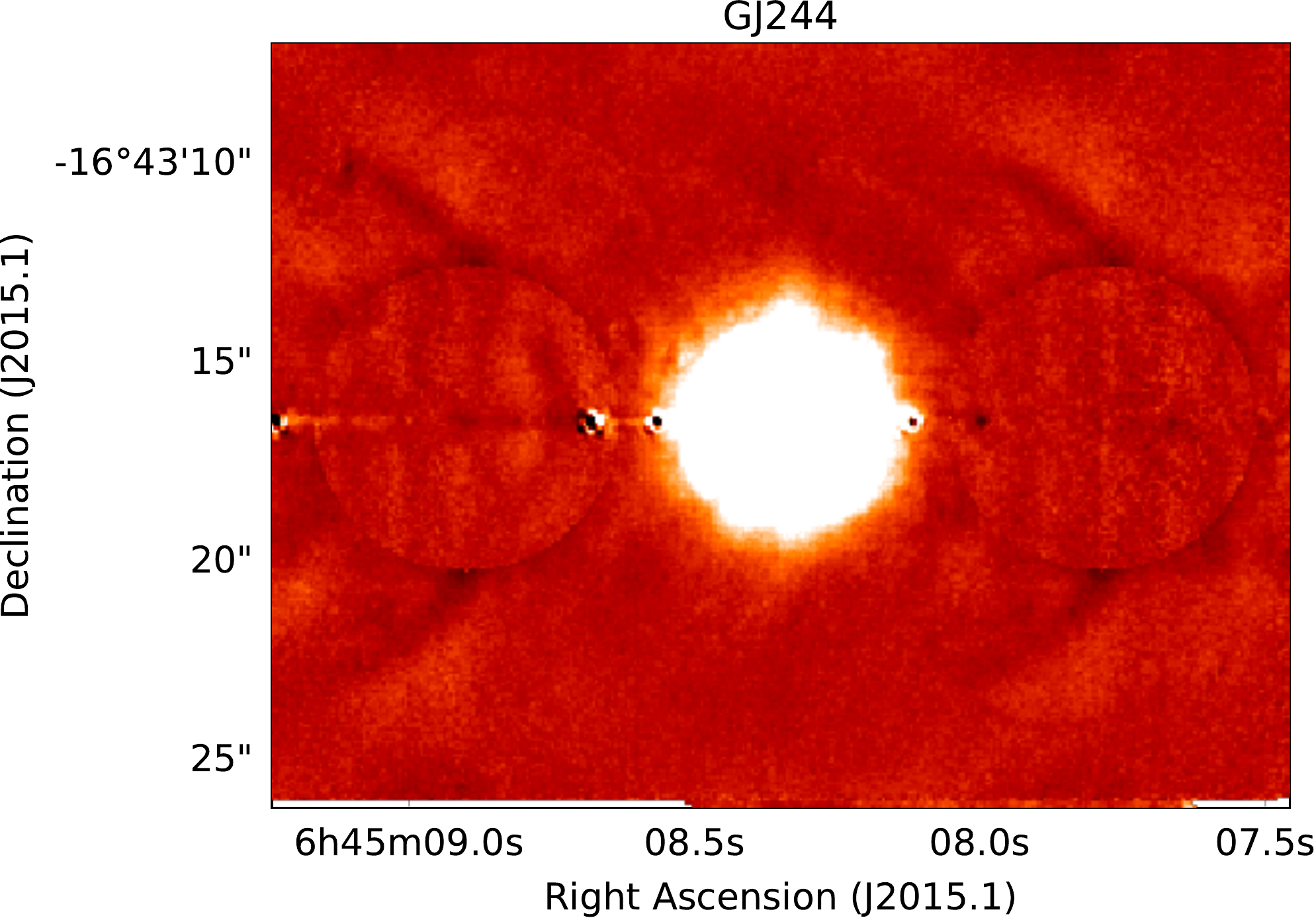}{0.31\textwidth}{}
          \fig{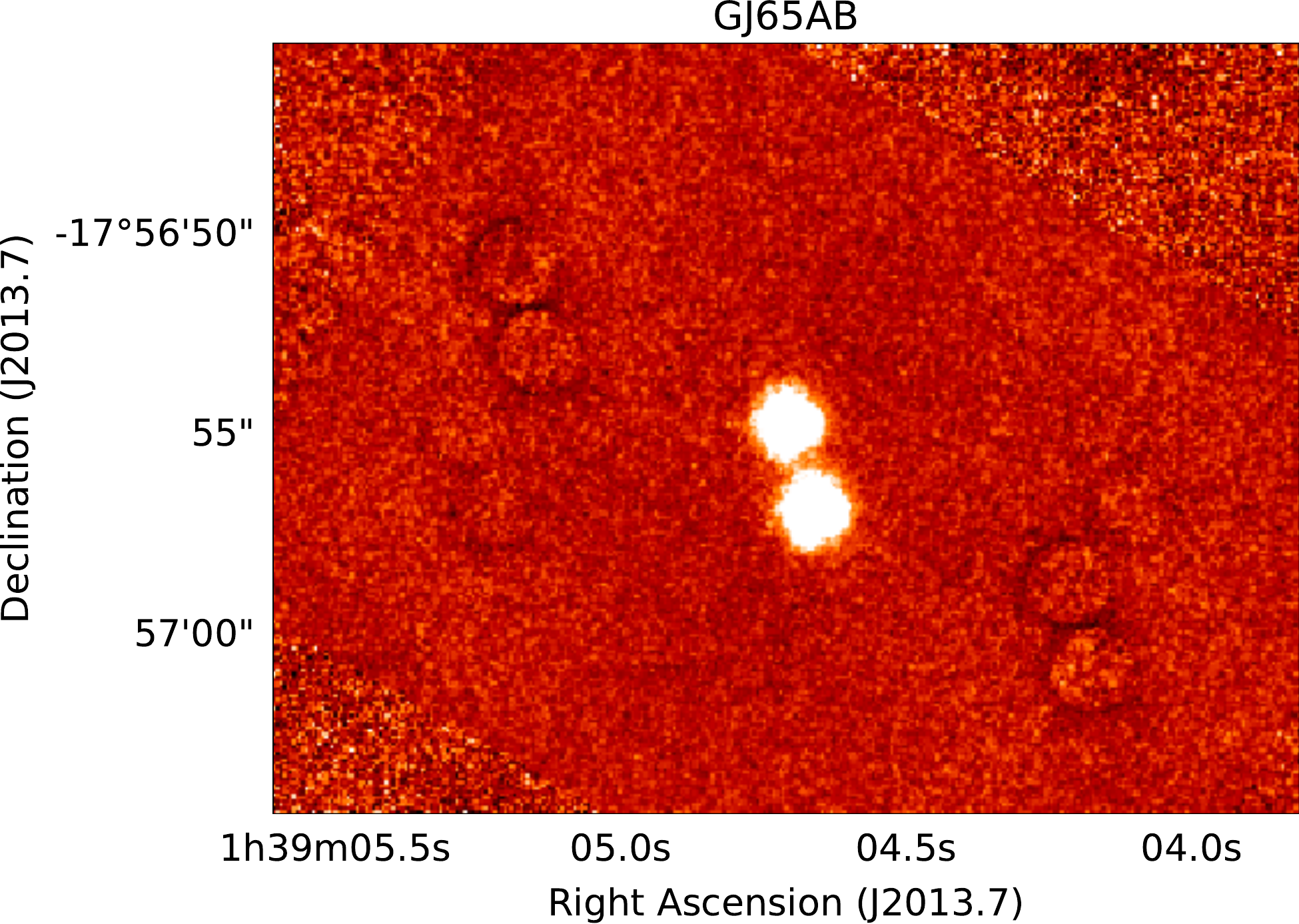}{0.31\textwidth}{}
          \fig{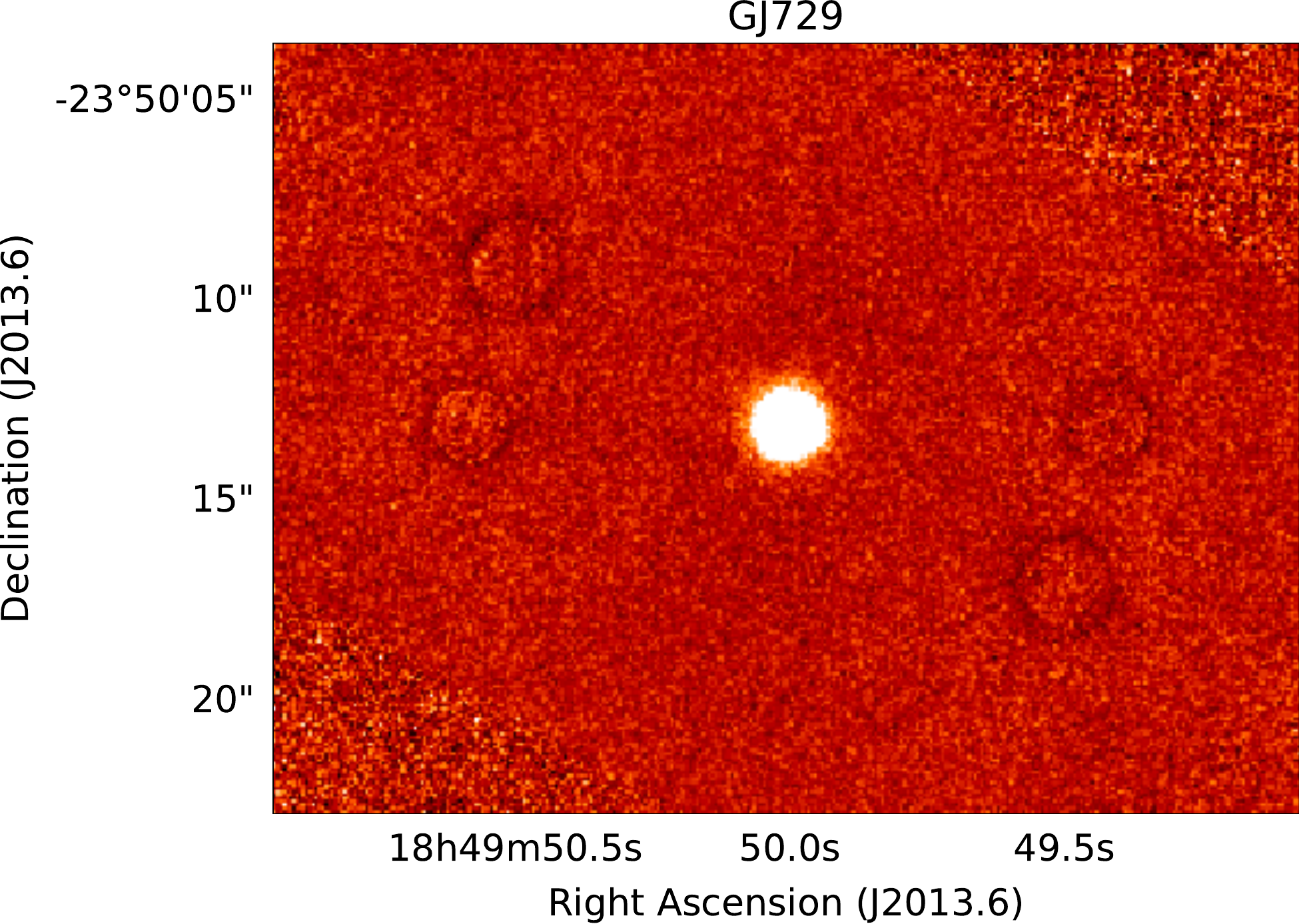}{0.31\textwidth}{}
          }
\vspace*{-0.3cm}
\gridline{\fig{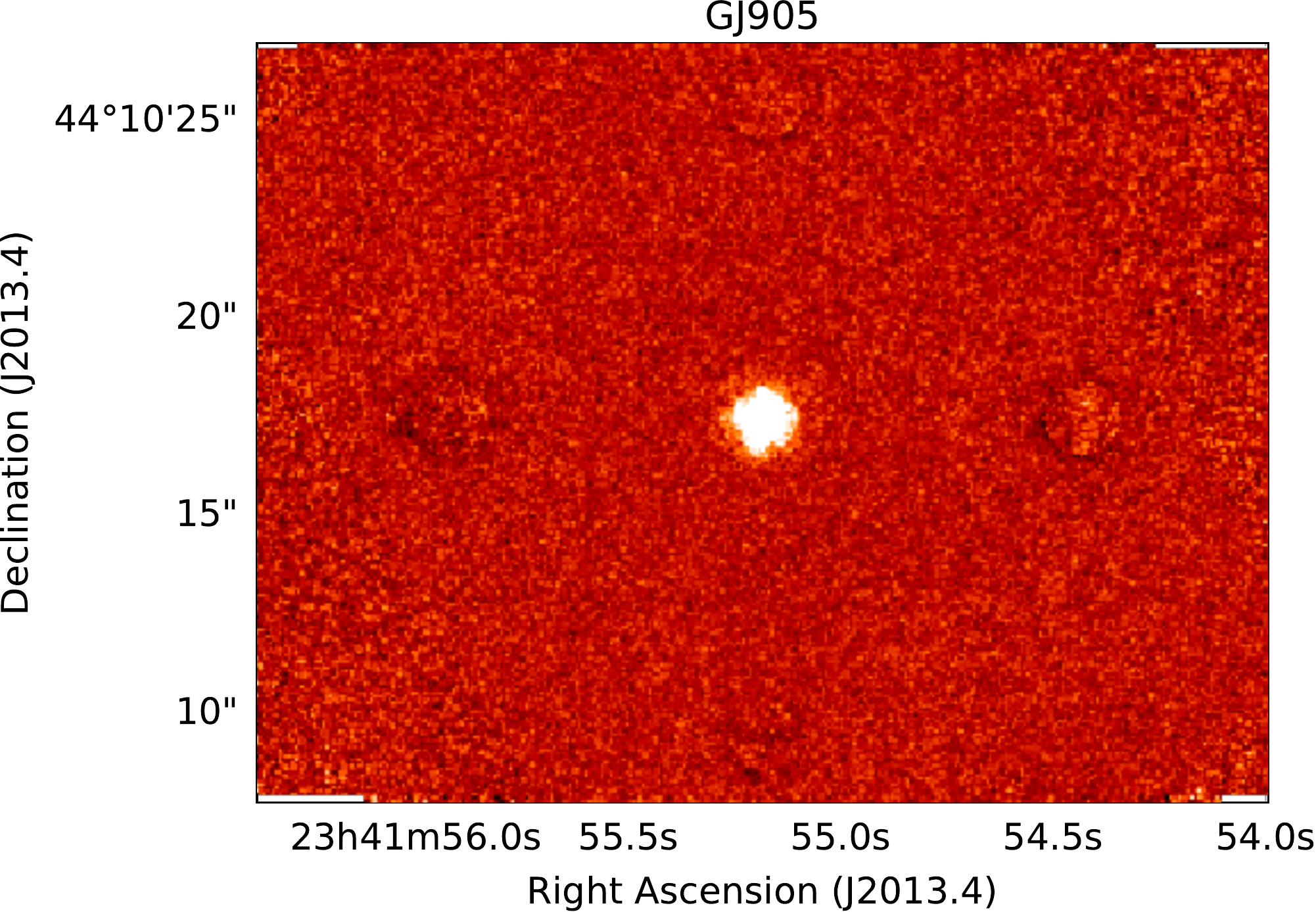}{0.31\textwidth}{}
          \fig{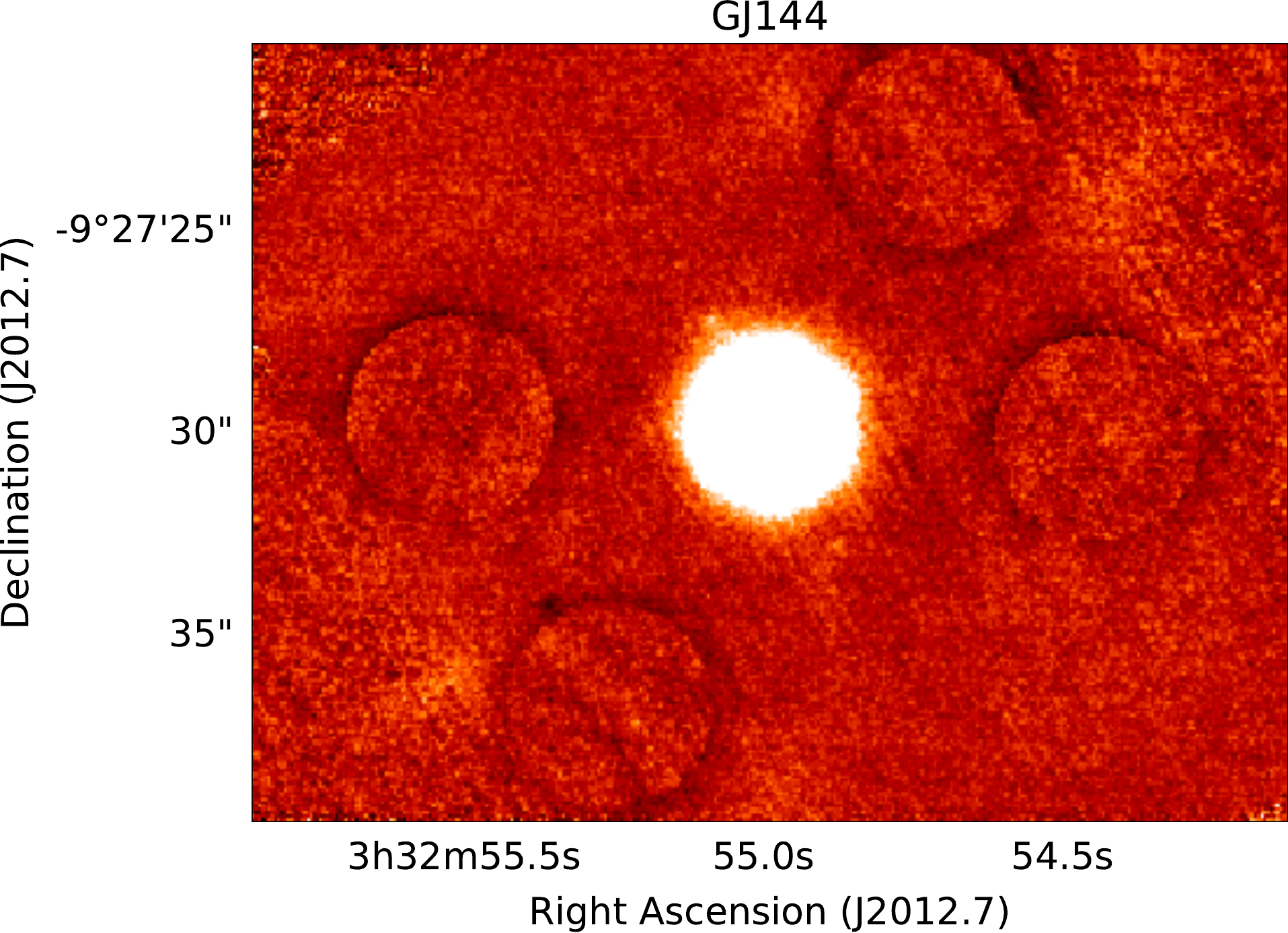}{0.31\textwidth}{}
          \fig{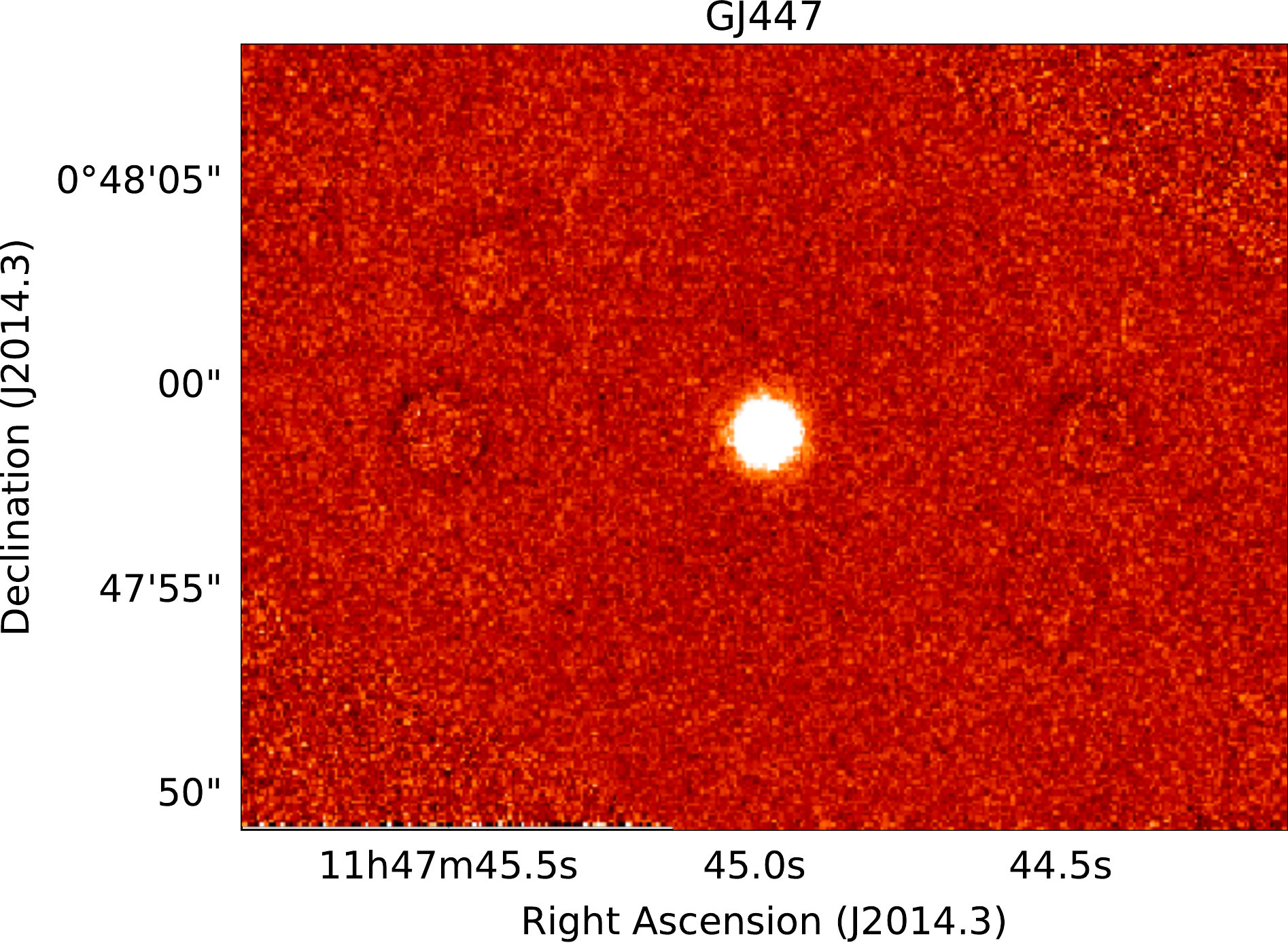}{0.31\textwidth}{}
          }
\vspace*{-0.3cm}
\gridline{\fig{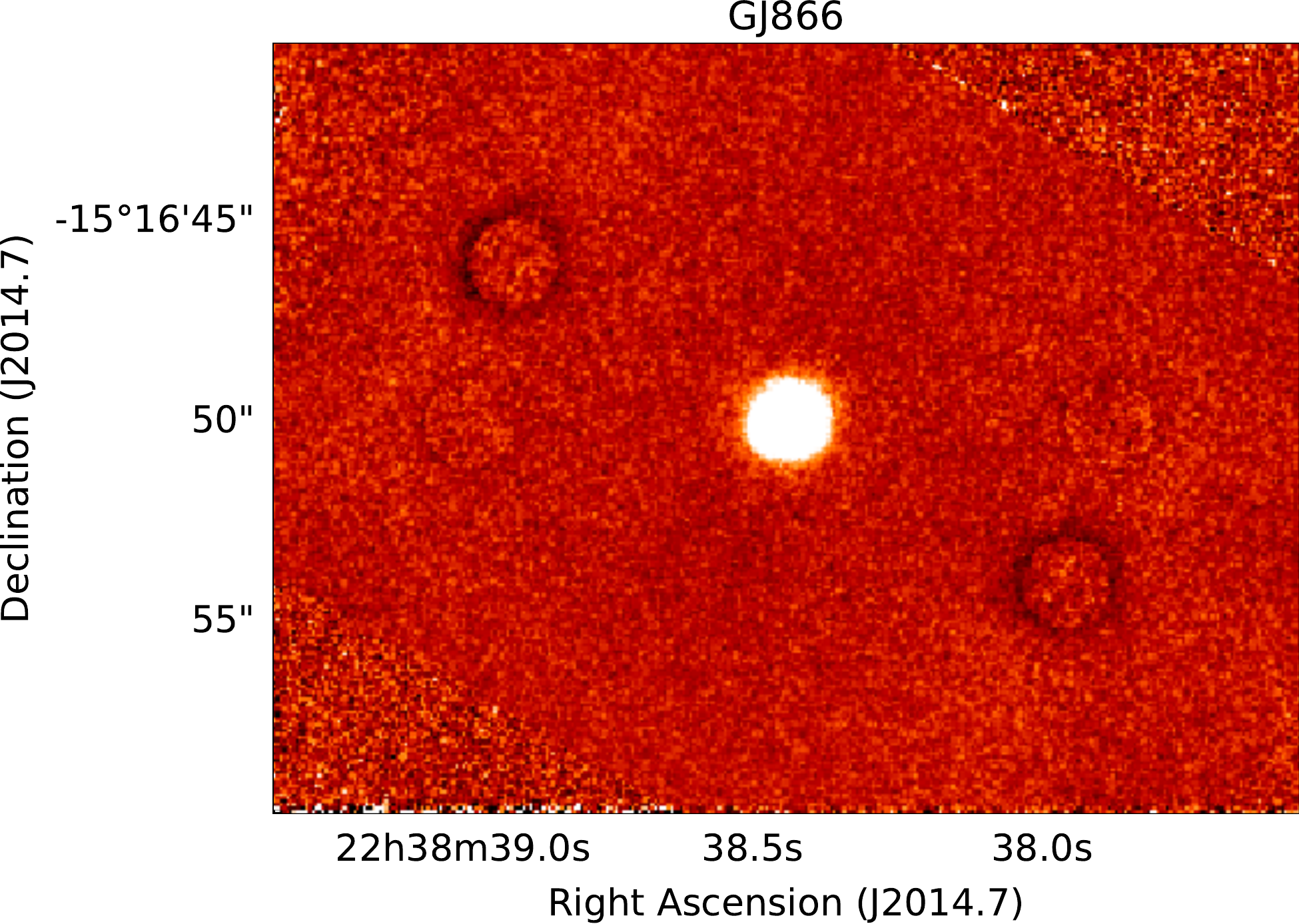}{0.31\textwidth}{}
          \fig{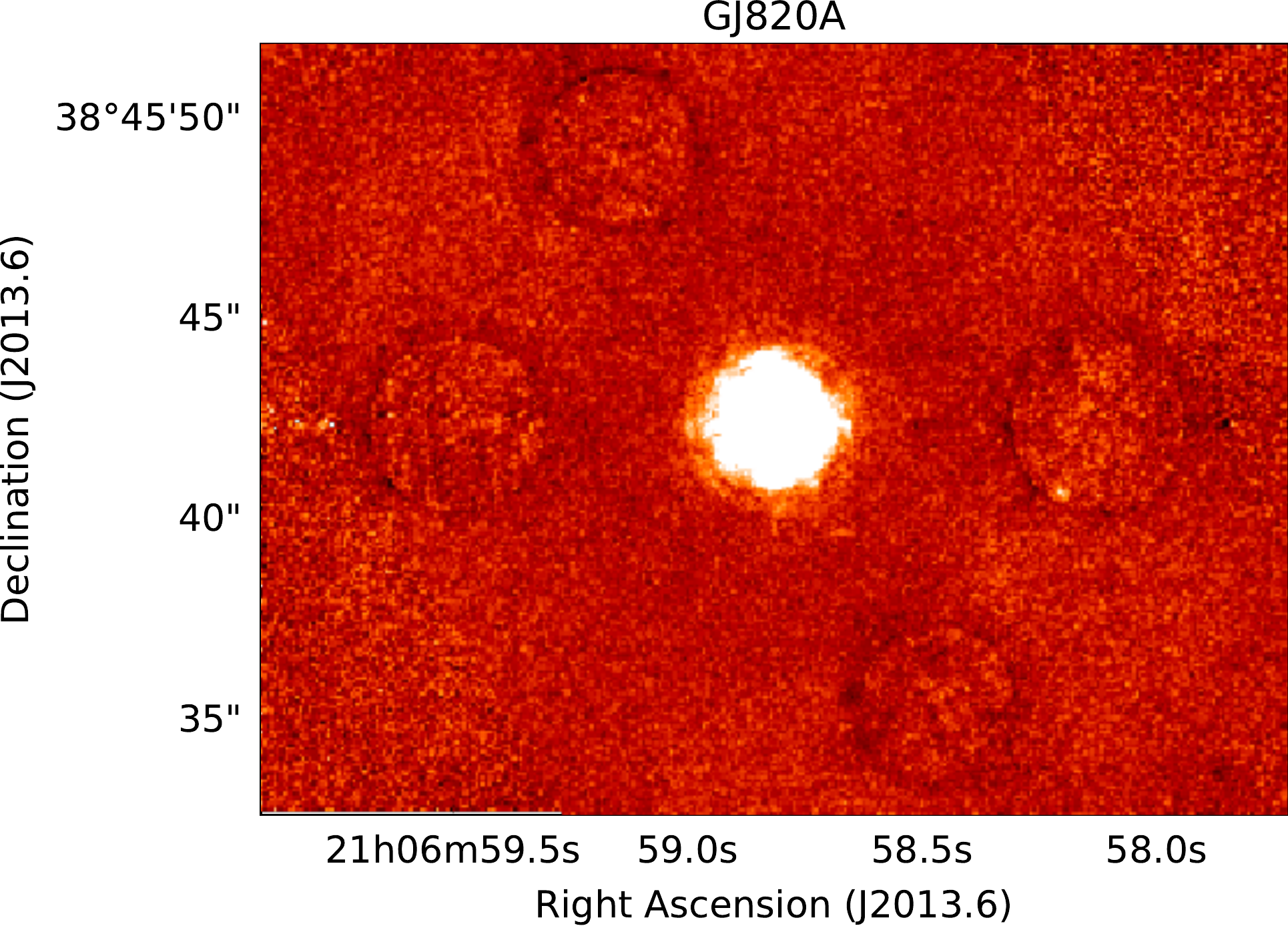}{0.31\textwidth}{}
          \fig{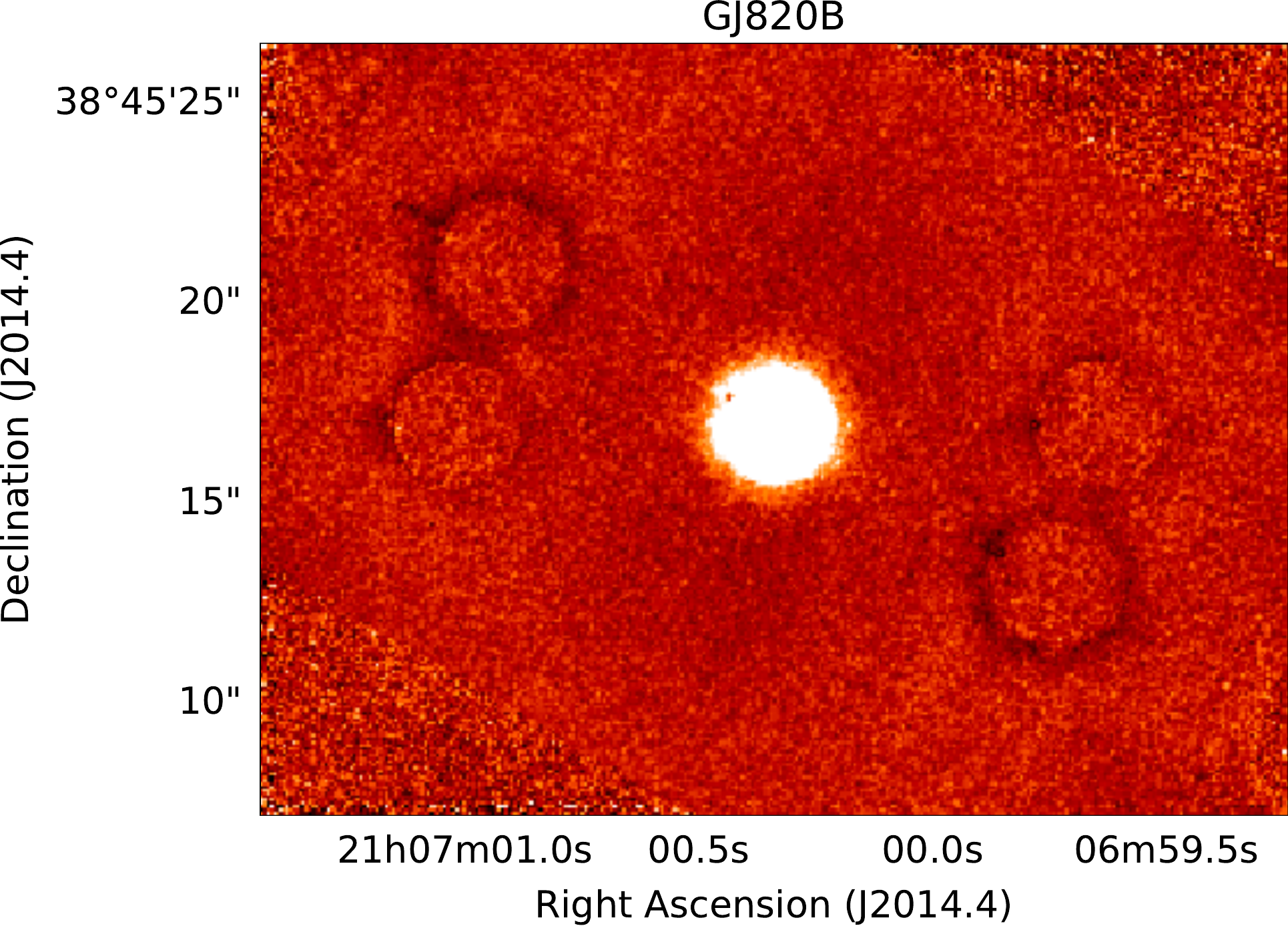}{0.31\textwidth}{}
          }
\vspace*{-0.3cm}
\gridline{\fig{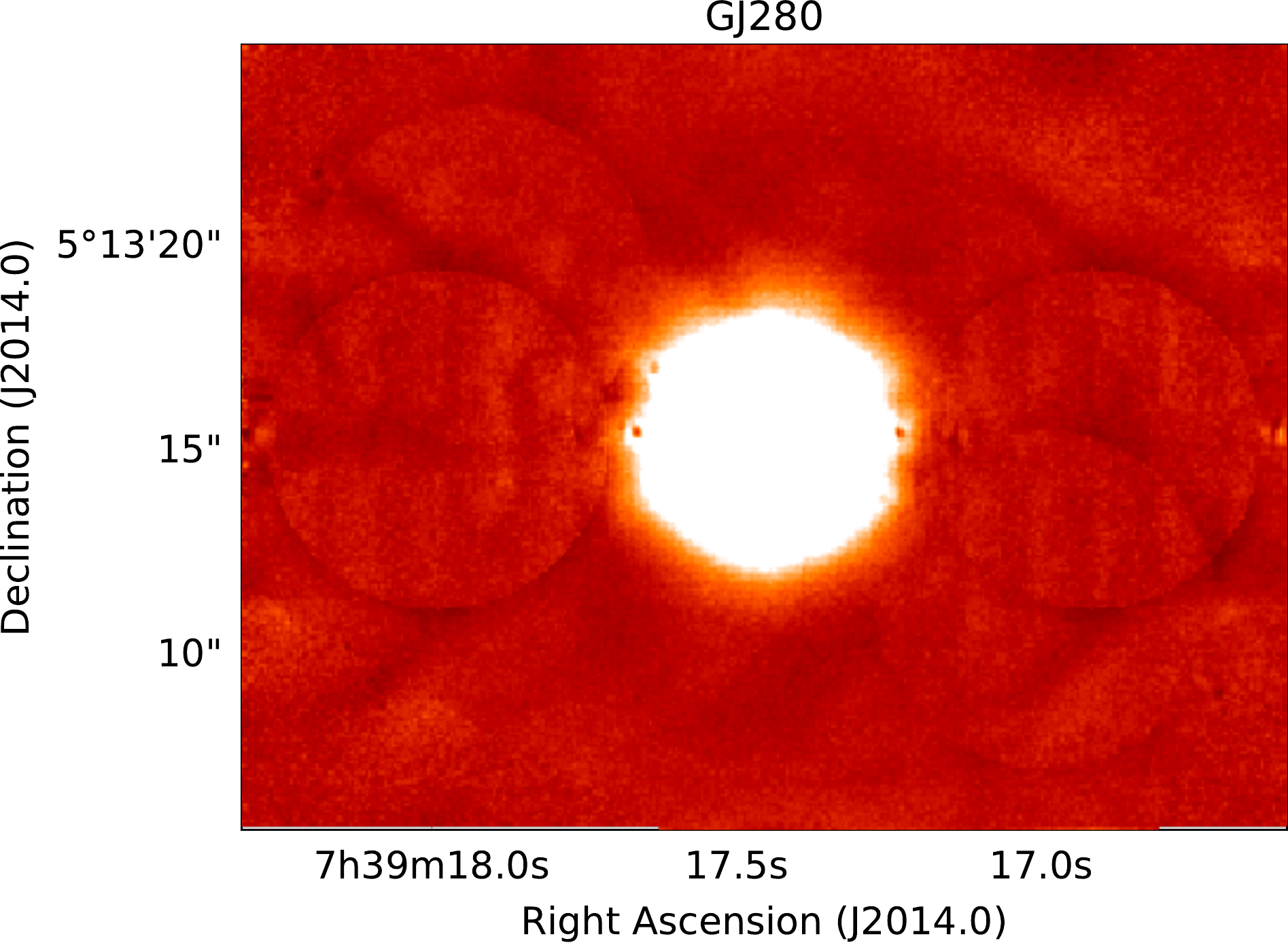}{0.31\textwidth}{}
          \fig{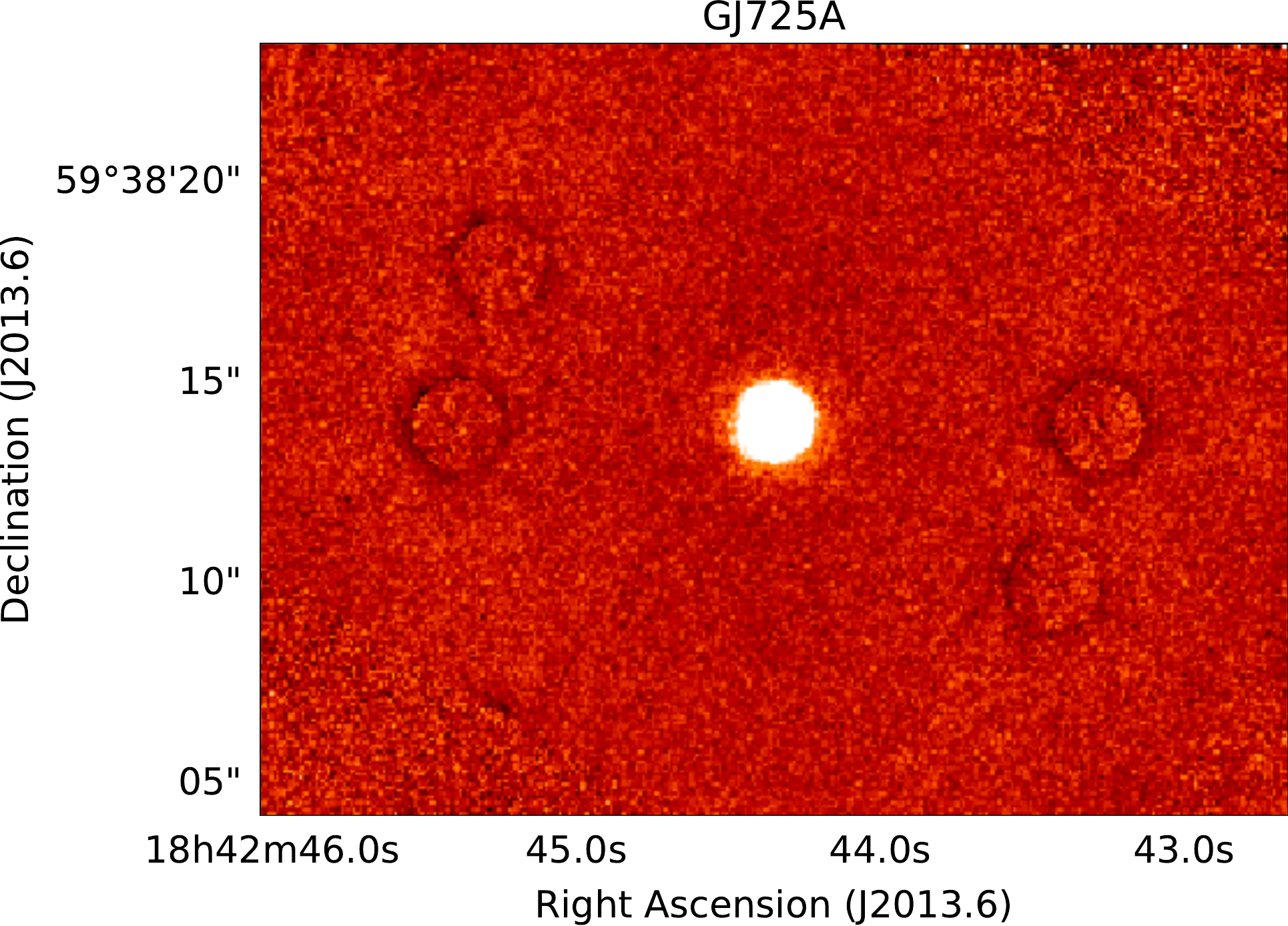}{0.31\textwidth}{}
          \fig{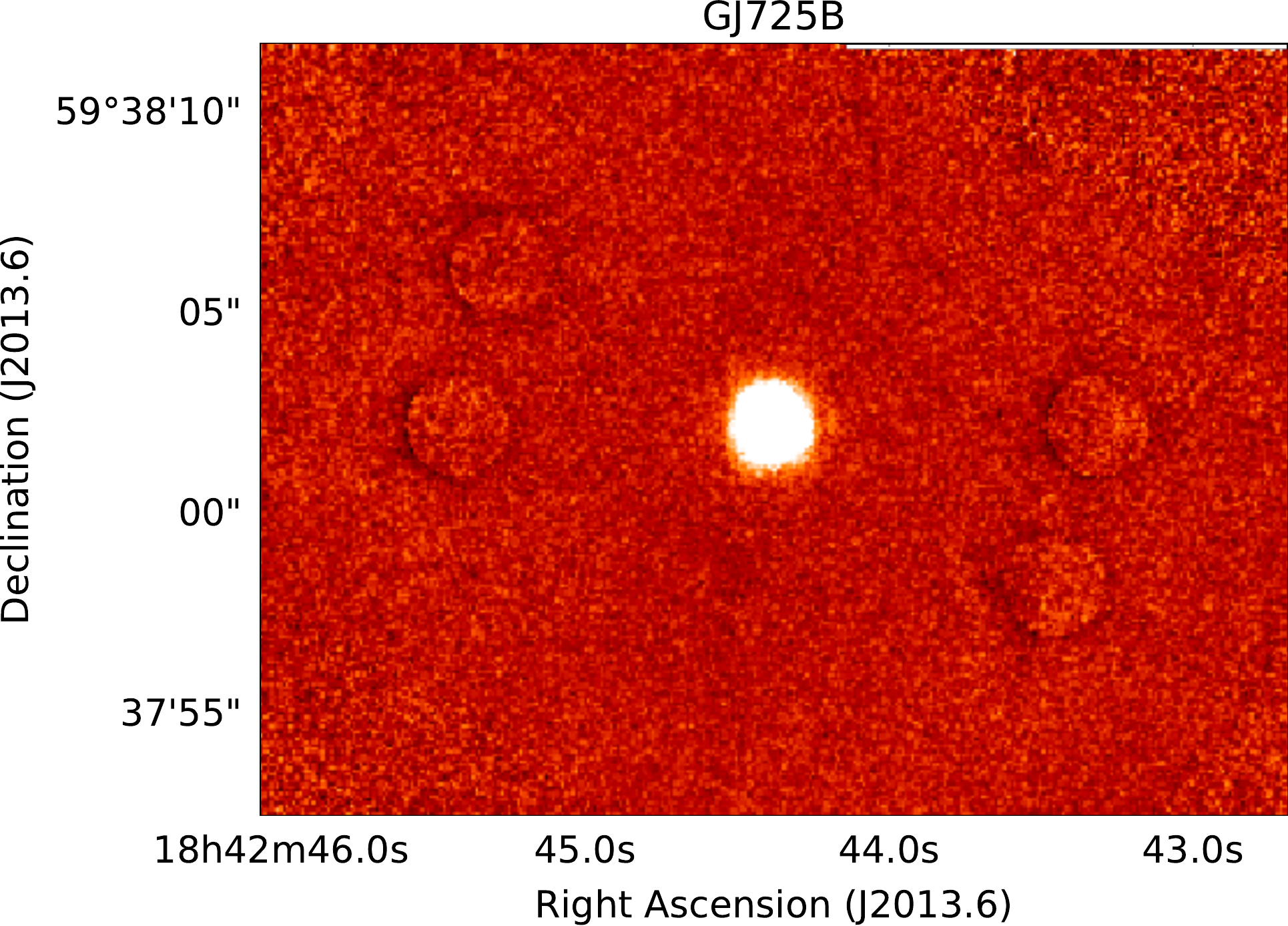}{0.31\textwidth}{}
          }          
\end{figure*}        

\begin{figure*}[h!]     
\gridline{\fig{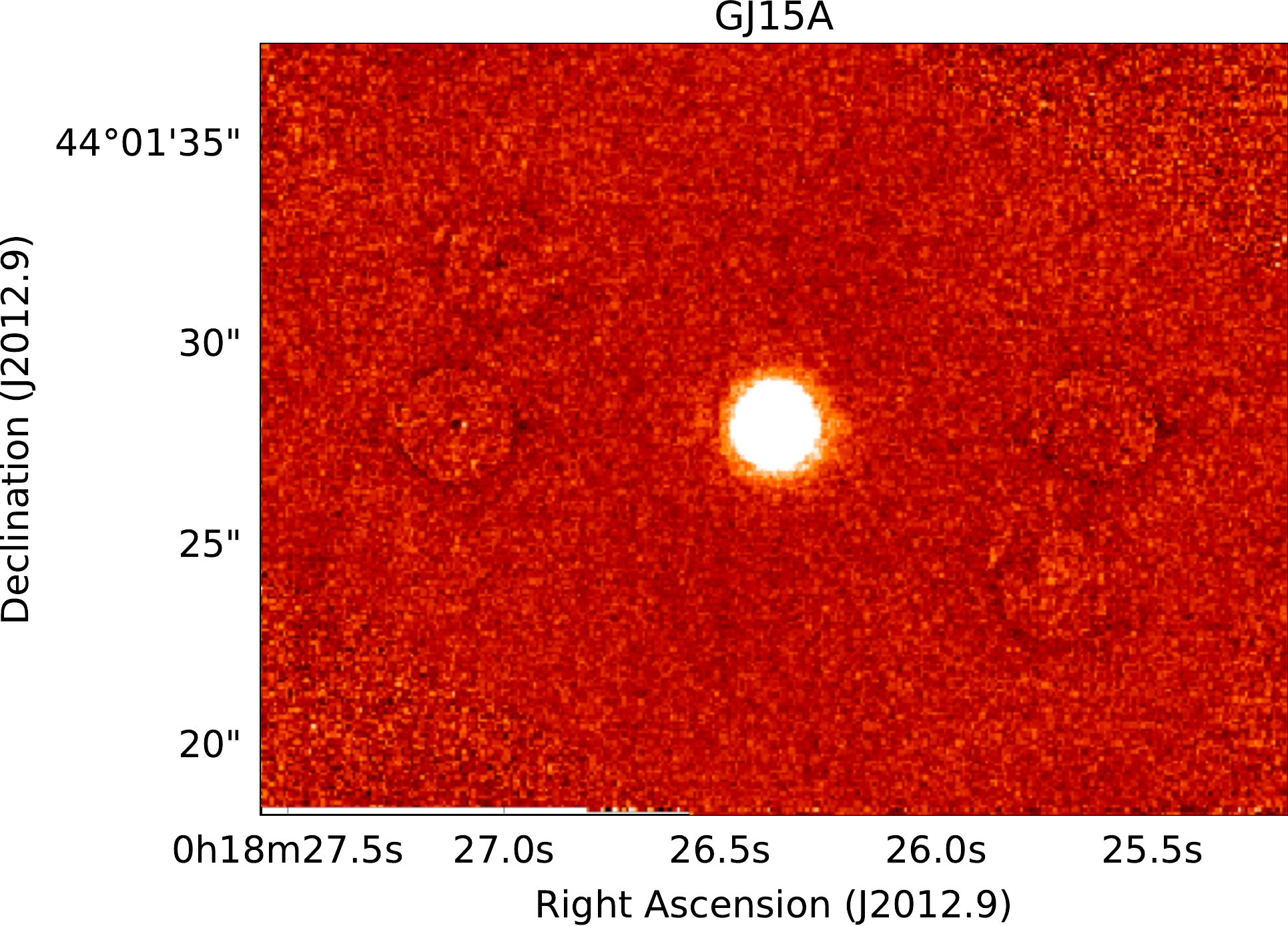}{0.31\textwidth}{}
          \fig{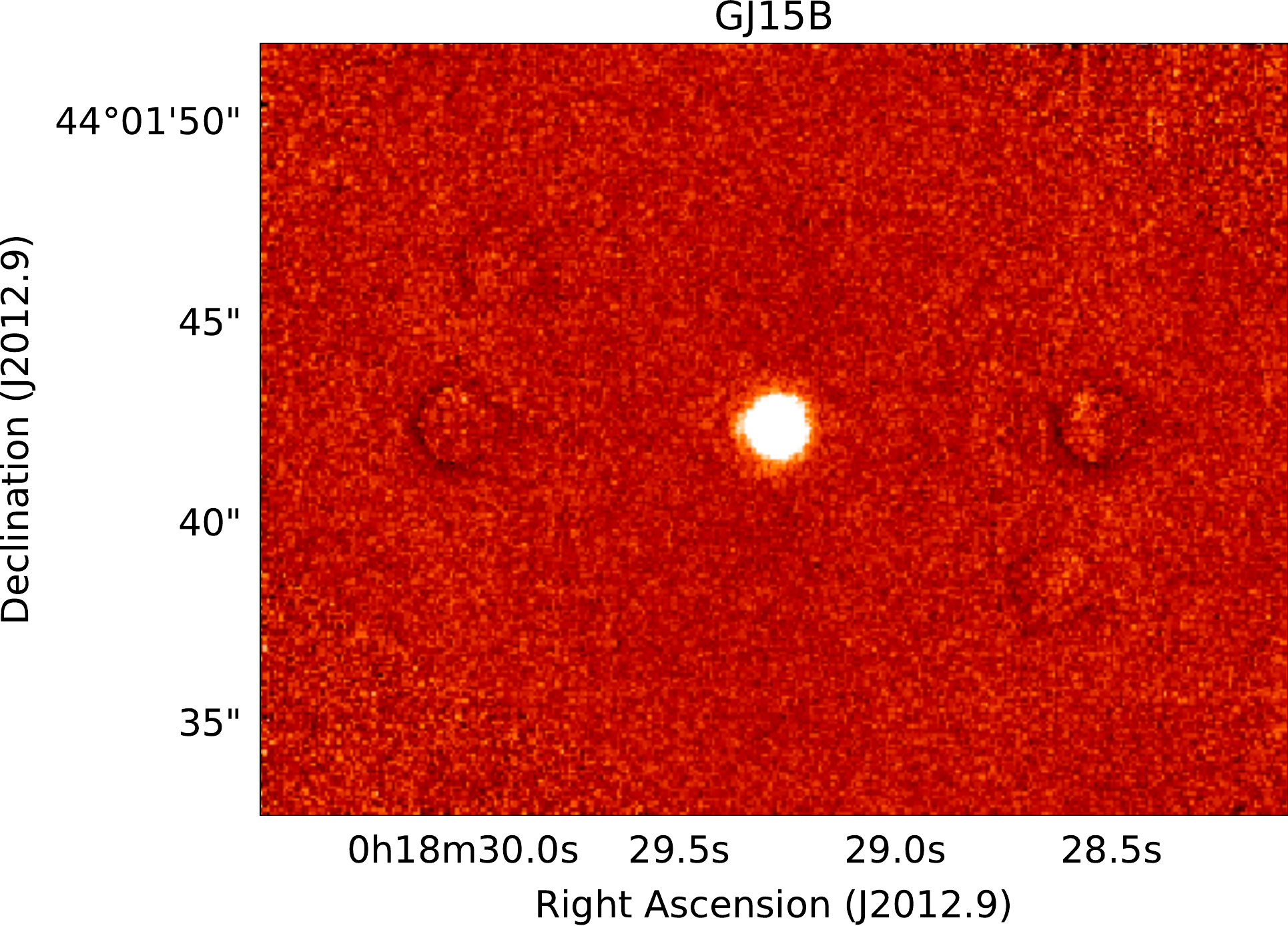}{0.31\textwidth}{}
          \fig{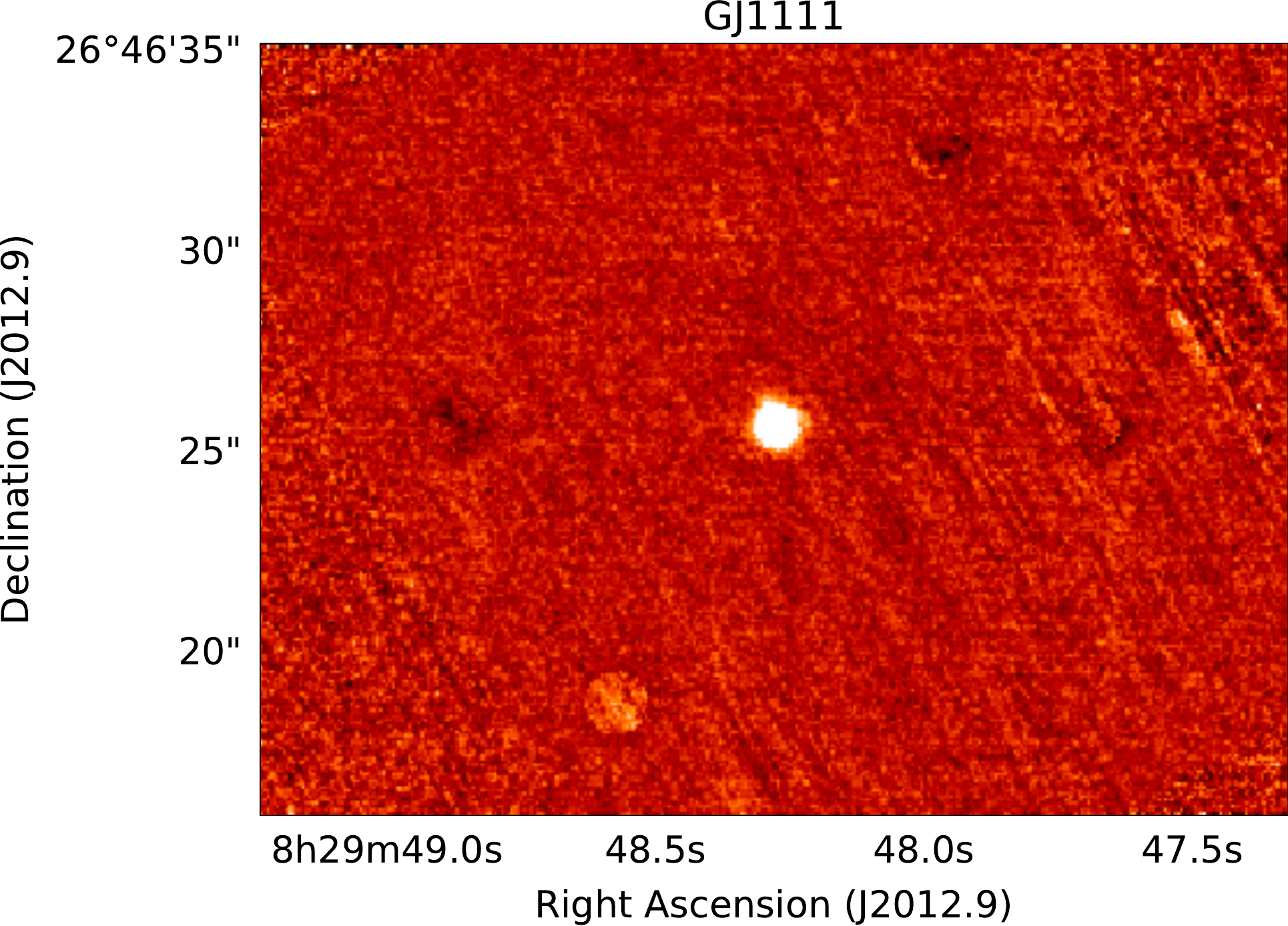}{0.31\textwidth}{}
          }
\vspace*{-0.3cm}
\gridline{\fig{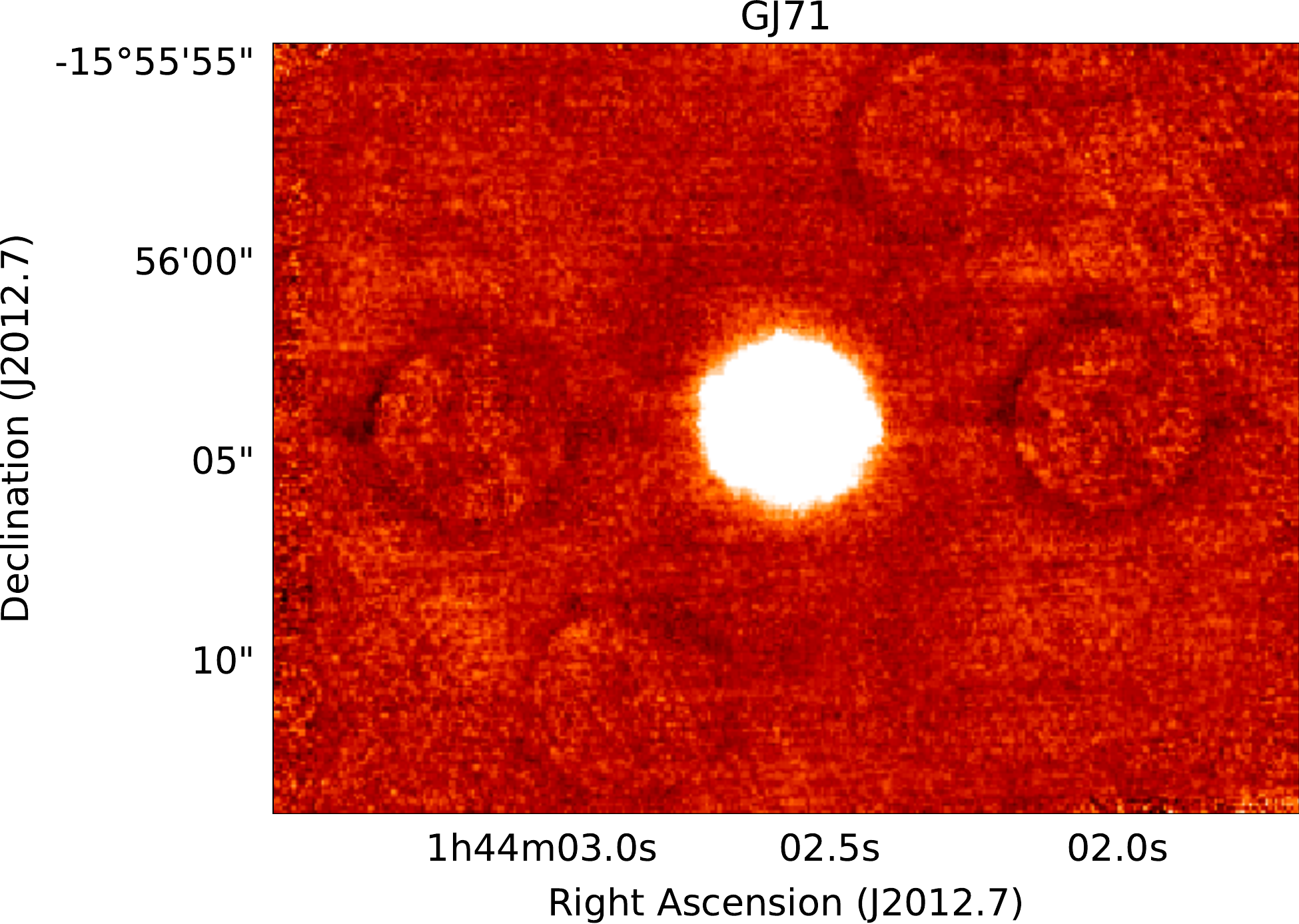}{0.31\textwidth}{}
          \fig{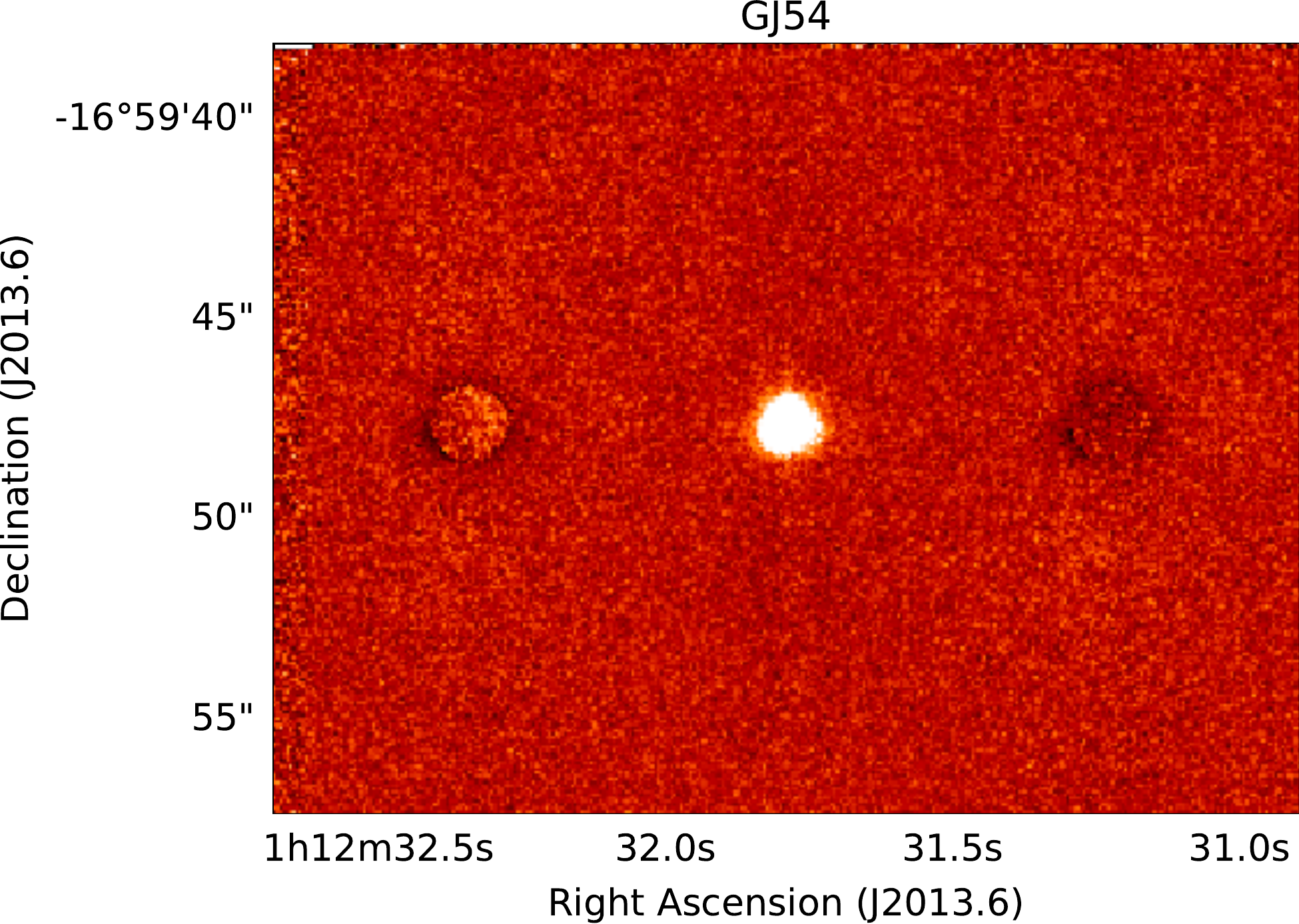}{0.31\textwidth}{}
          \fig{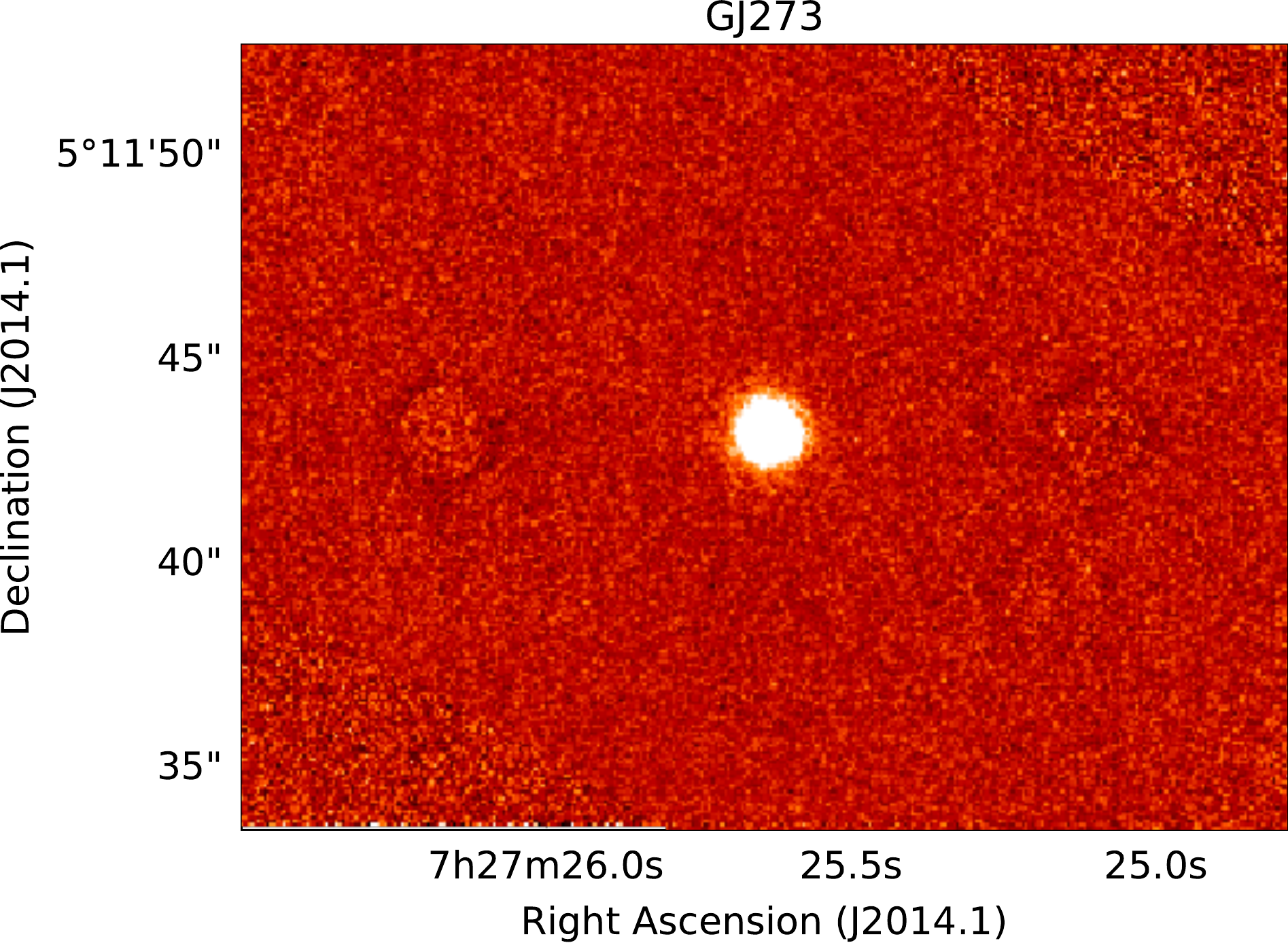}{0.31\textwidth}{}
          }
\vspace*{-0.3cm}
\gridline{\fig{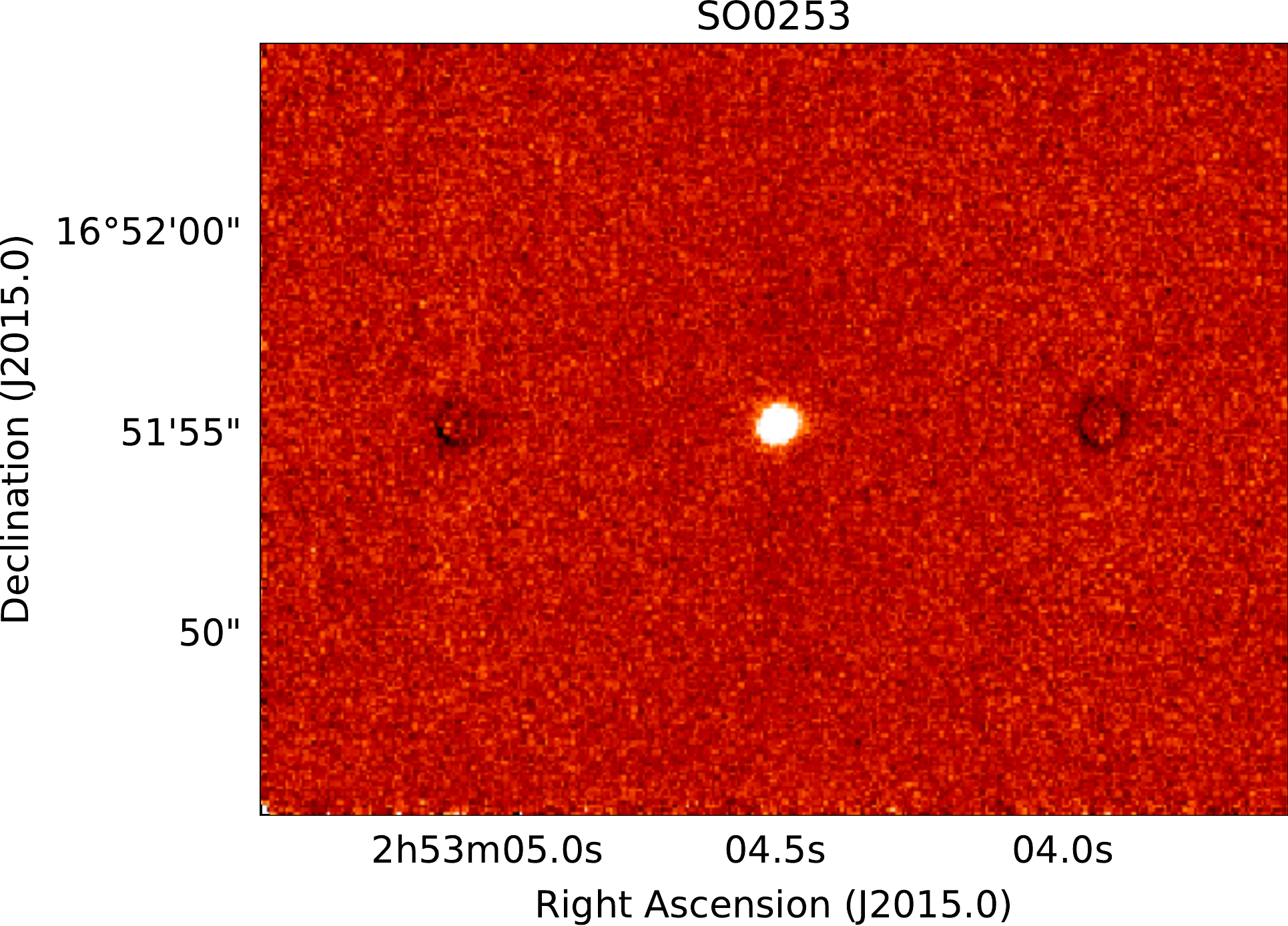}{0.31\textwidth}{}
          \fig{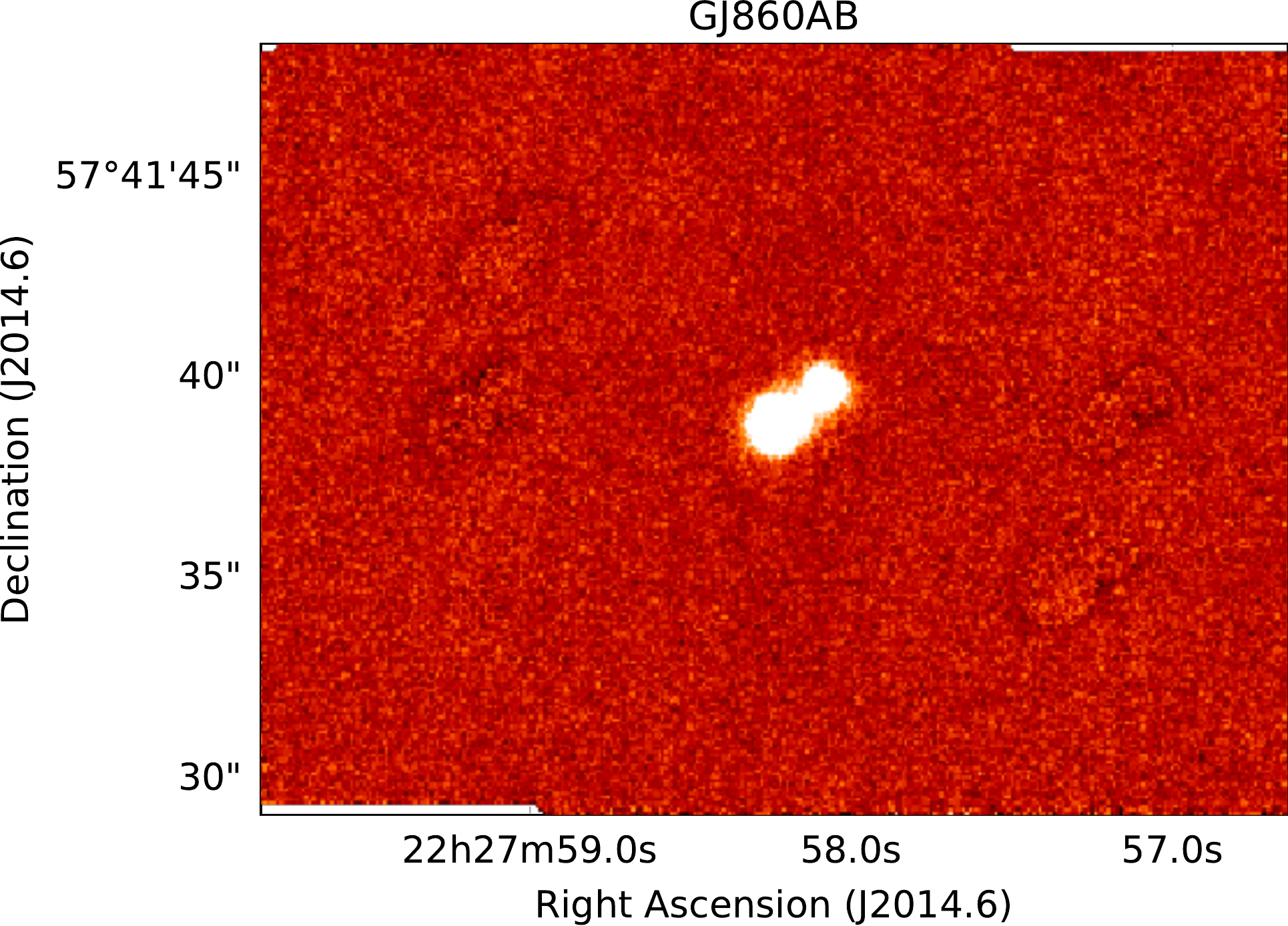}{0.31\textwidth}{}
          \fig{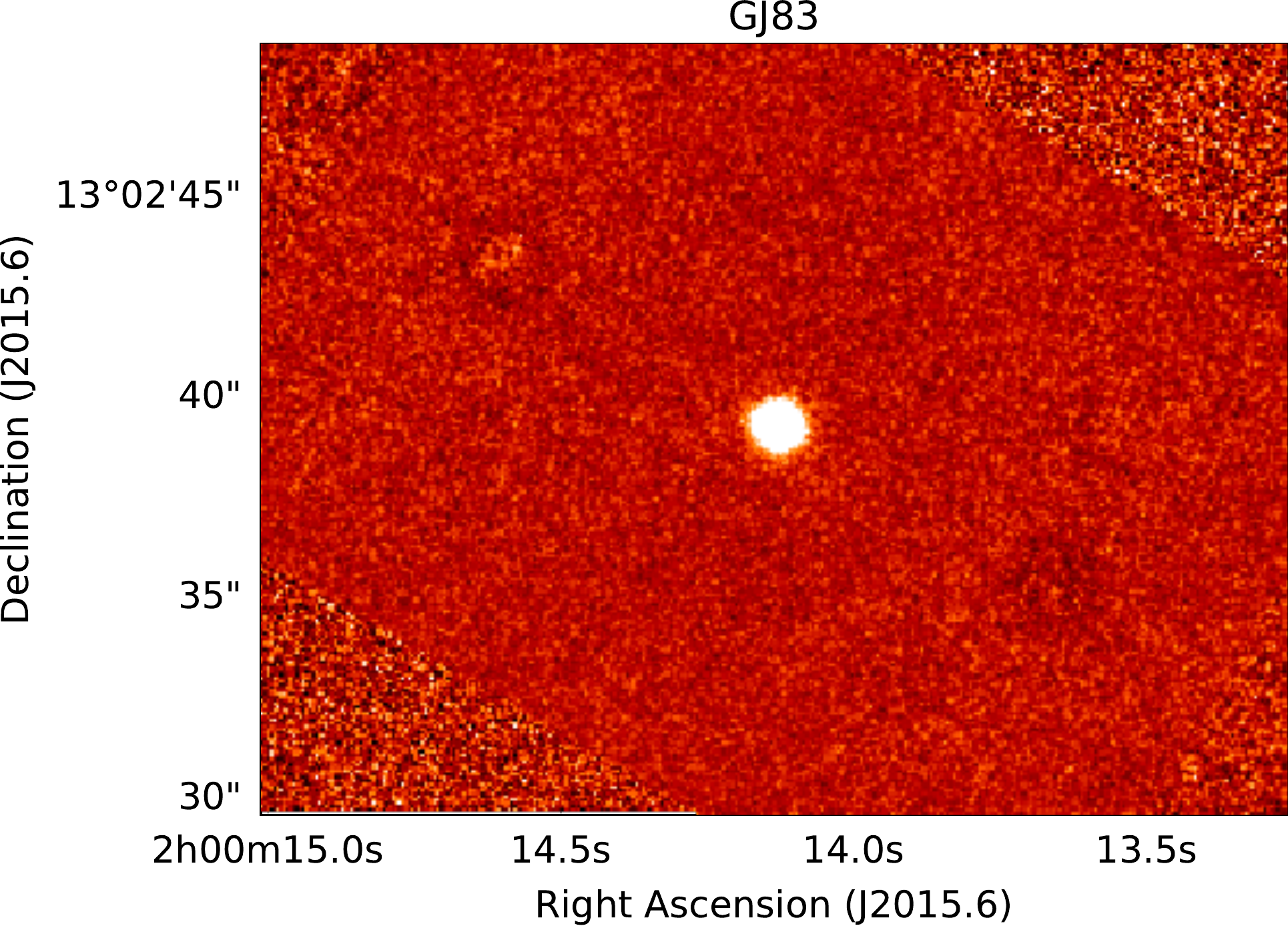}{0.31\textwidth}{}
          }
\vspace*{-0.3cm}
\gridline{\fig{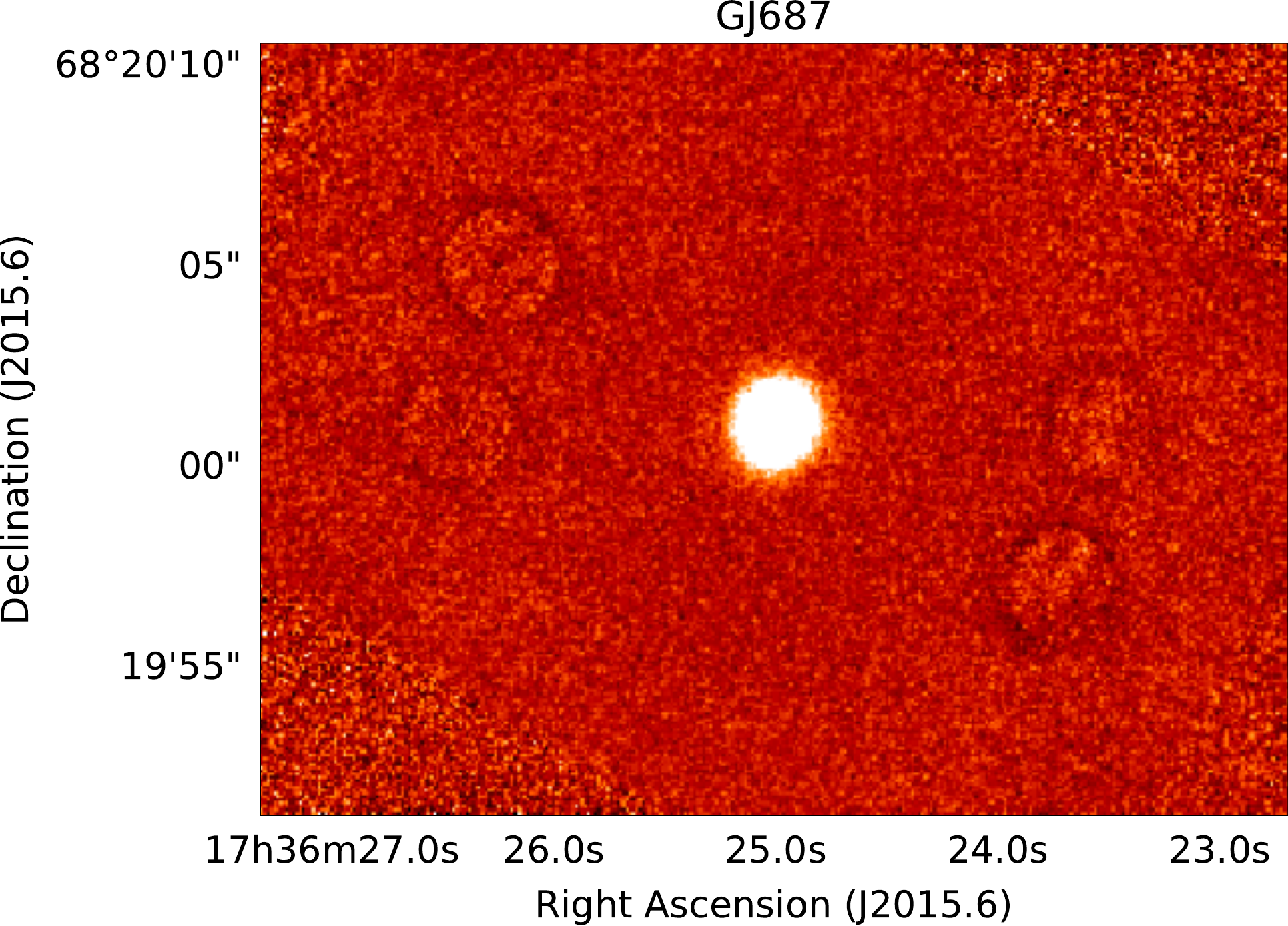}{0.31\textwidth}{}
          \fig{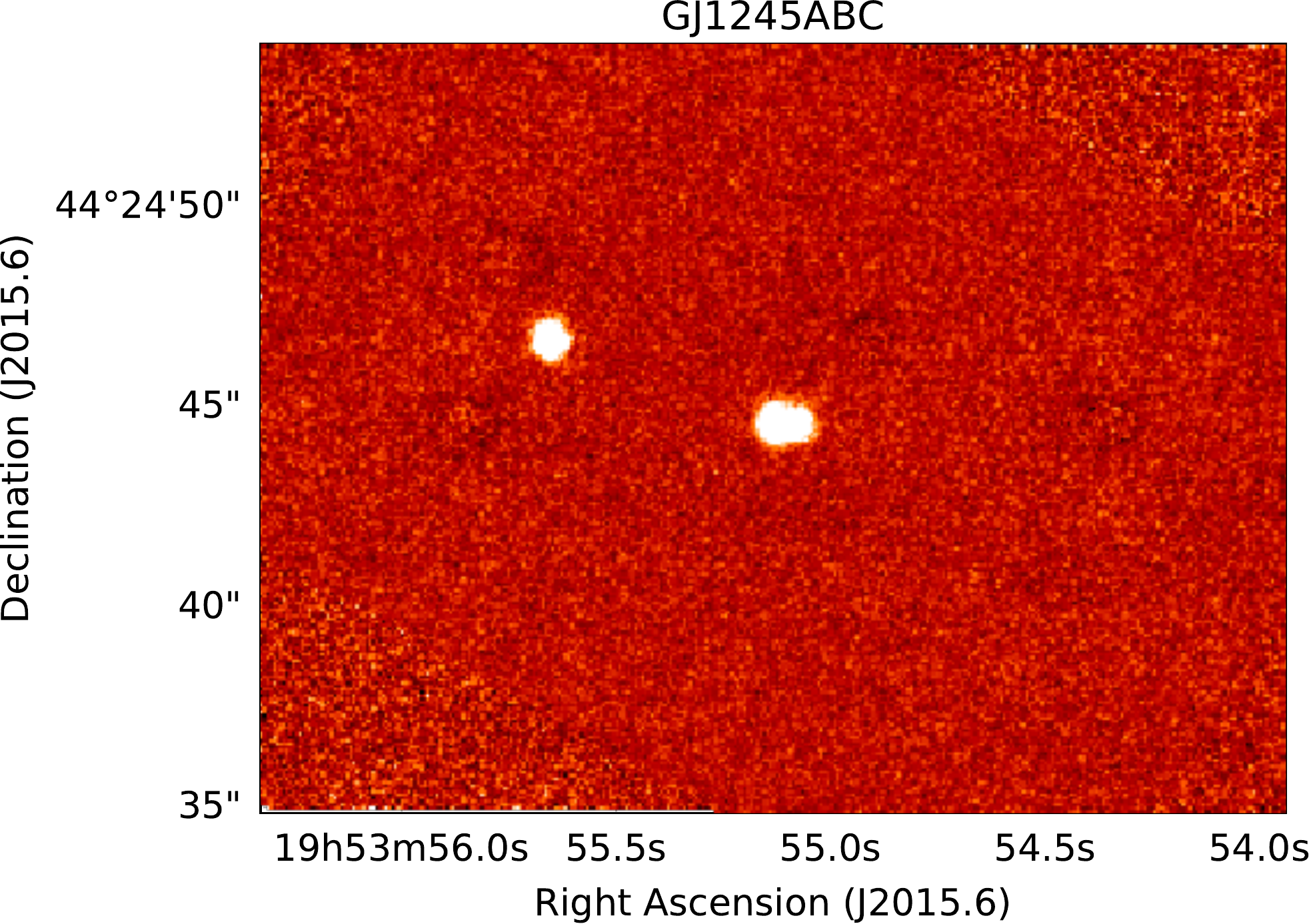}{0.31\textwidth}{}
          \fig{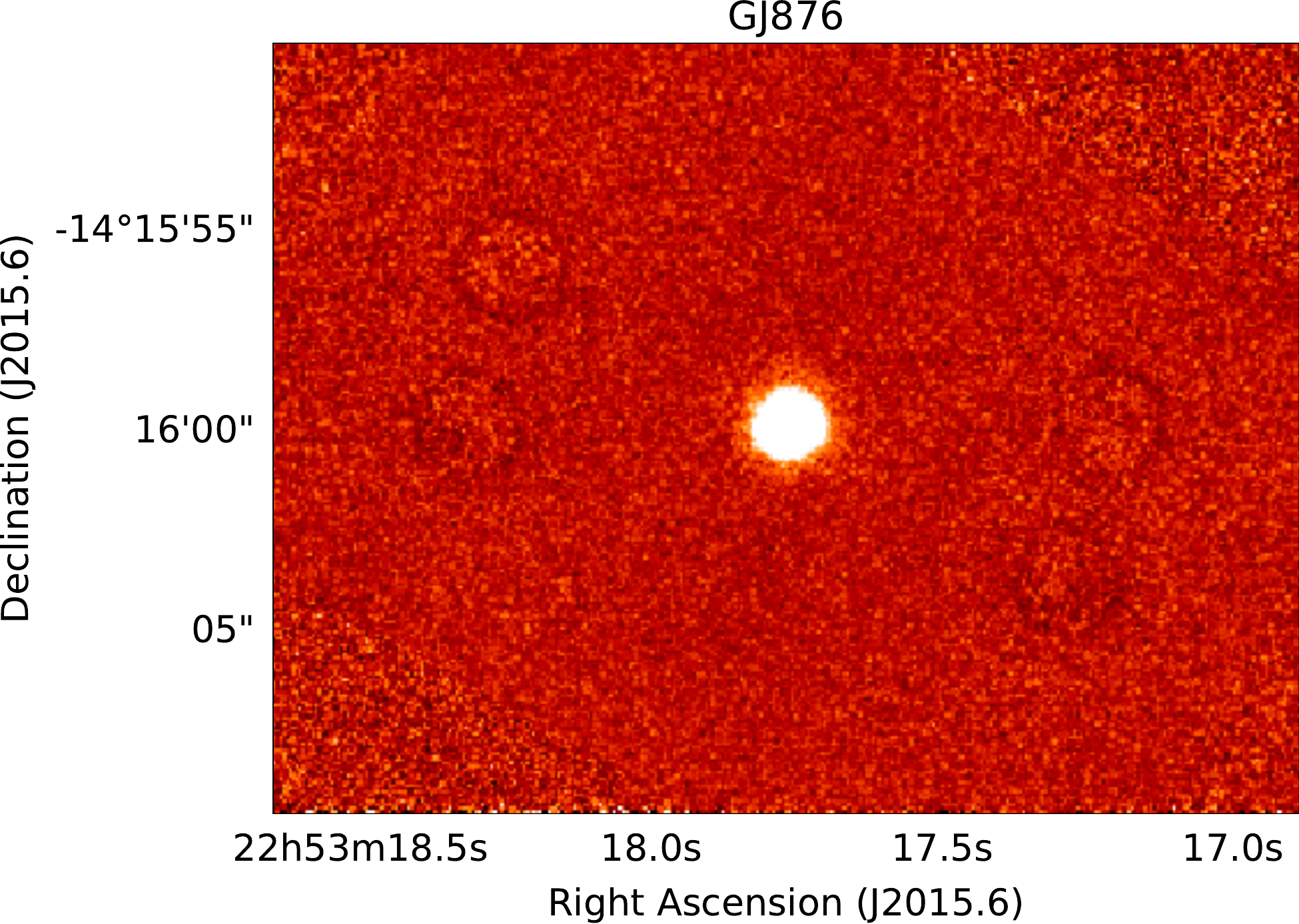}{0.31\textwidth}{}
          }
\vspace*{-0.3cm}
\gridline{\fig{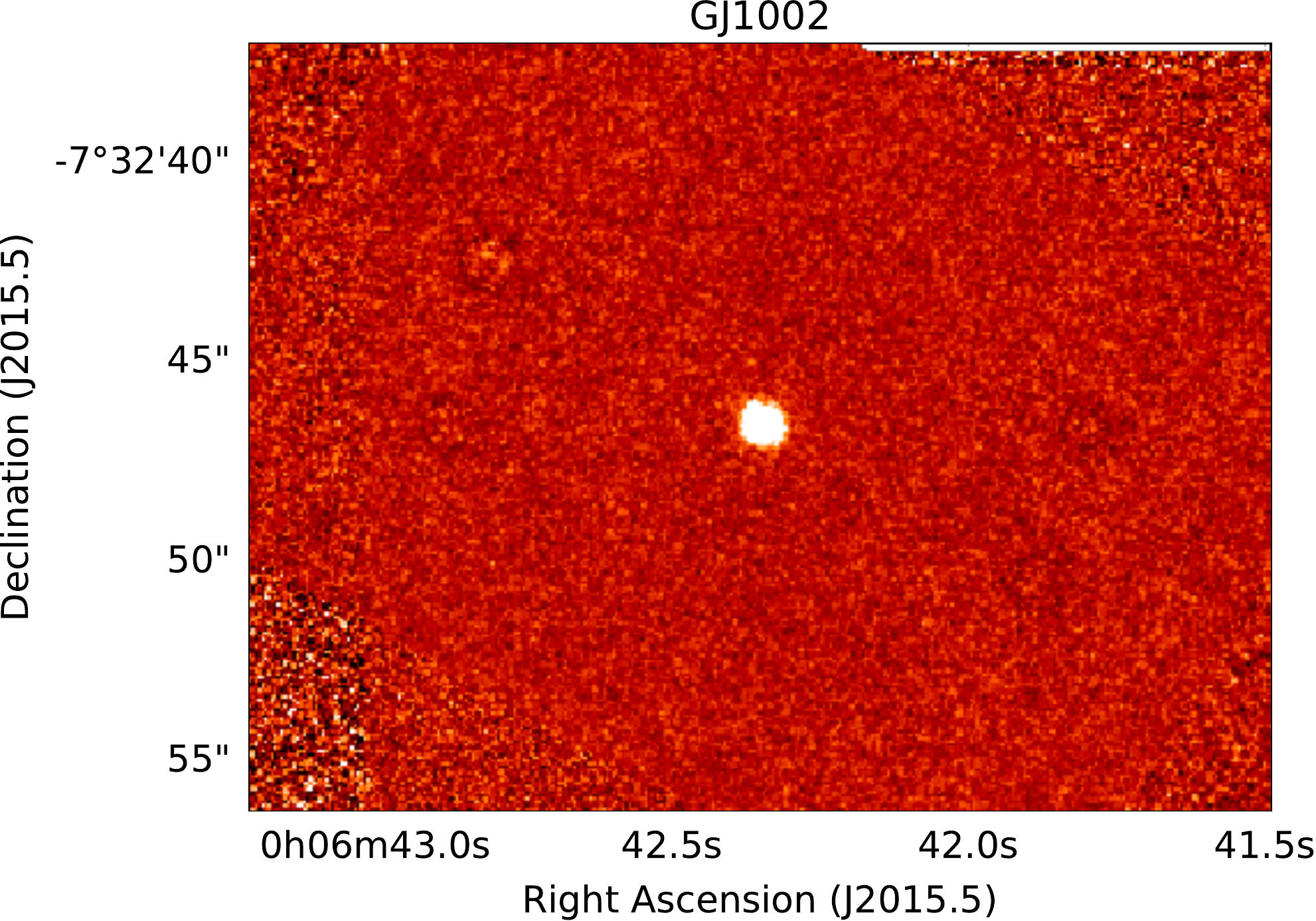}{0.31\textwidth}{}
          }
\vspace*{-0.3cm}
\caption{The final, deepest CanariCam Si-2 filter (8.7\,$\mu$m) images of the 
observed stars of the 5\,pc sample. Counts are in linear scale and in the 
$\pm$7$\sigma$ range relative to the zero background. Shown field of view is 
25.6$\times$19.2\,arcsec, orientation is North up and East to the left. The 
coordinates correspond to the first observing epoch, taking into account the 
proper motion.
\label{CCallimages}}
\end{figure*}

\begin{figure*}[h!]
\gridline{\fig{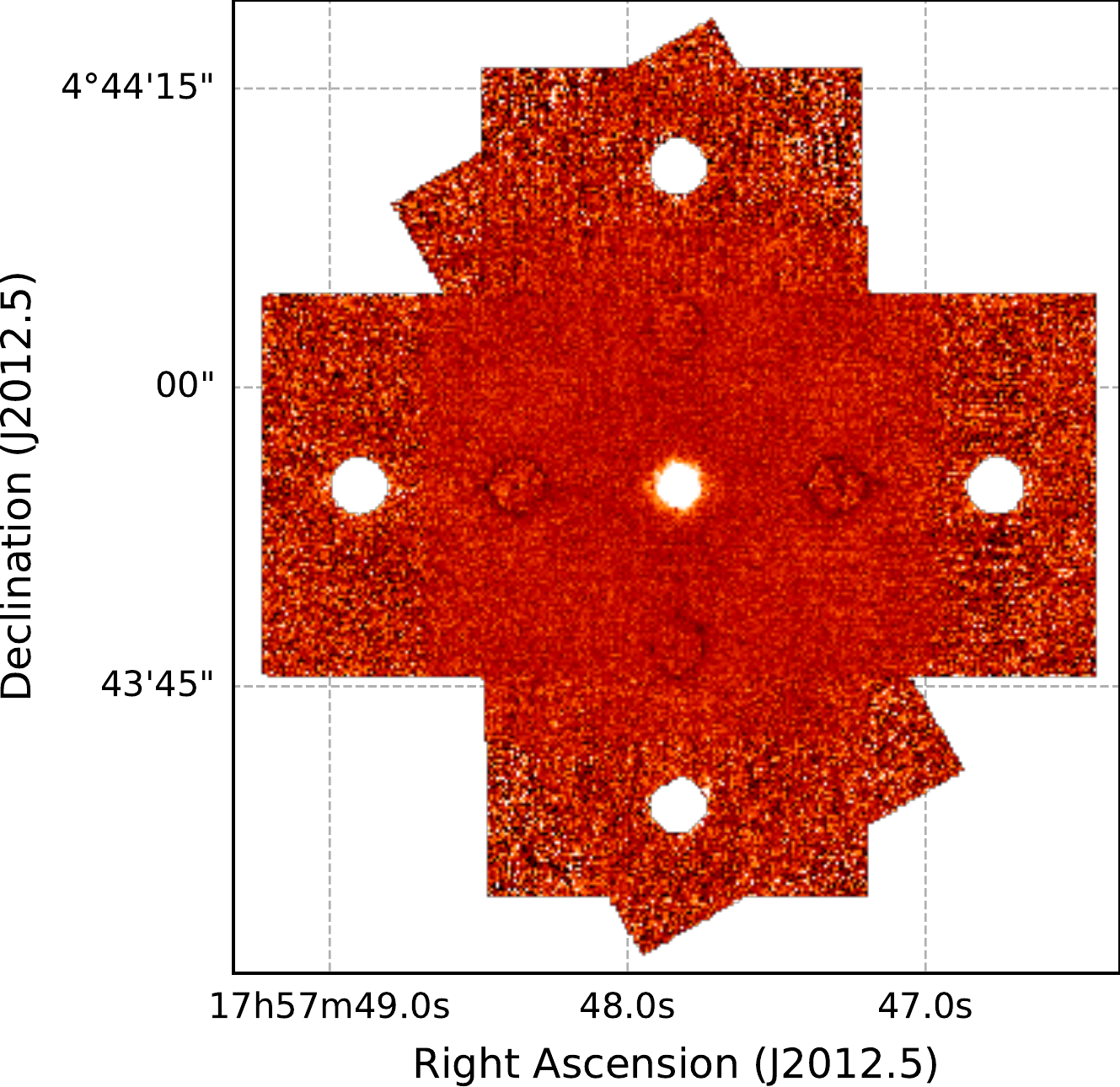}{0.32\textwidth}{(GJ 699)}
          \fig{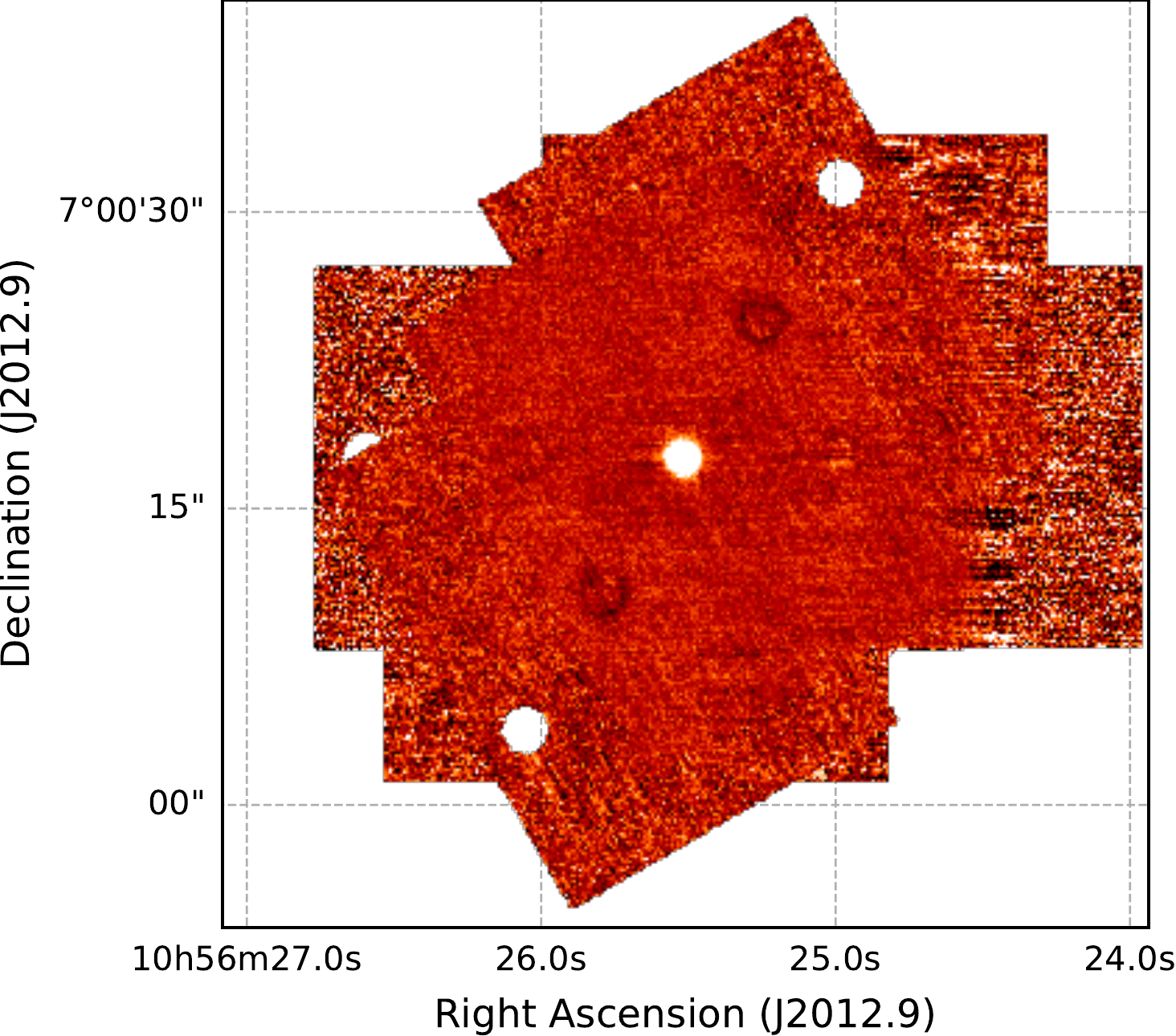}{0.32\textwidth}{(GJ 406)}
          \fig{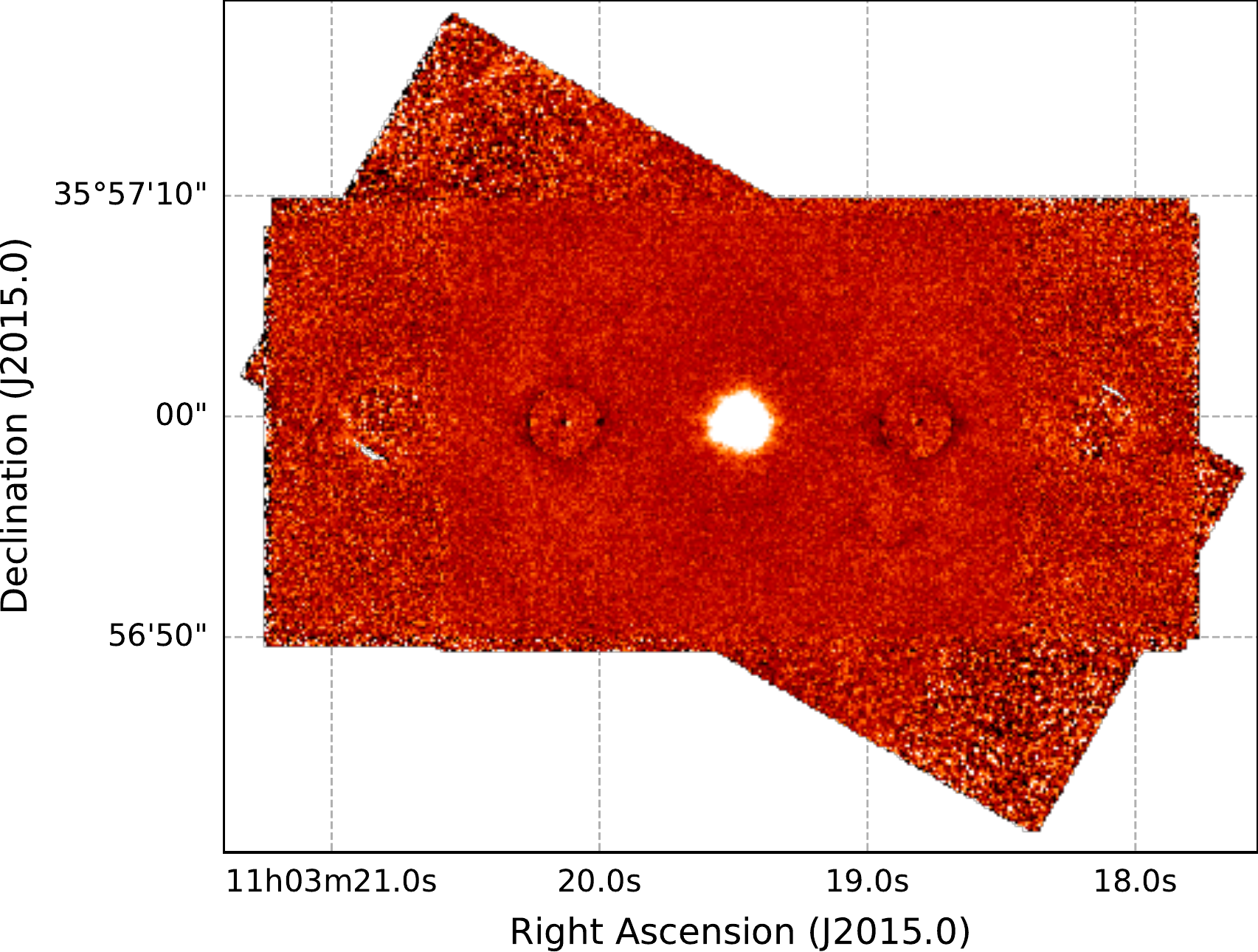}{0.32\textwidth}{(GJ 411)}
          }
\gridline{\fig{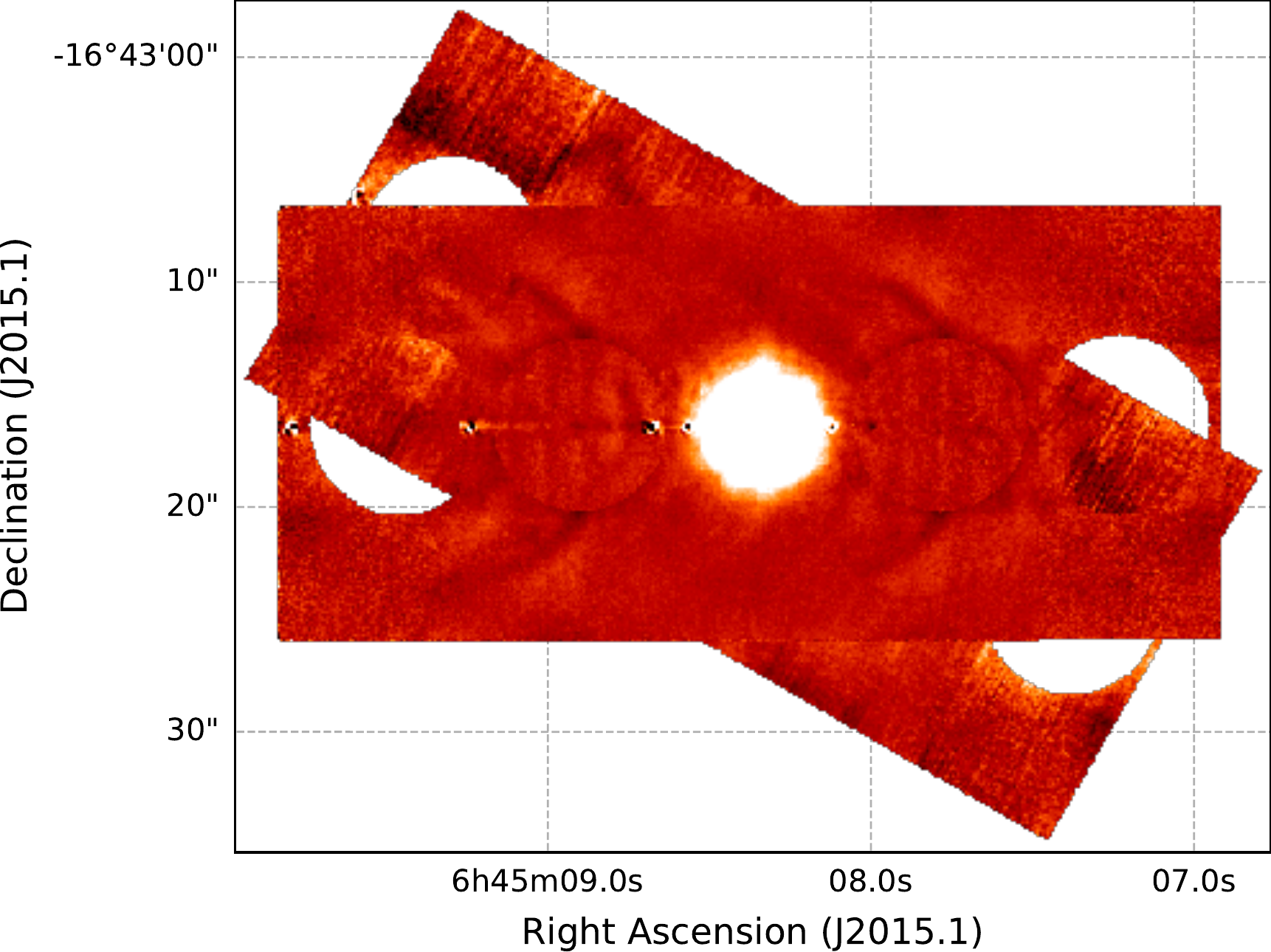}{0.32\textwidth}{(GJ 244)}
          \fig{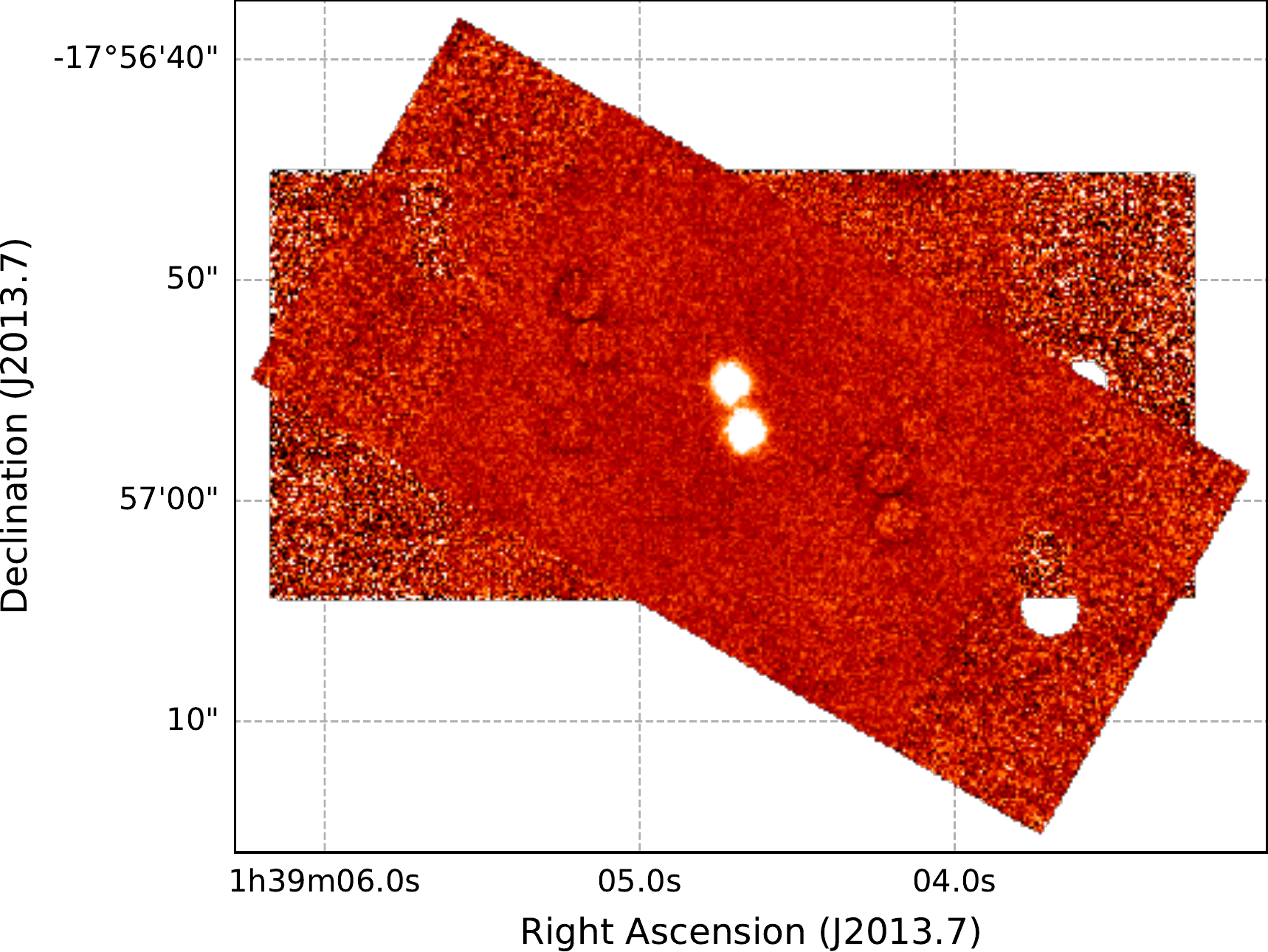}{0.32\textwidth}{(GJ 65AB)}
          \fig{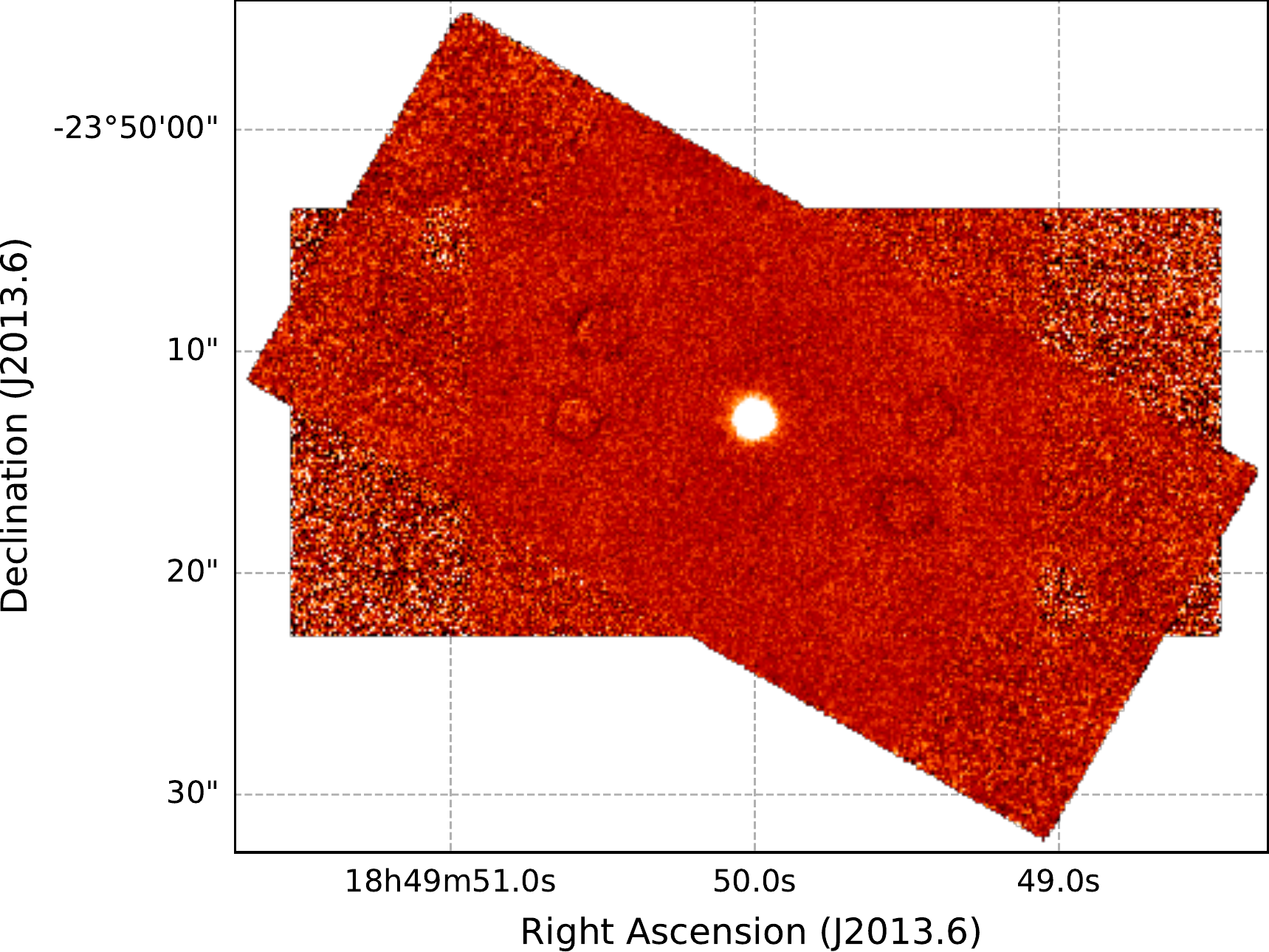}{0.32\textwidth}{(GJ 729)}
          }
\gridline{\fig{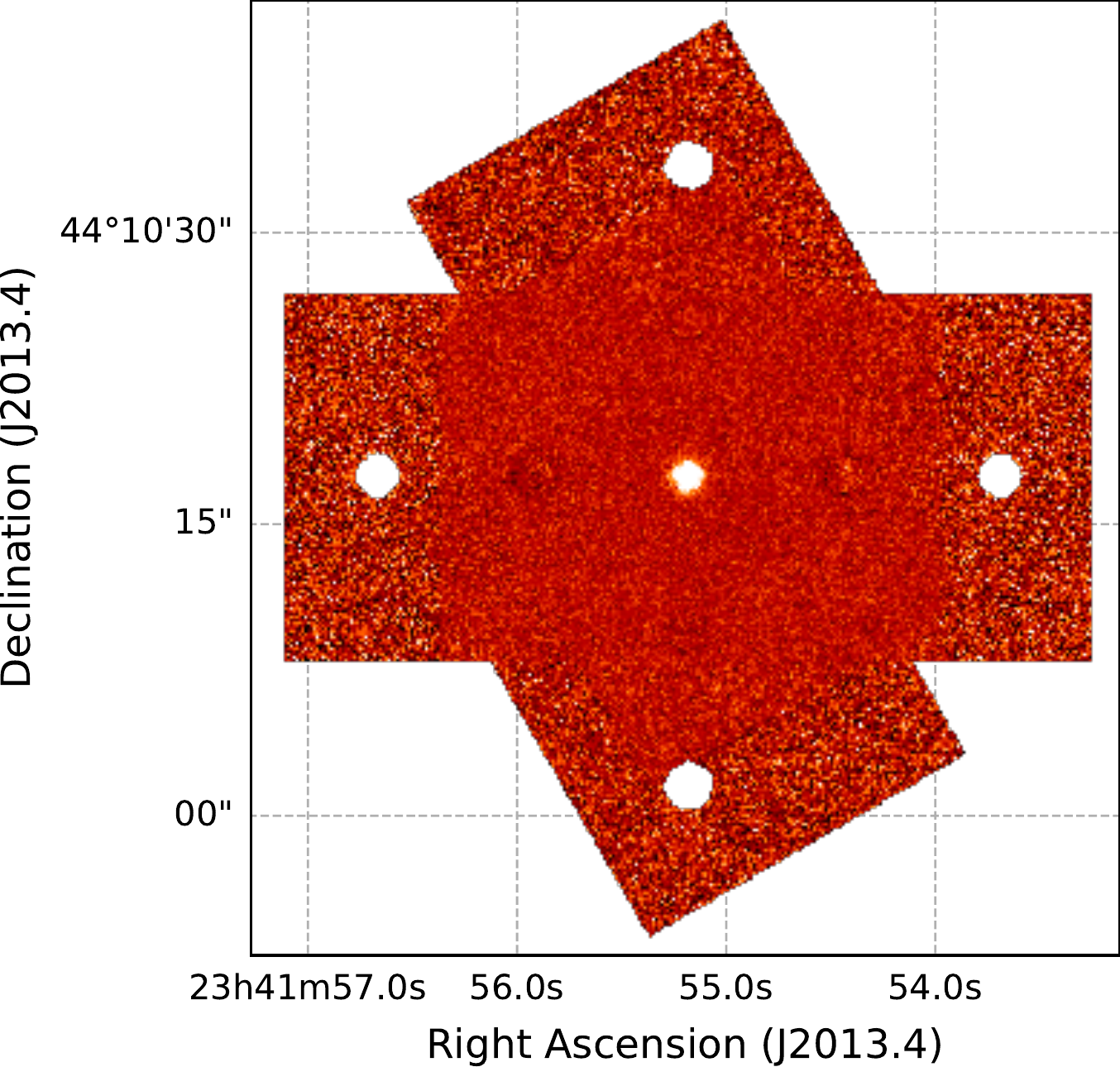}{0.32\textwidth}{(GJ 905)}
          \fig{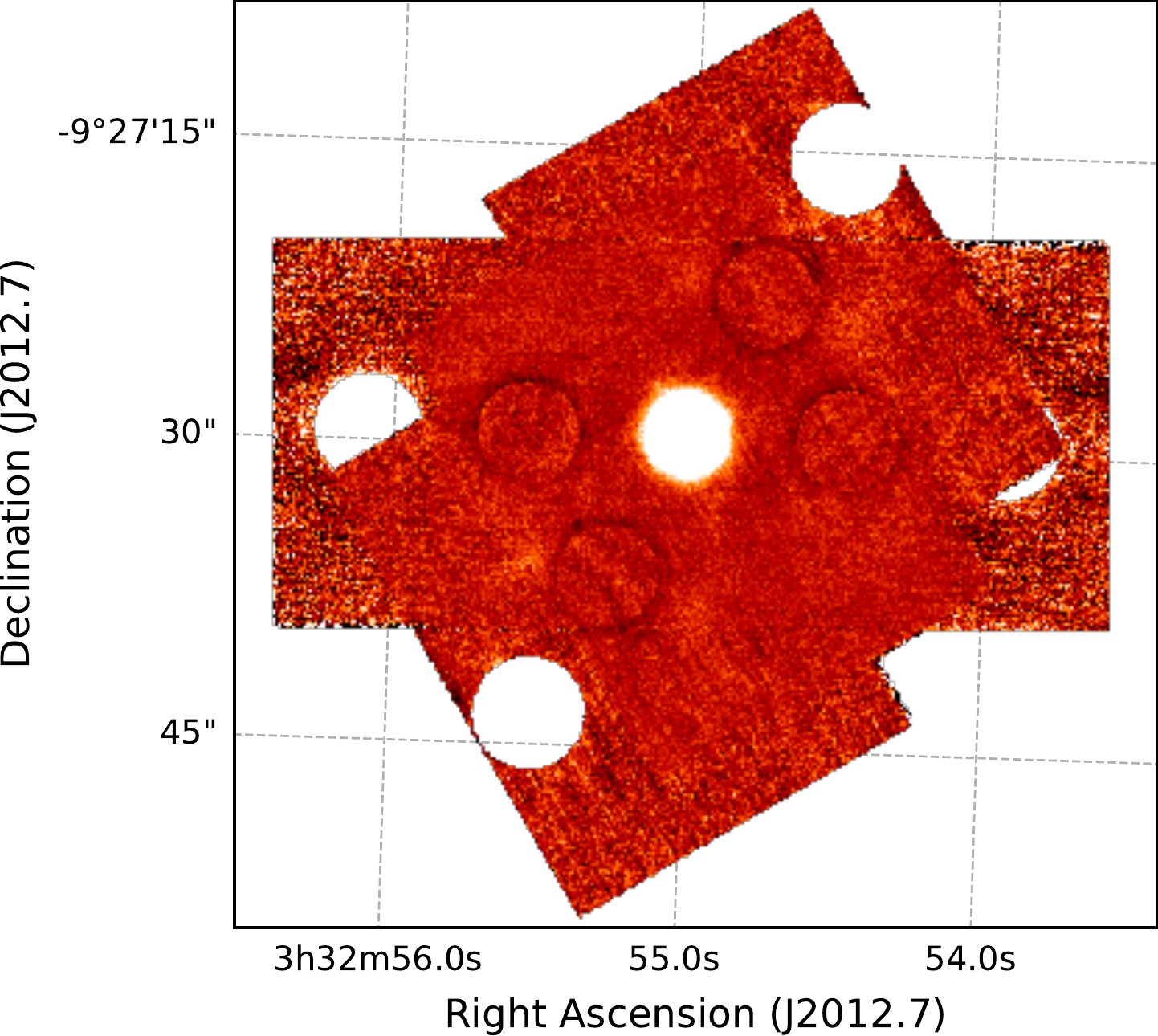}{0.32\textwidth}{(GJ 144)}
          \fig{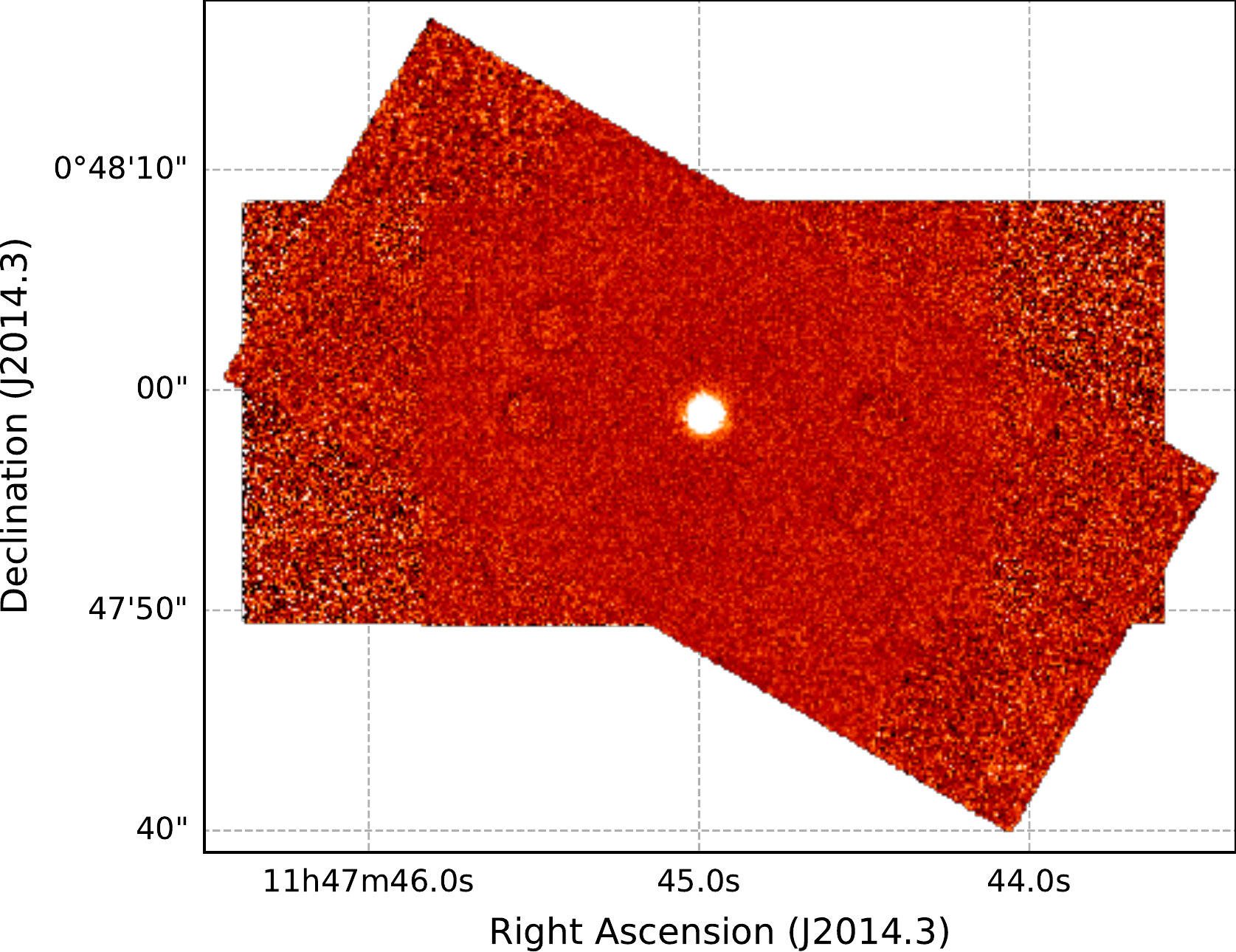}{0.32\textwidth}{(GJ 447)}
          }
\gridline{\fig{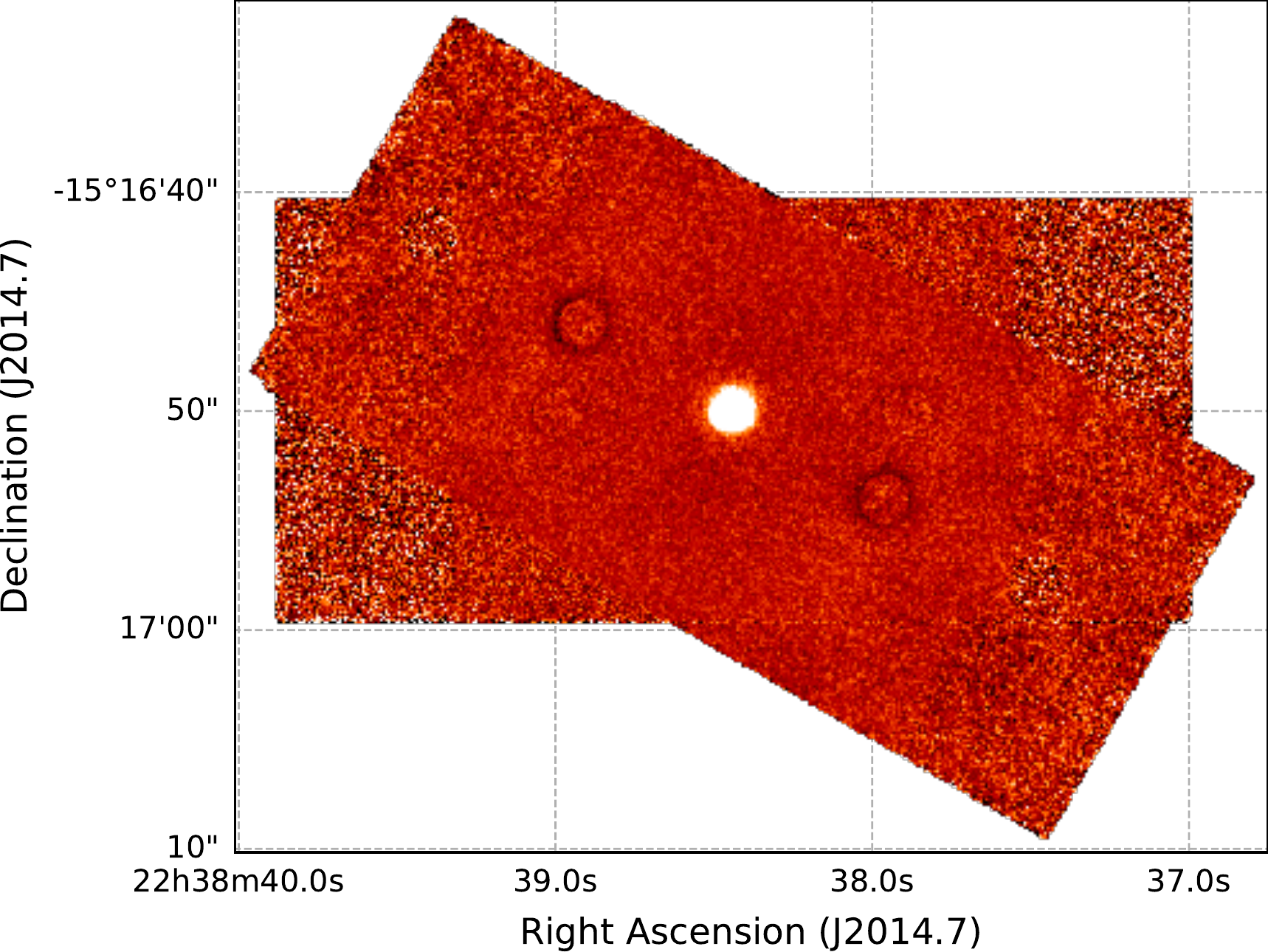}{0.32\textwidth}{(GJ 866)}
          \fig{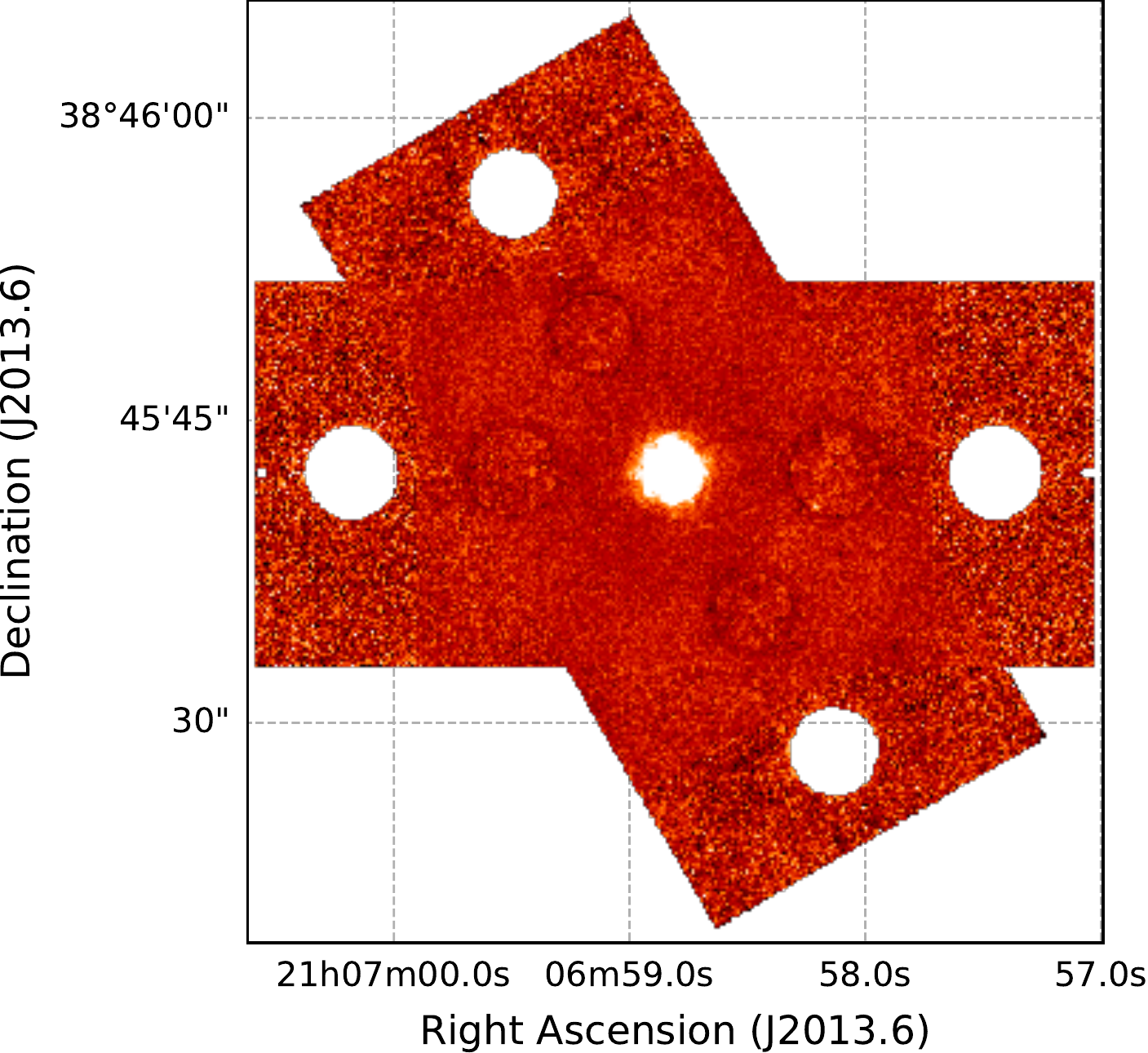}{0.32\textwidth}{(GJ 820A)}
          \fig{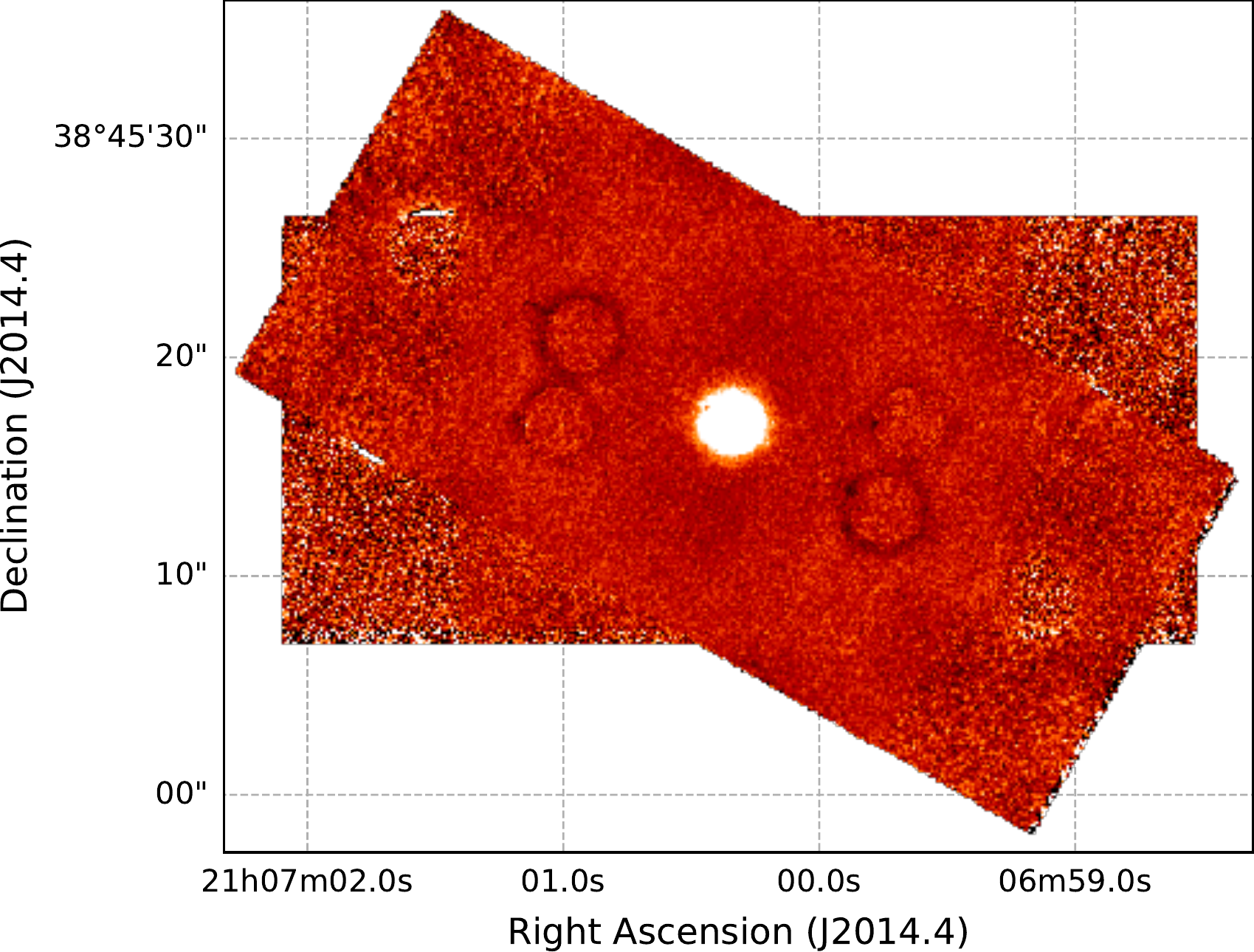}{0.32\textwidth}{(GJ 820B)}
          }          
\end{figure*}        

\begin{figure*}[h!]
\gridline{\fig{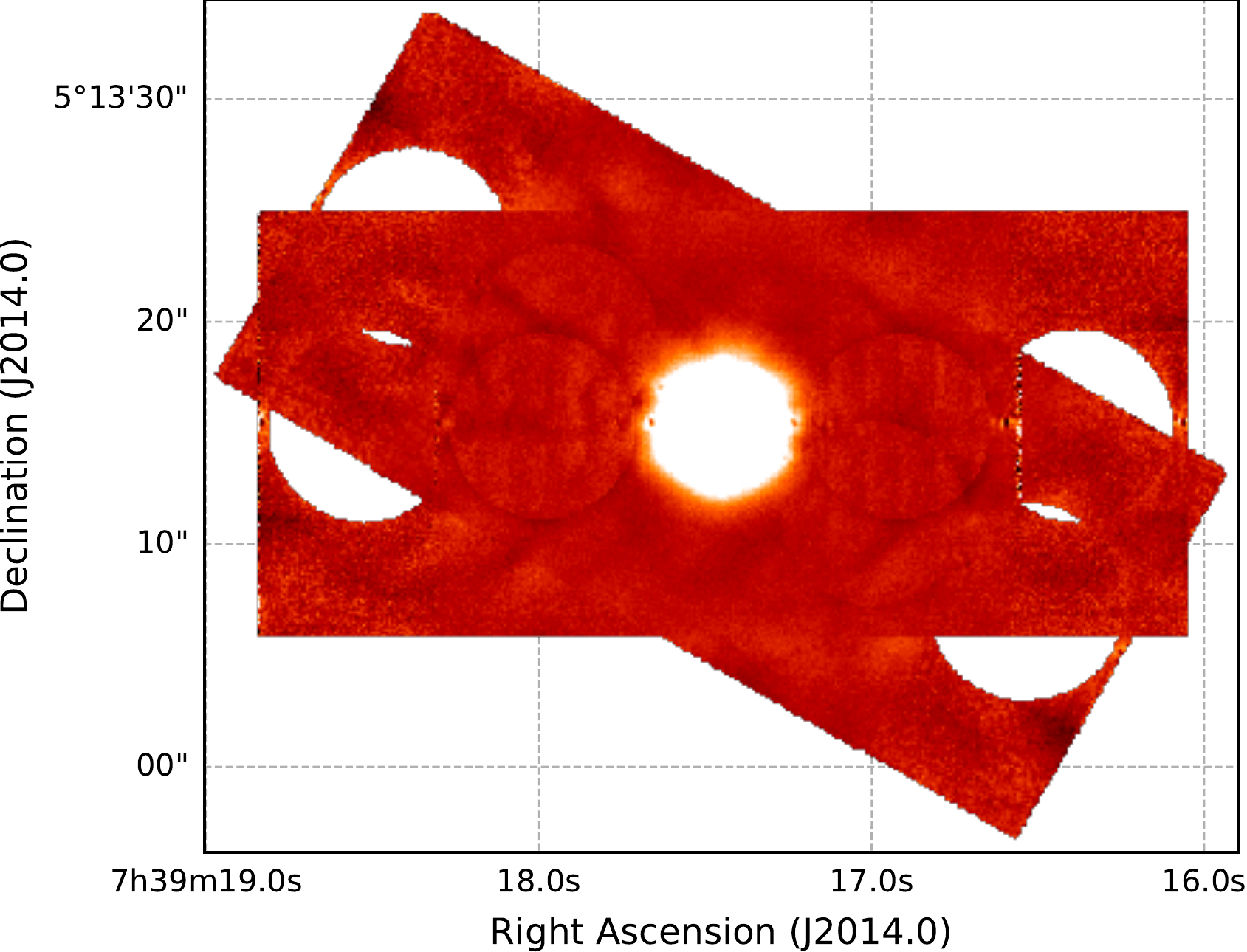}{0.32\textwidth}{(GJ 280)}
          \fig{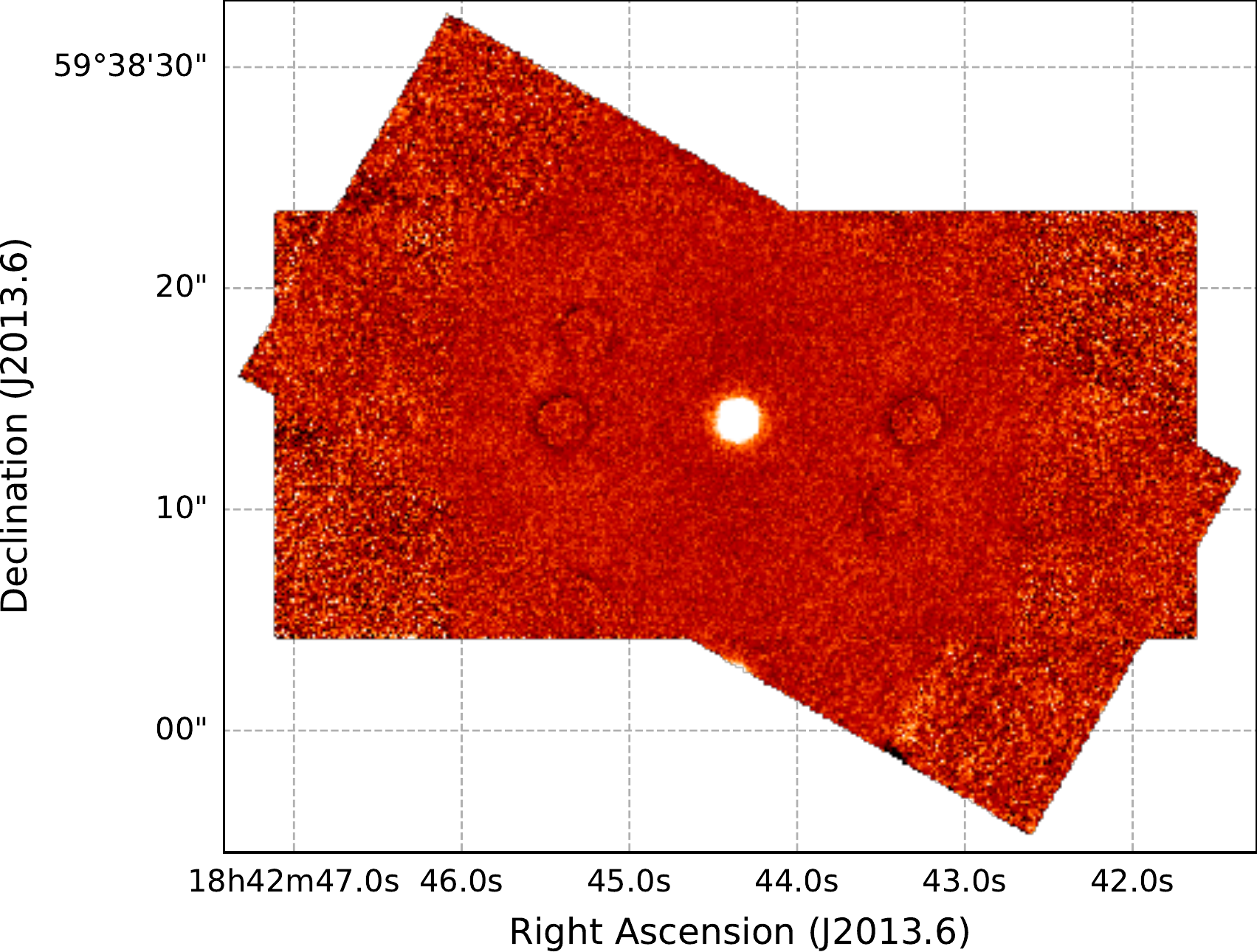}{0.32\textwidth}{(GJ 725A)}
          \fig{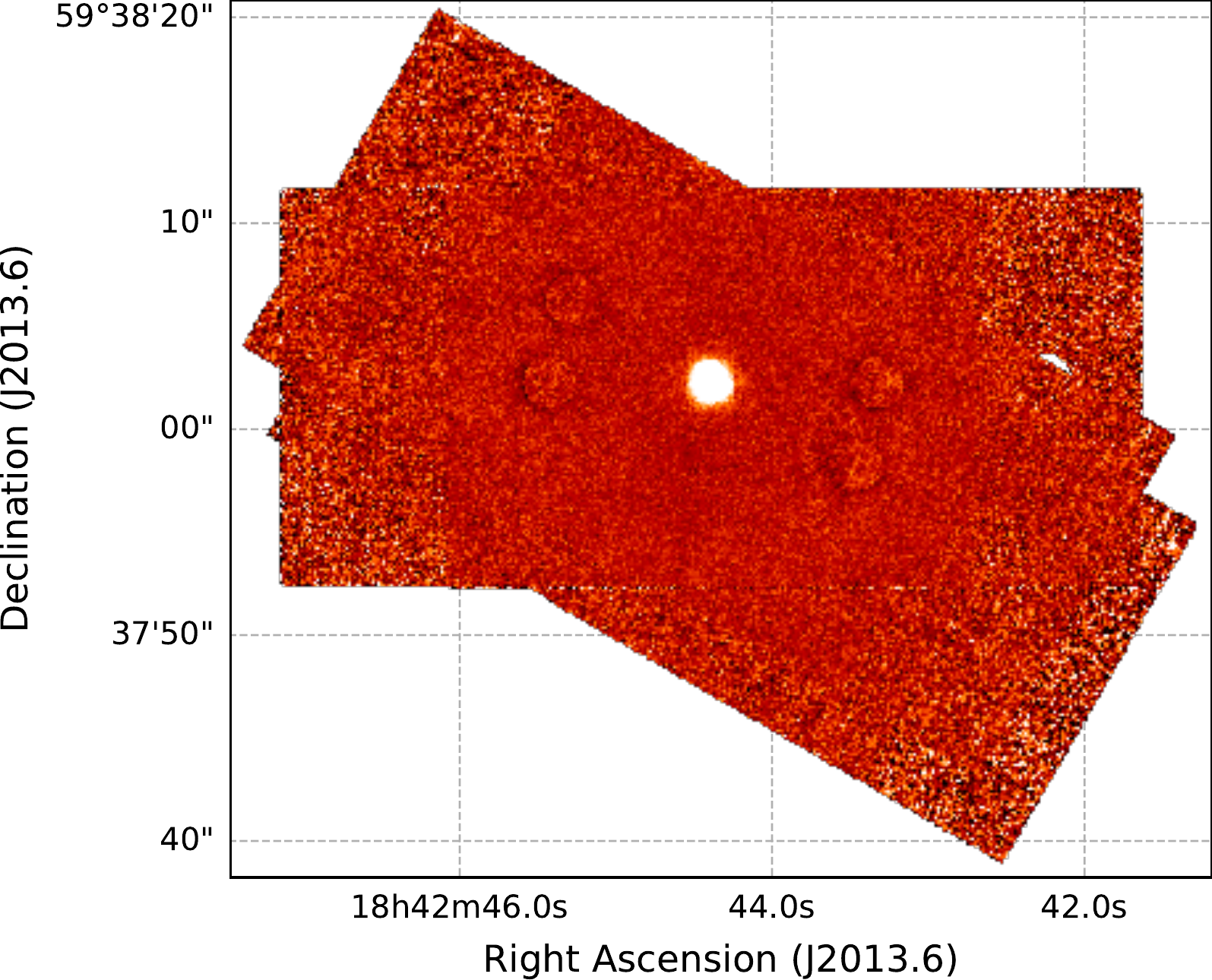}{0.32\textwidth}{(GJ 725B)}
          }
\gridline{\fig{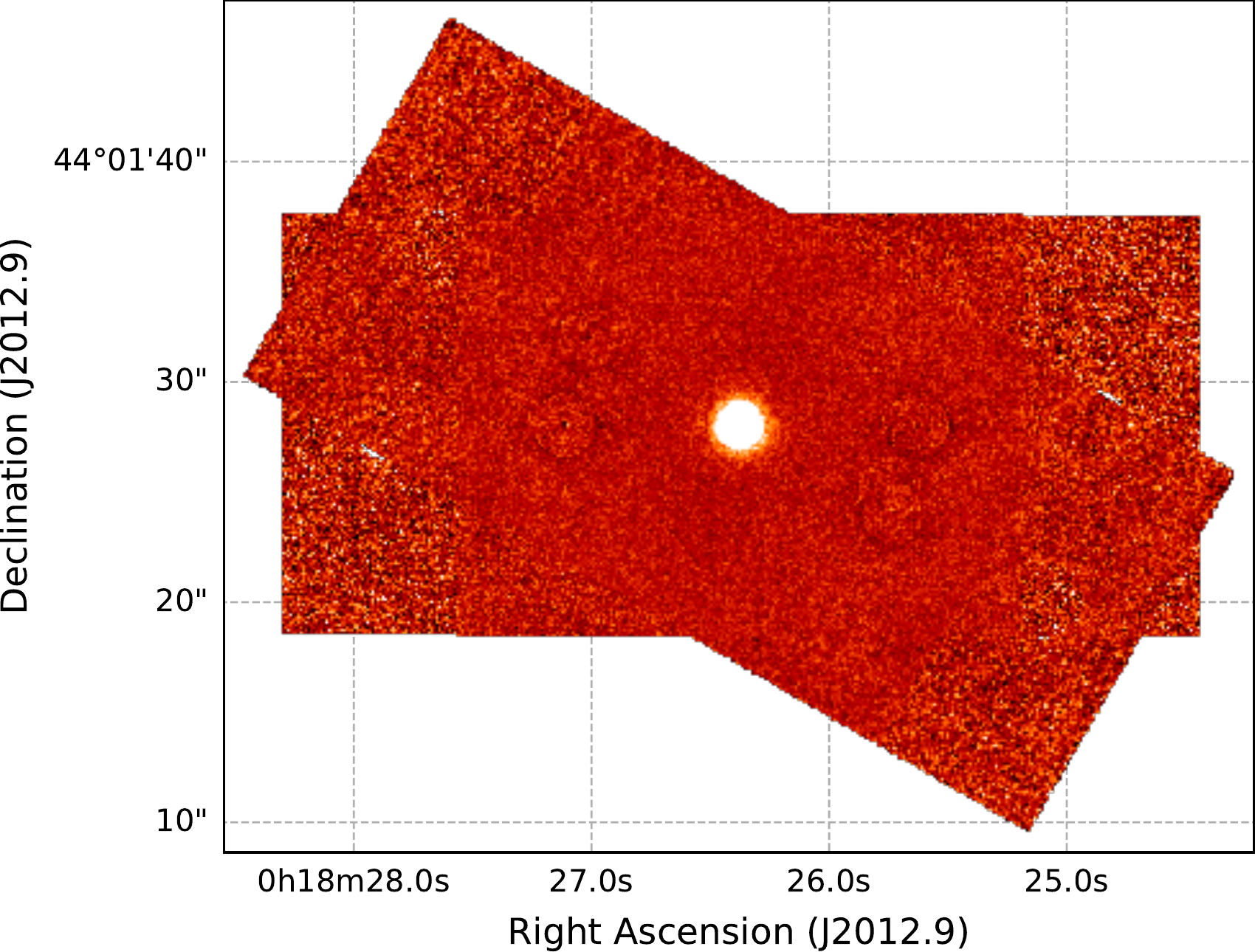}{0.32\textwidth}{(GJ 15A)}
          \fig{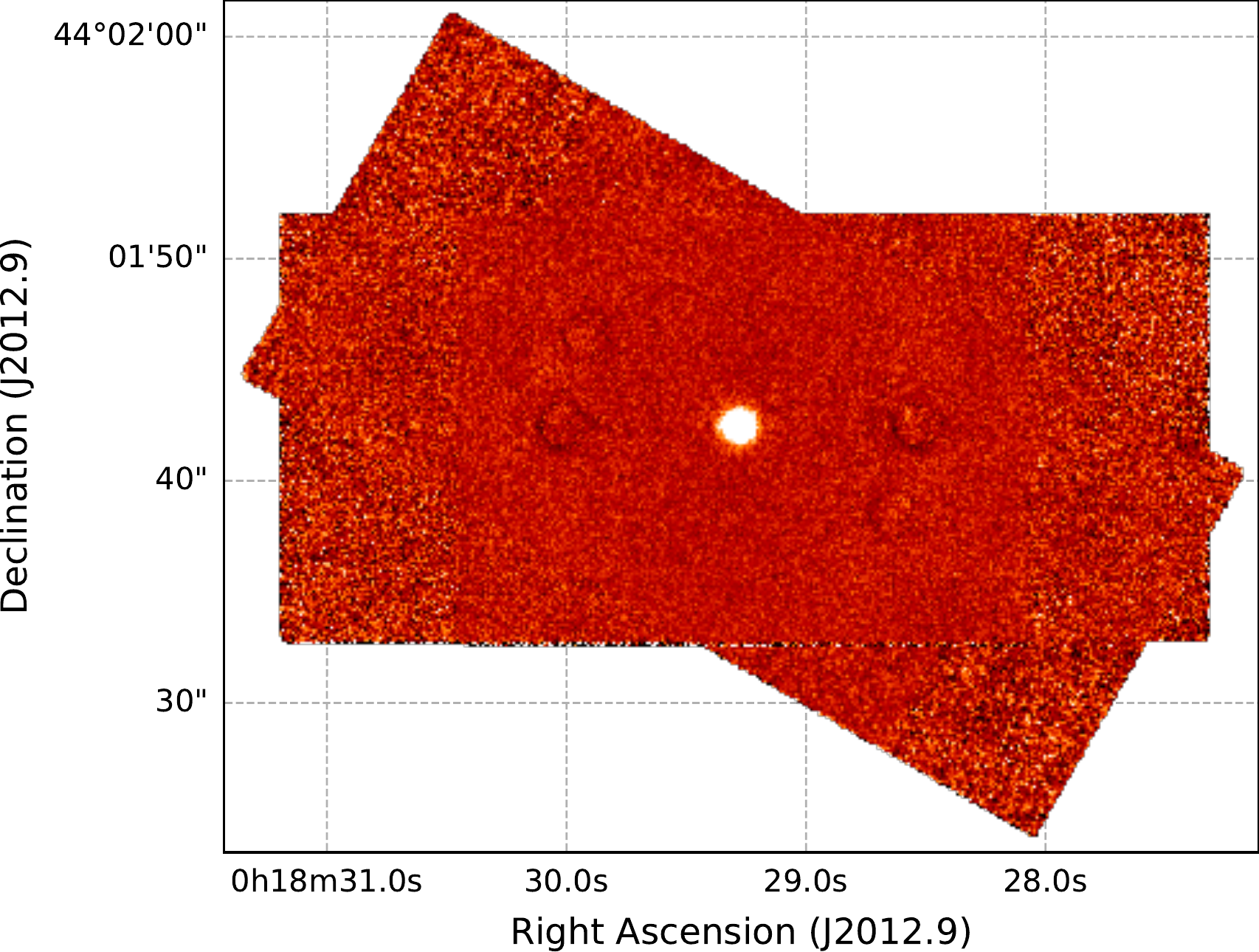}{0.32\textwidth}{(GJ 15B)}
          \fig{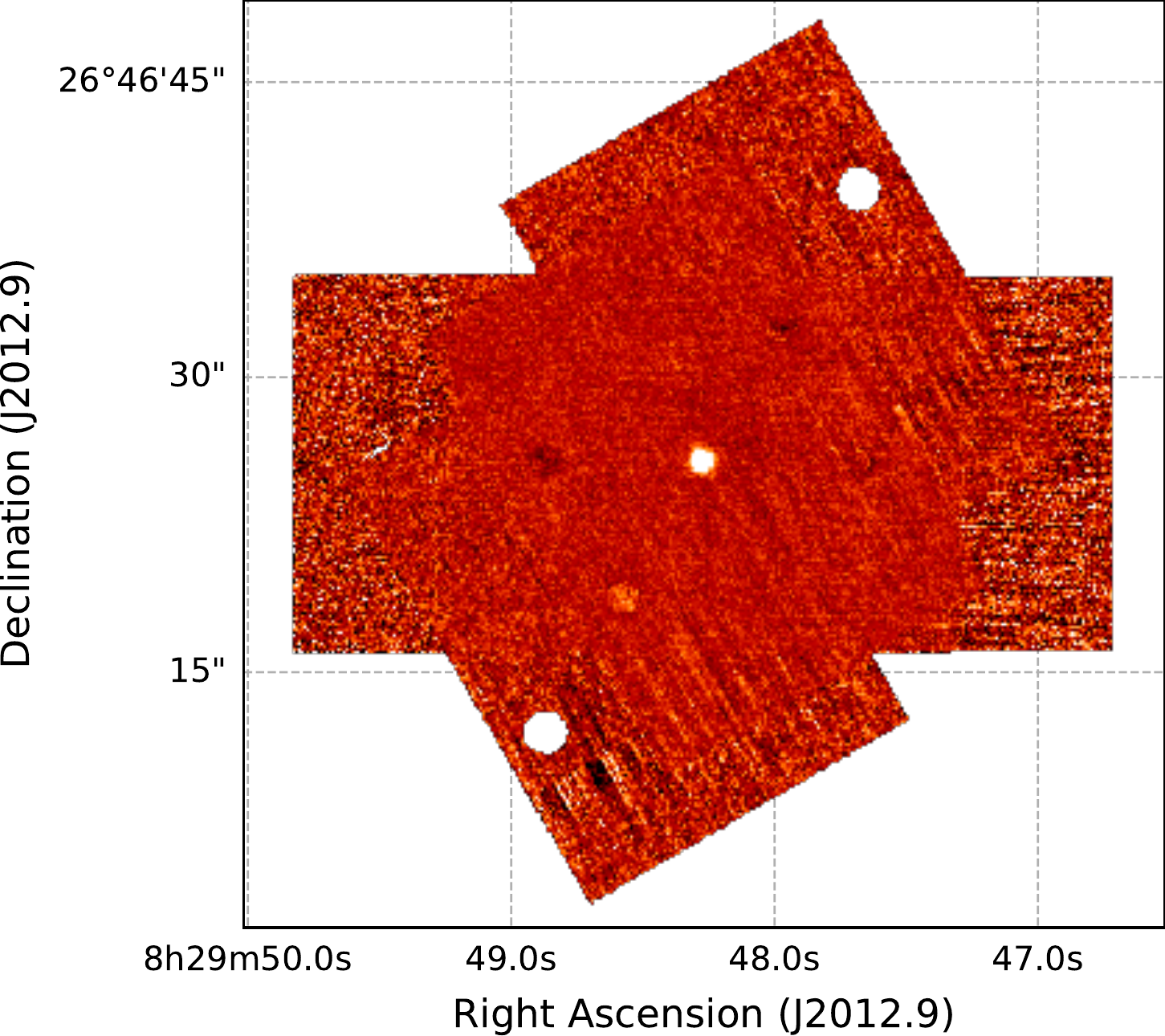}{0.32\textwidth}{(GJ 1111)}
          }
\gridline{\fig{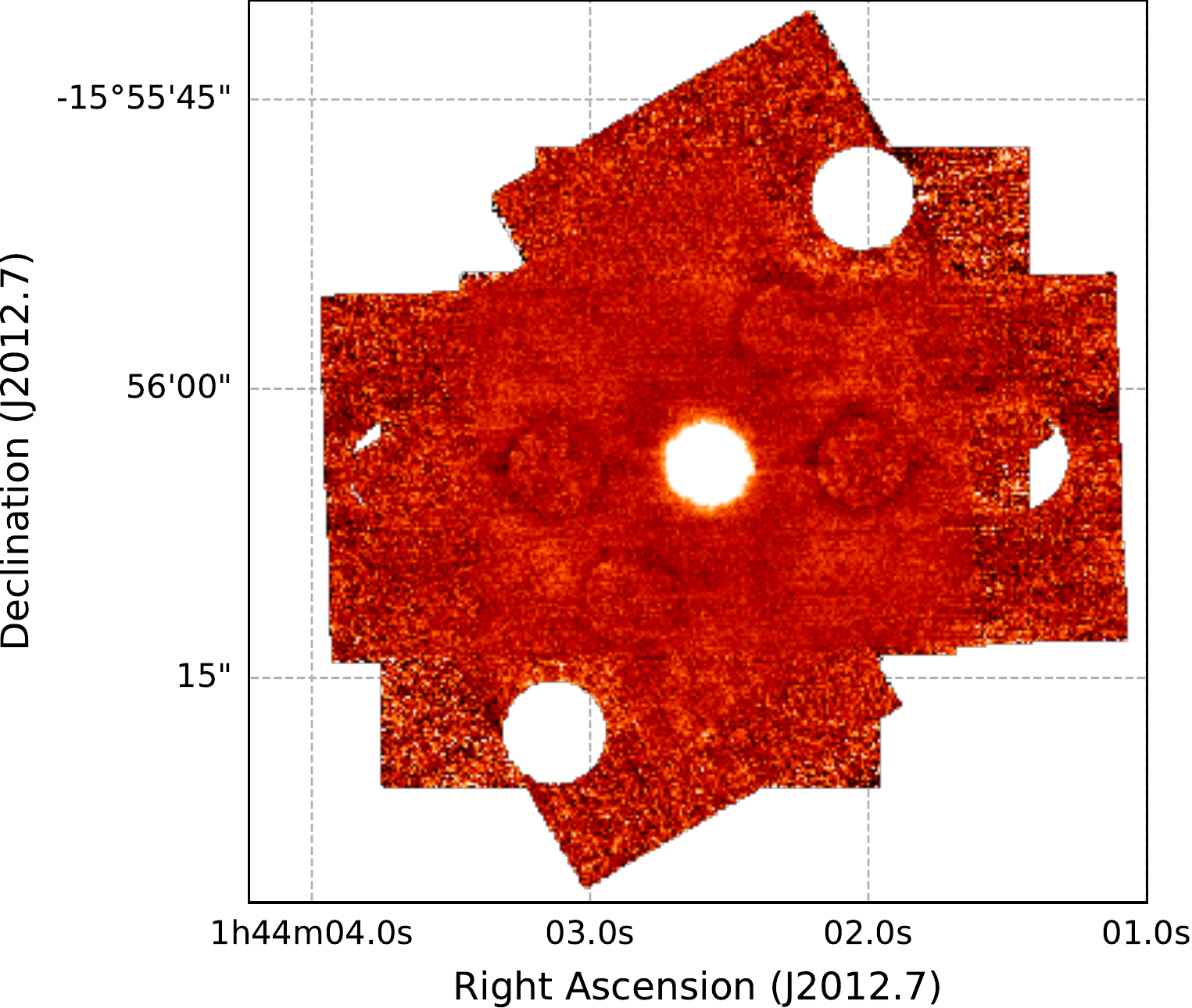}{0.32\textwidth}{(GJ 71)}
          \fig{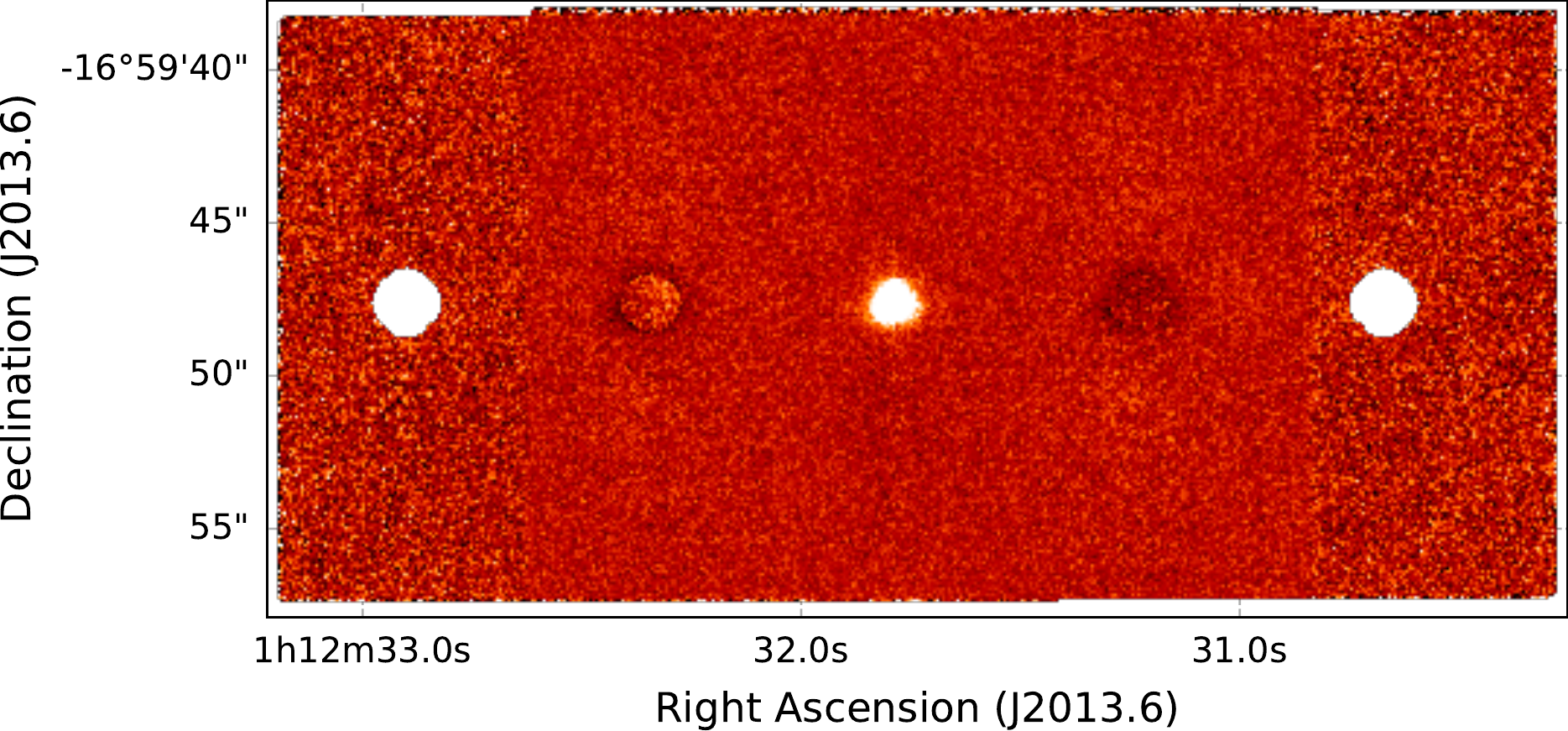}{0.32\textwidth}{(GJ 54.1)}
          \fig{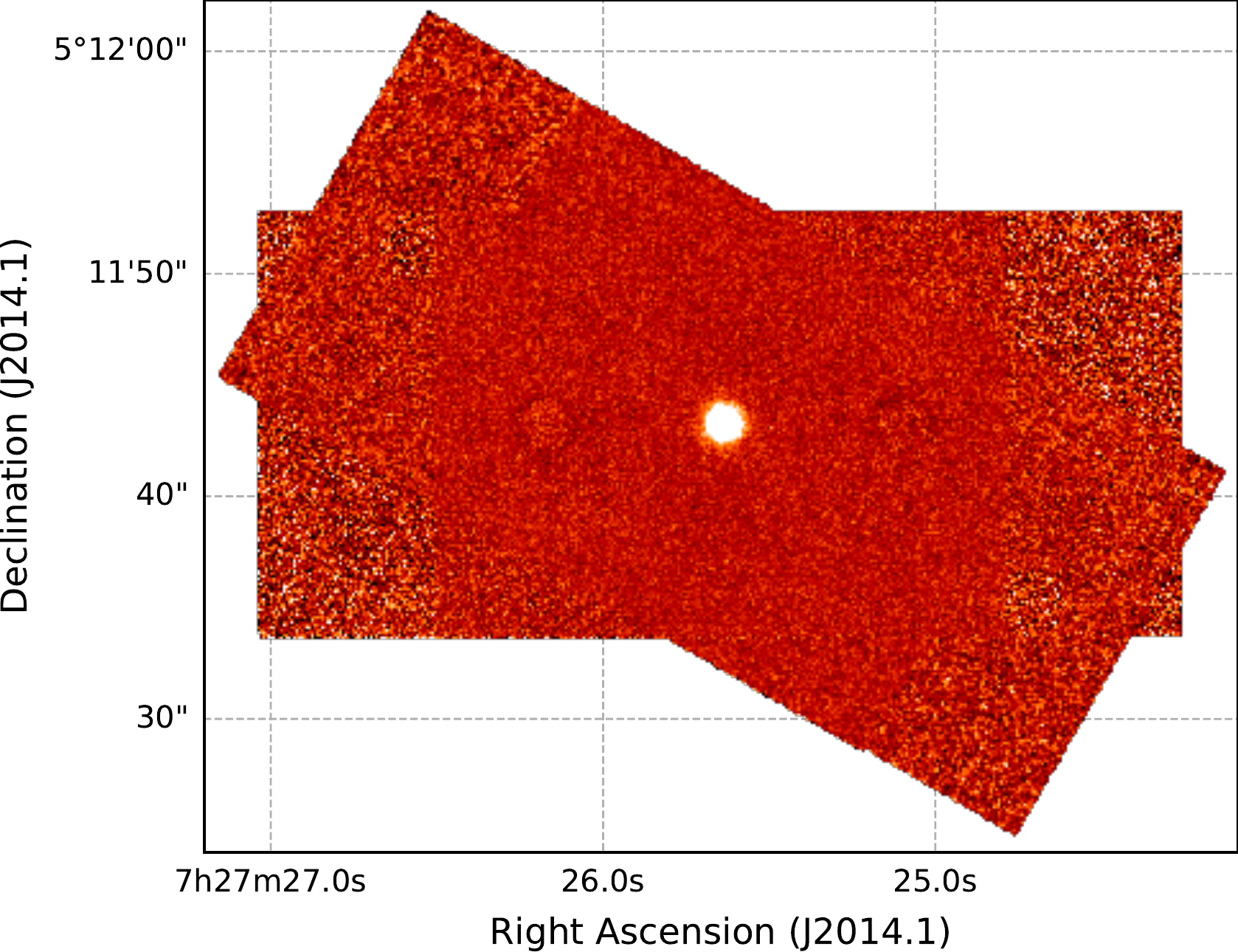}{0.32\textwidth}{(GJ 273)}
          }
\gridline{\fig{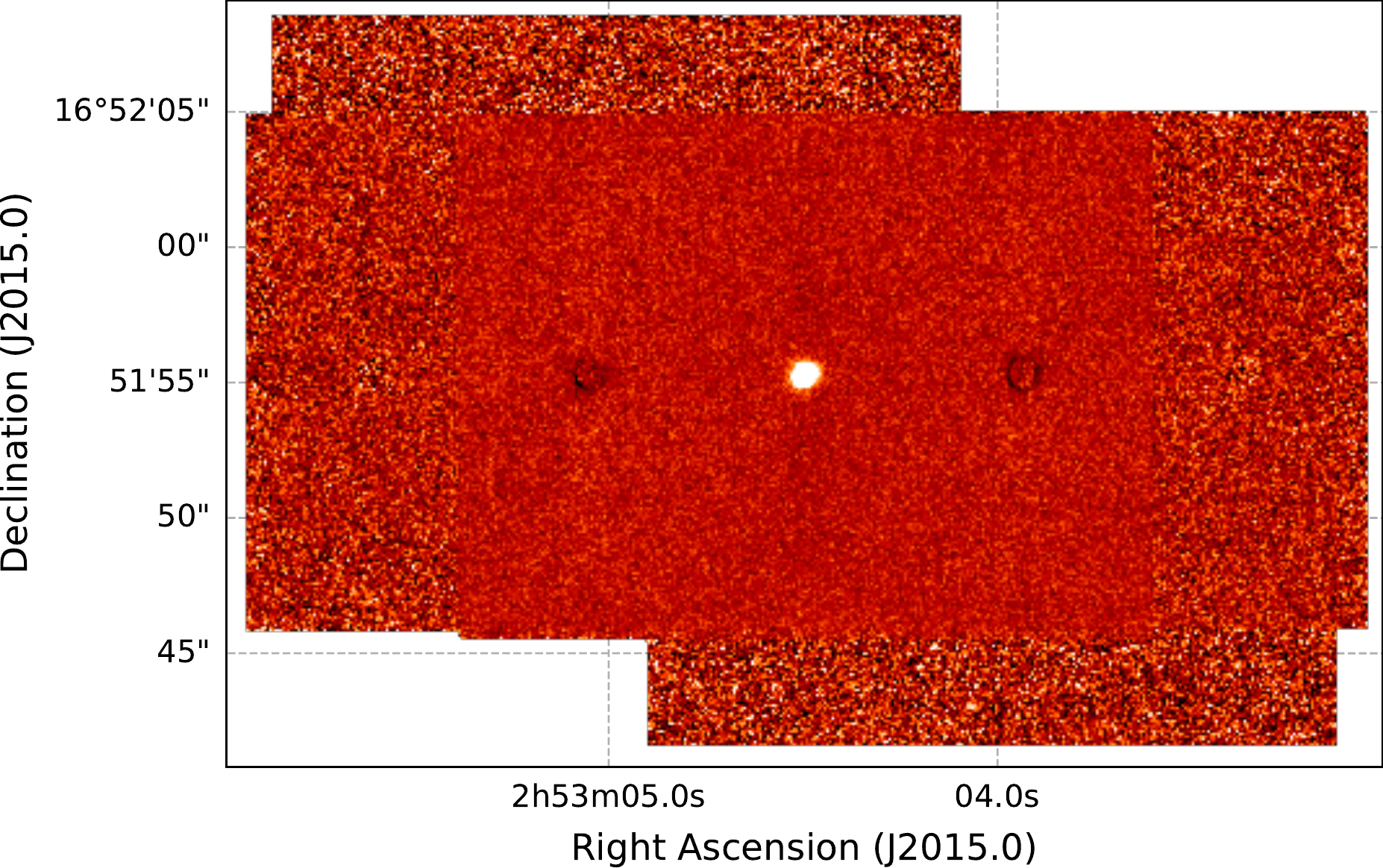}{0.32\textwidth}{(S0 0253+1652)}
          \fig{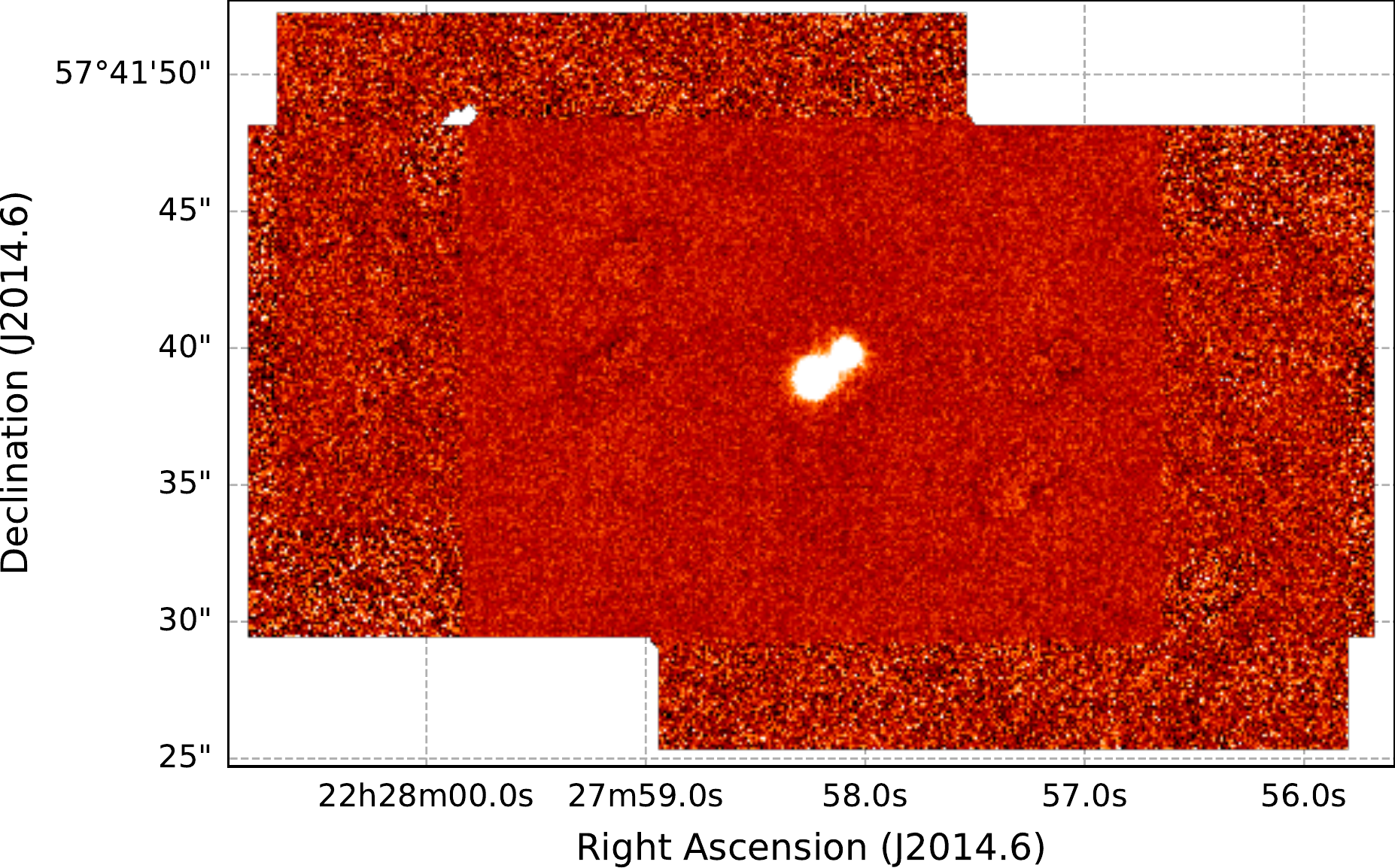}{0.32\textwidth}{(GJ 860AB)}
          \fig{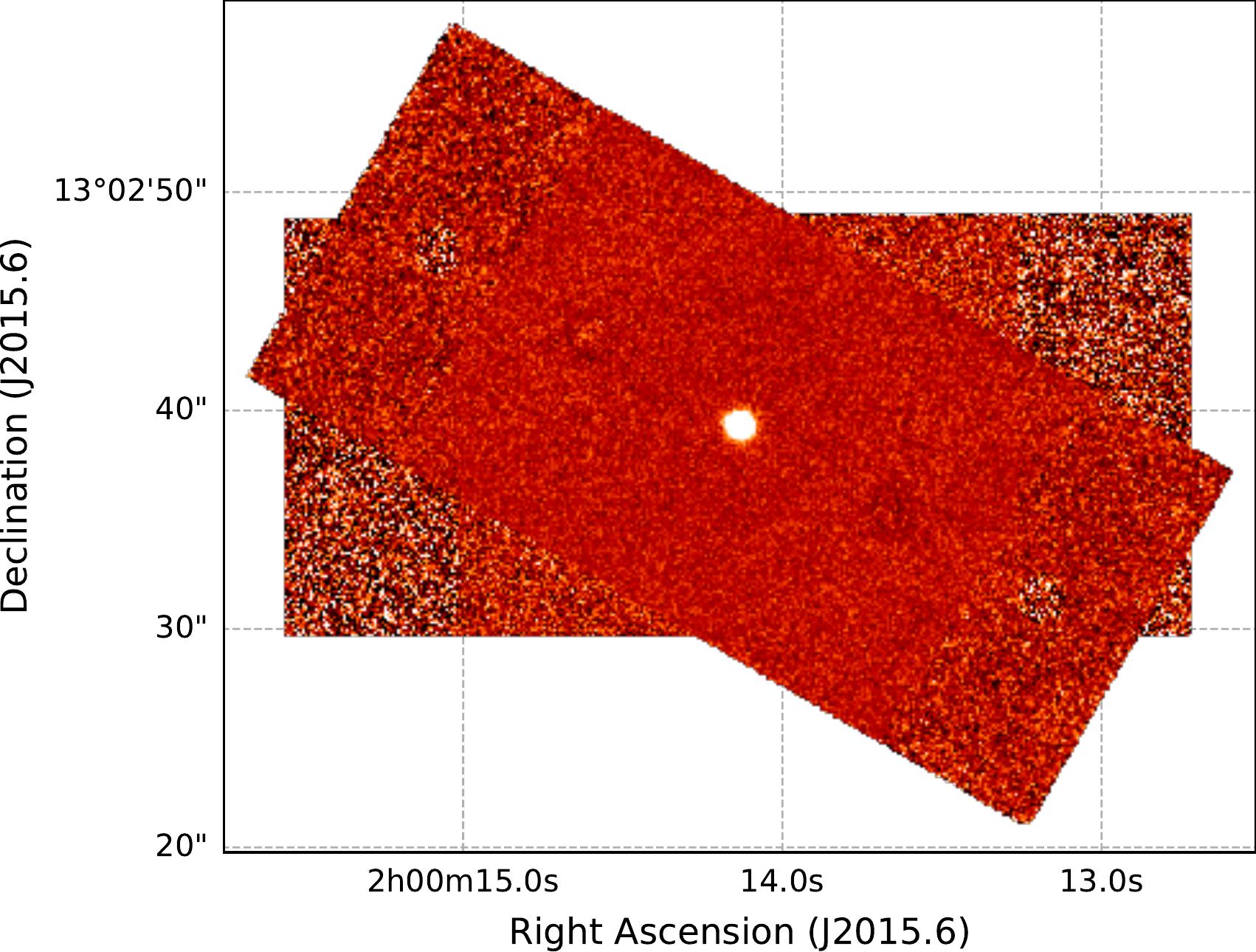}{0.32\textwidth}{(GJ 83.1)}
          }
\end{figure*}

\begin{figure*}[h!]
\gridline{\fig{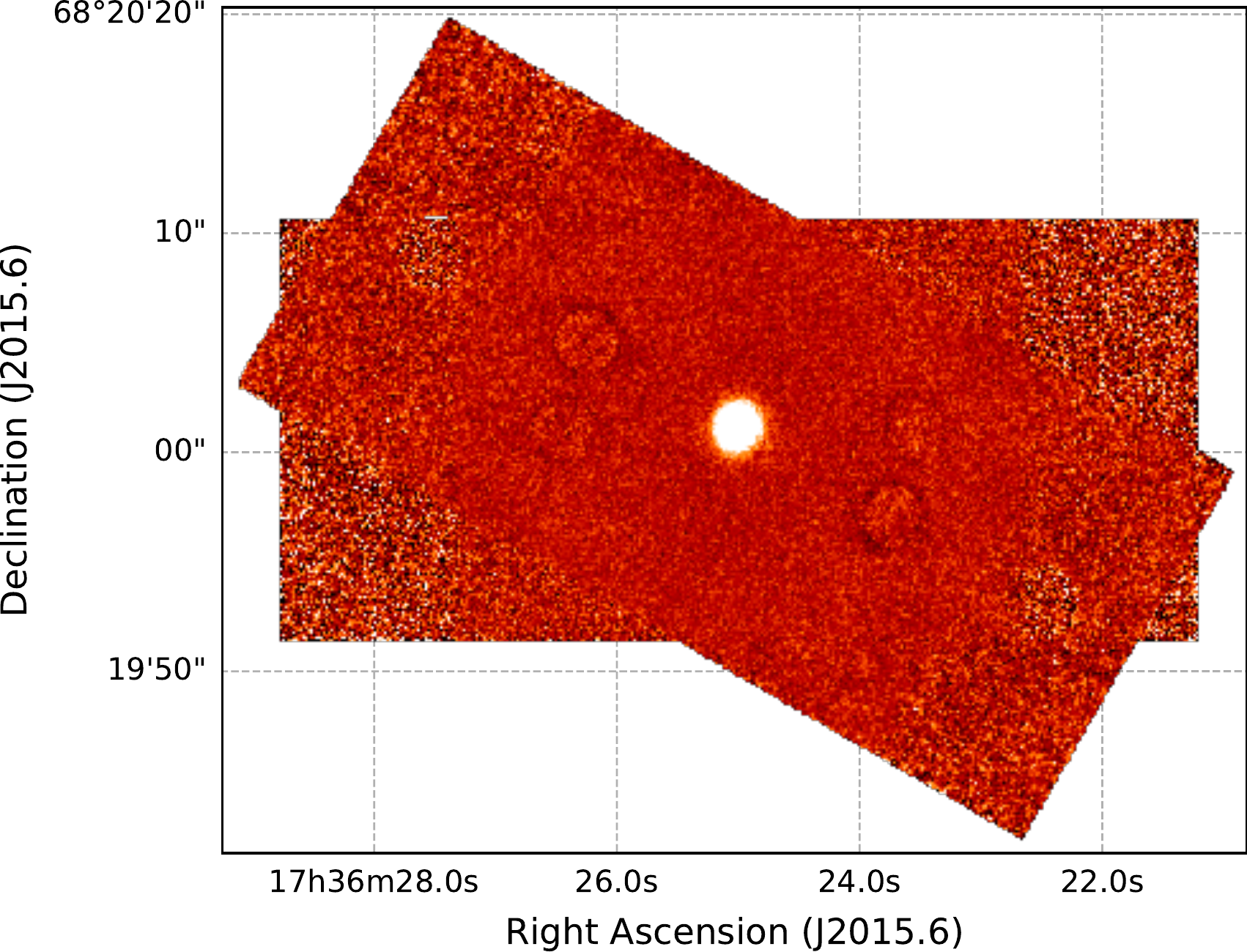}{0.32\textwidth}{(GJ 687)}
          \fig{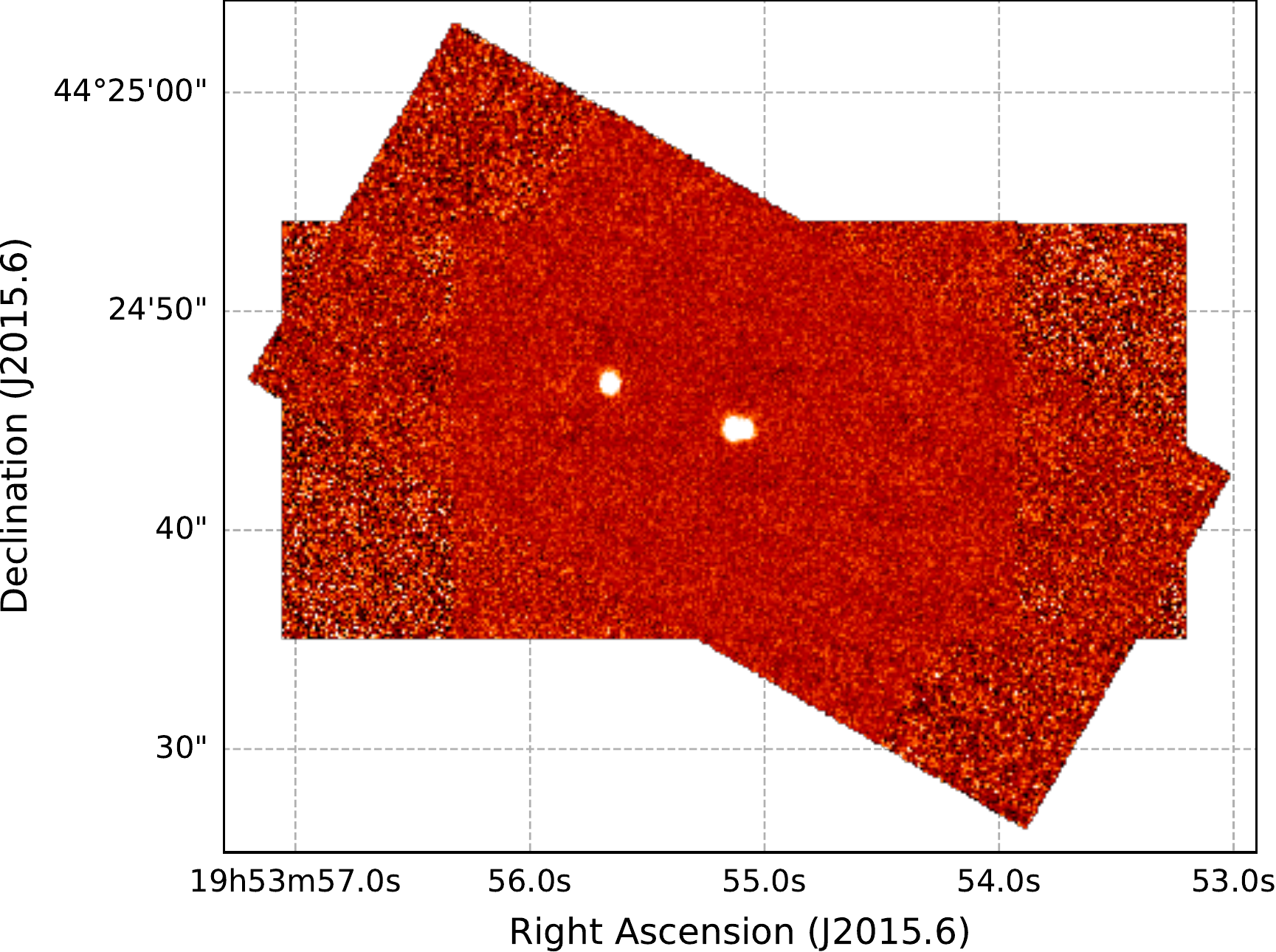}{0.32\textwidth}{(GJ 1245ABC)}
          \fig{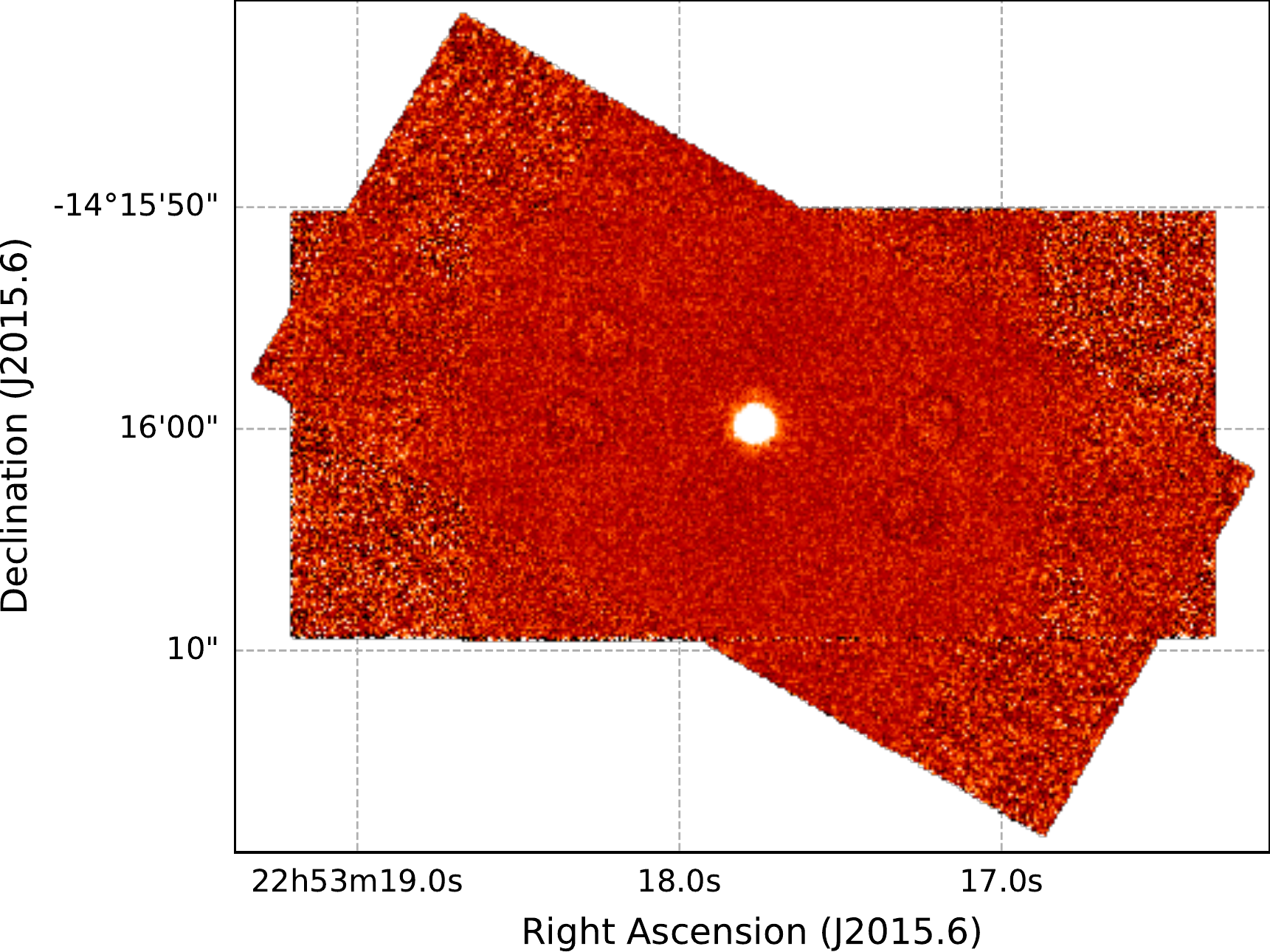}{0.32\textwidth}{(GJ 876)}
          }
\gridline{\fig{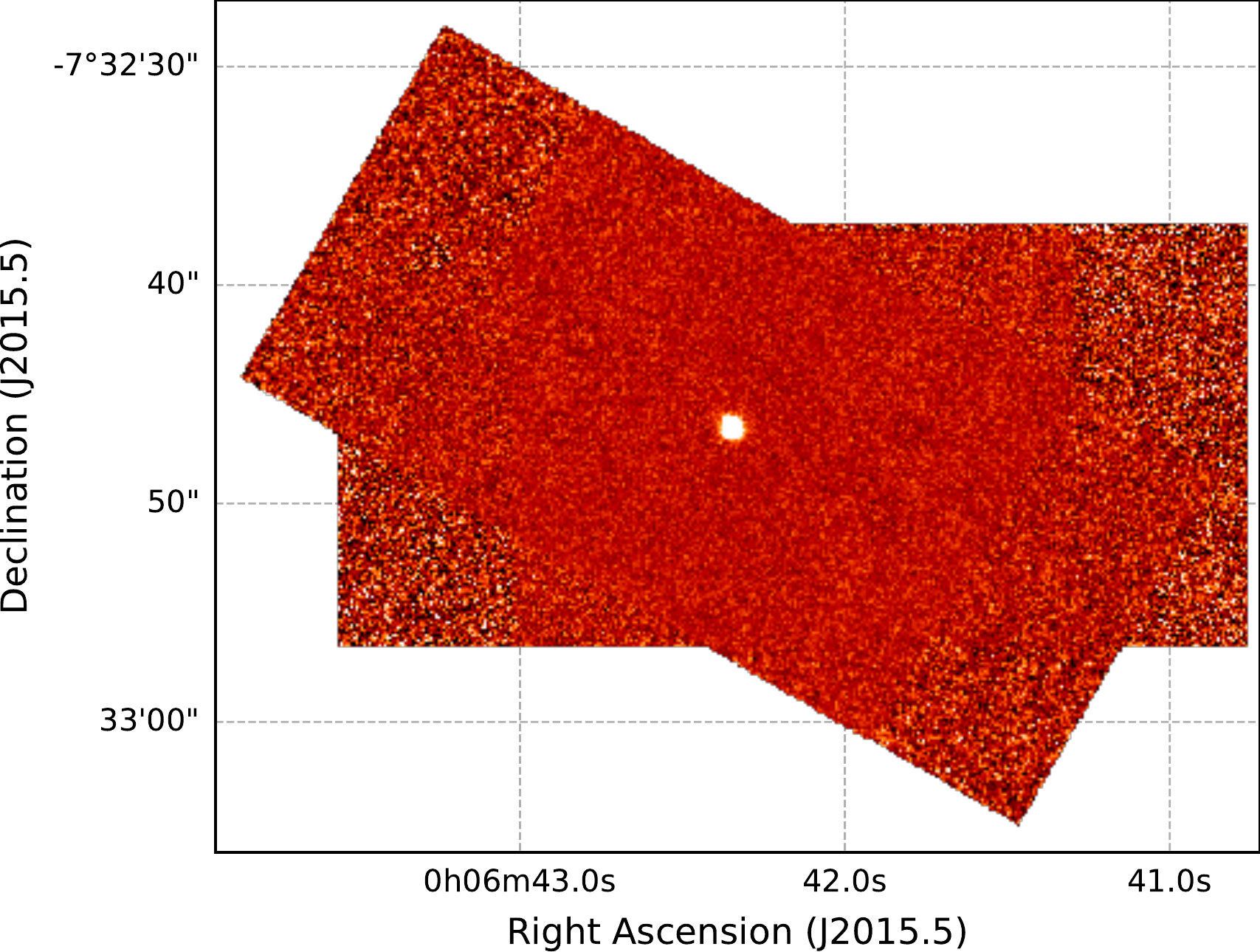}{0.32\textwidth}{(GJ 1002)}
          }
\caption{Same as Fig. B.1 but for full field of view mosaics.
\label{CCallimages_fullFOV}}
\end{figure*}

\clearpage
\bibliography{CC_survey_draft_v6.0}{}

\begin{thebibliography}{}
\expandafter\ifx\csname natexlab\endcsname\relax\def\natexlab#1{#1}\fi
\providecommand{\url}[1]{\href{#1}{#1}}
\providecommand{\dodoi}[1]{doi:~\href{http://doi.org/#1}{\nolinkurl{#1}}}
\providecommand{\doeprint}[1]{\href{http://ascl.net/#1}{\nolinkurl{http://ascl.net/#1}}}
\providecommand{\doarXiv}[1]{\href{https://arxiv.org/abs/#1}{\nolinkurl{https://arxiv.org/abs/#1}}}

\bibitem[{{Allard} {et~al.}(2001){Allard}, {Hauschildt}, {Alexander},
  {Tamanai}, \& {Schweitzer}}]{2001ApJ...556..357A}
{Allard}, F., {Hauschildt}, P.~H., {Alexander}, D.~R., {Tamanai}, A., \&
  {Schweitzer}, A. 2001, \apj, 556, 357, \dodoi{10.1086/321547}

\bibitem[{{Allard} {et~al.}(2012){Allard}, {Homeier}, {Freytag}, \&
  {Sharp}}]{2012EAS....57....3A}
{Allard}, F., {Homeier}, D., {Freytag}, B., \& {Sharp}, C.~M. 2012, in EAS
  Publications Series, Vol.~57, EAS Publications Series, ed. C.~{Reyl{\'e}},
  C.~{Charbonnel}, \& M.~{Schultheis}, 3--43, \dodoi{10.1051/eas/1257001}

\bibitem[{{Alonso-Floriano} {et~al.}(2015){Alonso-Floriano}, {Morales},
  {Caballero}, {Montes}, {Klutsch}, {Mundt}, {Cort{\'e}s-Contreras}, {Ribas},
  {Reiners}, {Amado}, {Quirrenbach}, \& {Jeffers}}]{2015AA...577A.128A}
{Alonso-Floriano}, F.~J., {Morales}, J.~C., {Caballero}, J.~A., {et~al.} 2015,
  \aap, 577, A128, \dodoi{10.1051/0004-6361/201525803}

\bibitem[{{Angus} {et~al.}(2015){Angus}, {Aigrain}, {Foreman-Mackey}, \&
  {McQuillan}}]{2015MNRAS.450.1787A}
{Angus}, R., {Aigrain}, S., {Foreman-Mackey}, D., \& {McQuillan}, A. 2015,
  \mnras, 450, 1787, \dodoi{10.1093/mnras/stv423}

\bibitem[{{Asensio-Torres} {et~al.}(2018){Asensio-Torres}, {Janson},
  {Bonavita}, {Desidera}, {Thalmann}, {Kuzuhara}, {Henning}, {Marzari},
  {Meyer}, {Calissendorff}, \& {Uyama}}]{2018AA...619A..43A}
{Asensio-Torres}, R., {Janson}, M., {Bonavita}, M., {et~al.} 2018, \aap, 619,
  A43, \dodoi{10.1051/0004-6361/201833349}

\bibitem[{{Astudillo-Defru} {et~al.}(2017{\natexlab{a}}){Astudillo-Defru},
  {D{\'\i}az}, {Bonfils}, {Almenara}, {Delisle}, {Bouchy}, {Delfosse},
  {Forveille}, {Lovis}, {Mayor}, {Murgas}, {Pepe}, {Santos}, {S{\'e}gransan},
  {Udry}, \& {W{\"u}nsche}}]{2017AA...605L..11A}
{Astudillo-Defru}, N., {D{\'\i}az}, R.~F., {Bonfils}, X., {et~al.}
  2017{\natexlab{a}}, \aap, 605, L11, \dodoi{10.1051/0004-6361/201731581}

\bibitem[{{Astudillo-Defru} {et~al.}(2017{\natexlab{b}}){Astudillo-Defru},
  {Forveille}, {Bonfils}, {S{\'e}gransan}, {Bouchy}, {Delfosse}, {Lovis},
  {Mayor}, {Murgas}, {Pepe}, {Santos}, {Udry}, \&
  {W{\"u}nsche}}]{2017AA...602A..88A}
{Astudillo-Defru}, N., {Forveille}, T., {Bonfils}, X., {et~al.}
  2017{\natexlab{b}}, \aap, 602, A88, \dodoi{10.1051/0004-6361/201630153}

\bibitem[{{Ayres}(1991)}]{1991ApJ...375..704A}
{Ayres}, T.~R. 1991, \apj, 375, 704, \dodoi{10.1086/170236}

\bibitem[{{Ballantyne} {et~al.}(2021){Ballantyne}, {Espaas}, {Norgrove},
  {Wootton}, {Harris}, {Pepper}, {Smith}, {Dommett}, \&
  {Parker}}]{2021MNRAS.tmp.2160B}
{Ballantyne}, H.~A., {Espaas}, T., {Norgrove}, B.~Z., {et~al.} 2021, \mnras,
  \dodoi{10.1093/mnras/stab2324}

\bibitem[{{Baraffe} {et~al.}(2003){Baraffe}, {Chabrier}, {Barman}, {Allard}, \&
  {Hauschildt}}]{2003A&A...402..701B}
{Baraffe}, I., {Chabrier}, G., {Barman}, T.~S., {Allard}, F., \& {Hauschildt},
  P.~H. 2003, \aap, 402, 701, \dodoi{10.1051/0004-6361:20030252}

\bibitem[{{Barnes} {et~al.}(2017){Barnes}, {Jeffers}, {Haswell}, {Jones},
  {Shulyak}, {Pavlenko}, \& {Jenkins}}]{2017MNRAS.471..811B}
{Barnes}, J.~R., {Jeffers}, S.~V., {Haswell}, C.~A., {et~al.} 2017, \mnras,
  471, 811, \dodoi{10.1093/mnras/stx1482}

\bibitem[{{Barnes}(2007)}]{2007ApJ...669.1167B}
{Barnes}, S.~A. 2007, \apj, 669, 1167, \dodoi{10.1086/519295}

\bibitem[{{Baron} {et~al.}(2019){Baron}, {Lafreni{\`e}re}, {Artigau},
  {Gagn{\'e}}, {Rameau}, {Delorme}, \& {Naud}}]{2019AJ....158..187B}
{Baron}, F., {Lafreni{\`e}re}, D., {Artigau}, {\'E}., {et~al.} 2019, \aj, 158,
  187, \dodoi{10.3847/1538-3881/ab4130}

\bibitem[{{Berdi{\~n}as} {et~al.}(2016){Berdi{\~n}as}, {Amado},
  {Anglada-Escud{\'e}}, {Rodr{\'\i}guez-L{\'o}pez}, \&
  {Barnes}}]{2016MNRAS.459.3551B}
{Berdi{\~n}as}, Z.~M., {Amado}, P.~J., {Anglada-Escud{\'e}}, G.,
  {Rodr{\'\i}guez-L{\'o}pez}, C., \& {Barnes}, J. 2016, \mnras, 459, 3551,
  \dodoi{10.1093/mnras/stw906}

\bibitem[{{Bergfors} {et~al.}(2013){Bergfors}, {Brandner}, {Daemgen}, {Biller},
  {Hippler}, {Janson}, {Kudryavtseva}, {Gei{\ss}ler}, {Henning}, \&
  {K{\"o}hler}}]{2013MNRAS.428..182B}
{Bergfors}, C., {Brandner}, W., {Daemgen}, S., {et~al.} 2013, \mnras, 428, 182,
  \dodoi{10.1093/mnras/sts019}

\bibitem[{{Biller} {et~al.}(2013){Biller}, {Liu}, {Wahhaj}, {Nielsen},
  {Hayward}, {Males}, {Skemer}, {Close}, {Chun}, {Ftaclas}, {Clarke}, {Thatte},
  {Shkolnik}, {Reid}, {Hartung}, {Boss}, {Lin}, {Alencar}, {de Gouveia Dal
  Pino}, {Gregorio-Hetem}, \& {Toomey}}]{2013ApJ...777..160B}
{Biller}, B.~A., {Liu}, M.~C., {Wahhaj}, Z., {et~al.} 2013, \apj, 777, 160,
  \dodoi{10.1088/0004-637X/777/2/160}

\bibitem[{{Bonavita} \& {Desidera}(2007)}]{2007AA...468..721B}
{Bonavita}, M., \& {Desidera}, S. 2007, \aap, 468, 721,
  \dodoi{10.1051/0004-6361:20066671}

\bibitem[{{Bonavita} {et~al.}(2016){Bonavita}, {Desidera}, {Thalmann},
  {Janson}, {Vigan}, {Chauvin}, \& {Lannier}}]{2016AA...593A..38B}
{Bonavita}, M., {Desidera}, S., {Thalmann}, C., {et~al.} 2016, \aap, 593, A38,
  \dodoi{10.1051/0004-6361/201628231}

\bibitem[{{Bond} {et~al.}(2015){Bond}, {Gilliland}, {Schaefer}, {Demarque},
  {Girard}, {Holberg}, {Gudehus}, {Mason}, {Kozhurina-Platais}, {Burleigh},
  {Barstow}, \& {Nelan}}]{2015ApJ...813..106B}
{Bond}, H.~E., {Gilliland}, R.~L., {Schaefer}, G.~H., {et~al.} 2015, \apj, 813,
  106, \dodoi{10.1088/0004-637X/813/2/106}

\bibitem[{{Bond} {et~al.}(2017){Bond}, {Schaefer}, {Gilliland}, {Holberg},
  {Mason}, {Lindenblad}, {Seitz-McLeese}, {Arnett}, {Demarque}, {Spada},
  {Young}, {Barstow}, {Burleigh}, \& {Gudehus}}]{2017ApJ...840...70B}
{Bond}, H.~E., {Schaefer}, G.~H., {Gilliland}, R.~L., {et~al.} 2017, \apj, 840,
  70, \dodoi{10.3847/1538-4357/aa6af8}

\bibitem[{{Bonfils} {et~al.}(2013){Bonfils}, {Delfosse}, {Udry}, {Forveille},
  {Mayor}, {Perrier}, {Bouchy}, {Gillon}, {Lovis}, {Pepe}, {Queloz}, {Santos},
  {S{\'e}gransan}, \& {Bertaux}}]{2013AA...549A.109B}
{Bonfils}, X., {Delfosse}, X., {Udry}, S., {et~al.} 2013, \aap, 549, A109,
  \dodoi{10.1051/0004-6361/201014704}

\bibitem[{{Bonfils} {et~al.}(2018){Bonfils}, {Astudillo-Defru}, {D{\'\i}az},
  {Almenara}, {Forveille}, {Bouchy}, {Delfosse}, {Lovis}, {Mayor}, {Murgas},
  {Pepe}, {Santos}, {S{\'e}gransan}, {Udry}, \&
  {W{\"u}nsche}}]{2018AA...613A..25B}
{Bonfils}, X., {Astudillo-Defru}, N., {D{\'\i}az}, R., {et~al.} 2018, \aap,
  613, A25, \dodoi{10.1051/0004-6361/201731973}

\bibitem[{{Bonnefoy} {et~al.}(2016){Bonnefoy}, {Zurlo}, {Baudino}, {Lucas},
  {Mesa}, {Maire}, {Vigan}, {Galicher}, {Homeier}, {Marocco}, {Gratton},
  {Chauvin}, {Allard}, {Desidera}, {Kasper}, {Moutou}, {Lagrange}, {Antichi},
  {Baruffolo}, {Baudrand }, {Beuzit}, {Boccaletti}, {Cantalloube}, {Carbillet},
  {Charton}, {Claudi}, {Costille}, {Dohlen}, {Dominik}, {Fantinel},
  {Feautrier}, {Feldt}, {Fusco}, {Gigan}, {Girard}, {Gluck}, {Gry}, {Henning},
  {Janson}, {Langlois}, {Madec}, {Magnard}, {Maurel}, {Mawet}, {Meyer},
  {Milli}, {Moeller-Nilsson}, {Mouillet}, {Pavlov}, {Perret}, {Pujet}, {Quanz},
  {Rochat}, {Rousset}, {Roux}, {Salasnich}, {Salter}, {Sauvage}, {Schmid},
  {Sevin}, {Soenke}, {Stadler}, {Turatto}, {Udry}, {Vakili}, {Wahhaj}, \&
  {Wildi}}]{2016A&A...587A..58B}
{Bonnefoy}, M., {Zurlo}, A., {Baudino}, J.~L., {et~al.} 2016, \aap, 587, A58,
  \dodoi{10.1051/0004-6361/201526906}

\bibitem[{{Bonnefoy} {et~al.}(2018){Bonnefoy}, {Perraut}, {Lagrange},
  {Delorme}, {Vigan}, {Line}, {Rodet}, {Ginski}, {Mourard}, {Marleau},
  {Samland}, {Tremblin}, {Ligi}, {Cantalloube}, {Molli{\`e}re}, {Charnay},
  {Kuzuhara}, {Janson}, {Morley}, {Homeier}, {D'Orazi}, {Klahr}, {Mordasini},
  {Lavie}, {Baudino}, {Beust}, {Peretti}, {Musso Bartucci}, {Mesa},
  {B{\'e}zard}, {Boccaletti}, {Galicher}, {Hagelberg}, {Desidera}, {Biller},
  {Maire}, {Allard}, {Borgniet}, {Lannier}, {Meunier}, {Desort}, {Alecian},
  {Chauvin}, {Langlois}, {Henning}, {Mugnier}, {Mouillet}, {Gratton}, {Brandt},
  {Mc Elwain}, {Beuzit}, {Tamura}, {Hori}, {Brandner}, {Buenzli}, {Cheetham},
  {Cudel}, {Feldt}, {Kasper}, {Keppler}, {Kopytova}, {Meyer}, {Perrot},
  {Rouan}, {Salter}, {Schmidt}, {Sissa}, {Zurlo}, {Wildi}, {Blanchard}, {De
  Caprio}, {Delboulb{\'e}}, {Maurel}, {Moulin}, {Pavlov}, {Rabou}, {Ramos},
  {Roelfsema}, {Rousset}, {Stadler}, {Rigal}, \& {Weber}}]{2018AA...618A..63B}
{Bonnefoy}, M., {Perraut}, K., {Lagrange}, A.~M., {et~al.} 2018, \aap, 618,
  A63, \dodoi{10.1051/0004-6361/201832942}

\bibitem[{{Bowler}(2016)}]{2016PASP..128j2001B}
{Bowler}, B.~P. 2016, \pasp, 128, 102001,
  \dodoi{10.1088/1538-3873/128/968/102001}

\bibitem[{{Bowler} {et~al.}(2015){Bowler}, {Liu}, {Shkolnik}, \&
  {Tamura}}]{2015ApJS..216....7B}
{Bowler}, B.~P., {Liu}, M.~C., {Shkolnik}, E.~L., \& {Tamura}, M. 2015, \apjs,
  216, 7, \dodoi{10.1088/0067-0049/216/1/7}

\bibitem[{{Brandt} {et~al.}(2014){Brandt}, {McElwain}, {Turner}, {Mede},
  {Spiegel}, {Kuzuhara}, {Schlieder}, {Wisniewski}, {Abe}, {Biller},
  {Brandner}, {Carson}, {Currie}, {Egner}, {Feldt}, {Golota}, {Goto}, {Grady},
  {Guyon}, {Hashimoto}, {Hayano}, {Hayashi}, {Hayashi}, {Henning}, {Hodapp},
  {Inutsuka}, {Ishii}, {Iye}, {Janson}, {Kandori}, {Knapp}, {Kudo}, {Kusakabe},
  {Kwon}, {Matsuo}, {Miyama}, {Morino}, {Moro-Mart{\'\i}n}, {Nishimura}, {Pyo},
  {Serabyn}, {Suto}, {Suzuki}, {Takami}, {Takato}, {Terada}, {Thalmann},
  {Tomono}, {Watanabe}, {Yamada}, {Takami}, {Usuda}, \&
  {Tamura}}]{2014ApJ...794..159B}
{Brandt}, T.~D., {McElwain}, M.~W., {Turner}, E.~L., {et~al.} 2014, \apj, 794,
  159, \dodoi{10.1088/0004-637X/794/2/159}

\bibitem[{{Burt} {et~al.}(2014){Burt}, {Vogt}, {Butler}, {Hanson}, {Meschiari},
  {Rivera}, {Henry}, \& {Laughlin}}]{2014ApJ...789..114B}
{Burt}, J., {Vogt}, S.~S., {Butler}, R.~P., {et~al.} 2014, \apj, 789, 114,
  \dodoi{10.1088/0004-637X/789/2/114}

\bibitem[{{Butler} {et~al.}(2017){Butler}, {Vogt}, {Laughlin}, {Burt},
  {Rivera}, {Tuomi}, {Teske}, {Arriagada}, {Diaz}, {Holden}, \&
  {Keiser}}]{2017AJ....153..208B}
{Butler}, R.~P., {Vogt}, S.~S., {Laughlin}, G., {et~al.} 2017, \aj, 153, 208,
  \dodoi{10.3847/1538-3881/aa66ca}

\bibitem[{{Caballero} {et~al.}(2016){Caballero}, {Cort{\'e}s-Contreras},
  {Alonso-Floriano}, {Montes}, {Quirrenbach}, {Amado}, {Ribas}, {Reiners},
  {Abellan}, {B{\'e}jar}, {Brinkm{\"o}ller}, {Czesla}, {Dorda}, {Gallardo},
  {Gonz{\'a}lez-{\'A}lvarez}, {Hidalgo}, {Holgado}, {Jeffers}, {Kim},
  {Klutsch}, {Lamert}, {Llamas}, {L{\'o}pez-Santiago},
  {Mart{\'\i}nez-Rodr{\'\i}guez}, {Morales}, {Mundt}, {Passegger},
  {Sch{\"o}fer}, {Seifert}, \& {Zechmeister}}]{2016csss.confE.148C}
{Caballero}, J.~A., {Cort{\'e}s-Contreras}, M., {Alonso-Floriano}, F.~J.,
  {et~al.} 2016, in 19th Cambridge Workshop on Cool Stars, Stellar Systems, and
  the Sun (CS19), Cambridge Workshop on Cool Stars, Stellar Systems, and the
  Sun, 148, \dodoi{10.5281/zenodo.60060}

\bibitem[{{Carson} {et~al.}(2011){Carson}, {Marengo}, {Patten}, {Luhman},
  {Sonnett}, {Hora}, {Schuster}, {Allen}, {Fazio}, {Stauffer}, \&
  {Schnupp}}]{2011ApJ...743..141C}
{Carson}, J.~C., {Marengo}, M., {Patten}, B.~M., {et~al.} 2011, \apj, 743, 141,
  \dodoi{10.1088/0004-637X/743/2/141}

\bibitem[{{Chauvin} {et~al.}(2015{\natexlab{a}}){Chauvin}, {Vigan}, {Bonnefoy},
  {Desidera}, {Bonavita}, {Mesa}, {Boccaletti}, {Buenzli}, {Carson}, {Delorme},
  {Hagelberg}, {Montagnier}, {Mordasini}, {Quanz}, {Segransan}, {Thalmann},
  {Beuzit}, {Biller}, {Covino}, {Feldt}, {Girard}, {Gratton}, {Henning},
  {Kasper}, {Lagrange}, {Messina}, {Meyer}, {Mouillet}, {Moutou}, {Reggiani},
  {Schlieder}, \& {Zurlo}}]{2015A&A...573A.127C}
{Chauvin}, G., {Vigan}, A., {Bonnefoy}, M., {et~al.} 2015{\natexlab{a}}, \aap,
  573, A127, \dodoi{10.1051/0004-6361/201423564}

\bibitem[{{Chauvin} {et~al.}(2015{\natexlab{b}}){Chauvin}, {Vigan}, {Bonnefoy},
  {Desidera}, {Bonavita}, {Mesa}, {Boccaletti}, {Buenzli}, {Carson}, {Delorme},
  {Hagelberg}, {Montagnier}, {Mordasini}, {Quanz}, {Segransan}, {Thalmann},
  {Beuzit}, {Biller}, {Covino}, {Feldt}, {Girard}, {Gratton}, {Henning},
  {Kasper}, {Lagrange}, {Messina}, {Meyer}, {Mouillet}, {Moutou}, {Reggiani},
  {Schlieder}, \& {Zurlo}}]{2015AA...573A.127C}
---. 2015{\natexlab{b}}, \aap, 573, A127, \dodoi{10.1051/0004-6361/201423564}

\bibitem[{{Chinchilla} {et~al.}(2021){Chinchilla}, {B{\'e}jar}, {Lodieu},
  {Zapatero Osorio}, \& {Gauza}}]{2021AA...645A..17C}
{Chinchilla}, P., {B{\'e}jar}, V.~J.~S., {Lodieu}, N., {Zapatero Osorio},
  M.~R., \& {Gauza}, B. 2021, \aap, 645, A17,
  \dodoi{10.1051/0004-6361/202038731}

\bibitem[{{Clanton} \& {Gaudi}(2014)}]{2014ApJ...791...91C}
{Clanton}, C., \& {Gaudi}, B.~S. 2014, \apj, 791, 91,
  \dodoi{10.1088/0004-637X/791/2/91}

\bibitem[{{Clanton} \& {Gaudi}(2016)}]{2016ApJ...819..125C}
---. 2016, \apj, 819, 125, \dodoi{10.3847/0004-637X/819/2/125}

\bibitem[{{Cohen} {et~al.}(2003){Cohen}, {Wheaton}, \&
  {Megeath}}]{2003AJ....126.1090C}
{Cohen}, M., {Wheaton}, W.~A., \& {Megeath}, S.~T. 2003, \aj, 126, 1090,
  \dodoi{10.1086/376474}

\bibitem[{{Correia} {et~al.}(2010){Correia}, {Couetdic}, {Laskar}, {Bonfils},
  {Mayor}, {Bertaux}, {Bouchy}, {Delfosse}, {Forveille}, {Lovis}, {Pepe},
  {Perrier}, {Queloz}, \& {Udry}}]{2010AA...511A..21C}
{Correia}, A.~C.~M., {Couetdic}, J., {Laskar}, J., {et~al.} 2010, \aap, 511,
  A21, \dodoi{10.1051/0004-6361/200912700}

\bibitem[{{Danielski} {et~al.}(2018){Danielski}, {Baudino}, {Lagage},
  {Boccaletti}, {Gastaud}, {Coulais}, \& {B{\'e}zard}}]{2018AJ....156..276D}
{Danielski}, C., {Baudino}, J.-L., {Lagage}, P.-O., {et~al.} 2018, \aj, 156,
  276, \dodoi{10.3847/1538-3881/aae651}

\bibitem[{{Delfosse} {et~al.}(1999){Delfosse}, {Forveille}, {Udry}, {Beuzit},
  {Mayor}, \& {Perrier}}]{1999AA...350L..39D}
{Delfosse}, X., {Forveille}, T., {Udry}, S., {et~al.} 1999, \aap, 350, L39.
\newblock \doarXiv{astro-ph/9909409}

\bibitem[{{Desidera} \& {Barbieri}(2007)}]{2007AA...462..345D}
{Desidera}, S., \& {Barbieri}, M. 2007, \aap, 462, 345,
  \dodoi{10.1051/0004-6361:20066319}

\bibitem[{{D{\'\i}az} {et~al.}(2019){D{\'\i}az}, {Delfosse}, {Hobson},
  {Boisse}, {Astudillo-Defru}, {Bonfils}, {Henry}, {Arnold}, {Bouchy},
  {Bourrier}, {Brugger}, {Dalal}, {Deleuil}, {Demangeon}, {Dolon}, {Dumusque},
  {Forveille}, {Hara}, {H{\'e}brard}, {Kiefer}, {Lopez}, {Mignon}, {Moreau},
  {Mousis}, {Moutou}, {Pepe}, {Perruchot}, {Richaud}, {Santerne}, {Santos},
  {Sottile}, {Stalport}, {S{\'e}gransan}, {Udry}, {Unger}, \&
  {Wilson}}]{2019AA...625A..17D}
{D{\'\i}az}, R.~F., {Delfosse}, X., {Hobson}, M.~J., {et~al.} 2019, \aap, 625,
  A17, \dodoi{10.1051/0004-6361/201935019}

\bibitem[{{Dieterich} {et~al.}(2012){Dieterich}, {Henry}, {Golimowski},
  {Krist}, \& {Tanner}}]{2012AJ....144...64D}
{Dieterich}, S.~B., {Henry}, T.~J., {Golimowski}, D.~A., {Krist}, J.~E., \&
  {Tanner}, A.~M. 2012, \aj, 144, 64, \dodoi{10.1088/0004-6256/144/2/64}

\bibitem[{{D{\'\i}ez Alonso} {et~al.}(2019){D{\'\i}ez Alonso}, {Caballero},
  {Montes}, {de Cos Juez}, {Dreizler}, {Dubois}, {Jeffers}, {Lalitha}, {Naves},
  {Reiners}, {Ribas}, {Vanaverbeke}, {Amado}, {B{\'e}jar},
  {Cort{\'e}s-Contreras}, {Herrero}, {Hidalgo}, {K{\"u}rster}, {Logie},
  {Quirrenbach}, {Rau}, {Seifert}, {Sch{\"o}fer}, \&
  {Tal-Or}}]{2019AA...621A.126D}
{D{\'\i}ez Alonso}, E., {Caballero}, J.~A., {Montes}, D., {et~al.} 2019, \aap,
  621, A126, \dodoi{10.1051/0004-6361/201833316}

\bibitem[{{Donahue} {et~al.}(1996){Donahue}, {Saar}, \&
  {Baliunas}}]{1996ApJ...466..384D}
{Donahue}, R.~A., {Saar}, S.~H., \& {Baliunas}, S.~L. 1996, \apj, 466, 384,
  \dodoi{10.1086/177517}

\bibitem[{{Durkan} {et~al.}(2016){Durkan}, {Janson}, \&
  {Carson}}]{2016ApJ...824...58D}
{Durkan}, S., {Janson}, M., \& {Carson}, J.~C. 2016, \apj, 824, 58,
  \dodoi{10.3847/0004-637X/824/1/58}

\bibitem[{{Dybczy{\'n}ski} \& {Kr{\'o}likowska}(2018)}]{2018AA...610L..11D}
{Dybczy{\'n}ski}, P.~A., \& {Kr{\'o}likowska}, M. 2018, \aap, 610, L11,
  \dodoi{10.1051/0004-6361/201732309}

\bibitem[{{Feng} {et~al.}(2017){Feng}, {Tuomi}, {Jones}, {Barnes},
  {Anglada-Escud{\'e}}, {Vogt}, \& {Butler}}]{2017AJ....154..135F}
{Feng}, F., {Tuomi}, M., {Jones}, H.~R.~A., {et~al.} 2017, \aj, 154, 135,
  \dodoi{10.3847/1538-3881/aa83b4}

\bibitem[{{Feng} {et~al.}(2020){Feng}, {Shectman}, {Clement}, {Vogt}, {Tuomi},
  {Teske}, {Burt}, {Crane}, {Holden}, {Wang}, {Thompson}, {D{\'\i}az}, \&
  {Butler}}]{2020ApJS..250...29F}
{Feng}, F., {Shectman}, S.~A., {Clement}, M.~S., {et~al.} 2020, \apjs, 250, 29,
  \dodoi{10.3847/1538-4365/abb139}

\bibitem[{{Fontanive} \& {Bardalez Gagliuffi}(2021)}]{2021FrASS...8...16F}
{Fontanive}, C., \& {Bardalez Gagliuffi}, D. 2021, Frontiers in Astronomy and
  Space Sciences, 8, 16, \dodoi{10.3389/fspas.2021.625250}

\bibitem[{{Fontanive} {et~al.}(2019){Fontanive}, {Rice}, {Bonavita}, {Lopez},
  {Mu{\v{z}}i{\'c}}, {}, \& {Biller}}]{2019MNRAS.485.4967F}
{Fontanive}, C., {Rice}, K., {Bonavita}, M., {et~al.} 2019, \mnras, 485, 4967,
  \dodoi{10.1093/mnras/stz671}

\bibitem[{{Fouqu{\'e}} {et~al.}(2018){Fouqu{\'e}}, {Moutou}, {Malo},
  {Martioli}, {Lim}, {Rajpurohit}, {Artigau}, {Delfosse}, {Donati},
  {Forveille}, {Morin}, {Allard}, {Delage}, {Doyon}, {H{\'e}brard}, \&
  {Neves}}]{2018MNRAS.475.1960F}
{Fouqu{\'e}}, P., {Moutou}, C., {Malo}, L., {et~al.} 2018, \mnras, 475, 1960,
  \dodoi{10.1093/mnras/stx3246}

\bibitem[{{Fulton} {et~al.}(2021){Fulton}, {Rosenthal}, {Hirsch}, {Isaacson},
  {Howard}, {Dedrick}, {Sherstyuk}, {Blunt}, {Petigura}, {Knutson}, {Behmard},
  {Chontos}, {Crepp}, {Crossfield}, {Dalba}, {Fischer}, {Henry}, {Kane},
  {Kosiarek}, {Marcy}, {Rubenzahl}, {Weiss}, \& {Wright}}]{2021ApJS..255...14F}
{Fulton}, B.~J., {Rosenthal}, L.~J., {Hirsch}, L.~A., {et~al.} 2021, \apjs,
  255, 14, \dodoi{10.3847/1538-4365/abfcc1}

\bibitem[{{Gaia Collaboration}(2018)}]{2018A&A...616A...1G}
{Gaia Collaboration}. 2018, \aap, 616, A1, \dodoi{10.1051/0004-6361/201833051}

\bibitem[{{Gaia Collaboration}(2021)}]{2021AA...649A...1G}
---. 2021, \aap, 649, A1, \dodoi{10.1051/0004-6361/202039657}

\bibitem[{{Galicher} {et~al.}(2016){Galicher}, {Marois}, {Macintosh},
  {Zuckerman}, {Barman}, {Konopacky}, {Song}, {Patience}, {Lafreni{\`e}re},
  {Doyon}, \& {Nielsen}}]{2016AA...594A..63G}
{Galicher}, R., {Marois}, C., {Macintosh}, B., {et~al.} 2016, \aap, 594, A63,
  \dodoi{10.1051/0004-6361/201527828}

\bibitem[{{Gauza} {et~al.}(2015{\natexlab{a}}){Gauza}, {B{\'e}jar},
  {P{\'e}rez-Garrido}, {Zapatero Osorio}, {Lodieu}, {Rebolo}, {Pall{\'e}}, \&
  {Nowak}}]{2015ApJ...804...96G}
{Gauza}, B., {B{\'e}jar}, V. J.~S., {P{\'e}rez-Garrido}, A., {et~al.}
  2015{\natexlab{a}}, \apj, 804, 96, \dodoi{10.1088/0004-637X/804/2/96}

\bibitem[{{Gauza} {et~al.}(2015{\natexlab{b}}){Gauza}, {B{\'e}jar}, {Rebolo},
  {{\'A}lvarez}, {Bihain}, {Zapatero Osorio}, {Caballero}, {Telesco}, \&
  {Packham}}]{2015MNRAS.452.1677G}
{Gauza}, B., {B{\'e}jar}, V.~J.~S., {Rebolo}, R., {et~al.} 2015{\natexlab{b}},
  \mnras, 452, 1677, \dodoi{10.1093/mnras/stv1350}

\bibitem[{{Gray} \& {Baliunas}(1995)}]{1995ApJ...441..436G}
{Gray}, D.~F., \& {Baliunas}, S.~L. 1995, \apj, 441, 436,
  \dodoi{10.1086/175368}

\bibitem[{{Hagelberg} {et~al.}(2020){Hagelberg}, {Engler}, {Fontanive},
  {Daemgen}, {Quanz}, {K{\"u}hn}, {Reggiani}, {Meyer}, {Jayawardhana}, \&
  {Kostov}}]{2020AA...643A..98H}
{Hagelberg}, J., {Engler}, N., {Fontanive}, C., {et~al.} 2020, \aap, 643, A98,
  \dodoi{10.1051/0004-6361/202039173}

\bibitem[{{Haghighipour}(2015)}]{2015enas.book.1942H}
{Haghighipour}, N. 2015, {Planets in Binary Star Systems}, ed. M.~{Gargaud},
  W.~M. {Irvine}, R.~{Amils}, I.~{Cleaves}, Henderson James~(Jim), D.~L.
  {Pinti}, J.~C. {Quintanilla}, D.~{Rouan}, T.~{Spohn}, S.~{Tirard}, \&
  M.~{Viso}, 1942--1944, \dodoi{10.1007/978-3-662-44185-5\_5285}

\bibitem[{{Hansen} \& {Murray}(2015)}]{2015MNRAS.448.1044H}
{Hansen}, B. M.~S., \& {Murray}, N. 2015, \mnras, 448, 1044,
  \dodoi{10.1093/mnras/stv049}

\bibitem[{{Hawley} {et~al.}(2014){Hawley}, {Davenport}, {Kowalski},
  {Wisniewski}, {Hebb}, {Deitrick}, \& {Hilton}}]{2014ApJ...797..121H}
{Hawley}, S.~L., {Davenport}, J. R.~A., {Kowalski}, A.~F., {et~al.} 2014, \apj,
  797, 121, \dodoi{10.1088/0004-637X/797/2/121}

\bibitem[{{He{\l}miniak} {et~al.}(2009){He{\l}miniak}, {Konacki}, {Kulkarni},
  \& {Eisner}}]{2009MNRAS.400..406H}
{He{\l}miniak}, K.~G., {Konacki}, M., {Kulkarni}, S.~R., \& {Eisner}, J. 2009,
  \mnras, 400, 406, \dodoi{10.1111/j.1365-2966.2009.15495.x}

\bibitem[{{Henry} {et~al.}(1997){Henry}, {Ianna}, {Kirkpatrick}, \&
  {Jahreiss}}]{1997AJ....114..388H}
{Henry}, T.~J., {Ianna}, P.~A., {Kirkpatrick}, J.~D., \& {Jahreiss}, H. 1997,
  \aj, 114, 388, \dodoi{10.1086/118482}

\bibitem[{{Henry} {et~al.}(2006){Henry}, {Jao}, {Subasavage}, {Beaulieu},
  {Ianna}, {Costa}, \& {M{\'e}ndez}}]{2006AJ....132.2360H}
{Henry}, T.~J., {Jao}, W.-C., {Subasavage}, J.~P., {et~al.} 2006, \aj, 132,
  2360, \dodoi{10.1086/508233}

\bibitem[{{Henry} {et~al.}(2018){Henry}, {Jao}, {Winters}, {Dieterich},
  {Finch}, {Ianna}, {Riedel}, {Silverstein}, {Subasavage}, \&
  {Vrijmoet}}]{2018AJ....155..265H}
{Henry}, T.~J., {Jao}, W.-C., {Winters}, J.~G., {et~al.} 2018, \aj, 155, 265,
  \dodoi{10.3847/1538-3881/aac262}

\bibitem[{{Holman} \& {Wiegert}(1999)}]{1999AJ....117..621H}
{Holman}, M.~J., \& {Wiegert}, P.~A. 1999, \aj, 117, 621,
  \dodoi{10.1086/300695}

\bibitem[{{Howard} {et~al.}(2014){Howard}, {Marcy}, {Fischer}, {Isaacson},
  {Muirhead}, {Henry}, {Boyajian}, {von Braun}, {Becker}, {Wright}, \&
  {Johnson}}]{2014ApJ...794...51H}
{Howard}, A.~W., {Marcy}, G.~W., {Fischer}, D.~A., {et~al.} 2014, \apj, 794,
  51, \dodoi{10.1088/0004-637X/794/1/51}

\bibitem[{{Janson} {et~al.}(2015){Janson}, {Quanz}, {Carson}, {Thalmann},
  {Lafreni{\`e}re}, \& {Amara}}]{2015AA...574A.120J}
{Janson}, M., {Quanz}, S.~P., {Carson}, J.~C., {et~al.} 2015, \aap, 574, A120,
  \dodoi{10.1051/0004-6361/201424944}

\bibitem[{{Jarrett} {et~al.}(2011){Jarrett}, {Cohen}, {Masci}, {Wright},
  {Stern}, {Benford}, {Blain}, {Carey}, {Cutri}, {Eisenhardt}, {Lonsdale},
  {Mainzer}, {Marsh}, {Padgett}, {Petty}, {Ressler}, {Skrutskie}, {Stanford},
  {Surace}, {Tsai}, {Wheelock}, \& {Yan}}]{2011ApJ...735..112J}
{Jarrett}, T.~H., {Cohen}, M., {Masci}, F., {et~al.} 2011, \apj, 735, 112,
  \dodoi{10.1088/0004-637X/735/2/112}

\bibitem[{{Johnson} {et~al.}(2010){Johnson}, {Aller}, {Howard}, \&
  {Crepp}}]{2010PASP..122..905J}
{Johnson}, J.~A., {Aller}, K.~M., {Howard}, A.~W., \& {Crepp}, J.~R. 2010,
  \pasp, 122, 905, \dodoi{10.1086/655775}

\bibitem[{{Kervella} {et~al.}(2019){Kervella}, {Arenou}, {Mignard}, \&
  {Th{\'e}venin}}]{2019AA...623A..72K}
{Kervella}, P., {Arenou}, F., {Mignard}, F., \& {Th{\'e}venin}, F. 2019, \aap,
  623, A72, \dodoi{10.1051/0004-6361/201834371}

\bibitem[{{Kervella} {et~al.}(2008){Kervella}, {M{\'e}rand}, {Pichon},
  {Th{\'e}venin}, {Heiter}, {Bigot}, {ten Brummelaar}, {McAlister}, {Ridgway},
  {Turner}, {Sturmann}, {Sturmann}, {Goldfinger}, \&
  {Farrington}}]{2008AA...488..667K}
{Kervella}, P., {M{\'e}rand}, A., {Pichon}, B., {et~al.} 2008, \aap, 488, 667,
  \dodoi{10.1051/0004-6361:200810080}

\bibitem[{{Kiraga} \& {Stepien}(2007)}]{2007AcA....57..149K}
{Kiraga}, M., \& {Stepien}, K. 2007, \actaa, 57, 149.
\newblock \doarXiv{0707.2577}

\bibitem[{{Knapp} \& {Nanson}(2018)}]{2018JDSO...14....3K}
{Knapp}, W., \& {Nanson}, J. 2018, Journal of Double Star Observations, 14, 3

\bibitem[{{Knutson} {et~al.}(2014){Knutson}, {Fulton}, {Montet}, {Kao}, {Ngo},
  {Howard}, {Crepp}, {Hinkley}, {Bakos}, {Batygin}, {Johnson}, {Morton}, \&
  {Muirhead}}]{2014ApJ...785..126K}
{Knutson}, H.~A., {Fulton}, B.~J., {Montet}, B.~T., {et~al.} 2014, \apj, 785,
  126, \dodoi{10.1088/0004-637X/785/2/126}

\bibitem[{{Kopparapu} {et~al.}(2016){Kopparapu}, {Wolf}, {Haqq-Misra}, {Yang},
  {Kasting}, {Meadows}, {Terrien}, \& {Mahadevan}}]{2016ApJ...819...84K}
{Kopparapu}, R.~k., {Wolf}, E.~T., {Haqq-Misra}, J., {et~al.} 2016, \apj, 819,
  84, \dodoi{10.3847/0004-637X/819/1/84}

\bibitem[{{Lafreni{\`e}re} {et~al.}(2007){Lafreni{\`e}re}, {Doyon}, {Marois},
  {Nadeau}, {Oppenheimer}, {Roche}, {Rigaut}, {Graham}, {Jayawardhana},
  {Johnstone}, {Kalas}, {Macintosh}, \& {Racine}}]{2007ApJ...670.1367L}
{Lafreni{\`e}re}, D., {Doyon}, R., {Marois}, C., {et~al.} 2007, \apj, 670,
  1367, \dodoi{10.1086/522826}

\bibitem[{{Lannier} {et~al.}(2016){Lannier}, {Delorme}, {Lagrange}, {Borgniet},
  {Rameau}, {Schlieder}, {Gagn{\'e}}, {Bonavita}, {Malo}, {Chauvin},
  {Bonnefoy}, \& {Girard}}]{2016AA...596A..83L}
{Lannier}, J., {Delorme}, P., {Lagrange}, A.~M., {et~al.} 2016, \aap, 596, A83,
  \dodoi{10.1051/0004-6361/201628237}

\bibitem[{{Launhardt} {et~al.}(2020){Launhardt}, {Henning}, {Quirrenbach},
  {S{\'e}gransan}, {Avenhaus}, {van Boekel}, {Brems}, {Cheetham}, {Cugno},
  {Girard}, {Godoy}, {Kennedy}, {Maire}, {Metchev}, {M{\"u}ller}, {Musso
  Barcucci}, {Olofsson}, {Pepe}, {Quanz}, {Queloz}, {Reffert}, {Rickman},
  {Ruh}, \& {Samland}}]{2020AA...635A.162L}
{Launhardt}, R., {Henning}, T., {Quirrenbach}, A., {et~al.} 2020, \aap, 635,
  A162, \dodoi{10.1051/0004-6361/201937000}

\bibitem[{{Leggett} {et~al.}(2012){Leggett}, {Saumon}, {Marley}, {Lodders},
  {Canty}, {Lucas}, {Smart}, {Tinney}, {Homeier}, {Allard}, {Burningham},
  {Day-Jones}, {Fegley}, {Ishii}, {Jones}, {Marocco}, {Pinfield}, \&
  {Tamura}}]{2012ApJ...748...74L}
{Leggett}, S.~K., {Saumon}, D., {Marley}, M.~S., {et~al.} 2012, \apj, 748, 74,
  \dodoi{10.1088/0004-637X/748/2/74}

\bibitem[{{Lestrade} {et~al.}(2006){Lestrade}, {Wyatt}, {Bertoldi}, {Dent}, \&
  {Menten}}]{2006AA...460..733L}
{Lestrade}, J.~F., {Wyatt}, M.~C., {Bertoldi}, F., {Dent}, W.~R.~F., \&
  {Menten}, K.~M. 2006, \aap, 460, 733, \dodoi{10.1051/0004-6361:20065873}

\bibitem[{{Liebert} {et~al.}(2005){Liebert}, {Young}, {Arnett}, {Holberg}, \&
  {Williams}}]{2005ApJ...630L..69L}
{Liebert}, J., {Young}, P.~A., {Arnett}, D., {Holberg}, J.~B., \& {Williams},
  K.~A. 2005, \apjl, 630, L69, \dodoi{10.1086/462419}

\bibitem[{{Lodieu} {et~al.}(2014){Lodieu}, {P{\'e}rez-Garrido}, {B{\'e}jar},
  {Gauza}, {Ruiz}, {Rebolo}, {Pinfield}, \& {Mart{\'\i}n}}]{2014AA...569A.120L}
{Lodieu}, N., {P{\'e}rez-Garrido}, A., {B{\'e}jar}, V.~J.~S., {et~al.} 2014,
  \aap, 569, A120, \dodoi{10.1051/0004-6361/201424210}

\bibitem[{{Luhman}(2014)}]{2014ApJ...786L..18L}
{Luhman}, K.~L. 2014, \apjl, 786, L18, \dodoi{10.1088/2041-8205/786/2/L18}

\bibitem[{{Luhman} \& {Esplin}(2016)}]{2016AJ....152...78L}
{Luhman}, K.~L., \& {Esplin}, T.~L. 2016, \aj, 152, 78,
  \dodoi{10.3847/0004-6256/152/3/78}

\bibitem[{{Lurie} {et~al.}(2015){Lurie}, {Davenport}, {Hawley}, {Wilkinson},
  {Wisniewski}, {Kowalski}, \& {Hebb}}]{2015ApJ...800...95L}
{Lurie}, J.~C., {Davenport}, J. R.~A., {Hawley}, S.~L., {et~al.} 2015, \apj,
  800, 95, \dodoi{10.1088/0004-637X/800/2/95}

\bibitem[{{MacDonald} {et~al.}(2018){MacDonald}, {Mullan}, \&
  {Dieterich}}]{2018ApJ...860...15M}
{MacDonald}, J., {Mullan}, D.~J., \& {Dieterich}, S. 2018, \apj, 860, 15,
  \dodoi{10.3847/1538-4357/aac2c0}

\bibitem[{{Mamajek} \& {Hillenbrand}(2008)}]{2008ApJ...687.1264M}
{Mamajek}, E.~E., \& {Hillenbrand}, L.~A. 2008, \apj, 687, 1264,
  \dodoi{10.1086/591785}

\bibitem[{{Mann} {et~al.}(2015){Mann}, {Feiden}, {Gaidos}, {Boyajian}, \& {von
  Braun}}]{2015ApJ...804...64M}
{Mann}, A.~W., {Feiden}, G.~A., {Gaidos}, E., {Boyajian}, T., \& {von Braun},
  K. 2015, \apj, 804, 64, \dodoi{10.1088/0004-637X/804/1/64}

\bibitem[{{Marcy} {et~al.}(2001){Marcy}, {Butler}, {Fischer}, {Vogt},
  {Lissauer}, \& {Rivera}}]{2001ApJ...556..296M}
{Marcy}, G.~W., {Butler}, R.~P., {Fischer}, D., {et~al.} 2001, \apj, 556, 296,
  \dodoi{10.1086/321552}

\bibitem[{{Marcy} {et~al.}(1998){Marcy}, {Butler}, {Vogt}, {Fischer}, \&
  {Lissauer}}]{1998ApJ...505L.147M}
{Marcy}, G.~W., {Butler}, R.~P., {Vogt}, S.~S., {Fischer}, D., \& {Lissauer},
  J.~J. 1998, \apjl, 505, L147, \dodoi{10.1086/311623}

\bibitem[{{Marois} {et~al.}(2008){Marois}, {Macintosh}, {Barman}, {Zuckerman},
  {Song}, {Patience}, {Lafreni{\`e}re}, \& {Doyon}}]{2008Sci...322.1348M}
{Marois}, C., {Macintosh}, B., {Barman}, T., {et~al.} 2008, Science, 322, 1348,
  \dodoi{10.1126/science.1166585}

\bibitem[{{Masset} \& {Papaloizou}(2003)}]{2003ApJ...588..494M}
{Masset}, F.~S., \& {Papaloizou}, J.~C.~B. 2003, \apj, 588, 494,
  \dodoi{10.1086/373892}

\bibitem[{{Mawet} {et~al.}(2019){Mawet}, {Hirsch}, {Lee}, {Ruffio}, {Bottom},
  {Fulton}, {Absil}, {Beichman}, {Bowler}, {Bryan}, {Choquet}, {Ciardi},
  {Christiaens}, {Defr{\`e}re}, {Gomez Gonzalez}, {Howard}, {Huby}, {Isaacson},
  {Jensen-Clem}, {Kosiarek}, {Marcy}, {Meshkat}, {Petigura}, {Reggiani},
  {Ruane}, {Serabyn}, {Sinukoff}, {Wang}, {Weiss}, \&
  {Ygouf}}]{2019AJ....157...33M}
{Mawet}, D., {Hirsch}, L., {Lee}, E.~J., {et~al.} 2019, \aj, 157, 33,
  \dodoi{10.3847/1538-3881/aaef8a}

\bibitem[{{McCarthy} {et~al.}(1988){McCarthy}, {Henry}, {Fleming}, {Saffer},
  {Liebert}, \& {Christou}}]{1988ApJ...333..943M}
{McCarthy}, D.~W., J., {Henry}, T.~J., {Fleming}, T.~A., {et~al.} 1988, \apj,
  333, 943, \dodoi{10.1086/166803}

\bibitem[{{Meshkat} {et~al.}(2015){Meshkat}, {Kenworthy}, {Reggiani}, {Quanz},
  {Mamajek}, \& {Meyer}}]{2015MNRAS.453.2533M}
{Meshkat}, T., {Kenworthy}, M.~A., {Reggiani}, M., {et~al.} 2015, \mnras, 453,
  2533, \dodoi{10.1093/mnras/stv1779}

\bibitem[{{Millholland} {et~al.}(2018){Millholland}, {Laughlin}, {Teske},
  {Butler}, {Burt}, {Holden}, {Vogt}, {Crane}, {Shectman}, \&
  {Thompson}}]{2018AJ....155..106M}
{Millholland}, S., {Laughlin}, G., {Teske}, J., {et~al.} 2018, \aj, 155, 106,
  \dodoi{10.3847/1538-3881/aaa894}

\bibitem[{{Moe} \& {Kratter}(2019)}]{2019arXiv191201699M}
{Moe}, M., \& {Kratter}, K.~M. 2019, arXiv e-prints, arXiv:1912.01699.
\newblock \doarXiv{1912.01699}

\bibitem[{{Montet} {et~al.}(2014){Montet}, {Crepp}, {Johnson}, {Howard}, \&
  {Marcy}}]{2014ApJ...781...28M}
{Montet}, B.~T., {Crepp}, J.~R., {Johnson}, J.~A., {Howard}, A.~W., \& {Marcy},
  G.~W. 2014, \apj, 781, 28, \dodoi{10.1088/0004-637X/781/1/28}

\bibitem[{{Morales} {et~al.}(2019){Morales}, {Mustill}, {Ribas}, {Davies},
  {Reiners}, {Bauer}, {Kossakowski}, {Herrero}, {Rodr{\'\i}guez},
  {L{\'o}pez-Gonz{\'a}lez}, {Rodr{\'\i}guez-L{\'o}pez}, {B{\'e}jar},
  {Gonz{\'a}lez-Cuesta}, {Luque}, {Pall{\'e}}, {Perger}, {Baroch}, {Johansen},
  {Klahr}, {Mordasini}, {Anglada-Escud{\'e}}, {Caballero},
  {Cort{\'e}s-Contreras}, {Dreizler}, {Lafarga}, {Nagel}, {Passegger},
  {Reffert}, {Rosich}, {Schweitzer}, {Tal-Or}, {Trifonov}, {Zechmeister},
  {Quirrenbach}, {Amado}, {Guenther}, {Hagen}, {Henning}, {Jeffers},
  {Kaminski}, {K{\"u}rster}, {Montes}, {Seifert}, {Abell{\'a}n}, {Abril},
  {Aceituno}, {Aceituno}, {Alonso-Floriano}, {Ammler-von Eiff}, {Antona},
  {Arroyo-Torres}, {Azzaro}, {Barrado}, {Becerril-Jarque}, {Ben{\'\i}tez},
  {Berdi{\~n}as}, {Bergond}, {Brinkm{\"o}ller}, {del Burgo}, {Burn},
  {Calvo-Ortega}, {Cano}, {C{\'a}rdenas}, {Cardona Guill{\'e}n}, {Carro},
  {Casal}, {Casanova}, {Casasayas-Barris}, {Chaturvedi}, {Cifuentes}, {Claret},
  {Colom{\'e}}, {Czesla}, {D{\'\i}ez-Alonso}, {Dorda}, {Emsenhuber},
  {Fern{\'a}ndez}, {Fern{\'a}ndez-Mart{\'\i}n}, {Ferro}, {Fuhrmeister},
  {Galad{\'\i}-Enr{\'\i}quez}, {Gallardo Cava}, {Garc{\'\i}a Vargas},
  {Garcia-Piquer}, {Gesa}, {Gonz{\'a}lez-{\'A}lvarez}, {Gonz{\'a}lez
  Hern{\'a}ndez}, {Gonz{\'a}lez-Peinado}, {Gu{\`a}rdia}, {Guijarro}, {de
  Guindos}, {Hatzes}, {Hauschildt}, {Hedrosa}, {Hermelo}, {Hern{\'a}ndez
  Arabi}, {Hern{\'a}ndez Otero}, {Hintz}, {Holgado}, {Huber}, {Huke},
  {Johnson}, {de Juan}, {Kehr}, {Kemmer}, {Kim}, {Kl{\"u}ter}, {Klutsch},
  {Labarga}, {Labiche}, {Lalitha}, {Lamp{\'o}n}, {Lara}, {Launhardt},
  {L{\'a}zaro}, {Lizon}, {Llamas}, {Lodieu}, {L{\'o}pez del Fresno}, {L{\'o}pez
  Salas}, {L{\'o}pez-Santiago}, {Mag{\'a}n Madinabeitia}, {Mall}, {Mancini},
  {Mandel}, {Marfil}, {Mar{\'\i}n Molina}, {Mart{\'\i}n},
  {Mart{\'\i}n-Fern{\'a}ndez}, {Mart{\'\i}n-Ruiz},
  {Mart{\'\i}nez-Rodr{\'\i}guez}, {Marvin}, {Mirabet}, {Moya}, {Naranjo},
  {Nelson}, {Nortmann}, {Nowak}, {Ofir}, {Pascual}, {Pavlov}, {Pedraz},
  {P{\'e}rez Medialdea}, {P{\'e}rez-Calpena}, {Perryman}, {Rabaza}, {Ram{\'o}n
  Ballesta}, {Rebolo}, {Redondo}, {Rix}, {Rodler}, {Rodr{\'\i}guez Trinidad},
  {Sabotta}, {Sadegi}, {Salz}, {S{\'a}nchez-Blanco}, {S{\'a}nchez Carrasco},
  {S{\'a}nchez-L{\'o}pez}, {Sanz-Forcada}, {Sarkis}, {Sarmiento},
  {Sch{\"a}fer}, {Schlecker}, {Schmitt}, {Sch{\"o}fer}, {Solano}, {Sota},
  {Stahl}, {Stock}, {Stuber}, {St{\"u}rmer}, {Su{\'a}rez}, {Tabernero},
  {Tulloch}, {Veredas}, {Vico-Linares}, {Vilardell}, {Wagner}, {Winkler},
  {Wolthoff}, {Yan}, \& {Zapatero Osorio}}]{2019Sci...365.1441M}
{Morales}, J.~C., {Mustill}, A.~J., {Ribas}, I., {et~al.} 2019, Science, 365,
  1441, \dodoi{10.1126/science.aax3198}

\bibitem[{{Murakami} {et~al.}(2007){Murakami}, {Baba}, {Barthel}, {Clements},
  {Cohen}, {Doi}, {Enya}, {Figueredo}, {Fujishiro}, {Fujiwara}, {Fujiwara},
  {Garcia-Lario}, {Goto}, {Hasegawa}, {Hibi}, {Hirao}, {Hiromoto}, {Hong},
  {Imai}, {Ishigaki}, {Ishiguro}, {Ishihara}, {Ita}, {Jeong}, {Jeong},
  {Kaneda}, {Kataza}, {Kawada}, {Kawai}, {Kawamura}, {Kessler}, {Kester},
  {Kii}, {Kim}, {Kim}, {Kobayashi}, {Koo}, {Kwon}, {Lee}, {Lorente}, {Makiuti},
  {Matsuhara}, {Matsumoto}, {Matsuo}, {Matsuura}, {M{\"U}ller}, {Murakami},
  {Nagata}, {Nakagawa}, {Naoi}, {Narita}, {Noda}, {Oh}, {Ohnishi}, {Ohyama},
  {Okada}, {Okuda}, {Oliver}, {Onaka}, {Ootsubo}, {Oyabu}, {Pak}, {Park},
  {Pearson}, {Rowan-Robinson}, {Saito}, {Sakon}, {Salama}, {Sato}, {Savage},
  {Serjeant}, {Shibai}, {Shirahata}, {Sohn}, {Suzuki}, {Takagi}, {Takahashi},
  {Tanab{\'E}}, {Takeuchi}, {Takita}, {Thomson}, {Uemizu}, {Ueno}, {Usui},
  {Verdugo}, {Wada}, {Wang}, {Watabe}, {Watarai}, {White}, {Yamamura},
  {Yamauchi}, \& {Yasuda}}]{2007PASJ...59S.369M}
{Murakami}, H., {Baba}, H., {Barthel}, P., {et~al.} 2007, \pasj, 59, S369,
  \dodoi{10.1093/pasj/59.sp2.S369}

\bibitem[{{Nakajima} {et~al.}(1995){Nakajima}, {Oppenheimer}, {Kulkarni},
  {Golimowski}, {Matthews}, \& {Durrance}}]{1995Natur.378..463N}
{Nakajima}, T., {Oppenheimer}, B.~R., {Kulkarni}, S.~R., {et~al.} 1995, \nat,
  378, 463, \dodoi{10.1038/378463a0}

\bibitem[{{Naoz}(2016)}]{2016ARAA..54..441N}
{Naoz}, S. 2016, \araa, 54, 441, \dodoi{10.1146/annurev-astro-081915-023315}

\bibitem[{{Ngo} {et~al.}(2016){Ngo}, {Knutson}, {Hinkley}, {Bryan}, {Crepp},
  {Batygin}, {Crossfield}, {Hansen}, {Howard}, {Johnson}, {Mawet}, {Morton},
  {Muirhead}, \& {Wang}}]{2016ApJ...827....8N}
{Ngo}, H., {Knutson}, H.~A., {Hinkley}, S., {et~al.} 2016, \apj, 827, 8,
  \dodoi{10.3847/0004-637X/827/1/8}

\bibitem[{{Nielsen} \& {Close}(2010)}]{2010ApJ...717..878N}
{Nielsen}, E.~L., \& {Close}, L.~M. 2010, \apj, 717, 878,
  \dodoi{10.1088/0004-637X/717/2/878}

\bibitem[{{Nielsen} {et~al.}(2019){Nielsen}, {De Rosa}, {Macintosh}, {Wang},
  {Ruffio}, {Chiang}, {Marley}, {Saumon}, {Savransky}, {Ammons}, {Bailey},
  {Barman}, {Blain}, {Bulger}, {Burrows}, {Chilcote}, {Cotten}, {Czekala},
  {Doyon}, {Duch{\^e}ne}, {Esposito}, {Fabrycky}, {Fitzgerald}, {Follette},
  {Fortney}, {Gerard}, {Goodsell}, {Graham}, {Greenbaum}, {Hibon}, {Hinkley},
  {Hirsch}, {Hom}, {Hung}, {Dawson}, {Ingraham}, {Kalas}, {Konopacky},
  {Larkin}, {Lee}, {Lin}, {Maire}, {Marchis}, {Marois}, {Metchev},
  {Millar-Blanchaer}, {Morzinski}, {Oppenheimer}, {Palmer}, {Patience},
  {Perrin}, {Poyneer}, {Pueyo}, {Rafikov}, {Rajan}, {Rameau}, {Rantakyr{\"o}},
  {Ren}, {Schneider}, {Sivaramakrishnan}, {Song}, {Soummer}, {Tallis},
  {Thomas}, {Ward-Duong}, \& {Wolff}}]{2019AJ....158...13N}
{Nielsen}, E.~L., {De Rosa}, R.~J., {Macintosh}, B., {et~al.} 2019, \aj, 158,
  13, \dodoi{10.3847/1538-3881/ab16e9}

\bibitem[{{Oppenheimer} {et~al.}(2001){Oppenheimer}, {Golimowski}, {Kulkarni},
  {Matthews}, {Nakajima}, {Creech-Eakman}, \& {Durrance}}]{2001AJ....121.2189O}
{Oppenheimer}, B.~R., {Golimowski}, D.~A., {Kulkarni}, S.~R., {et~al.} 2001,
  \aj, 121, 2189, \dodoi{10.1086/319941}

\bibitem[{{Pavlenko} {et~al.}(2006){Pavlenko}, {Jones}, {Lyubchik}, {Tennyson},
  \& {Pinfield}}]{2006AA...447..709P}
{Pavlenko}, Y.~V., {Jones}, H.~R.~A., {Lyubchik}, Y., {Tennyson}, J., \&
  {Pinfield}, D.~J. 2006, \aap, 447, 709, \dodoi{10.1051/0004-6361:20052979}

\bibitem[{{Pinamonti} {et~al.}(2018){Pinamonti}, {Damasso}, {Marzari},
  {Sozzetti}, {Desidera}, {Maldonado}, {Scandariato}, {Affer}, {Lanza},
  {Bignamini}, {Bonomo}, {Borsa}, {Claudi}, {Cosentino}, {Giacobbe},
  {Gonz{\'a}lez-{\'A}lvarez}, {Gonz{\'a}lez Hern{\'a}ndez}, {Gratton}, {Leto},
  {Malavolta}, {Martinez Fiorenzano}, {Micela}, {Molinari}, {Pagano}, {Pedani},
  {Perger}, {Piotto}, {Rebolo}, {Ribas}, {Su{\'a}rez Mascare{\~n}o}, \&
  {Toledo-Padr{\'o}n}}]{2018AA...617A.104P}
{Pinamonti}, M., {Damasso}, M., {Marzari}, F., {et~al.} 2018, \aap, 617, A104,
  \dodoi{10.1051/0004-6361/201732535}

\bibitem[{{Piskorz} {et~al.}(2015){Piskorz}, {Knutson}, {Ngo}, {Muirhead},
  {Batygin}, {Crepp}, {Hinkley}, \& {Morton}}]{2015ApJ...814..148P}
{Piskorz}, D., {Knutson}, H.~A., {Ngo}, H., {et~al.} 2015, \apj, 814, 148,
  \dodoi{10.1088/0004-637X/814/2/148}

\bibitem[{{Pozuelos} {et~al.}(2020){Pozuelos}, {Su{\'a}rez}, {de El{\'\i}a},
  {Berdi{\~n}as}, {Bonfanti}, {Dugaro}, {Gillon}, {Jehin}, {G{\"u}nther}, {Van
  Grootel}, {Garcia}, {Thuillier}, {Delrez}, \&
  {Rod{\'o}n}}]{2020AA...641A..23P}
{Pozuelos}, F.~J., {Su{\'a}rez}, J.~C., {de El{\'\i}a}, G.~C., {et~al.} 2020,
  \aap, 641, A23, \dodoi{10.1051/0004-6361/202038047}

\bibitem[{{Quanz} {et~al.}(2015){Quanz}, {Crossfield}, {Meyer}, {Schmalzl}, \&
  {Held}}]{2015IJAsB..14..279Q}
{Quanz}, S.~P., {Crossfield}, I., {Meyer}, M.~R., {Schmalzl}, E., \& {Held}, J.
  2015, International Journal of Astrobiology, 14, 279,
  \dodoi{10.1017/S1473550414000135}

\bibitem[{{Reiners} {et~al.}(2007){Reiners}, {Schmitt}, \&
  {Liefke}}]{2007AA...466L..13R}
{Reiners}, A., {Schmitt}, J.~H.~M.~M., \& {Liefke}, C. 2007, \aap, 466, L13,
  \dodoi{10.1051/0004-6361:20077095}

\bibitem[{{Reiners} {et~al.}(2018){Reiners}, {Zechmeister}, {Caballero},
  {Ribas}, {Morales}, {Jeffers}, {Sch{\"o}fer}, {Tal-Or}, {Quirrenbach},
  {Amado}, {Kaminski}, {Seifert}, {Abril}, {Aceituno}, {Alonso-Floriano},
  {Ammler-von Eiff}, {Antona}, {Anglada-Escud{\'e}}, {Anwand-Heerwart},
  {Arroyo-Torres}, {Azzaro}, {Baroch}, {Barrado}, {Bauer}, {Becerril},
  {B{\'e}jar}, {Ben{\'\i}tez}, {Berdinas̃}, {Bergond}, {Bl{\"u}mcke},
  {Brinkm{\"o}ller}, {del Burgo}, {Cano}, {C{\'a}rdenas V{\'a}zquez}, {Casal},
  {Cifuentes}, {Claret}, {Colom{\'e}}, {Cort{\'e}s-Contreras}, {Czesla},
  {D{\'\i}ez-Alonso}, {Dreizler}, {Feiz}, {Fern{\'a}ndez}, {Ferro},
  {Fuhrmeister}, {Galad{\'\i}-Enr{\'\i}quez}, {Garcia-Piquer}, {Garc{\'\i}a
  Vargas}, {Gesa}, {G{\'o}mez Galera}, {Gonz{\'a}lez Hern{\'a}ndez},
  {Gonz{\'a}lez-Peinado}, {Gr{\"o}zinger}, {Grohnert}, {Gu{\`a}rdia},
  {Guenther}, {Guijarro}, {de Guindos}, {Guti{\'e}rrez-Soto}, {Hagen},
  {Hatzes}, {Hauschildt}, {Hedrosa}, {Helmling}, {Henning}, {Hermelo},
  {Hern{\'a}ndez Arab{\'\i}}, {Hern{\'a}ndez Casta{\~n}o}, {Hern{\'a}ndez
  Hernando}, {Herrero}, {Huber}, {Huke}, {Johnson}, {de Juan}, {Kim}, {Klein},
  {Kl{\"u}ter}, {Klutsch}, {K{\"u}rster}, {Lafarga}, {Lamert}, {Lamp{\'o}n},
  {Lara}, {Laun}, {Lemke}, {Lenzen}, {Launhardt}, {L{\'o}pez del Fresno},
  {L{\'o}pez-Gonz{\'a}lez}, {L{\'o}pez-Puertas}, {L{\'o}pez Salas},
  {L{\'o}pez-Santiago}, {Luque}, {Mag{\'a}n Madinabeitia}, {Mall}, {Mancini},
  {Mand el}, {Marfil}, {Mar{\'\i}n Molina}, {Maroto Fern{\'a}ndez},
  {Mart{\'\i}n}, {Mart{\'\i}n-Ruiz}, {Marvin}, {Mathar}, {Mirabet}, {Montes},
  {Moreno-Raya}, {Moya}, {Mundt}, {Nagel}, {Naranjo}, {Nortmann}, {Nowak},
  {Ofir}, {Oreiro}, {Pall{\'e}}, {Pand uro}, {Pascual}, {Passegger}, {Pavlov},
  {Pedraz}, {P{\'e}rez-Calpena}, {P{\'e}rez Medialdea}, {Perger}, {Perryman},
  {Pluto}, {Rabaza}, {Ram{\'o}n}, {Rebolo}, {Redondo}, {Reffert}, {Reinhart},
  {Rhode}, {Rix}, {Rodler}, {Rodr{\'\i}guez}, {Rodr{\'\i}guez-L{\'o}pez},
  {Rodr{\'\i}guez Trinidad}, {Rohloff}, {Rosich}, {Sadegi},
  {S{\'a}nchez-Blanco}, {S{\'a}nchez Carrasco}, {S{\'a}nchez-L{\'o}pez},
  {Sanz-Forcada}, {Sarkis}, {Sarmiento}, {Sch{\"a}fer}, {Schmitt}, {Schiller},
  {Schweitzer}, {Solano}, {Stahl}, {Strachan}, {St{\"u}rmer}, {Su{\'a}rez},
  {Tabernero}, {Tala}, {Trifonov}, {Tulloch}, {Ulbrich}, {Veredas}, {Vico
  Linares}, {Vilardell}, {Wagner}, {Winkler}, {Wolthoff}, {Xu}, {Yan}, \&
  {Zapatero Osorio}}]{2018AA...612A..49R}
{Reiners}, A., {Zechmeister}, M., {Caballero}, J.~A., {et~al.} 2018, \aap, 612,
  A49, \dodoi{10.1051/0004-6361/201732054}

\bibitem[{{Reyl{\'e}} {et~al.}(2021){Reyl{\'e}}, {Jardine}, {Fouqu{\'e}},
  {Caballero}, {Smart}, \& {Sozzetti}}]{2021AA...650A.201R}
{Reyl{\'e}}, C., {Jardine}, K., {Fouqu{\'e}}, P., {et~al.} 2021, \aap, 650,
  A201, \dodoi{10.1051/0004-6361/202140985}

\bibitem[{{Ribas} {et~al.}(2018){Ribas}, {Tuomi}, {Reiners}, {Butler},
  {Morales}, {Perger}, {Dreizler}, {Rodr{\'\i}guez-L{\'o}pez}, {Gonz{\'a}lez
  Hern{\'a}ndez}, {Rosich}, {Feng}, {Trifonov}, {Vogt}, {Caballero}, {Hatzes},
  {Herrero}, {Jeffers}, {Lafarga}, {Murgas}, {Nelson}, {Rodr{\'\i}guez},
  {Strachan}, {Tal-Or}, {Teske}, {Toledo-Padr{\'o}n}, {Zechmeister},
  {Quirrenbach}, {Amado}, {Azzaro}, {B{\'e}jar}, {Barnes}, {Berdi{\~n}as},
  {Burt}, {Coleman}, {Cort{\'e}s-Contreras}, {Crane}, {Engle}, {Guinan},
  {Haswell}, {Henning}, {Holden}, {Jenkins}, {Jones}, {Kaminski}, {Kiraga},
  {K{\"u}rster}, {Lee}, {L{\'o}pez-Gonz{\'a}lez}, {Montes}, {Morin}, {Ofir},
  {Pall{\'e}}, {Rebolo}, {Reffert}, {Schweitzer}, {Seifert}, {Shectman},
  {Staab}, {Street}, {Su{\'a}rez Mascare{\~n}o}, {Tsapras}, {Wang}, \&
  {Anglada-Escud{\'e}}}]{2018Natur.563..365R}
{Ribas}, I., {Tuomi}, M., {Reiners}, A., {et~al.} 2018, Nature, 563, 365,
  \dodoi{10.1038/s41586-018-0677-y}

\bibitem[{{Rivera} {et~al.}(2010){Rivera}, {Laughlin}, {Butler}, {Vogt},
  {Haghighipour}, \& {Meschiari}}]{2010ApJ...719..890R}
{Rivera}, E.~J., {Laughlin}, G., {Butler}, R.~P., {et~al.} 2010, \apj, 719,
  890, \dodoi{10.1088/0004-637X/719/1/890}

\bibitem[{{Robertson} {et~al.}(2020){Robertson}, {Stefansson}, {Mahadevan},
  {Endl}, {Cochran}, {Beard}, {Bender}, {Diddams}, {Duong}, {Ford}, {Fredrick},
  {Halverson}, {Hearty}, {Holcomb}, {Juan}, {Kanodia}, {Lubin}, {Metcalf},
  {Monson}, {Ninan}, {Palafoutas}, {Ramsey}, {Roy}, {Schwab}, {Terrien}, \&
  {Wright}}]{2020ApJ...897..125R}
{Robertson}, P., {Stefansson}, G., {Mahadevan}, S., {et~al.} 2020, \apj, 897,
  125, \dodoi{10.3847/1538-4357/ab989f}

\bibitem[{{Rosenthal} {et~al.}(2021){Rosenthal}, {Fulton}, {Hirsch},
  {Isaacson}, {Howard}, {Dedrick}, {Sherstyuk}, {Blunt}, {Petigura}, {Knutson},
  {Behmard}, {Chontos}, {Crepp}, {Crossfield}, {Dalba}, {Fischer}, {Henry},
  {Kane}, {Kosiarek}, {Marcy}, {Rubenzahl}, {Weiss}, \&
  {Wright}}]{2021ApJS..255....8R}
{Rosenthal}, L.~J., {Fulton}, B.~J., {Hirsch}, L.~A., {et~al.} 2021, \apjs,
  255, 8, \dodoi{10.3847/1538-4365/abe23c}

\bibitem[{{Sabotta} {et~al.}(2021){Sabotta}, {Schlecker}, {Chaturvedi},
  {Guenther}, {Mu{\~n}oz Rodr{\'\i}guez}, {Mu{\~n}oz S{\'a}nchez}, {Caballero},
  {Shan}, {Reffert}, {Ribas}, {Reiners}, {Hatzes}, {Amado}, {Klahr}, {Morales},
  {Quirrenbach}, {Henning}, {Dreizler}, {Pall{\'e}}, {Perger}, {Azzaro},
  {Jeffers}, {Kaminski}, {K{\"u}rster}, {Lafarga}, {Montes}, {Passegger}, \&
  {Zechmeister}}]{2021arXiv210703802S}
{Sabotta}, S., {Schlecker}, M., {Chaturvedi}, P., {et~al.} 2021, arXiv
  e-prints, arXiv:2107.03802.
\newblock \doarXiv{2107.03802}

\bibitem[{{Skemer} \& {Close}(2011)}]{2011ApJ...730...53S}
{Skemer}, A.~J., \& {Close}, L.~M. 2011, \apj, 730, 53,
  \dodoi{10.1088/0004-637X/730/1/53}

\bibitem[{{Skemer} {et~al.}(2016){Skemer}, {Morley}, {Zimmerman}, {Skrutskie},
  {Leisenring}, {Buenzli}, {Bonnefoy}, {Bailey}, {Hinz}, {Defr{\'e}re},
  {Esposito}, {Apai}, {Biller}, {Brandner}, {Close}, {Crepp}, {De Rosa},
  {Desidera}, {Eisner}, {Fortney}, {Freedman}, {Henning}, {Hofmann},
  {Kopytova}, {Lupu}, {Maire}, {Males}, {Marley}, {Morzinski}, {Oza},
  {Patience}, {Rajan}, {Rieke}, {Schertl}, {Schlieder}, {Stone}, {Su}, {Vaz},
  {Visscher}, {Ward-Duong}, {Weigelt}, \& {Woodward}}]{2016ApJ...817..166S}
{Skemer}, A.~J., {Morley}, C.~V., {Zimmerman}, N.~T., {et~al.} 2016, \apj, 817,
  166, \dodoi{10.3847/0004-637X/817/2/166}

\bibitem[{{Southworth} {et~al.}(2020){Southworth}, {Bohn}, {Kenworthy},
  {Ginski}, \& {Mancini}}]{2020AA...635A..74S}
{Southworth}, J., {Bohn}, A.~J., {Kenworthy}, M.~A., {Ginski}, C., \&
  {Mancini}, L. 2020, \aap, 635, A74, \dodoi{10.1051/0004-6361/201937334}

\bibitem[{{Stock} {et~al.}(2020{\natexlab{a}}){Stock}, {Nagel}, {Kemmer},
  {Passegger}, {Reffert}, {Quirrenbach}, {Caballero}, {Czesla}, {B{\'e}jar},
  {Cardona}, {D{\'\i}ez-Alonso}, {Herrero}, {Lalitha}, {Schlecker}, {Tal-Or},
  {Rodr{\'\i}guez}, {Rodr{\'\i}guez-L{\'o}pez}, {Ribas}, {Reiners}, {Amado},
  {Bauer}, {Bluhm}, {Cort{\'e}s-Contreras}, {Gonz{\'a}lez-Cuesta}, {Dreizler},
  {Hatzes}, {Henning}, {Jeffers}, {Kaminski}, {K{\"u}rster}, {Lafarga},
  {L{\'o}pez-Gonz{\'a}lez}, {Montes}, {Morales}, {Pedraz}, {Sch{\"o}fer},
  {Schweitzer}, {Trifonov}, {Zapatero Osorio}, \&
  {Zechmeister}}]{2020AA...643A.112S}
{Stock}, S., {Nagel}, E., {Kemmer}, J., {et~al.} 2020{\natexlab{a}}, \aap, 643,
  A112, \dodoi{10.1051/0004-6361/202038820}

\bibitem[{{Stock} {et~al.}(2020{\natexlab{b}}){Stock}, {Kemmer}, {Reffert},
  {Trifonov}, {Kaminski}, {Dreizler}, {Quirrenbach}, {Caballero}, {Reiners},
  {Jeffers}, {Anglada-Escud{\'e}}, {Ribas}, {Amado}, {Barrado}, {Barnes},
  {Bauer}, {Berdi{\~n}as}, {B{\'e}jar}, {Coleman}, {Cort{\'e}s-Contreras},
  {D{\'\i}ez-Alonso}, {Dom{\'\i}nguez-Fern{\'a}ndez}, {Espinoza}, {Haswell},
  {Hatzes}, {Henning}, {Jenkins}, {Jones}, {Kossakowski}, {K{\"u}rster},
  {Lafarga}, {Lee}, {L{\'o}pez Gonz{\'a}lez}, {Montes}, {Morales}, {Morales},
  {Pall{\'e}}, {Pedraz}, {Rodr{\'\i}guez}, {Rodr{\'\i}guez-L{\'o}pez}, \&
  {Zechmeister}}]{2020AA...636A.119S}
{Stock}, S., {Kemmer}, J., {Reffert}, S., {et~al.} 2020{\natexlab{b}}, \aap,
  636, A119, \dodoi{10.1051/0004-6361/201936732}

\bibitem[{{Stone} {et~al.}(2018){Stone}, {Skemer}, {Hinz}, {Bonavita},
  {Kratter}, {Maire}, {Defrere}, {Bailey}, {Spalding}, {Leisenring},
  {Desidera}, {Bonnefoy}, {Biller}, {Woodward}, {Henning}, {Skrutskie},
  {Eisner}, {Crepp}, {Patience}, {Weigelt}, {De Rosa}, {Schlieder}, {Brandner},
  {Apai}, {Su}, {Ertel}, {Ward-Duong}, {Morzinski}, {Schertl}, {Hofmann},
  {Close}, {Brems}, {Fortney}, {Oza}, {Buenzli}, \&
  {Bass}}]{2018AJ....156..286S}
{Stone}, J.~M., {Skemer}, A.~J., {Hinz}, P.~M., {et~al.} 2018, \aj, 156, 286,
  \dodoi{10.3847/1538-3881/aaec00}

\bibitem[{{Su{\'a}rez Mascare{\~n}o} {et~al.}(2015){Su{\'a}rez Mascare{\~n}o},
  {Rebolo}, {Gonz{\'a}lez Hern{\'a}ndez}, \& {Esposito}}]{2015MNRAS.452.2745S}
{Su{\'a}rez Mascare{\~n}o}, A., {Rebolo}, R., {Gonz{\'a}lez Hern{\'a}ndez},
  J.~I., \& {Esposito}, M. 2015, \mnras, 452, 2745,
  \dodoi{10.1093/mnras/stv1441}

\bibitem[{{Su{\'a}rez Mascare{\~n}o} {et~al.}(2017){Su{\'a}rez Mascare{\~n}o},
  {Rebolo}, {Gonz{\'a}lez Hern{\'a}ndez}, \& {Esposito}}]{2017MNRAS.468.4772S}
---. 2017, \mnras, 468, 4772, \dodoi{10.1093/mnras/stx771}

\bibitem[{{Su{\'a}rez Mascare{\~n}o} {et~al.}(2018){Su{\'a}rez Mascare{\~n}o},
  {Rebolo}, {Gonz{\'a}lez Hern{\'a}ndez}, {Toledo-Padr{\'o}n}, {Perger},
  {Ribas}, {Affer}, {Micela}, {Damasso}, {Maldonado}, {Gonz{\'a}lez-Alvarez},
  {Leto}, {Pagano}, {Scandariato}, {Sozzetti}, {Lanza}, {Malavolta}, {Claudi},
  {Cosentino}, {Desidera}, {Giacobbe}, {Maggio}, {Rainer}, {Esposito},
  {Benatti}, {Pedani}, {Morales}, {Herrero}, {Lafarga}, {Rosich}, \&
  {Pinamonti}}]{2018AA...612A..89S}
{Su{\'a}rez Mascare{\~n}o}, A., {Rebolo}, R., {Gonz{\'a}lez Hern{\'a}ndez},
  J.~I., {et~al.} 2018, \aap, 612, A89, \dodoi{10.1051/0004-6361/201732143}

\bibitem[{{Telesco} {et~al.}(2008){Telesco}, {Packham}, {Ftaclas}, {Hough},
  {Moerchen}, {Hanna}, {Julian}, {Varosi}, {Julian}, {Bennett}, {Murphey},
  {Reyes}, \& {Warner}}]{2008SPIE.7014E..0RT}
{Telesco}, C.~M., {Packham}, C., {Ftaclas}, C., {et~al.} 2008, in Society of
  Photo-Optical Instrumentation Engineers (SPIE) Conference Series, Vol. 7014,
  \procspie, 70140R, \dodoi{10.1117/12.787697}

\bibitem[{{Thebault} \& {Haghighipour}(2015)}]{2015pes..book..309T}
{Thebault}, P., \& {Haghighipour}, N. 2015, {Planet Formation in Binaries},
  309--340, \dodoi{10.1007/978-3-662-45052-9\_13}

\bibitem[{{Trifonov} {et~al.}(2018){Trifonov}, {K{\"u}rster}, {Zechmeister},
  {Tal-Or}, {Caballero}, {Quirrenbach}, {Amado}, {Ribas}, {Reiners}, {Reffert},
  {Dreizler}, {Hatzes}, {Kaminski}, {Launhardt}, {Henning}, {Montes},
  {B{\'e}jar}, {Mundt}, {Pavlov}, {Schmitt}, {Seifert}, {Morales}, {Nowak},
  {Jeffers}, {Rodr{\'\i}guez-L{\'o}pez}, {del Burgo}, {Anglada-Escud{\'e}},
  {L{\'o}pez-Santiago}, {Mathar}, {Ammler-von Eiff}, {Guenther}, {Barrado},
  {Gonz{\'a}lez Hern{\'a}ndez}, {Mancini}, {St{\"u}rmer}, {Abril}, {Aceituno},
  {Alonso-Floriano}, {Antona}, {Anwand-Heerwart}, {Arroyo-Torres}, {Azzaro},
  {Baroch}, {Bauer}, {Becerril}, {Ben{\'\i}tez}, {Berdi{\~n}as}, {Bergond},
  {Bl{\"u}mcke}, {Brinkm{\"o}ller}, {Cano}, {C{\'a}rdenas V{\'a}zquez},
  {Casal}, {Cifuentes}, {Claret}, {Colom{\'e}}, {Cort{\'e}s-Contreras},
  {Czesla}, {D{\'\i}ez-Alonso}, {Feiz}, {Fern{\'a}ndez}, {Ferro},
  {Fuhrmeister}, {Galad{\'\i}-Enr{\'\i}quez}, {Garcia-Piquer}, {Garc{\'\i}a
  Vargas}, {Gesa}, {G{\'o}mez Galera}, {Gonz{\'a}lez-Peinado}, {Gr{\"o}zinger},
  {Grohnert}, {Gu{\`a}rdia}, {Guijarro}, {de Guindos}, {Guti{\'e}rrez-Soto},
  {Hagen}, {Hauschildt}, {Hedrosa}, {Helmling}, {Hermelo}, {Hern{\'a}ndez
  Arab{\'\i}}, {Hern{\'a}ndez Casta{\~n}o}, {Hern{\'a}ndez Hernando},
  {Herrero}, {Huber}, {Huke}, {Johnson}, {de Juan}, {Kim}, {Klein},
  {Kl{\"u}ter}, {Klutsch}, {Lafarga}, {Lamp{\'o}n}, {Lara}, {Laun}, {Lemke},
  {Lenzen}, {L{\'o}pez del Fresno}, {L{\'o}pez-Gonz{\'a}lez},
  {L{\'o}pez-Puertas}, {L{\'o}pez Salas}, {Luque}, {Mag{\'a}n Madinabeitia},
  {Mall}, {Mandel}, {Marfil}, {Mar{\'\i}n Molina}, {Maroto Fern{\'a}ndez},
  {Mart{\'\i}n}, {Mart{\'\i}n-Ruiz}, {Marvin}, {Mirabet}, {Moya},
  {Moreno-Raya}, {Nagel}, {Naranjo}, {Nortmann}, {Ofir}, {Oreiro}, {Pall{\'e}},
  {Panduro}, {Pascual}, {Passegger}, {Pedraz}, {P{\'e}rez-Calpena}, {P{\'e}rez
  Medialdea}, {Perger}, {Perryman}, {Pluto}, {Rabaza}, {Ram{\'o}n}, {Rebolo},
  {Redondo}, {Reinhardt}, {Rhode}, {Rix}, {Rodler}, {Rodr{\'\i}guez},
  {Rodr{\'\i}guez Trinidad}, {Rohloff}, {Rosich}, {Sadegi},
  {S{\'a}nchez-Blanco}, {S{\'a}nchez Carrasco}, {S{\'a}nchez-L{\'o}pez},
  {Sanz-Forcada}, {Sarkis}, {Sarmiento}, {Sch{\"a}fer}, {Schiller},
  {Sch{\"o}fer}, {Schweitzer}, {Solano}, {Stahl}, {Strachan}, {Su{\'a}rez},
  {Tabernero}, {Tala}, {Tulloch}, {Veredas}, {Vico Linares}, {Vilardell},
  {Wagner}, {Winkler}, {Wolthoff}, {Xu}, {Yan}, \& {Zapatero
  Osorio}}]{2018AA...609A.117T}
{Trifonov}, T., {K{\"u}rster}, M., {Zechmeister}, M., {et~al.} 2018, \aap, 609,
  A117, \dodoi{10.1051/0004-6361/201731442}

\bibitem[{{Tuomi} {et~al.}(2013){Tuomi}, {Jones}, {Jenkins}, {Tinney},
  {Butler}, {Vogt}, {Barnes}, {Wittenmyer}, {O'Toole}, {Horner}, {Bailey},
  {Carter}, {Wright}, {Salter}, \& {Pinfield}}]{2013AA...551A..79T}
{Tuomi}, M., {Jones}, H.~R.~A., {Jenkins}, J.~S., {et~al.} 2013, \aap, 551,
  A79, \dodoi{10.1051/0004-6361/201220509}

\bibitem[{{Uyama} {et~al.}(2017){Uyama}, {Hashimoto}, {Kuzuhara}, {Mayama},
  {Akiyama}, {Currie}, {Livingston}, {Kudo}, {Kusakabe}, {Abe}, {Brandner},
  {Brandt}, {Carson}, {Egner}, {Feldt}, {Goto}, {Grady}, {Guyon}, {Hayano},
  {Hayashi}, {Hayashi}, {Henning}, {Hodapp}, {Ishii}, {Iye}, {Janson},
  {Kandori}, {Knapp}, {Kwon}, {Matsuo}, {Mcelwain}, {Miyama}, {Morino},
  {Moro-Martin}, {Nishimura}, {Pyo}, {Serabyn}, {Suenaga}, {Suto}, {Suzuki},
  {Takahashi}, {Takami}, {Takato}, {Terada}, {Thalmann}, {Turner}, {Watanabe},
  {Wisniewski}, {Yamada}, {Takami}, {Usuda}, \& {Tamura}}]{2017AJ....153..106U}
{Uyama}, T., {Hashimoto}, J., {Kuzuhara}, M., {et~al.} 2017, \aj, 153, 106,
  \dodoi{10.3847/1538-3881/153/3/106}

\bibitem[{{Vigan} {et~al.}(2017){Vigan}, {Bonavita}, {Biller}, {Forgan},
  {Rice}, {Chauvin}, {Desidera}, {Meunier}, {Delorme}, {Schlieder}, {Bonnefoy},
  {Carson}, {Covino}, {Hagelberg}, {Henning}, {Janson}, {Lagrange}, {Quanz},
  {Zurlo}, {Beuzit}, {Boccaletti}, {Buenzli}, {Feldt}, {Girard}, {Gratton},
  {Kasper}, {Le Coroller}, {Mesa}, {Messina}, {Meyer}, {Montagnier},
  {Mordasini}, {Mouillet}, {Moutou}, {Reggiani}, {Segransan}, \&
  {Thalmann}}]{2017AA...603A...3V}
{Vigan}, A., {Bonavita}, M., {Biller}, B., {et~al.} 2017, \aap, 603, A3,
  \dodoi{10.1051/0004-6361/201630133}

\bibitem[{{Vigan} {et~al.}(2021){Vigan}, {Fontanive}, {Meyer}, {Biller},
  {Bonavita}, {Feldt}, {Desidera}, {Marleau}, {Emsenhuber}, {Galicher}, {Rice},
  {Forgan}, {Mordasini}, {Gratton}, {Le Coroller}, {Maire}, {Cantalloube},
  {Chauvin}, {Cheetham}, {Hagelberg}, {Lagrange}, {Langlois}, {Bonnefoy},
  {Beuzit}, {Boccaletti}, {D'Orazi}, {Delorme}, {Dominik}, {Henning}, {Janson},
  {Lagadec}, {Lazzoni}, {Ligi}, {Menard}, {Mesa}, {Messina}, {Moutou},
  {M{\"u}ller}, {Perrot}, {Samland}, {Schmid}, {Schmidt}, {Sissa}, {Turatto},
  {Udry}, {Zurlo}, {Abe}, {Antichi}, {Asensio-Torres}, {Baruffolo}, {Baudoz},
  {Baudrand}, {Bazzon}, {Blanchard}, {Bohn}, {Brown Sevilla}, {Carbillet},
  {Carle}, {Cascone}, {Charton}, {Claudi}, {Costille}, {De Caprio},
  {Delboulb{\'e}}, {Dohlen}, {Engler}, {Fantinel}, {Feautrier}, {Fusco},
  {Gigan}, {Girard}, {Giro}, {Gisler}, {Gluck}, {Gry}, {Hubin}, {Hugot},
  {Jaquet}, {Kasper}, {Le Mignant}, {Llored}, {Madec}, {Magnard}, {Martinez},
  {Maurel}, {M{\"o}ller-Nilsson}, {Mouillet}, {Moulin}, {Orign{\'e}}, {Pavlov},
  {Perret}, {Petit}, {Pragt}, {Puget}, {Rabou}, {Ramos}, {Rickman}, {Rigal},
  {Rochat}, {Roelfsema}, {Rousset}, {Roux}, {Salasnich}, {Sauvage}, {Sevin},
  {Soenke}, {Stadler}, {Suarez}, {Wahhaj}, {Weber}, \&
  {Wildi}}]{2021AA...651A..72V}
{Vigan}, A., {Fontanive}, C., {Meyer}, M., {et~al.} 2021, \aap, 651, A72,
  \dodoi{10.1051/0004-6361/202038107}

\bibitem[{{Wang} {et~al.}(2015){Wang}, {Fischer}, {Horch}, \&
  {Xie}}]{2015ApJ...806..248W}
{Wang}, J., {Fischer}, D.~A., {Horch}, E.~P., \& {Xie}, J.-W. 2015, \apj, 806,
  248, \dodoi{10.1088/0004-637X/806/2/248}

\bibitem[{{Wargelin} {et~al.}(2008){Wargelin}, {Kashyap}, {Drake},
  {Garc{\'\i}a-Alvarez}, \& {Ratzlaff}}]{2008ApJ...676..610W}
{Wargelin}, B.~J., {Kashyap}, V.~L., {Drake}, J.~J., {Garc{\'\i}a-Alvarez}, D.,
  \& {Ratzlaff}, P.~W. 2008, \apj, 676, 610, \dodoi{10.1086/528702}

\bibitem[{{Winn} \& {Fabrycky}(2015)}]{2015ARAA..53..409W}
{Winn}, J.~N., \& {Fabrycky}, D.~C. 2015, \araa, 53, 409,
  \dodoi{10.1146/annurev-astro-082214-122246}

\bibitem[{{Wittenmyer} {et~al.}(2006){Wittenmyer}, {Endl}, {Cochran}, {Hatzes},
  {Walker}, {Yang}, \& {Paulson}}]{2006AJ....132..177W}
{Wittenmyer}, R.~A., {Endl}, M., {Cochran}, W.~D., {et~al.} 2006, \aj, 132,
  177, \dodoi{10.1086/504942}

\bibitem[{{Wittenmyer} {et~al.}(2016){Wittenmyer}, {Butler}, {Tinney},
  {Horner}, {Carter}, {Wright}, {Jones}, {Bailey}, \&
  {O'Toole}}]{2016ApJ...819...28W}
{Wittenmyer}, R.~A., {Butler}, R.~P., {Tinney}, C.~G., {et~al.} 2016, \apj,
  819, 28, \dodoi{10.3847/0004-637X/819/1/28}

\bibitem[{{Wittenmyer} {et~al.}(2020){Wittenmyer}, {Wang}, {Horner}, {Butler},
  {Tinney}, {Carter}, {Wright}, {Jones}, {Bailey}, {O'Toole}, \&
  {Johns}}]{2020MNRAS.492..377W}
{Wittenmyer}, R.~A., {Wang}, S., {Horner}, J., {et~al.} 2020, \mnras, 492, 377,
  \dodoi{10.1093/mnras/stz3436}

\bibitem[{{Woitas} {et~al.}(2000){Woitas}, {Leinert}, {Jahrei{\ss}}, {Henry},
  {Franz}, \& {Wasserman}}]{2000AA...353..253W}
{Woitas}, J., {Leinert}, C., {Jahrei{\ss}}, H., {et~al.} 2000, \aap, 353, 253.
\newblock \doarXiv{astro-ph/9910411}

\bibitem[{{Wright} {et~al.}(2010){Wright}, {Eisenhardt}, {Mainzer}, {Ressler},
  {Cutri}, {Jarrett}, {Kirkpatrick}, {Padgett}, {McMillan}, {Skrutskie},
  {Stanford}, {Cohen}, {Walker}, {Mather}, {Leisawitz}, {Gautier}, {McLean},
  {Benford}, {Lonsdale}, {Blain}, {Mendez}, {Irace}, {Duval}, {Liu}, {Royer},
  {Heinrichsen}, {Howard}, {Shannon}, {Kendall}, {Walsh}, {Larsen}, {Cardon},
  {Schick}, {Schwalm}, {Abid}, {Fabinsky}, {Naes}, \&
  {Tsai}}]{2010AJ....140.1868W}
{Wright}, E.~L., {Eisenhardt}, P. R.~M., {Mainzer}, A.~K., {et~al.} 2010, \aj,
  140, 1868, \dodoi{10.1088/0004-6256/140/6/1868}

\bibitem[{{Yee} {et~al.}(2017){Yee}, {Petigura}, \& {von
  Braun}}]{2017ApJ...836...77Y}
{Yee}, S.~W., {Petigura}, E.~A., \& {von Braun}, K. 2017, \apj, 836, 77,
  \dodoi{10.3847/1538-4357/836/1/77}

\bibitem[{{Zechmeister} {et~al.}(2019){Zechmeister}, {Dreizler}, {Ribas},
  {Reiners}, {Caballero}, {Bauer}, {B{\'e}jar}, {Gonz{\'a}lez-Cuesta},
  {Herrero}, {Lalitha}, {L{\'o}pez-Gonz{\'a}lez}, {Luque}, {Morales},
  {Pall{\'e}}, {Rodr{\'\i}guez}, {Rodr{\'\i}guez L{\'o}pez}, {Tal-Or},
  {Anglada-Escud{\'e}}, {Quirrenbach}, {Amado}, {Abril}, {Aceituno},
  {Aceituno}, {Alonso-Floriano}, {Ammler-von Eiff}, {Antona Jim{\'e}nez},
  {Anwand-Heerwart}, {Arroyo-Torres}, {Azzaro}, {Baroch}, {Barrado},
  {Becerril}, {Ben{\'\i}tez}, {Berdi{\~n}as}, {Bergond}, {Bluhm},
  {Brinkm{\"o}ller}, {del Burgo}, {Calvo Ortega}, {Cano}, {Cardona
  Guill{\'e}n}, {Carro}, {C{\'a}rdenas V{\'a}zquez}, {Casal},
  {Casasayas-Barris}, {Casanova}, {Chaturvedi}, {Cifuentes}, {Claret},
  {Colom{\'e}}, {Cort{\'e}s-Contreras}, {Czesla}, {D{\'\i}ez-Alonso}, {Dorda},
  {Fern{\'a}ndez}, {Fern{\'a}ndez-Mart{\'\i}n}, {Fuhrmeister}, {Fukui},
  {Galad{\'\i}-Enr{\'\i}quez}, {Gallardo Cava}, {Garcia de la Fuente},
  {Garcia-Piquer}, {Garc{\'\i}a Vargas}, {Gesa}, {G{\'o}ngora Rueda},
  {Gonz{\'a}lez-{\'A}lvarez}, {Gonz{\'a}lez Hern{\'a}ndez},
  {Gonz{\'a}lez-Peinado}, {Gr{\"o}zinger}, {Gu{\`a}rdia}, {Guijarro}, {de
  Guindos}, {Hatzes}, {Hauschildt}, {Hedrosa}, {Helmling}, {Henning},
  {Hermelo}, {Hern{\'a}ndez Arabi}, {Hern{\'a}ndez Casta{\~n}o}, {Hern{\'a}ndez
  Otero}, {Hintz}, {Huke}, {Huber}, {Jeffers}, {Johnson}, {de Juan},
  {Kaminski}, {Kemmer}, {Kim}, {Klahr}, {Klein}, {Kl{\"u}ter}, {Klutsch},
  {Kossakowski}, {K{\"u}rster}, {Labarga}, {Lafarga}, {Llamas}, {Lamp{\'o}n},
  {Lara}, {Launhardt}, {L{\'a}zaro}, {Lodieu}, {L{\'o}pez del Fresno},
  {L{\'o}pez-Puertas}, {L{\'o}pez Salas}, {L{\'o}pez-Santiago}, {Mag{\'a}n
  Madinabeitia}, {Mall}, {Mancini}, {Mand el}, {Marfil}, {Mar{\'\i}n Molina},
  {Maroto Fern{\'a}ndez}, {Mart{\'\i}n}, {Mart{\'\i}n-Fern{\'a}ndez},
  {Mart{\'\i}n-Ruiz}, {Marvin}, {Mirabet}, {Monta{\~n}{\'e}s-Rodr{\'\i}guez},
  {Montes}, {Moreno-Raya}, {Nagel}, {Naranjo}, {Narita}, {Nortmann}, {Nowak},
  {Ofir}, {Oshagh}, {Panduro}, {Parviainen}, {Pascual}, {Passegger}, {Pavlov},
  {Pedraz}, {P{\'e}rez-Calpena}, {P{\'e}rez Medialdea}, {Perger}, {Perryman},
  {Rabaza}, {Ram{\'o}n Ballesta}, {Rebolo}, {Redondo}, {Reffert}, {Reinhardt},
  {Rhode}, {Rix}, {Rodler}, {Rodr{\'\i}guez Trinidad}, {Rosich}, {Sadegi},
  {S{\'a}nchez-Blanco}, {S{\'a}nchez Carrasco}, {S{\'a}nchez-L{\'o}pez},
  {Sanz-Forcada}, {Sarkis}, {Sarmiento}, {Sch{\"a}fer}, {Schmitt},
  {Sch{\"o}fer}, {Schweitzer}, {Seifert}, {Shulyak}, {Solano}, {Sota}, {Stahl},
  {Stock}, {Strachan}, {Stuber}, {St{\"u}rmer}, {Su{\'a}rez}, {Tabernero},
  {Tala Pinto}, {Trifonov}, {Veredas}, {Vico Linares}, {Vilardell}, {Wagner},
  {Wolthoff}, {Xu}, {Yan}, \& {Zapatero Osorio}}]{2019AA...627A..49Z}
{Zechmeister}, M., {Dreizler}, S., {Ribas}, I., {et~al.} 2019, \aap, 627, A49,
  \dodoi{10.1051/0004-6361/201935460}

\end{thebibliography}
\bibliographystyle{aasjournal}


\end{document}